\documentclass[fleqn,useAMS,usenatbib]{mnras}
\usepackage{mathptmx}
\usepackage{color}
\usepackage{graphics,graphicx,amssymb,amsmath,natbib}
\usepackage{deluxetable}
\usepackage{float}
\usepackage{color}
\usepackage{lscape,graphicx}
\usepackage{amsmath}
\usepackage{amssymb}
\usepackage{multirow}
\usepackage{afterpage}
\usepackage{threeparttable}
\usepackage{subfigure}
\usepackage{rotating}
\usepackage{lscape}
\usepackage{caption}
\usepackage{verbatim}
\usepackage{morefloats}
\usepackage[T1]{fontenc}
\usepackage{ae,aecompl}
\usepackage{graphicx}
\usepackage{epstopdf}
\usepackage{amsmath}
\usepackage{amssymb}
\usepackage{enumerate}
\usepackage{dblfloatfix}
\usepackage{lipsum}

\newcommand{\avg}[1]{\left\langle #1 \right\rangle}

\def\cha    {{\em Chandra}\/}

\def\xmm        {{\em XMM-Newton}\/}
\def\XMM        {{\em XMM}\/}
\def\pla        {{\em Planck}\/}

\def\rosat      {{\em ROSAT}\/}

\def\ga         {{CGCG~254-021}\/}
\def\max         {{maxBCG}\/}
\def\red         {{redMaPPer}\/}
\def\sdss        {{\em SDSS}\/}

\title[X-ray emission from richest \max\ clusters]{X-ray scaling relations from a complete sample of the richest \max\ clusters}
\author
[Ge et al.]{Chong Ge$^{1}$\thanks{chong.ge@uah.edu}, Ming Sun$^{1}$\thanks{ming.sun@uah.edu}, Eduardo Rozo$^{2}$, Neelima Sehgal$^{3}$, Alexey Vikhlinin$^{4}$,
\newauthor
William Forman$^{4}$, Christine Jones$^{4}$, and Daisuke Nagai$^{5,6,7}$\\
$^{1}$Department of Physics and Astronomy, University of Alabama in Huntsville, Huntsville, AL 35899, USA\\
$^{2}$Department of Physics, University of Arizona, Tucson, AZ 85721, USA\\
$^{3}$Physics and Astronomy Department, Stony Brook University, Stony Brook, NY 11794, USA\\
$^{4}$Harvard-Smithsonian Center for Astrophysics, 60 Garden Street, Cambridge, MA 02138, USA\\
$^{5}$Department of Physics, Yale University, New Haven, CT 06520, USA\\
$^{6}$Department of Astronomy, Yale University, New Haven, CT 06520, USA\\
$^{7}$Yale Center for Astronomy \& Astrophysics, Yale University, New Haven, CT 06520, USA\\
}
\begin{document}
\date{Accepted. Received; in original form}

\pubyear{2018}

\maketitle

\label{firstpage}

\begin{abstract}
We use a complete sample of 38 richest \max\ clusters to study the ICM-galaxy scaling relations and the halo mass selection properties of the \max\ algorithm, based on X-ray and optical observations. The clusters are selected from the two largest bins of optical richness in the \pla\ stacking work with the \max\ richness $N_{200} \geq 78$. We analyze their \cha\ and \xmm\ data to derive the X-ray properties of the ICM. 
We then use the distribution of $P(X|N)$, $X=T_X,\ L_X,\ Y_X$, to study the mass selection $P(M|N)$ of \max. Compared with previous works based on the whole richness sample, a significant fraction of blended systems with boosted richness is skewed into this richest sample. Parts of the blended halos are picked apart by the \red, an updated red-sequence cluster finding algorithm with lower mass scatter. Moreover, all the optical blended halos are resolved as individual X-ray halos, following the established $L_X-T_X$ and $L_X-Y_X$ relations.
We further discuss that the discrepancy between ICM-galaxy scaling relations, especially for future blind stacking, can come from several factors, including miscentering, projection, contamination of low mass systems, mass bias and covariance bias. We also evaluate the fractions of relaxed and cool core clusters in our sample. Both are smaller than those from SZ or X-ray selected samples. Moreover, disturbed clusters show a higher level of mass bias than relaxed clusters.
\end{abstract}


\section{Introduction}
\label{s:int}
The spatial distribution and number density of massive dark matter halos are very sensitive to the underlying cosmological parameters. 
These halos host clusters of galaxies, and they have been used to study cosmology for over 20 yr (e.g. \citealt{1991ApJ...372..410H}; \citealt{1998MNRAS.298.1145E}; \citealt{2002ApJ...567..716R}; \citealt{2009ApJ...692.1060V}; \citealt{2010ApJ...708..645R}; \citealt{2011ApJ...732...44S}; \citealt{2011ARA&A..49..409A}; \citealt{2013PhR...530...87W}). The key questions to study cosmology with clusters are how to find them and how to measure their masses, which can be measured by the velocity dispersion of cluster galaxies and gravitational lensing,  both of them are observationally expensive. They are not efficient methods for finding clusters, though they are important tools for direct mass calibration. These measurements find that most cluster mass is in dark matter, which cannot be detected directly, thus more conveniently the cluster finding and mass measuring are tied to baryons in clusters.

A small fraction of cluster baryons cooled and formed stars. Clusters can be selected optically by looking for overdense regions of galaxies.
The optical identifications have resulted in large samples of clusters, benefiting from large optical surveys like \sdss\ (e.g. \citealt{2007ApJ...660..239K}; \citealt{2014ApJ...785..104R}).
Mass proxies in optical are generally the total number or luminosity of member galaxies but such scaling relations
suffer from large fractional scatter (e.g. $\sigma_{m|n}=0.45$; \citealt{2009ApJ...699..768R}).

Most of the cluster baryons are in the hot intracluster medium (ICM). The ICM emits X-rays via thermal bremsstrahlung and line emission, which are proportional to the density square of the gas that traces the deep potential well of the cluster. Thus, the X-ray detection is very effective to find clusters and relatively free from projection contamination. Meanwhile, the cluster mass can be derived from the distribution of their X-ray emitting gas under the assumption of hydrostatic equilibrium (HE). A more economical way is using X-ray mass proxies like $T_X$, $L_X$ and $Y_X$ through scaling relations (e.g. \citealt{1986MNRAS.222..323K}; \citealt{2009AA...498..361P}; \citealt{2009ApJ...692.1033V}; \citealt{2009ApJ...693.1142S}; \citealt{2010MNRAS.406.1773M}; \citealt{2011ApJ...727L..49S}; \citealt{2012NJPh...14d5004S}). These X-ray mass proxies, usually considered robust and low scattered, can also be calibrated with the robust mass indicators like weak lensing mass (e.g. $\sigma_{m|t}=0.26$; \citealt{2016MNRAS.463.3582M}). The ICM scaling relations are also important to understand the key baryon physics that governs galaxy formation, e.g. cooling, efficiency of star formation, and stellar or AGN feedback (e.g. \citealt{2005ApJ...625..588K}; \citealt{2006MNRAS.365...11C}; \citealt{2008MNRAS.390.1399B}; \citealt{2014MNRAS.441.1270L}; \citealt{2017MNRAS.470..166H}; \citealt{2018MNRAS.474.4089T}).

The X-ray emitting hot electrons also boost microwave background radiation (CMB) photons to higher energy through inverse Compton scattering (Sunyaev-Zel'dovich --- SZ effect; \citealt{2002ARA&A..40..643C}). The SZ signal, which is proportional to gas density and temperature, is a promising way to find clusters and complementary to X-ray measurements. Tremendous progress has been made in the past 10 yr with the SZ surveys from \pla, the Atacama Cosmology Telescope ({\em ACT}), and the South Pole Telescope ({\em SPT}). Hundreds of new clusters have been discovered (e.g. \citealt{2016A&A...594A..27P}). Scaling relations between the SZ observables and the other observables in X-ray and optical have been established (e.g. \citealt{2011A&A...536A..11P}). The total SZ flux has been considered a robust mass proxy, as it measures the total thermal energy of the ICM. However, the new SZ data have also raised new puzzles. One that has received much attention is the mismatch between the stacked \pla\ SZ fluxes and the model expectations for the optically selected clusters. All the model SZ fluxes from optically selected clusters overpredict the observed SZ fluxes from \pla\ \citep{2011A&A...536A..12P}, {\em ACT} (\citealt{2013ApJ...767...38S}), and {\em SPT} (\citealt{2017MNRAS.468.3347S}). Possible solutions include: 1) miscentering because of the offset between the optical center and the ICM peak; 2) contamination of optical richness estimates from line-of-sight projections; 3) contamination of the SZ flux from radio galaxies, infrared galaxies, Galactic emission, and the SZ background from unresolved clusters, groups, and the intergalactic medium; 4) underestimate of the true mass by the X-ray HE mass because of additional non-thermal pressure from e.g. gas bulk motion and turbulence; 5) Property covariance and X-ray selection bias (e.g. \citealt{2008ApJ...675.1106R}; \citealt{2011A&A...536A..12P}; \citealt{2012ApJ...757....1B}; \citealt{2013ApJ...767...38S}; \citealt{2014MNRAS.441.3562E}; \citealt{2014MNRAS.438...78R}; \citealt{2017MNRAS.468.3347S}).
Among them, the hydrostatic mass bias seems the dominate source for the discrepancy with a typical value $1-b\sim0.75$ (e.g. \citealt{2017MNRAS.465.3361H}; \citealt{2017A&A...604A..89P}; \citealt{2018arXiv180405873M}).

While the previous studies addressing this puzzle require the calibration of the true mass and assume pressure templates, we bypass this intermediate step to directly compare the X-ray properties with the optical richness, which further allows a direct comparison of the pressure content from X-rays and SZ in the future. 
We target the most massive \max\ clusters as the discrepancy exists in all richness bins. These clusters should have the strongest X-ray and SZ signals. It is natural to study them directly to examine the reasons for the discrepancy. There are 38 clusters in the two richest bins (Fig.~1 in \citealt{2011A&A...536A..12P} and Table~\ref{t:smp}). When we first started this project, 31 of them had existing \cha\ or \XMM\ data, and were mostly X-ray selected. The remaining seven clusters are generally X-ray faint from the \rosat\ All Sky Survey (RASS). We proposed new \XMM\ observations (proposal id 74251 and 76159, PI: M. Sun) on these seven clusters (shown in Table~\ref{t:obs}) to have full X-ray coverage of this sample. 
This sample also provides an extremal study of halo mass selection properties of the maxBCG algorithm. What is being indirectly tested here (ignoring $z$-dependence) is the conditional probability $P(M_{500}|N_{200})$, which is the probability of a mass $M_{500}$ halo with given $N_{200}$ galaxies, where $N_{200}$ is the \max\ richness. The probability is typically assumed to be a log-normal distribution that characterized by a mean relation and an rms scatter. What is actually measured is $P(X|N_{200})$, with $X=\{L_X, T_X, Y_X\}$.
In addition to the \max\ catalogue, we also use the \red\ catalogue \citep{2014ApJ...785..104R}, which uses a similar but different richness measure with different halo selection properties from \max.

We organise the paper as follows: Section~\ref{s:smp} includes the cluster sample and catalogues; Section~\ref{s:obs} details our X-ray data reduction and analysis; Section~\ref{s:res} presents results based on X-ray data and optical catalogues; Section~\ref{s:dis} is the discussion; and Section~\ref{s:sum} is our conclusions.
We use the standard cold dark matter cosmology with $H_0 = 70 {\rm\ km\ s^{-1}\ Mpc^{-1}}$, $\Omega_{m}$ = 0.3, and $\Omega_{\Lambda}$ = 0.7.

\begin{table*}
\centering
\caption{Optical and SZ properties of the richest \max\ clusters}
\begin{tabular}{lcccccccccc} \hline \hline
Cluster & RA$^a$ & DEC$^a$ & $z_{\rm photo}^a$ & $N_{200}^a$ & \red$^b$ & $\lambda^b$ & offset$^b$ & \pla$^c$ & SNR$^c$ & offset$^c$\\
&		&		&						&				&	RM	&				&	arcmin	&	PSZ2	&		& arcmin \\ \hline	 
A2142	&	239.58334	&	27.23342	&	0.103	&	188	&	J155820.0+271400.3	&	169.8	$\pm$	4.2	&	0.00	&	G044.20+48.66	&	28.4	&	1.56	\\
J150	&	150.68705	&	20.49151	&	0.297	&	164	&	J100214.1+203216.6	&	151.4	$\pm$	5.8	&	7.72	&	G213.30+50.99	&	4.5	&	3.91	\\
	&		&		&		&		&	J100311.3+203230.2	&	54.7	$\pm$	4.4	&	6.88	&		&		&		\\
A1689	&	197.87292	&	-1.34110	&	0.189	&	156	&	J131129.5-012028.0	&	164.7	$\pm$	4.2	&	0.00	&	G313.33+61.13	&	16.7	&	0.81	\\
A1443	&	180.31985	&	23.10909	&	0.262	&	139	&	J120112.2+230557.3	&	131.3	$\pm$	5.9	&	1.20	&	G229.74+77.96	&	12.0	&	0.87	\\
A781	&	140.10742	&	30.49406	&	0.292	&	126	&	J092052.5+302740.3	&	128.6	$\pm$	5.7	&	6.09	&	G195.60+44.06	&	7.6	&	0.72	\\
	&		&		&		&		&	J092030.0+302946.8	&	38.7	$\pm$	5.2	&	0.92	&		&		&		\\
A1986	&	223.37531	&	21.87434	&	0.116	&	120	&	J145308.4+215339.3	&	68.8	$\pm$	3.2	&	5.18	&	-	&	-	&	-	\\
A1882	&	213.78496	&	-0.49325	&	0.135	&	115	&	J141424.1-002239.5	&	105.1	$\pm$	3.8	&	13.07	&	-	&	-	&	-	\\
A1758	&	203.16008	&	50.55992	&	0.284	&	114	&	J133238.4+503336.0	&	152.9	$\pm$	5.7	&	0.00	&	G107.10+65.32	&	16.5	&	2.59	\\
	&		&		&		&		&	J133224.5+502425.8	&	51.9	$\pm$	4.2	&	9.43	&		&		&		\\
A1760	&	203.53616	&	20.24806	&	0.176	&	113	&	J133431.5+201217.7	&	73.4	$\pm$	3.7	&	5.94	&	G000.13+78.04	&	9.3	&	1.35	\\
	&		&		&		&		&	J133359.1+201801.3	&	26.5	$\pm$	3.7	&	3.85	&		&		&		\\
A1622	&	192.38517	&	49.89549	&	0.275	&	107	&	J124922.1+494742.1	&	129.3	$\pm$	5.6	&	6.25	&	G123.66+67.25	&	5.7	&	2.02	\\
A750	&	137.21745	&	11.02649	&	0.173	&	107	&	J090912.2+105824.9	&	174.7	$\pm$	4.9	&	5.84	&	G218.81+35.51	&	8.4	&	4.84	\\
A1682	&	196.69040	&	46.55854	&	0.235	&	106	&	J130650.0+463333.4	&	122.4	$\pm$	5.3	&	0.74	&	G114.99+70.36	&	10.0	&	2.52	\\
A1246	&	171.01653	&	21.49114	&	0.186	&	103	&	J112358.8+212849.7	&	113.6	$\pm$	4.1	&	1.36	&	G224.00+69.33	&	8.8	&	0.67	\\
A1961	&	221.07852	&	31.28601	&	0.240	&	101	&	J144431.8+311336.0	&	114.8	$\pm$	4.6	&	4.51	&	G049.18+65.05	&	6.1	&	3.79	\\
A2034	&	227.54883	&	33.48646	&	0.122	&	99	&	J151011.7+332911.3	&	99.8	$\pm$	3.7	&	0.00	&	G053.53+59.52	&	15.9	&	0.42	\\
A655	&	126.37104	&	47.13348	&	0.135	&	99	&	J082529.1+470800.9	&	131.6	$\pm$	4.8	&	0.01	&	G172.63+35.15	&	7.4	&	1.19	\\
A1914	&	216.48612	&	37.81645	&	0.167	&	98	&	J142556.7+374859.2	&	103.3	$\pm$	4.0	&	0.00	&	G067.17+67.46	&	17.8	&	0.81	\\
Z5247	&	188.57278	&	9.76624	&	0.243	&	98	&	J123424.1+094715.5	&	93.0	$\pm$	4.2	&	2.08	&	G289.13+72.19	&	8.8	&	2.71	\\
A657	&	125.83033	&	15.96272	&	0.157	&	94	&	J082319.3+155745.8	&	82.1	$\pm$	3.9	&	0.00	&	-	&	-	&	-	\\
J229	&	229.35068	&	-0.73816	&	0.119	&	93	&	J151721.9-004256.3	&	58.1	$\pm$	3.0	&	1.47	&	-	&	-	&	-	\\
A1423	&	179.32219	&	33.61093	&	0.219	&	91	&	J115715.2+333638.5	&	86.0	$\pm$	4.2	&	0.45	&	G180.60+76.65	&	11.0	&	0.56	\\
A801	&	142.01889	&	20.52921	&	0.208	&	91	&	J092804.5+203145.1	&	93.0	$\pm$	4.5	&	0.00	&	G209.53+43.32	&	4.8	&	2.32	\\
A773	&	139.47261	&	51.72704	&	0.224	&	90	&	J091753.4+514337.5	&	153.6	$\pm$	4.7	&	0.00	&	G166.09+43.38	&	12.4	&	0.51	\\
A1576	&	189.24684	&	63.18658	&	0.294	&	89	&	J123658.6+631114.1	&	102.6	$\pm$	4.8	&	0.08	&	G125.71+53.86	&	11.4	&	0.60	\\
A2631	&	354.41554	&	0.27138	&	0.286	&	88	&	J233739.7+001616.9	&	126.3	$\pm$	4.9	&	0.00	&	G087.03-57.37	&	11.2	&	1.00	\\
A1703	&	198.77182	&	51.81738	&	0.286	&	87	&	J131505.2+514902.8	&	152.5	$\pm$	4.9	&	0.00	&	G114.31+64.89	&	11.1	&	0.96	\\
A2219	&	250.08254	&	46.71148	&	0.235	&	85	&	J164019.8+464241.5	&	199.5	$\pm$	5.3	&	0.00	&	G072.62+41.46	&	27.4	&	0.48	\\
A1319	&	174.05437	&	40.04327	&	0.292	&	85	&	J113613.0+400235.8	&	155.1	$\pm$	5.9	&	0.00	&	G168.33+69.73	&	5.6	&	3.60	\\
A1560	&	188.47087	&	15.19464	&	0.275	&	84	&	J123416.1+151508.4	&	136.9	$\pm$	6.1	&	6.57	&	G283.26+77.37	&	6.0	&	1.53	\\
J175	&	175.57289	&	10.29192	&	0.127	&	84	&	J114207.4+100857.6	&	38.4	$\pm$	2.6	&	8.91	&	-	&	-	&	-	\\
	&		&		&		&		&	J114215.3+102649.4	&	23.5	$\pm$	2.4	&	9.32	&		&		&		\\
J249	&	249.89850	&	47.07569	&	0.227	&	84	&	J163938.0+470310.4	&	61.8	$\pm$	3.4	&	1.43	&	-	&	-	&	-	\\
A1201	&	168.22709	&	13.43584	&	0.176	&	83	&	J111250.1+132830.4	&	78.3	$\pm$	3.5	&	2.59	&	G238.69+63.26	&	7.3	&	3.01	\\
A2009	&	225.08133	&	21.36941	&	0.165	&	82	&	J150019.5+212209.9	&	91.8	$\pm$	3.6	&	0.00	&	G028.89+60.13	&	7.5	&	0.79	\\
A2111	&	234.91872	&	34.42425	&	0.243	&	81	&	J153940.5+342527.3	&	151.5	$\pm$	5.2	&	0.00	&	G054.99+53.41	&	8.4	&	0.94	\\
A815	&	143.03590	&	29.06782	&	0.281	&	80	&	J093208.6+290404.1	&	92.2	$\pm$	5.9	&	0.00	&	-	&	-	&	-	\\
	&		&		&		&		&	J093251.2+290318.0	&	33.3	$\pm$	3.3	&	9.33	&		&		&		\\
Z1450	&	223.05481	&	37.87803	&	0.284	&	79	&	J145213.1+375251.9	&	101.9	$\pm$	4.5	&	0.18	&	-	&	-	&	-	\\
A1765	&	204.20577	&	10.44002	&	0.159	&	78	&	J133649.4+102624.1	&	82.7	$\pm$	3.7	&	0.00	&	-	&	-	&	-	\\
A1902	&	215.41700	&	37.29131	&	0.165	&	78	&	J142140.1+371728.7	&	82.5	$\pm$	3.8	&	0.00	&	G066.68+68.44	&	7.1	&	0.90	\\
\hline
\end{tabular}
\begin{tablenotes}
      \item {\sl Note.}
$^a$: From \max\ catalogue \citep{2007ApJ...660..239K}; RA and DEC mark the cluster center defined as the BCG location; $z_{\rm photo}$ is the photometric redshift; $N_{200}$ is the \max\ richness given by the number of E/S0 ridgeline galaxies more luminous than $0.4L^{\ast}$ within a scaled radius $R_{200}$.
$^b$: From \red\ catalogue \citep{2014ApJ...785..104R}; $\lambda$ is the \red\ richness defined as the sum of the membership probabilities over all galaxies; offset relatives to the \max\ center. 
$^c$: From the second \pla\ catalogue of SZ sources \citep{2016A&A...594A..27P}; the significance of the \pla\ detection and the offset to \max\ center.
    \end{tablenotes} 
\label{t:smp}
\end{table*}

\section{Cluster sample}
\label{s:smp}
Table~\ref{t:smp} lists the 38 richest \max\ clusters with optical and SZ properties assembled from \max, \red, and PSZ2 catalogues, which are introduced briefly below. We use $N$ as a general representative of optical richness, particularly $N_{200}$ for \max\ and $\lambda$ for \red.

\subsection{\max\ catalogue}
The \max\ cluster catalogue \citep{2007ApJ...660..239K} contains 13823 clusters selected from \sdss\ Data Release 5 (DR5), with photometric redshifts of $0.1 - 0.3$ over approximately volume-limited 7500 deg$^2$ of sky. Each cluster is selected as a spatial overdensity of red-sequence galaxies that form a tight E/S0 ridgeline in the color-magnitude diagram. The cluster center is on the brightest cluster galaxy (BCG) with the highest likelihood. The tight relation between the ridgeline color and redshift also provides an accurate photometric
redshift estimate for each cluster ($\Delta_z \simeq 0.01$). The cluster richness $N_{200}$ is defined as the number of red-sequence galaxies with rest-frame i-band luminosity $L_i > 0.4L_{\ast}$ within a projected scaled radius $R_{200}$, interior to which the density of galaxies with $-24 < M_r < -16$ is 200 times the mean density of such kind of galaxies. Applying the cluster selection algorithm to mock catalogues suggests that the catalogue is $\sim$90\% pure and $\sim$85\% complete for clusters with masses $\geq 10^{14}M_{\odot}$. The \max\ cluster mass has been calibrated (\citealt{2007arXiv0709.1159J}; \citealt{2009ApJ...699..768R}) with the $M_{500c}-N_{200}$ relation.

\subsection{\red\ catalogue}
The \red\ cluster catalogue (v6.3; \citealt{2014ApJ...785..104R}) includes 26111 clusters identified from \sdss\ DR8,  with photometric redshifts of $0.08 - 0.60$ (volume-limited in $z \leqslant 0.35$), covering nearly 10000 deg$^2$ of sky. Each cluster is identified as an overdensity of red-sequence galaxies, which relies on iteratively self-training a model of the red-sequence as a function of redshift based on the red galaxies with known spectroscopic redshifts. Then this model is used on photometric data to assign membership probabilities to galaxies with luminosities $\geq 0.2L_{\ast}$ in the cluster vicinity. The cluster richness $\lambda$ is the sum of the membership probabilities of the galaxies within a radius $R_{\lambda}$, which scales with richness as $R_{\lambda} = (\lambda/100)^{0.2}h^{-1}$ Mpc. The cluster photometric redshift is evaluated by simultaneously fitting all high probability cluster members with a single red sequence model, with an accurate of 0.005-0.01. The cluster center is on the central galaxy chosen with a probabilistic approach that weights not just galaxy luminosity, but also local galaxy density, as well as the consistency to cluster redshift. The completeness is $\gtrsim 99$\% at $\lambda > 30$ and z < 0.3, and the purity is $> 95$\% at all richness and redshift. The \red\ cluster mass has been calibrated by \cite{2017MNRAS.466.3103S} with the $M_{200m}-\lambda$ relation. We convert $M_{200m}$ to $M_{500c}$ assuming a cluster mass distribution of Navarro-Frenk-White (NFW; \citealt{1997ApJ...490..493N}) profile with a typical concentration of 6 (e.g. \citealt{2007ApJ...664..123B}) and a median $z=0.23$ for the sample. The ratio of $M_{200m}$/$M_{500c}$ is 1.67.

Compared with \max, some main improvements of \red\ include (i) using multicolor filter rather than single-color; (ii) the aperture used to estimate cluster richness is scaled with richness rather than the fixed scaled radius; (iii) the cluster center is on a weighted position rather than simply the BCG. These updates reduce the scatter at fixed richness (e.g. \citealt{2009ApJ...703..601R}).

\subsection{PSZ2 catalogue}
The second \pla\ Catalogue of Sunyaev-Zel'dovich Sources (PSZ2; \citealt{2016A&A...594A..27P}) exploits the 29 month full-mission data. It contains 1653 candidate clusters with a signal-to-noise ratio (SNR) above 4.5 and distributes across 83.6\% of the sky. 
Among them, more than 1203 are confirmed to be clusters with identified counterparts in external optical or X-ray samples or by dedicated follow-ups, and with a purity larger than 83\%.
The median redshift is $z \sim 0.2$ and the farthest clusters are at $z \lesssim 1.0$.

\begin{table*}
\caption{X-ray observations of the richest \max\ clusters}
\tabcolsep=0.10cm
\begin{tabular}{lccccc} \hline \hline
Cluster & $N_{\rm H}$ & \cha\ id & \cha\ time & \xmm\ id & \xmm\ time \\
& $10^{20}{\rm\ cm}^{-2}$ & & ks  &	 & ks\\	\hline	 
A2142	&	4.36	& (1196 1228 5005 7692) & (231.2) &		 (0111870101 0111870401) 0674560201 &	51.5, 52.9, 41.1/108.7	\\
 	&	 	&	(15186 16564 16565)	&		 	&	  &	 	\\
J150	&	2.59	&	-	&		-	&	0742510101$^\ast$ 0761590101$^\ast$ &	7.9, 7.8, 5.7/43.6	\\
A1689	&	1.98	& 540 1663 5004 6930 7289 7701	&	196.7/199.8	&	0093030101	&	34.4, 35.3, 26.3/39.8	\\
A1443	&	2.41	& 11762 16279 &	25.5/26.1	&	-	&		-	\\
A781	&	1.75	&	(534 15128)	&	(45.6)	& 0150620201 0401170101 0401170201 &	73.1, 74.8, 58.2/105.2	\\
A1986	&	3.18	&	17474	&	5.0/5.1	&	-	&		-	\\
A1882	&	3.76	&	12904  12905  12906  12907  12908    	&	392.2/397.9	&	0145480101 0762870501	&	30.2, 30.9, 20.0/45.6	\\
 & & 12909  12910  12911  12912  17149 & & & \\
  & & 17150  17151  17671 & & & \\
A1758	&	1.06	& (2213) 7710 13997 15538 15540	&	154.7/215.8	&	(0142860201) & (57.2)\\
A1760	&	1.87	&	17159	&	7.0/7.1	&	-	&		-	\\
A1622	&	1.19	&	11763 (17154)	&	12.9/27.1	&	-	&		-	\\
A750	&	3.66	&	924 7699	&	34.5/34.9	&	0605000901 	0673850901 & 21.0, 21.2, 9.7/45.7	\\
A1682	&	1.07	& 	3244 11725 &	29.4/30.0	&	-	&		-	\\
A1246	&	1.66	&	11770	&	5.0/5.1	&	-	&		-	\\
A1961	&	1.21	&	11764	&	6.9/7.0	&	-	&		-	\\
A2034	&	1.62	& (2204 7695 12885 12886 13192 13193) & (259.1)	& 0149880101 0303930101	0303930201 & 18.8, 19.5, 11.1/75.4 \\
A655	&	4.39	&	15159	&	7.9/8.1	&	-	&		-	\\
A1914	&	1.10	& (542) 3593 12197 12892 12893 12894 &	38.6/47.2 & 0112230201 & 19.1, 20.3, 8.8/25.8	\\
Z5247	&	1.73	&	539 11727	&	29.3/30.5	&	(0673851101 0673852101) &		(46.1)	\\
A657	&	3.38	&	-	&		-	&	0742510401$^\ast$	&	14.9, 15.7, 8.4/18.0	\\
J229	&	6.09	&	-	&		-	& 0201902001 0201902101	0761590301$^\ast$	&	76.0, 76.6, 58.4/83.8\\
A1423	&	1.93	&	538 11724 &	35.5/36.0	&	-	&		-	\\
A801	&	4.37	&	11767	&	6.6/6.7	&	-	&		-	\\
A773	&	1.34	&	533 3588 5006 13592 	&	60.2/	61.3	&	0084230601	&	12.5, 14.3, 14.2/25.2	\\
 	&	 	&	13594 13593 13591	&		 	&	  &	 	\\
A1576	&	1.12	&	7938 15127	&	43.4/45.0	&	(0402250101 0402250601)	&		(28.8)	\\
A2631	&	3.96	&	3248 11728	&	26.0/26.3	&	(0042341301)	&		(14.0)	\\
A1703	&	1.39	& 15123	(16126)	& 29.1/78.8	& (0653530101 0653530201 0653530301) & (135.5)\\
& & & & (0653530401 0653530501 0653530601)  & \\
& & & & (0653530701 0653530801 0653530901) & \\
A2219	&	1.87	& (896) 7892 13988 14431  &	152.5/197.4	& (0112231801 0112231901) 0605000501 	&	12.0, 12.8, 6.8/55.1	\\
 	&	 	&	14451 14355 14356	&		 	&	  &	 	\\
A1319	&	2.20	& 11760 17153 &	23.8/24.1	&	-	&		-	\\
A1560	&	2.63	&	11761 (17155) &	12.1/20.8	&	(0404120101)	&	(31.9)	\\
J175	&	4.51	&	-	&		-	&	0655380101	&	4.2, 4.3, 3.1/13.9	\\
J249	&	1.94	&	-	&		-	&	0761590401$^\ast$	&	20.2, 20.7, 13.7/23.0	\\
A1201	&	1.66	&	(4216 7697) 9616	&	47.4/93.2	&	(0500760101) &		(51.8)	\\
A2009	&	3.89	&	10438	&	19.9/20.1	&	(0673851001 0673852201) 0693011001 &	10.6, 12.3, 6.1/	60.2	\\
& & & & (0693011301)  & \\
A2111	&	2.00	&	544 11726	&	31.2/31.6	&	(0673850601 0673852301)	&		(40.5)	\\
A815	&	1.98	&	-	&		-	&	0742510701$^\ast$	&	16.1, 16.6, 12.5/20.0	\\
Z1450	&	1.19	&	-	&		-	&	0761590601$^\ast$	&	19.6, 19.9, 6.3/23.0	\\
A1765	&	2.33	&	-	&		-	&	0761590701$^\ast$	&	21.7, 21.4, 15.6/	24.0	\\
A1902	&	1.02	&	16151	&	5.0/5.1	&	-	&		-	\\
\hline	 
\end{tabular}
\begin{tablenotes}
      \item {\sl Note.}
The total Galactic column density of Hydrogen is from \cite{2013MNRAS.431..394W}. The obsids and their total exposure time in brackets are not included in the data analysis of this work. The exposure time is the clean/total time for \cha\ or clean MOS1, MOS2, pn/total time for \xmm. The obsids with $^\ast$ are clusters from our \XMM\ project (proposal id 74251 and 76159, PI: M. Sun).
    \end{tablenotes} 
\label{t:obs}
\end{table*}

\section{Data analysis}
\label{s:obs}
Table~\ref{t:obs} provides the detail of observations from \cha\ and \xmm\ for this sample. We present here the data reduction procedure, the derivation of X-ray properties, and how we deal with complicated multiple cluster systems.

\subsection{\cha}
We process the \cha\ ACIS data with the \cha\ Interactive Analysis of Observation (CIAO, version 4.9) and calibration database (CALDB, version 4.7.3), following the procedures in \cite{2015MNRAS.450.2261M}. 
We reprocess the level-1 event files using {\tt acis\_process\_events} tool to check for the presence of cosmic ray background events,
correct for spatial gain variations due to charge transfer inefficiency (CTI), and recompute the event grades. We then filter the data to include the standard events grades 0, 2, 3, 4, and 6 only. Most observations were taken in very faint (VFAINT) mode, and in this case we applied VFAINT cleaning to both the cluster and stowed background observations. The light curve is then created with {\tt dmextract} and filtered with {\tt deflare} to exclude intervals of deviating more than $3\sigma$ of the mean value. Then we filter the ACIS event files in 0.3-12 keV to obtain a level-2 event file. Point sources and extended substructures are detected and removed using {\tt wavedetect}, which provides candidate point
sources, and the result is then checked through visual inspection. We produce the X-ray images from the level-2 event file and then create an exposure-corrected image from a set of observations using {\tt merge\_obs} (Fig.~\ref{fig:smp}). We then measure the surface brightness radial profile $S_X$ from the exposure-corrected images. We applied a direct subtraction of the cosmic X-ray (CXB)+particle+readout artifact backgrounds. For the particle background modeling, we use the stowed background scaled with the count rate in the 9.5-12 keV band, where the \cha\ effective area is negligible and the flux is dominated by the particle background. In order to measure the CXB, we considered the region where the CXB is more dominant than the cluster emission, which can be determined from the flattened portion at the outer radial profile.

The spectra and response files of ACIS (0.7-7 keV) are extracted using {\tt specextract} and fitted with {\tt XSPEC} package \citep[][version 12.9.1]{1996ASPC..101...17A}. We adopt the {\tt APEC} emissivity model \citep{2012ApJ...756..128F} to fit the on-cluster emission and the AtomDB (version 3.0.8) database of atomic data, 
the solar abundance tables are adopted from \cite{2009ARA&A..47..481A} and fixed to $0.3\ Z_{\odot}$, the redshift is fixed to the optical spectroscopy redshift of BCG.
We apply the Tuebingen-Boulder absorption model ({\tt TBABS}) for X-ray absorption by the interstellar medium (ISM), with fixed hydrogen column density $N_{\rm H}$ to the Galactic value from the NHtot tool \citep{2013MNRAS.431..394W}.
The off-cluster background spectra are extracted from cluster emission insignificant regions ($\gtrsim R_{100}$) of the same exposure.

\subsection{\xmm}
We reduce the \xmm\ MOS and pn data using the Extended Source Analysis Software (ESAS; \citealt{2008A&A...478..575K}; \citealt{2008A&A...478..615S}), as integrated into the \xmm\ Science Analysis System (version 15.0.0) with the associated Current Calibration Files (CCF), following the procedures in \cite{2016MNRAS.459..366G}.
We reproduce the raw event files from MOS and pn CCDs using tasks {\tt emchain} and {\tt epchain}, respectively. The solar soft proton flares are filtered out with {\tt mos-filter} and {\tt pn-filter} through the light curve screening to obtain the clean event files. The MOS CCDs that are damaged or in the anomalous state are excluded in downstream processing. The point sources are detected by {\tt cheese} and checked with visual inspection and then excluded. We use {\tt mos-spectra} and {\tt pn-spectra} to produce event images and exposure maps, as well as to extract spectra and response files. The instrumental background images and spectra are modeled with  {\tt mos$\_$back} and {\tt pn$\_$back}. 
We combine the event images, background images, and exposure maps from MOS and pn with {\tt comb}, and combine the images from multiple observations with {\tt merge\_comp\_xmm}. We use {\tt adapt} to produce the final background subtracted, exposure corrected, and smoothed images (Fig.~\ref{fig:smp}). The surface brightness profiles are extracted from exposure corrected images of combined MOS1/MOS2 and pn separately. The derived X-ray properties are evaluated with error weight means of MOS and pn.  

The spectra of MOS (0.3-11.0 keV) and pn (0.4-11.0 keV) are fitted jointly.
The on-cluster spectra are also fitted with {\tt APEC} model.
The same solar abundance table (also fixed to $0.3\ Z_{\odot}$), AtomDB version and the Galactic absorption model as in the \cha\ analysis are used.
The background spectra consist of mainly four components: CXB, quiescent particle background (QPB), residual soft proton (SP), and solar wind charge exchange (SWCX). They are fit simultaneously with the on-cluster emission. The CXB is modeled with three components: an unabsorbed thermal emission ($E \sim 0.1$ keV) from the local hot bubble or heliosphere; an absorbed thermal emission ($E \sim 0.25$ keV) from the Galactic halo and/or intergalactic medium; and an absorbed power-law emission ($\Gamma \sim 1.46$) from an unresolved background of cosmological sources. An off-cluster ROSAT all-sky survey (RASS) spectrum is also extracted from a $1^{\circ}-2^{\circ}$ annulus surrounding the cluster and joint fit with other spectra to constrain the contribution of CXB. The QPB originates from the interaction of cosmic rays with detectors. The QPB continuum is subtracted as background spectra in {\tt XSPEC}, while its bright instrumental fluorescent lines vary from observation to observation. Thus, they are not included in the QPB model spectra and are individually fit by Gaussian models. The residual SP may still exist after the light-curve screening. As they are not X-ray photons and not folded through the instrumental effective area, SP is modeled by a power law with diagonal response matrices supplied with the ESAS calibration files. The SWCX process may produce additional emission lines in the observed spectra, and they are modeled with the Gaussian components. 

\subsection{$T_X$, $L_X$, and $Y_X$}
The spectroscopic X-ray temperature $T_X$ is measured in 0.15 - 0.75 $R_{500}$. The inner boundary of 0.15 $R_{500}$ is chosen to exclude the central cool core (CC) and the outer boundary of 0.75 $R_{500}$ is limited by the quality of the spectroscopic data. The X-ray peak or centroid of a cluster is the position where the derivatives of the surface brightness variation along two orthogonal (e.g. X and Y) directions become zero.
$R_{500}$ is estimated from the $M-T_X$ relation \citep{2009ApJ...693.1142S} iteratively. In fact, the $R_{500}$ is dependent on temperature with a power-law index of 0.55 and the $R_{500}$ typically converges in three iterations. 

The bolometric X-ray luminosity $L_X$ is derived within $R_{500}$. We sum the count rates from surface brightness profiles in the 0.7-2 keV for \cha\ (maximize the SNR of cluster emission and minimize the dependency of the cooling function on the temperature and metallicity) and 0.7-1.3 keV for \xmm\ (maximize the SNR of cluster emission and minimize the contamination of SWCX lines below 0.7 keV and instrumental lines above 1.3 keV). Regions masked for point sources, chip gaps and bad pixels are added back in this process. We then use the best-fit spectral model to convert the count rate to bolometric luminosity.
For some clusters with shallow data, surface brightness profiles can not be robustly constrained to $R_{500}$. We extrapolate the surface brightness profile using a power law measured from the profile at large radii. The typical slope is about -3, which corresponds to $\beta=2/3$ for a $\beta$ model and is typical for cluster density profiles around $R_{500}$ \citep{2015MNRAS.450.2261M}. The extrapolated correction factors ([data+extrapolation]/data) are always smaller than 1.5.

The X-ray Compton parameter $Y_X$, $M_{gas}\ (<R_{500}) \times T_X\ (0.15-1.0\ R_{500})$, is also derived. We convert our $T_X\ (0.15-0.75\ R_{500})$ to $T_X\ (0.15-1.0\ R_{500})$ with the correction factor of 0.96 based on \cite{2015MNRAS.450.2261M}. The gas mass $M_{gas}$ within $R_{500}$ is summed from the gas density profile, which is derived from deprojecting the surface brightness profile. For clusters with surface brightness profiles not reaching $R_{500}$, we also extrapolate the gas density profiles using a power-law model measured from the profile at large radii, similar to what we did for luminosity extrapolation. The resulting $Y_X$ value is only included in the analysis if the extrapolated correction factor of $M_{gas}$ ([data+extrapolation]/data) is smaller than 1.5. Table~\ref{t:xray} includes the results of X-ray properties.

\subsection{Multiple cluster systems}
Most clusters in the sample are mergers. Some clusters with multiple components, e.g. J150, A781, A750 and A1319, require special attention as the association of the X-ray emission with the optical cluster becomes ambiguous. We apply different strategies to deal with such systems.

For systems of one optical cluster corresponding to multiple X-ray clusters (e.g. J150, A750, A1319), we use two different methods to associate the X-ray emission with the optical cluster and study the scaling relations.

(i) Add. The X-ray properties are derived individually for each cluster after masking the $R_{500}$ region of the nearby cluster. Then we  derive $L_X$-weighted $T_X$, add $L_X$ or $Y_X$ together, and then assign to the corresponding optical counterpart from \max\ or \red\ catalogue. For instance, A750 is a close pair composed of A750E and A750W (Fig.~\ref{fig:smp}, Appendix~\ref{app:id}). Both \max\ and \red\ catalogue mix them as one cluster. We add the $L_X$ values from A750E and A750W, as well as their $Y_X$ values. The $T_X$ is from the $L_X$-weighted $T_X$ of A750E and A750W, which should be similar to the emission-weighted.

(ii) Mix. The multiple subclusters are treated as one cluster. We center the cluster on the peak of the X-ray brightest subcluster and derived the X-ray properties within $R_{500}$ determined from the $M-N$ relation based on the optical catalogue, under the assumption of spherical symmetry. In the case of A750, we center the cluster in A750E and do not mask out A750W.

We compare the results in Table~\ref{t:asi} from these two methods and find that the $L_X$ are comparable from add and mix, while the $Y_X$ from add is smaller than the $Y_X$ from mix. For clusters in the early merge stage (J150, A750, A1319 and J175), we use the add method. For clusters in the late merge stage (Z5247, A1560 and A815) where the individual $R_{500}$ regions heavily overlap, we use the mix method. We emphasize that switching the method for multiple cluster systems in Table~\ref{t:asi} does not affect any of our conclusions for the whole sample.

For systems of one X-ray cluster corresponding to multiple optical clusters (e.g. A781, A1760, A2631), we match the X-ray cluster with the richest optical cluster.

{\setlength{\tabcolsep}{0.5em}
\begin{landscape}
\begin{deluxetable}{lcccccccccccccc}
\tablecolumns{11}
  \tablecaption{X-ray properties}
  \tablewidth{0pt}
  \tablehead{
  \colhead{Cluster} &
  \colhead{RA} &
  \colhead{DEC} &
  \colhead{$z_{\rm spec}$} &
  \colhead{$T_X$} &
  \colhead{$L_{\rm X}$} &
  \colhead{$Y_{\rm X}$} &
  \colhead{$R_{500}$} &
  \colhead{offset$^{N}$} &
  \colhead{offset$^{\lambda}$} &  
  \colhead{$M_{500}$} &   
  \colhead{$M_{500}^{N}$} &
  \colhead{$M_{500}^{\lambda}$} &
  \colhead{opt$^{N}$ } &
  \colhead{ opt$^{\lambda}$} \\
 & & & & keV & $10^{44}\rm{erg\ s}^{-1}$ & $10^{14}M_{\odot}$ keV  & Mpc & kpc & kpc &  $10^{14}M_{\odot}$ & $10^{14}M_{\odot}$ & $10^{14}M_{\odot}$ & &
}   
\startdata
A2142	&	239.58563	&	27.22967	&	0.091	&	8.20	$\pm$	0.05	(X)	&	32.24	$\pm$	0.06	&	11.44	$\pm$	0.33	&	1.43	&	25.7	&	25.8	&	9.1	&	12.2	&	12.6	&	R	&	R	\\
J150M	&	150.61145	&	20.51861	&	0.320	&	3.63	$\pm$	0.48	(X)	&	3.24	$\pm$	0.12	&	0.98	$\pm$	0.15	&	0.81	&	1273.2	&	888.0	&	2.1	&	10.6	&	10.9	&	M, P	&	M, P	\\
J150E	&	150.68539	&	20.49089	&	0.309	&	2.44	$\pm$	0.35	(X)	&	1.08	$\pm$	0.05	&	0.28	$\pm$	0.05	&	0.65	&		&		&	1.1	&		&		&		&		\\
J150W	&	150.55545	&	20.53728	&	0.320	&	3.11	$\pm$	0.71	(X)	&	1.00	$\pm$	0.09	&	0.55	$\pm$	0.14	&	0.74	&		&		&	1.6	&		&		&		&		\\
A1689	&	197.87247	&	-1.34129	&	0.183	&	10.70	$\pm$	0.26	(C)	&	41.73	$\pm$	1.06	&	13.89	$\pm$	0.63	&	1.58	&	5.4	&	5.8	&	13.4	&	10.0	&	12.2	&	R	&	R	\\
	&		&		&		&	8.01	$\pm$	0.13	(X)	&	33.59	$\pm$	0.11	&	8.86	$\pm$	0.37	&	1.34	&		&		&	8.3	&		&		&		&		\\
A1443	&	180.30410	&	23.10630	&	0.264	&	8.78	$\pm$	0.60	(C)	&	19.97	$\pm$	2.08	&	11.58	$\pm$	1.02	&	1.35	&	216.3	&	112.2	&	9.3	&	8.9	&	9.1	&	M	&	M	\\
A781	&	140.11223	&	30.49477	&	0.299	&	5.55	$\pm$	0.17	(X)	&	10.03	$\pm$	0.08	&	3.88	$\pm$	0.22	&	1.03	&	67.3	&	179.0	&	4.3	&	8.0	&	1.9	&	M, P	&	M, P	\\
A781M	&	140.22101	&	30.46535	&	0.293	&	3.12	$\pm$	0.14	(X)	&	0.75	$\pm$	0.06	&	2.32	$\pm$	0.15	&	0.75	&		&	71.7	&	1.7	&		&	8.8	&		&	M, P	\\
A781E	&	140.29418	&	30.46433	&	0.429	&	3.34	$\pm$	0.34	(X)	&	3.26	$\pm$	0.09	&	0.83	$\pm$	0.10	&	0.72	&		&	84.5	&	1.7	&		&	1.9	&		&	M, P	\\
A781W	&	139.89517	&	30.53303	&	0.427	&	3.24	$\pm$	0.24	(X)	&	2.81	$\pm$	0.06	&	0.62	$\pm$	0.06	&	0.71	&		&		&	1.6	&		&		&		&		\\
A1986	&	223.28780	&	21.89560	&	0.117	&	3.65	$\pm$	0.47	(C)	&	2.16	$\pm$	0.38	&	1.09	$\pm$	0.17	&	0.90	&	639.5	&	23.5	&	2.4	&	7.6	&	3.9	&	M	&	R	\\
A1882E	&	213.77818	&	-0.49216	&	0.139	&	3.35	$\pm$	0.40	(C)	&	0.85	$\pm$	0.10	&				&	0.85	&	13.0	&		&	2.0	&	7.3	&		&	P	&		\\
	&		&		&		&	2.44	$\pm$	0.17	(X)	&	0.68	$\pm$	0.10	&				&	0.72	&		&		&	1.2	&		&		&		&		\\
A1882W	&	213.60034	&	-0.37877	&	0.138	&	1.91	$\pm$	0.16	(X)	&	0.33	$\pm$	0.01	&				&	0.63	&		&	10.0	&	0.8	&		&	6.8	&		&	P	\\
A1882M	&	213.74050	&	-0.34915	&	0.139	&	1.37	$\pm$	0.08	(X)	&	0.27	$\pm$	0.01	&				&	0.52	&		&		&	0.5	&		&		&		&		\\
A1882N	&	213.53597	&	-0.27103	&	0.140	&	1.36	$\pm$	0.14	(X)	&	0.11	$\pm$	0.01	&				&	0.52	&		&		&	0.5	&		&		&		&		\\
A1758N	&	203.19958	&	50.53972	&	0.279	&	10.15	$\pm$	0.43	(C)	&	17.01	$\pm$	2.31	&	11.76	$\pm$	0.77	&	1.45	&	491.0	&	492.0	&	11.7	&	7.2	&	11.1	&	M, P	&	M, P	\\
A1758S	&	203.12317	&	50.40931	&	0.273	&	5.87	$\pm$	0.24	(C)	&	11.61	$\pm$	0.67	&	4.40	$\pm$	0.29	&	1.08	&		&	203.7	&	4.7	&		&	2.7	&		&	M, P	\\
A1760	&	203.53705	&	20.23874	&	0.173	&	6.71	$\pm$	0.92	(C)	&	7.10	$\pm$	1.02	&	5.41	$\pm$	0.84	&	1.23	&	138.4	&	1001.5	&	6.2	&	7.1	&	4.3	&	M	&	M	\\
A1622	&	192.42163	&	49.87525	&	0.283	&	4.70	$\pm$	0.89	(C)	&	2.84	$\pm$	0.14	&	1.31	$\pm$	0.36	&	0.95	&	477.3	&	604.1	&	3.3	&	6.7	&	8.9	&	M, P	&	M, P	\\
A750E	&	137.30282	&	10.97472	&	0.176	&	5.02	$\pm$	0.26	(C)	&	7.44	$\pm$	0.56	&	2.52	$\pm$	0.19	&	1.05	&		&	23.7	&	3.9	&		&	13.1	&		&	P	\\
	&		&		&		&	4.76	$\pm$	0.16	(X)	&	5.83	$\pm$	0.06	&	2.04	$\pm$	0.12	&	1.02	&		&		&	3.6	&		&		&		&		\\
A750W	&	137.21662	&	11.02747	&	0.164	&	3.36	$\pm$	0.18	(X)	&	1.96	$\pm$	0.04	&	0.86	$\pm$	0.07	&	0.84	&	13.0	&		&	2.0	&	6.7	&		&	P	&		\\
A1682	&	196.70864	&	46.55945	&	0.218	&	7.90	$\pm$	0.67	(C)	&	10.42	$\pm$	0.52	&	5.38	$\pm$	0.67	&	1.31	&	159.7	&	3.6	&	8.0	&	6.7	&	8.3	&	M	&	R	\\
A1246	&	170.98878	&	21.48200	&	0.193	&	6.23	$\pm$	0.66	(C)	&	9.77	$\pm$	0.49	&	3.93	$\pm$	0.60	&	1.17	&	312.3	&	68.4	&	5.5	&	6.5	&	7.5	&	M	&	M	\\
A1961	&	221.13280	&	31.22670	&	0.233	&	4.12	$\pm$	1.28	(C)	&	3.63	$\pm$	0.48	&	1.35	$\pm$	0.44	&	0.91	&	1005.8	&	3.8	&	2.7	&	6.3	&	7.6	&	M	&	R	\\
A2034	&	227.55129	&	33.50973	&	0.112	&	5.98	$\pm$	0.17	(X)	&	8.82	$\pm$	0.06	&	3.96	$\pm$	0.20	&	1.19	&	174.0	&	174.0	&	5.3	&	6.2	&	6.3	&	M	&	M	\\
A655	&	126.37410	&	47.13130	&	0.129	&	4.31	$\pm$	0.35	(C)	&	3.83	$\pm$	0.55	&	1.91	$\pm$	0.20	&	0.98	&	25.0	&	25.4	&	3.1	&	6.2	&	9.1	&	R	&	R	\\
A1914	&	216.50380	&	37.82557	&	0.170	&	9.33	$\pm$	0.66	(C)	&	33.10	$\pm$	1.57	&	11.41	$\pm$	0.94	&	1.47	&	174.9	&	174.9	&	10.8	&	6.1	&	6.6	&	M	&	M	\\
	&		&		&		&	7.84	$\pm$	0.19	(X)	&	29.39	$\pm$	0.14	&	8.71	$\pm$	0.39	&	1.34	&		&		&	8.1	&		&		&		&		\\
Z5247	&	188.57520	&	9.77170	&	0.231	&	5.91	$\pm$	0.64	(C)	&	5.85	$\pm$	0.66	&	3.68	$\pm$	0.71	&	1.11	&	79.1	&	35.3	&	4.9	&	6.1	&	5.8	&	M, P	&	P	\\
A657	&	125.83066	&	15.96316	&	0.153	&	2.89	$\pm$	0.17	(X)	&	1.28	$\pm$	0.02	&	0.53	$\pm$	0.05	&	0.78	&	5.2	&	5.2	&	1.6	&	5.9	&	4.9	&	R	&	R	\\
J229	&	229.33991	&	-0.71679	&	0.114	&	2.50	$\pm$	0.12	(X)	&	0.61	$\pm$	0.01	&	0.29	$\pm$	0.03	&	0.73	&	178.3	&	12.5	&	1.3	&	5.8	&	3.1	&	M, P	&	P	\\
A2051	&	229.18317	&	-0.97167	&	0.120	&	3.78	$\pm$	0.13	(X)	&	2.28	$\pm$	0.03	&	1.18	$\pm$	0.08	&	0.92	&		&		&	2.5	&		&		&		&		\\
A2051S	&	229.24147	&	-1.11056	&	0.117	&	2.45	$\pm$	0.26	(X)	&	0.27	$\pm$	0.01	&				&	0.73	&	4.6	&	233.6	&	1.2	&	1.1	&	2.8	&	R	&	M	\\
A2051N	&	229.10108	&	-0.83036	&	0.376	&	3.43	$\pm$	0.52	(X)	&	1.70	$\pm$	0.13	&	0.68	$\pm$	0.14	&	0.76	&		&	6.2	&	1.8	&		&	1.1	&		&		\\
A2050	&	229.07404	&	0.08937	&	0.121	&	4.46	$\pm$	0.12	(X)	&	4.05	$\pm$	0.04	&	1.69	$\pm$	0.10	&	1.01	&	5.9	&	5.8	&	3.3	&	4.4	&	4.6	&	R	&	R	\\
A1423	&	179.32222	&	33.61106	&	0.214	&	6.99	$\pm$	0.48	(C)	&	11.60	$\pm$	0.58	&	5.68	$\pm$	0.64	&	1.23	&	1.6	&	93.5	&	6.5	&	5.7	&	5.2	&	R	&	M	\\
A801	&	142.01788	&	20.52813	&	0.192	&	4.31	$\pm$	0.64	(C)	&	4.95	$\pm$	0.67	&	1.74	$\pm$	0.29	&	0.95	&	22.2	&	22.3	&	3.0	&	5.7	&	5.8	&	R	&	R	\\
A773	&	139.47280	&	51.72909	&	0.217	&	8.47	$\pm$	0.52	(C)	&	18.68	$\pm$	0.19	&	8.79	$\pm$	0.69	&	1.36	&	26.0	&	25.3	&	9.0	&	5.6	&	11.1	&	R	&	R	\\
	&		&		&		&	6.53	$\pm$	0.17	(X)	&	14.68	$\pm$	0.11	&	5.31	$\pm$	0.27	&	1.18	&		&		&	5.8	&		&		&		&		\\
A1576	&	189.24296	&	63.18712	&	0.301	&	7.99	$\pm$	0.55	(C)	&	14.44	$\pm$	2.17	&	7.95	$\pm$	0.87	&	1.26	&	29.4	&	9.2	&	7.8	&	5.5	&	6.6	&	R	&	R	\\
A2631	&	354.41084	&	0.26799	&	0.277	&	7.83	$\pm$	0.54	(C)	&	20.24	$\pm$	1.01	&	8.18	$\pm$	0.96	&	1.26	&	87.9	&	87.8	&	7.6	&	5.5	&	8.6	&	M	&	M	\\
A1703	&	198.77059	&	51.81964	&	0.282	&	9.63	$\pm$	0.75	(C)	&	21.80	$\pm$	1.09	&	9.81	$\pm$	1.19	&	1.41	&	36.9	&	35.8	&	10.7	&	5.4	&	11.0	&	R	&	R	\\
A2219	&	250.08197	&	46.71113	&	0.225	&	12.59	$\pm$	0.29	(C)	&	61.02	$\pm$	3.05	&	26.43	$\pm$	0.99	&	1.69	&	6.7	&	7.3	&	17.2	&	5.3	&	15.6	&	R	&	R	\\
A1319M	&	174.05510	&	40.04070	&	0.290	&	2.95	$\pm$	0.59	(C)	&	2.00	$\pm$	0.39	&	0.57	$\pm$	0.13	&	0.73	&	41.2	&	41.2	&	1.5	&	5.3	&	11.3	&	P	&	P	\\
A1319NW	&	173.99769	&	40.08607	&	0.295	&	2.95	$\pm$	0.44	(C)	&	2.59	$\pm$	0.49	&				&	0.73	&		&		&	1.5	&		&		&		&		\\
A1560	&	188.47084	&	15.19453	&	0.284	&	5.18	$\pm$	0.77	(C)	&	4.95	$\pm$	0.25	&	2.15	$\pm$	0.48	&	1.00	&	1.8	&	308.8	&	3.8	&	5.2	&	9.6	&	P	&	M, P	\\
J175N	&	175.56923	&	10.45006	&	0.117	&	1.66	$\pm$	0.32	(X)	&	0.20	$\pm$	0.01	&				&	0.59	&	1205.2	&	47.3	&	0.6	&	5.2	&	1.0	&	M, P	&	P	\\
J175S	&	175.52926	&	10.14652	&	0.119	&	1.42	$\pm$	0.40	(X)	&	0.14	$\pm$	0.02	&				&	0.54	&		&	24.8	&	0.5	&	-	&	1.8	&		&	P	\\
J249SW	&	249.90682	&	47.05049	&	0.227	&	3.51	$\pm$	0.28	(X)	&	1.76	$\pm$	0.04	&	0.71	$\pm$	0.07	&	0.83	&	338.3	&	33.9	&	2.1	&	5.2	&	3.4	&	M, P	&	P	\\
J249NE	&	250.03190	&	47.16119	&	0.224	&	2.67	$\pm$	0.64	(X)	&	0.77	$\pm$	0.03	&	0.30	$\pm$	0.08	&	0.72	&		&		&	1.3	&	-	&	-	&		&		\\
A1201	&	168.22715	&	13.43506	&	0.168	&	6.12	$\pm$	0.30	(C)	&	7.54	$\pm$	0.38	&	3.83	$\pm$	0.26	&	1.17	&	8.1	&	452.8	&	5.4	&	5.1	&	4.6	&	R	&	M	\\
A2009	&	225.08160	&	21.36920	&	0.153	&	6.30	$\pm$	0.37	(C)	&	14.62	$\pm$	0.84	&	4.21	$\pm$	0.31	&	1.20	&	3.2	&	3.3	&	5.7	&	5.1	&	5.7	&	R	&	R	\\
	&		&		&		&	5.91	$\pm$	0.21	(X)	&	14.04	$\pm$	0.10	&	3.96	$\pm$	0.24	&	1.16	&		&		&	5.1	&		&		&		&		\\
A2111	&	234.92407	&	34.41675	&	0.228	&	8.57	$\pm$	0.59	(C)	&	12.87	$\pm$	0.73	&	8.60	$\pm$	0.08	&	1.36	&	114.4	&	114.4	&	9.1	&	5.0	&	10.9	&	M	&	M	\\
A815	&	143.03589	&	29.06782	&	0.300	&	2.58	$\pm$	0.30	(X)	&	1.47	$\pm$	0.05	&				&	0.68	&	0.2	&	0.0	&	1.2	&	4.9	&	5.7	&	P	&	P	\\
Z1450	&	223.05214	&	37.88133	&	0.282	&	2.30	$\pm$	0.40	(X)	&	1.48	$\pm$	0.05	&	0.34	$\pm$	0.07	&	0.64	&	60.2	&	28.6	&	1.0	&	4.9	&	6.5	&	M, P	&	P	\\
A1765	&	204.20616	&	10.44110	&	0.155	&	2.22	$\pm$	0.16	(X)	&	0.49	$\pm$	0.02	&	0.20	$\pm$	0.02	&	0.67	&	11.1	&	11.1	&	1.0	&	4.8	&	5.0	&	P	&	P	\\
A1902	&	215.41900	&	37.29160	&	0.163	&	7.21	$\pm$	1.24	(C)	&	9.72	$\pm$	1.31	&	4.21	$\pm$	0.79	&	1.28	&	16.3	&	16.3	&	7.1	&	4.8	&	5.0	&	R	&	R	\\
\enddata
\begin{tablenotes}
      \item {\sl Note.}
The individual cluster name is marked the same as in Fig.~\ref{fig:smp}.  
RA and DEC are the X-ray peak position.   
$z_{\rm spec}$ is the spectroscope redshift of BCG.
$T_X$ is the X-ray temperature within 0.15-0.75 $R_{500}$.
$L_X$ is the bolometric luminosity within $R_{500}$.
$Y_X$ is the X-ray Compton parameter, $Y_X = M_{gas}\ (<R_{500}) \times T_X\ (0.15-1.0\ R_{500})$, and is not listed if the extrapolation factor for $M_{\rm gas}$ is larger than 1.5.
$R_{500}$ and $M_{500}$ are estimated from the $M-T_X$ relation \citep{2009ApJ...693.1142S}.
offset$^{N}$ or offset$^{\lambda}$ is the offset between X-ray peak and \max\ or \red\ center, respectively.
$M_{500}^{N}$ and $M_{500}^{\lambda}$ is from $M-N$ relation \citep{2009ApJ...699..768R} for \max\ and $M-\lambda$ relation (\citealt{2017MNRAS.466.3103S}, with erratum in \citealt{2018MNRAS.480.5385S} applied below) for \red, respectively. We apply $\sim -10\%$ and $\sim -3\%$ correction for the \max\ and \red\ masses because we cite $\tilde{a}_{m|n}$ from $M-N$ relation and correct to $a_{m|n}$ with the relation of $\tilde{a}_{m|n}=a_{m|n}+\frac{1}{2}\sigma_{m|n}^2$ (Appendix~\ref{app:msr}).
The label of ``M'' means clusters with {\bf M}iscentering problems (the offset between X-ray peak and optical center larger than 50 kpc), ``P'' means clusters with {\bf P}rojection problems (another X-ray cluster in a 2 Mpc radius) and ``R'' means {\bf R}elaxed cluster without both miscentering and projection problems.
    \end{tablenotes} 
\label{t:xray}
\end{deluxetable}
\end{landscape}

\begin{table*}
\caption{Assignment table for multiple cluster system}
\tabcolsep=0.33cm
\begin{tabular}{lccc} \hline \hline
Cluster	&	Subclusters	& \max\ $T_X, L_X, Y_X$	&	\red\ $T_X, L_X, Y_X$ \\	\hline
J150 & J150E, J150M, J150W (add) & $3.29\pm0.33,\ 5.32\pm0.16,\ 1.81\pm0.21$ & $3.29\pm0.33,\ 5.32\pm0.16,\ 1.81\pm0.21$	\\	
& J150E, J150M, J150W (mix) & $2.89\pm0.28,\ 5.54\pm0.15,\ 2.95\pm0.35$ & $2.94\pm0.29,\ 5.47\pm0.14,\ 2.88\pm0.35$	\\	
A750 & A750E, A750W (add) & $4.41\pm0.13,\ 7.79\pm0.07,\ 2.90\pm0.68$ & $4.41\pm0.13,\ 7.79\pm0.07,\ 2.90\pm0.68$\\
 & A750E, A750W (mix) & $4.01\pm0.11,\ 7.16\pm0.07,\ 3.88\pm0.21$ & $3.66\pm0.12,\ 7.35\pm0.07,\ 4.09\pm0.24$\\
Z5247 & Z5247NE, Z5247SW (mix) & $6.03\pm1.23$, $7.52\pm0.74$, $5.58\pm1.79$ & $6.03\pm1.23$, $7.52\pm0.74$, $5.58\pm1.79$\\
A1319 & A1319M, A1319NW, A1319SW (add) & $2.95\pm0.36,\ 4.59\pm0.63$, - & $2.95\pm0.36,\ 4.59\pm0.63$, -\\
A1560 & A1560NW, A1560SE (mix) & $5.31\pm0.71,\ 4.33\pm0.33,\ 2.24\pm0.42$ & $5.31\pm0.71,\ 4.33\pm0.33,\ 2.24\pm0.42$ \\
J175 & J175N, J175S (add) & $1.56\pm0.25,\ 0.34\pm0.02$, -  & -\\
A815 & A815N, A815S (mix) & $2.63\pm0.30,\ 1.57\pm0.06,$ - & $2.58\pm0.30,\ 1.47\pm0.05,$ -\\
\hline
\end{tabular}
\begin{tablenotes}
\item {\sl Note.} For systems of one optical cluster corresponding to multiple X-ray clusters, we use two different methods to assign the X-ray properties for clusters in different merge stages. The add method adds $L_X$ or $Y_X$ of multiple clusters, the $T_X$ is from the $L_X$ weighted $T_X$ of individual cluster. The mix method treats the multiple clusters as one cluster and uses the same routines for the single cluster, then derives the X-ray properties centering on the brightest X-ray peak within $R_{500}$ from $M-N$ relations.
\end{tablenotes} 
\label{t:asi}
\end{table*}

\section{Results}
\label{s:res}
We first study the scaling relations between X-ray and optical properties as $T_X-N$, $L_X-N$, and $Y_X-N$ relations. We then study the X-ray scaling relations as $L_X-T_X$ and $L_X-Y_X$ relations. Finally we classify the clusters using quantitative X-ray morphology parameters.

\subsection{X-ray --- optical scaling relations}

\begin{figure*}
 	\begin{center}
\includegraphics[width=0.495\textwidth,keepaspectratio=true,clip=true]{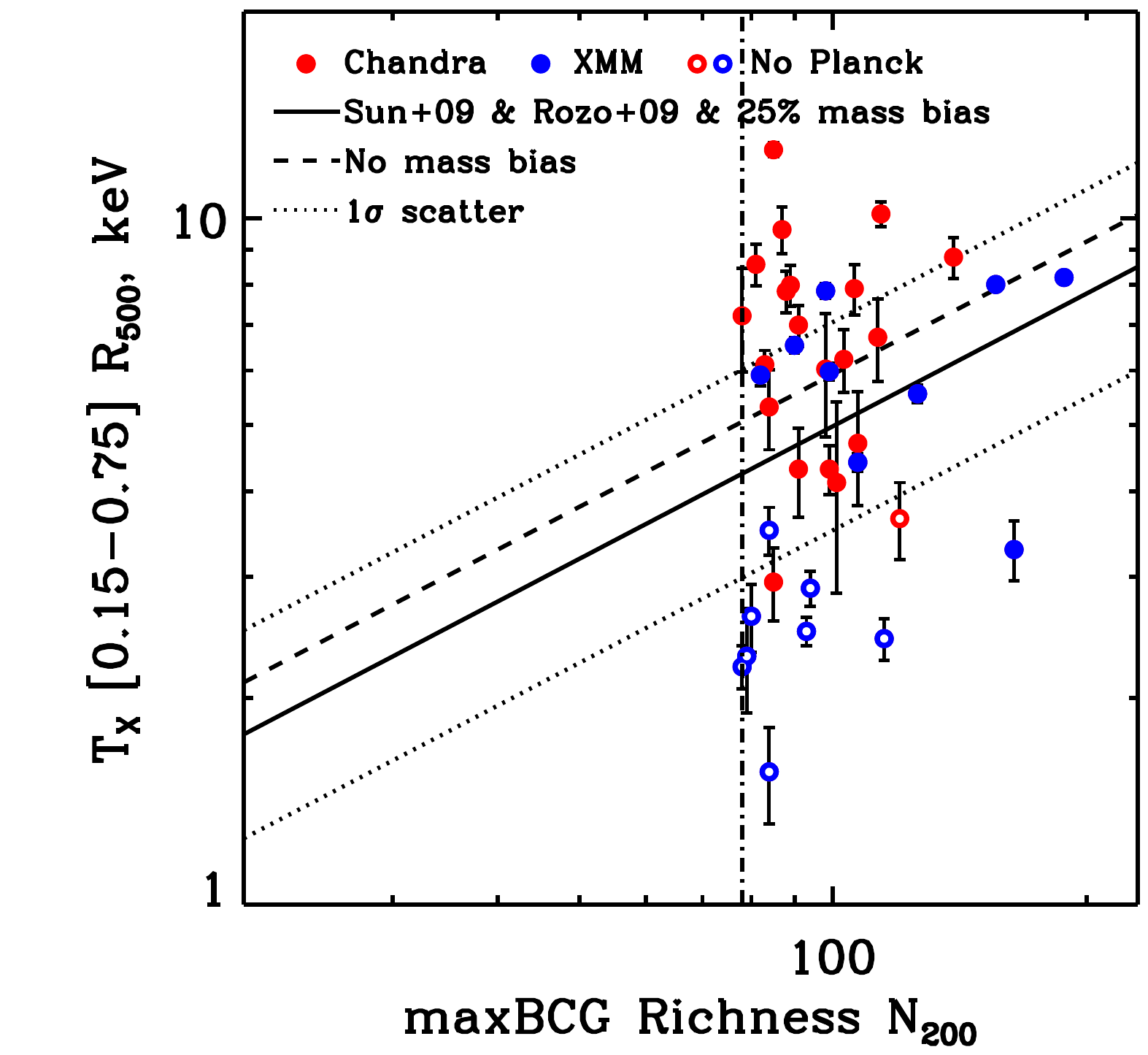}
\includegraphics[width=0.495\textwidth,keepaspectratio=true,clip=true]{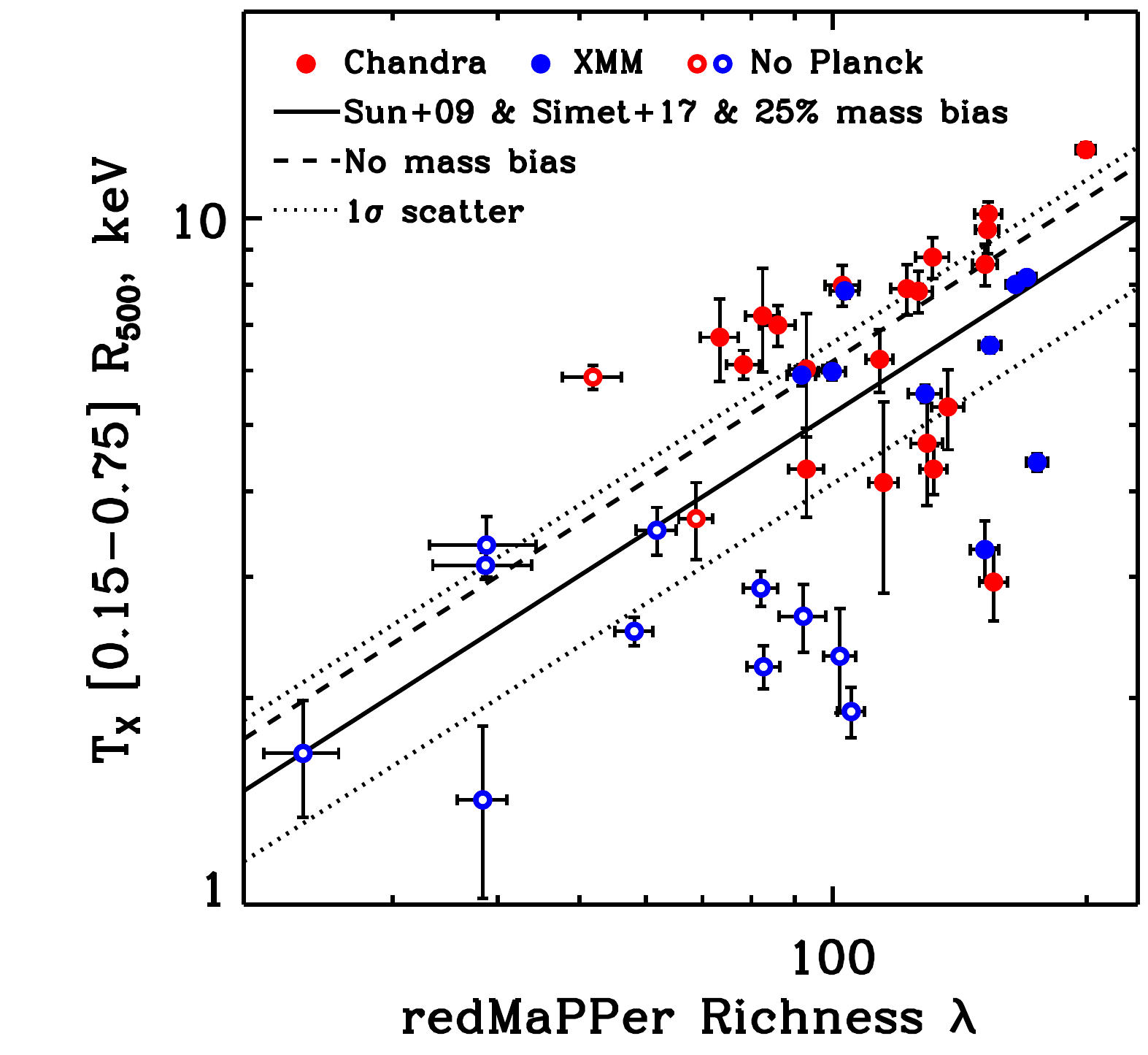}
 	\end{center}
	\caption{
		{\sl Left}: the $T_X-N_{200}$ relation for the richest \max\ sample, with $T_X$ measured from \cha\ (red) or \xmm\ (blue). The empty circles are clusters without \pla\ SZ detection. The dashed line are the expected $T_X-N_{200}$ relations from $M-T_X$ relation (\citealt{2009ApJ...693.1142S}) and weak-lensing $M-N_{200}$ relation \citep{2009ApJ...699..768R} on the full \max\ sample (see Appendix~\ref{app:msr} for the detail and bias correction). The solid line shows 25\% hydrostatic bias (the X-ray mass is 25\% lower than the weak lensing mass), and the dotted line shows expected $1\sigma$ scatter from multivariate scaling relation. The dash-dotted line shows the richness threshold of $N_{200} = 78$.
{\sl Right}: the $T_X-\lambda$ relation for the same sample. The dashed line is from X-ray $M-T_X$ relations and weak-lensing $M-\lambda$ relation (\citealt{2017MNRAS.466.3103S}).
There are more \XMM\ outliers than \cha, which reflects the fact that many \XMM\ clusters were selected for observations with non-X-ray reasons, including seven clusters in our new \XMM\ project. Moreover, most of these low $T_X$ outliers have no \pla\ SZ detection. We also note that the highest $T_X$ system, A2219, shifts from $N_{200}$=85 to $\lambda$=199.5, more in line with the overall population.
	}
	 \label{fig:T-N}
\end{figure*}

As the $T_X$ is a robust mass proxy with low scatter, the temperature-richness relation $P(T_X|N_{200})$ offers a insight into halo mass selection properties of the \max\ algorithm.
Fig.~\ref{fig:T-N} left panel shows the $T_X-N_{200}$ relation for this sample, along with the expected relation from the X-ray $M-T_X$ relation and the weak-lensing $M-N_{200}$ relation using the multivariate scaling relation in Appendix~\ref{app:msr}. 
While the richness threshold of 78 (dash-dotted line) corresponds to $\sim$ 5 keV, a wide temperature range of from $\sim$ 2 keV to $\sim$ 10 keV is observed from the data, with approximately half of the systems cooler than 5 keV. 
The temperature distribution is not symmetric around the expected relation (dashed line), but with more outliers towards lower temperatures.
Indeed, 29 clusters in the sample are detected in the PSZ2 catalogue and the \max\ clusters without \pla\ detections all have low temperatures (Fig.~\ref{fig:T-N}). 
Since the optical richness $N_{\rm gal}$ is a `projected measure' with only moderate redshift resolution, the mapping of $M$ to $N_{\rm gal}$ is skewed to high $N_{\rm gal}$ when surrounding filamentary projections are favorable. Explicit simulation studies (\citealt{2007MNRAS.382.1738C}; \citealt{2011MNRAS.413..301N}; \citealt{2011ApJ...740...53R}) have shown that $P(N_{\rm gal}|M)$ is asymmetric, with a tail extending to high richness.  While the overall amplitude of this asymmetry may be modest (e.g. involving ~10\% of all halos), it will be significantly larger when one selects only the highest richness systems from a large population. As we study only the richest 0.3\% of the \max\ catalog, the fraction of projected or `blended' halos \citep{2007MNRAS.382.1738C} could be boosted by a factor of $\sim 2$. Fig.~\ref{fig:richness} and detail in Fig.~\ref{fig:smp} show that 14/38 clusters are in multi systems, which could potentially boost the richness estimate. In fact, 5/38 ($\sim$13\%) \max\ clusters are picked apart by \red. Moreover, the $P(T_X|\lambda)$ relation in Fig.~\ref{fig:T-N} right panel shows more low mass clusters are identified by \red\ and the whole sample disperses along the mean relation with lower scatter. This demonstrates that the update algorithm of \red\ improves the richness estimate, which is less affected by the projection and with lower scatter. These improvements are mostly due to the multicolor filter \citep{2009ApJ...703..601R}.

Fig.~\ref{fig:L-N} shows the $L_X-N$ relation for \max\ (left panel) and \red\ (right panel). The dashed line shows the expected $L_X-N$ relation combined from the $L_X-M$ relation and the $M-N_{200}$ or the $M-\lambda$ relation. As miscentering and substructure can also contribute to the mismatch problem, we attempt to classify clusters in this sample. The red dots (marked as `M' in Table~\ref{t:xray}) are clusters with miscentering problems (the offset between X-ray peak and optical center larger than 50 kpc), with most of them representing merging clusters (e.g. \citealt{2010A&A...513A..37H}). The blue dots (marked as `P' in Table~\ref{t:xray}) are clusters with projection problems (another nearby X-ray cluster with similar redshift in a 2 Mpc projected radius). The half-red half-blue dots (marked as `M, P' in Table~\ref{t:xray}) are with both miscentering and projection problems. The grey dots (marked as `R' in Table~\ref{t:xray}) are clusters without miscentering and projection problems. The $L_X-N$ distribution is similar to the $T_X-N$ distribution. Similarly, there are more outliers towards lower X-ray luminosity.

Fig.~\ref{fig:Y-N} shows the $Y_X-N$ relation for \max\ (left panel) and \red\ (right panel). The $Y_X-N$ distribution is similar to the previous two distributions.
The SZ signal $Y_{SZ}$ is characterized by the integrated Compton parameter $y_c=\int n_e T_e dl$. As an X-ray analogue of the integrated Compton parameter, $Y_X$ is closely related to $Y_{SZ}$ with a linear relation (\citealt{2014MNRAS.438...62R}). Therefore, Fig.~\ref{fig:Y-N} may be converted to the $Y_{SZ}-N$ relation, which also recovers previous stacking results of lower observed SZ signal compared with the model prediction (e.g. \citealt{2011A&A...536A..12P}; \citealt{2013ApJ...767...38S}; \citealt{2017MNRAS.468.3347S}). Moreover, we convert the \pla\ mean stacking $Y_{SZ}$ \citep{2011A&A...536A..12P} in the two richest \max\ bins to the expected $Y_X$, using the $D_A^2Y_{SZ}-CY_{X}$ relation (\citealt{2014MNRAS.438...62R}; more detail in Appendix~\ref{app:msr}). We also evaluate the observed mean $Y_X$ (with \xmm\ based $Y_X$ converted to \cha\ using the cross-calibration in Appendix~\ref{app:cc}) in the same two bins. The expected $Y_X$ and observed $Y_X$ are shown as orange and green bowties in Fig.~\ref{fig:Y-N}. They are consistent with each other, and both are lower than the model prediction from the combination of X-ray mass proxy with optical weak-lensing mass calibration, which is shown as a dashed line.

\subsection{X-ray scaling relation}
X-ray imaging is good at identifying individual halos. Each extended X-ray source would map cleanly onto a single halo, while optical clusters offer a dirtier mapping to halos mainly due to projection effect. So long as the hot gas properties of the involved halos are not influenced by the projection effect in optical, we expected that the low-mass halos associated with blended (projection-dominated) optical systems will have X-ray properties comparable to those of the overall massive halo population. If $M$ is lower so will be $L_X$, $T_X$, and $Y_X$.
Figs.~\ref{fig:L-T} and \ref{fig:L-Y} show the $L_X-T_X$ relation and the $L_X-Y_X$ relation for this sample, respectively. Established relations from \cite{2009ApJ...692.1033V} (\cha\ calibration) and \cite{2009AA...498..361P} (\xmm\ calibration) are also shown.
With the cross-calibration derived in Appendix~\ref{app:cc}, we also rescale all \cha\ properties to \xmm. 
Indeed, we find that these optically selected clusters follow the `normal' $L_X-T_X$ and $L_X-Y_X$ relations calibrated from X-ray selected samples, `normal' means the behavior of the underlying true population of massive halos.

\begin{figure*}
 	\begin{center}
\includegraphics[width=0.495\textwidth,keepaspectratio=true,clip=true]{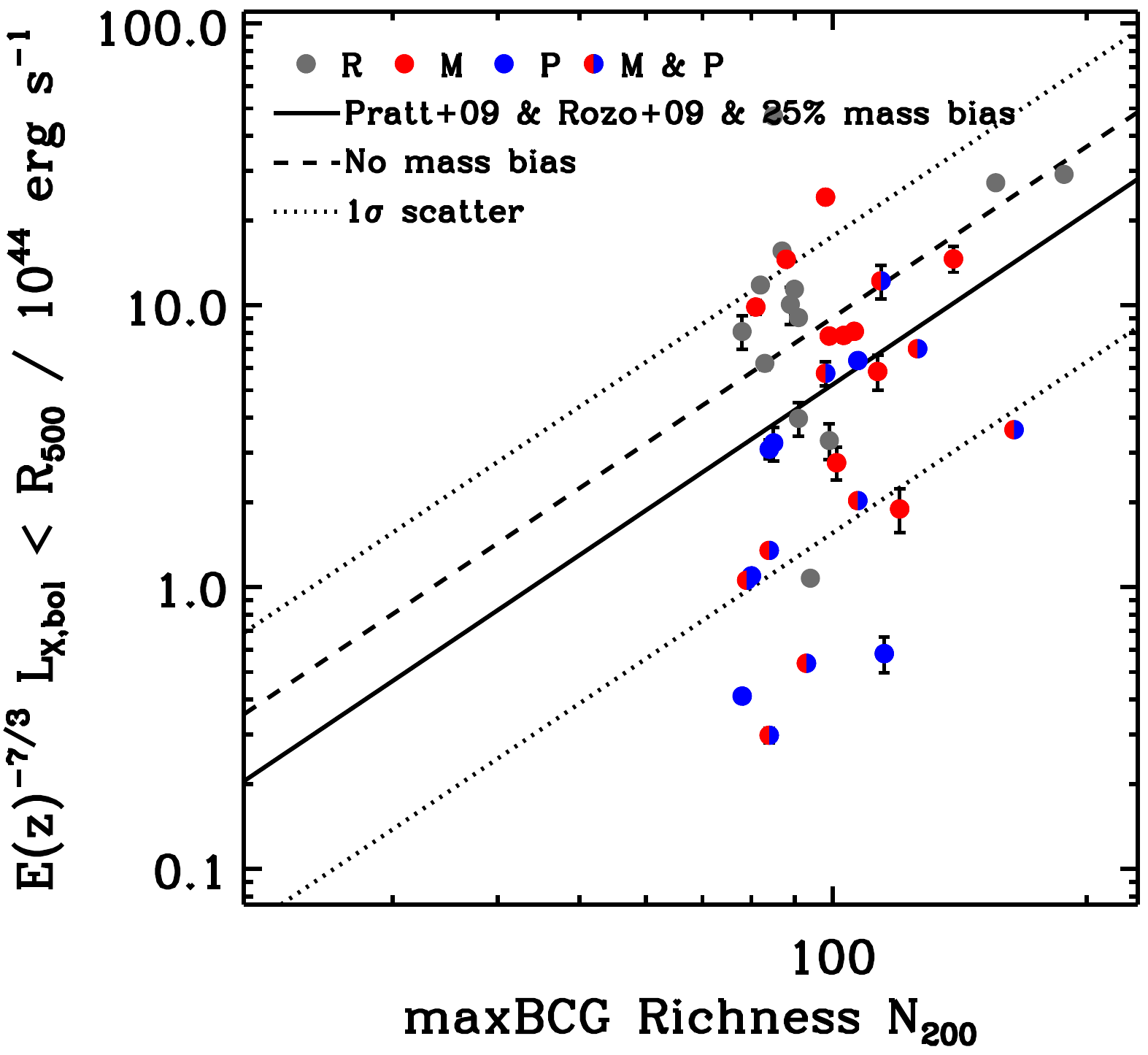}
\includegraphics[width=0.495\textwidth,keepaspectratio=true,clip=true]{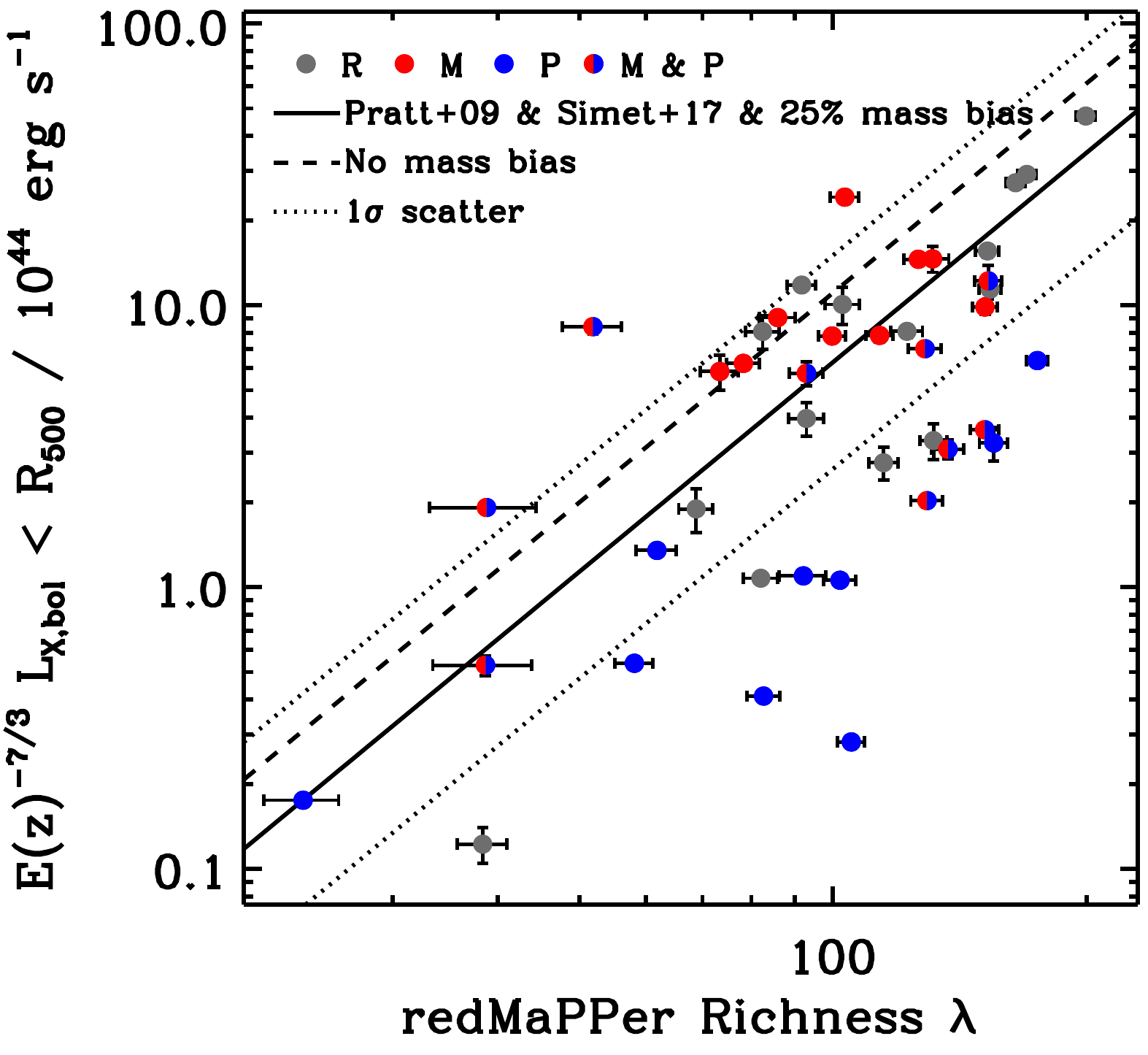}
 	\end{center}
	\caption{
{\sl Left}: the bolometric $L_X-N_{200}$ relation for the sample. Red dots are clusters with {\bf M}iscentering problems (the offset between the X-ray peak and the optical center larger than 50 kpc), with most of them are merging clusters. Blue dots are clusters with {\bf P}rojection problems (another X-ray cluster within 2 Mpc, which corresponds to 9 arcmin at median redshift = 0.23 for the \max\ sample, comparable to \pla's resolution). Half-red half-blue dots are clusters with both {\bf M}iscentering and {\bf P}rojection problems. Grey dots are {\bf R}elaxed clusters without miscentering and projection problems. The dashed line shows the expected $L_X-N_{200}$ relation from the X-ray $L_X-M$ relation (\citealt{2009AA...498..361P}) and the weak-lensing $M-N_{200}$ relation \citep{2009ApJ...699..768R} on the full \max\ sample (Appendix~\ref{app:msr}). The solid line and the dotted line show 25\% hydrostatic bias and expected $1\sigma$ scatter, respectively.
{\sl Right}: $L_X-\lambda$ relation for the sample. The $L_X-M$ relation (\citealt{2009AA...498..361P}) and the $M-\lambda$ relation \citep{2017MNRAS.466.3103S} are combined.
These two plots show the roles of miscentering and projection on the scatter of the optical richness relations, while the cluster mass bias can also reduce the discrepancy.
	}
	 \label{fig:L-N}
\end{figure*}

\begin{figure*}
 	\begin{center}
\includegraphics[width=0.495\textwidth,keepaspectratio=true,clip=true]{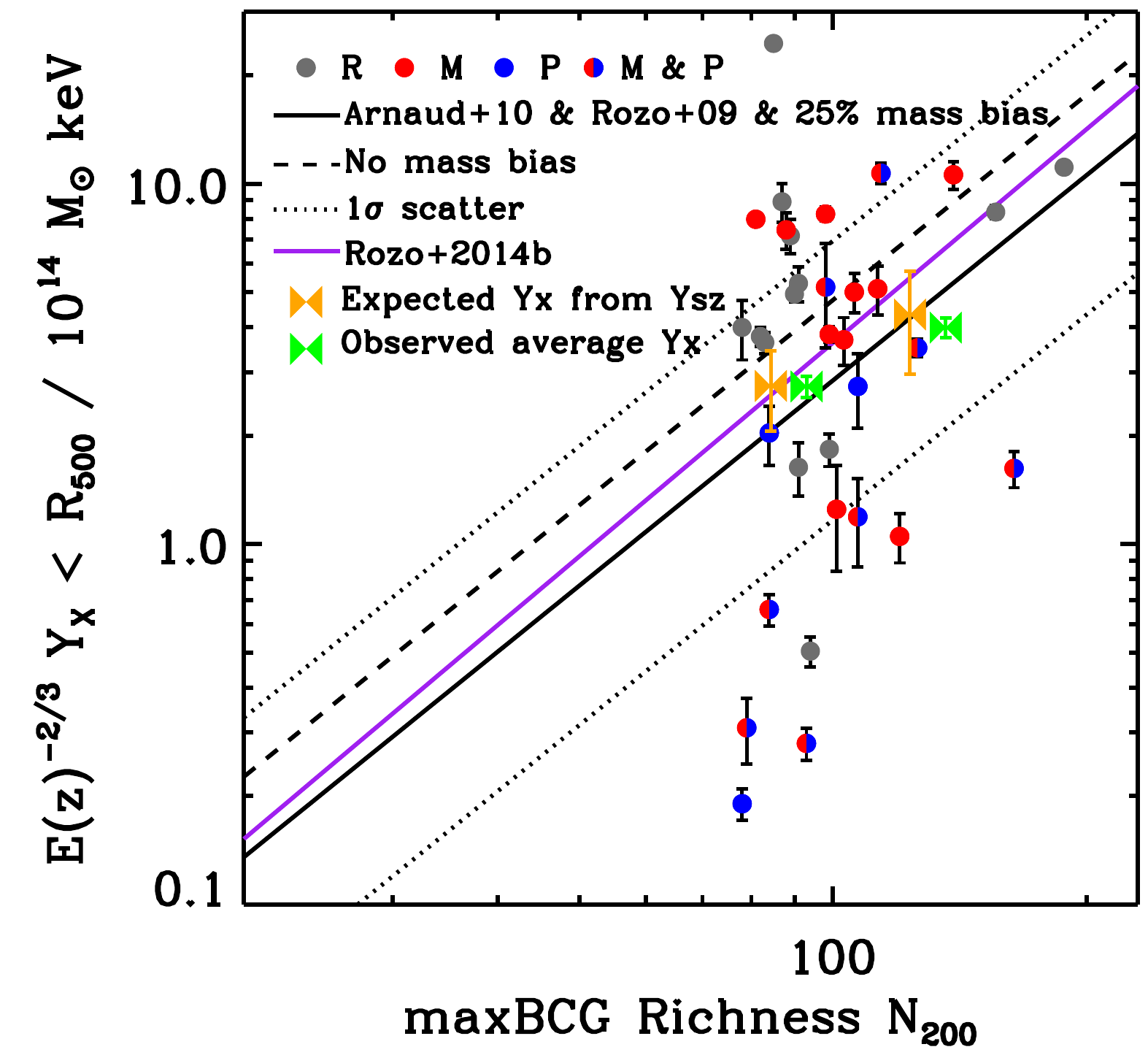}
\includegraphics[width=0.495\textwidth,keepaspectratio=true,clip=true]{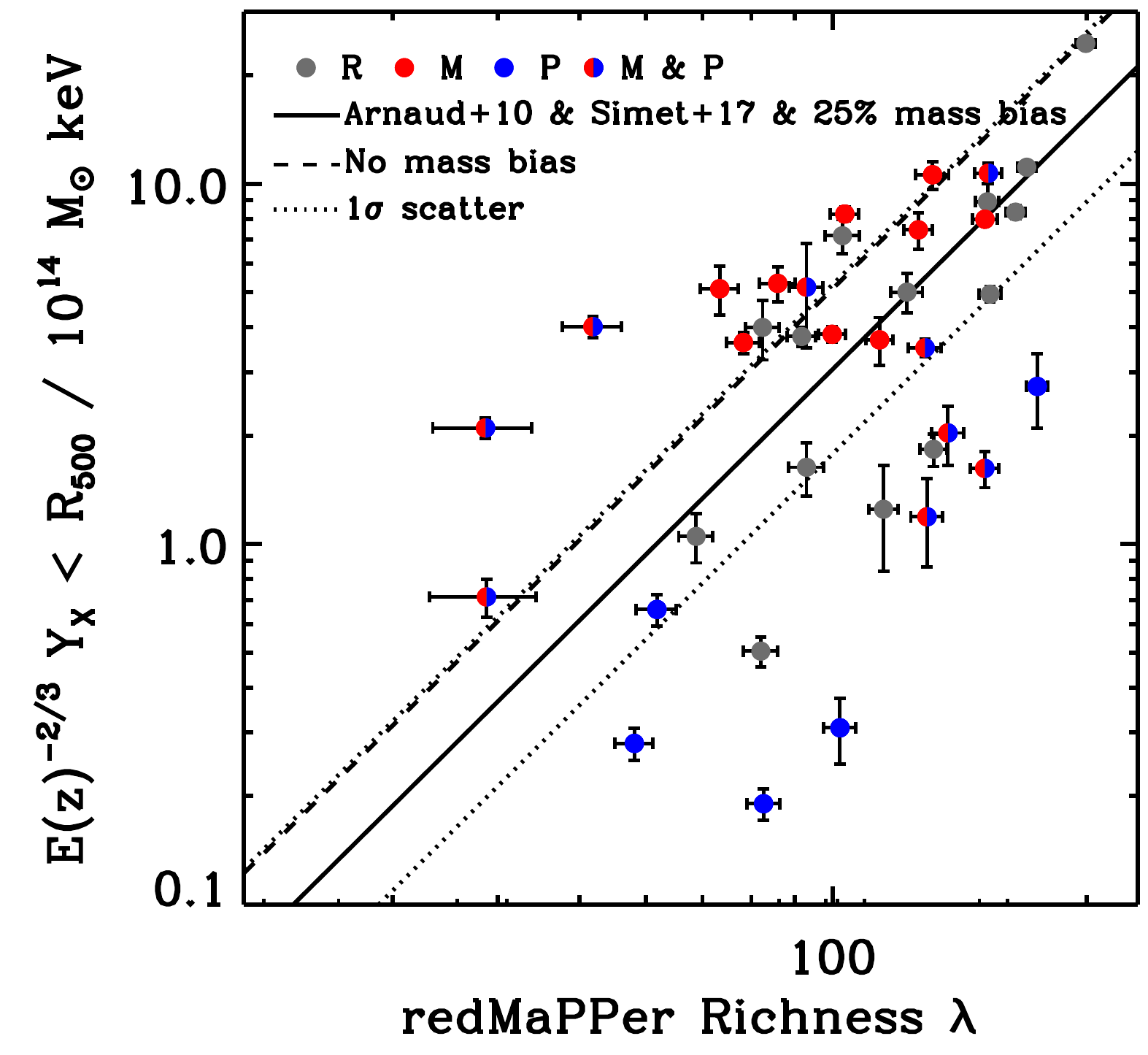}
 	\end{center}
	\caption{The relations between $Y_X$ and the optical richness for the sample, with the same symbol set as in Fig.~\ref{fig:L-N}. Orange bowties are the expected $Y_X$, which is inferred from the \pla\ mean staking $Y_{SZ}$ in the two richest \max\ bins \citep{2011A&A...536A..12P}. The $D_A^2Y_{SZ}-CY_{X}$ relation from \citet{2014MNRAS.438...62R} is assumed. Green bowties are the observed mean $Y_X$ in the same two bins. The bowties are misplaced to the average richness of each bin to avoid overlapping. The expected $Y_X$ and observed $Y_X$ are consistent with each other, and both of them are lower than the model prediction shown by the dashed line. The black dashed line combines the $M-Y_X$ relation (\citealt{2010AA...517A..92A}) and the $M-N_{200}$ relation \citep{2009ApJ...699..768R} on the full \max\ sample (Appendix~\ref{app:msr}) for the {\sl Left}, and the $M-Y_X$ relation (\citealt{2010AA...517A..92A}) and the $M-\lambda$ relation \citep{2017MNRAS.466.3103S} for the {\sl Right}. We also plot the predicted $Y_X-N_{200}$ from \citet{2014MNRAS.438...78R} as a purple line in the left panel. Most relaxed clusters are more close to the expected scaling relation.
	}
	 \label{fig:Y-N}
\end{figure*}

\begin{figure*}
 	\begin{center}
\includegraphics[width=0.495\textwidth,keepaspectratio=true,clip=true]{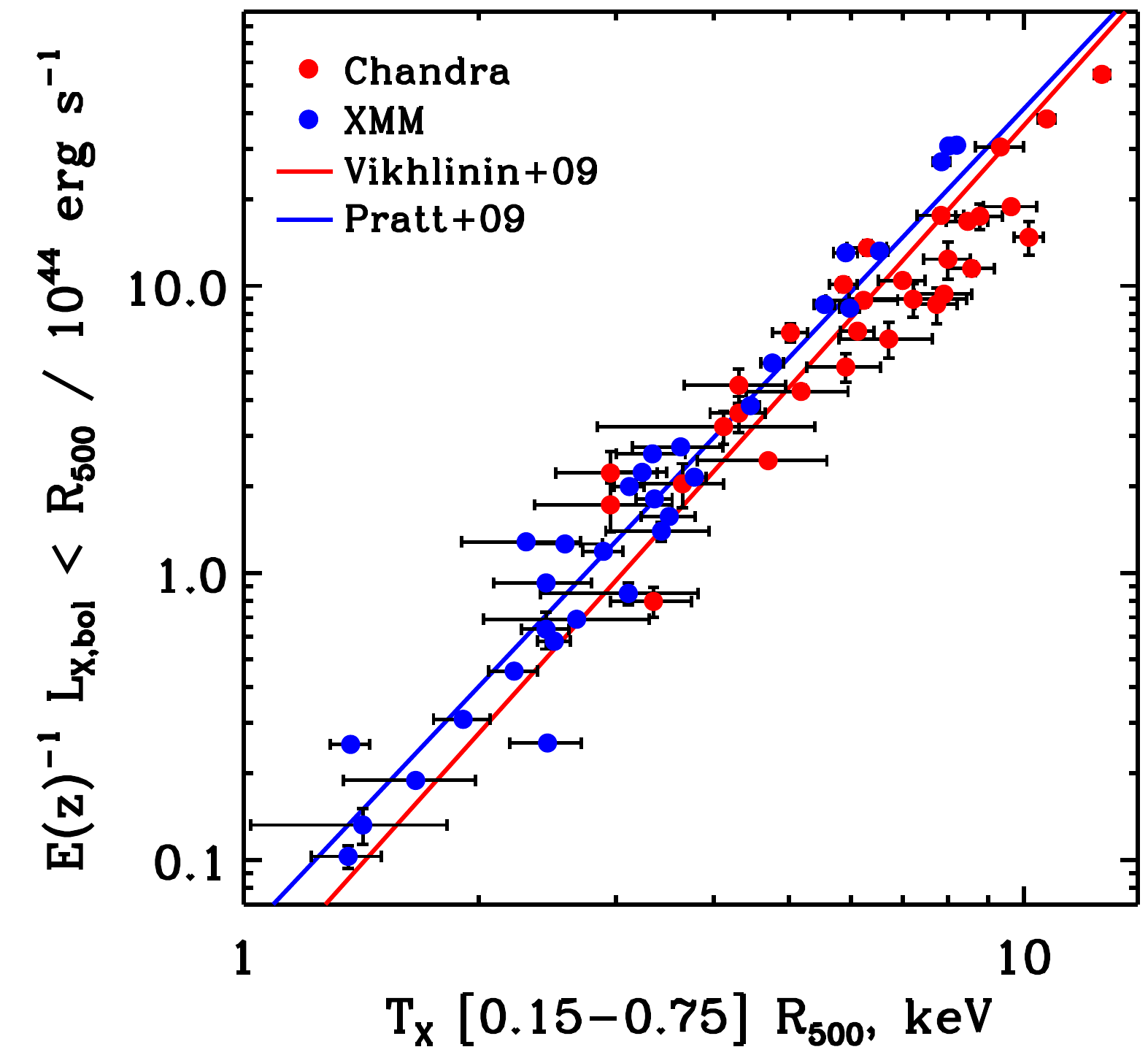}
\includegraphics[width=0.495\textwidth,keepaspectratio=true,clip=true]{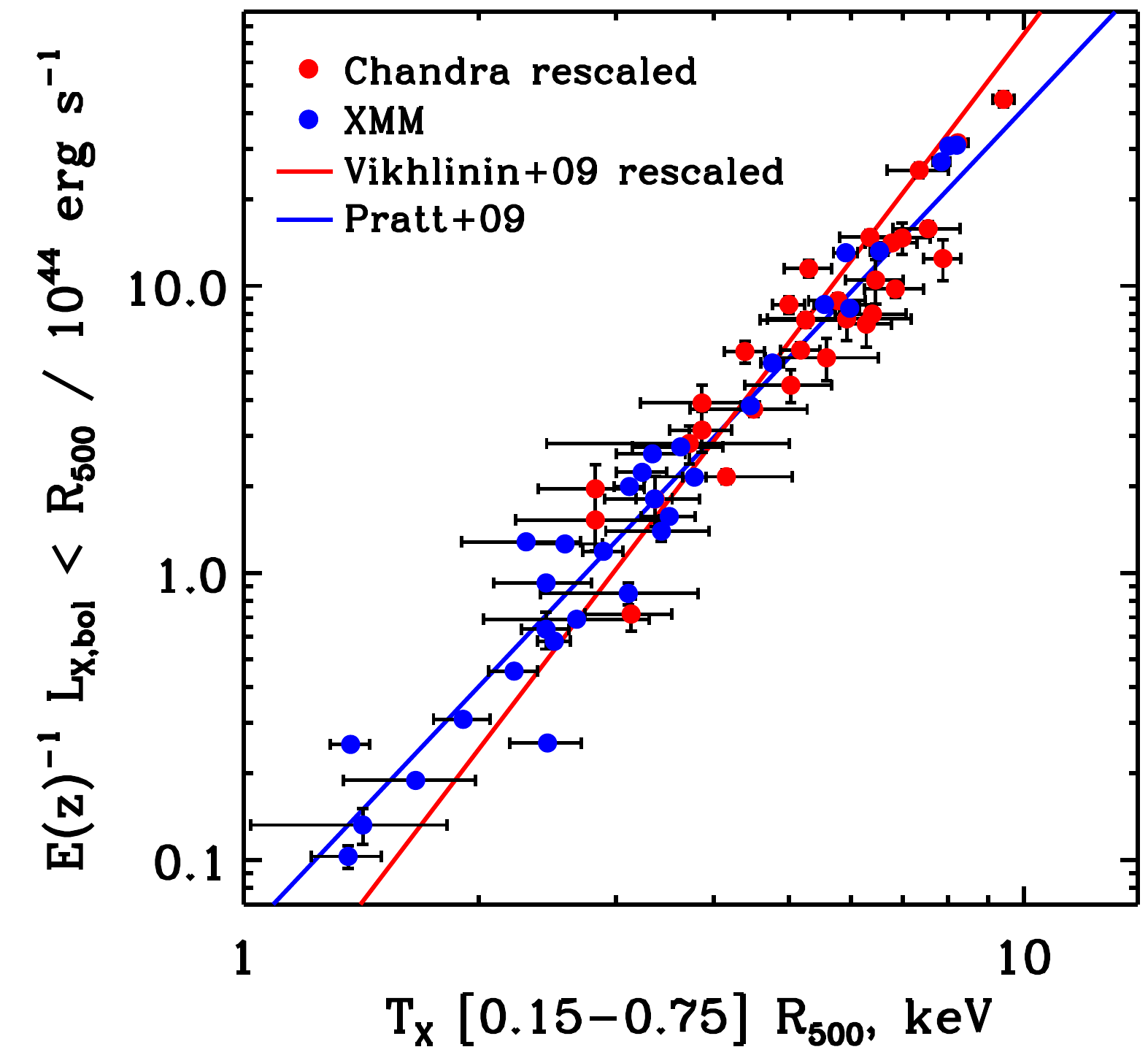}
 	\end{center}
	\caption{
{\sl Left}: the $L_X-T_X$ relation for individual clusters from Table~\ref{t:xray} (red: \cha, blue: \xmm). The red line is the $L_X-T_X$ relation \protect\citep{2009ApJ...692.1033V} based on \cha\ calibration, while the blue line \protect\citep{2009AA...498..361P} is based on \xmm\ calibration.
{\sl Right}: \cha\ data ($L_X$ and $T_X$) and $L_X-T_X$ relation are rescaled to \xmm\ with the in-house cross-calibration in Appendix~\ref{app:cc}. These clusters follow the normal $L_X-T_X$ relation with less scatters than those in Fig.~\ref{fig:L-N}.
	}
	 \label{fig:L-T}
\end{figure*}

\begin{figure*}
 	\begin{center}
\includegraphics[width=0.495\textwidth,keepaspectratio=true,clip=true]{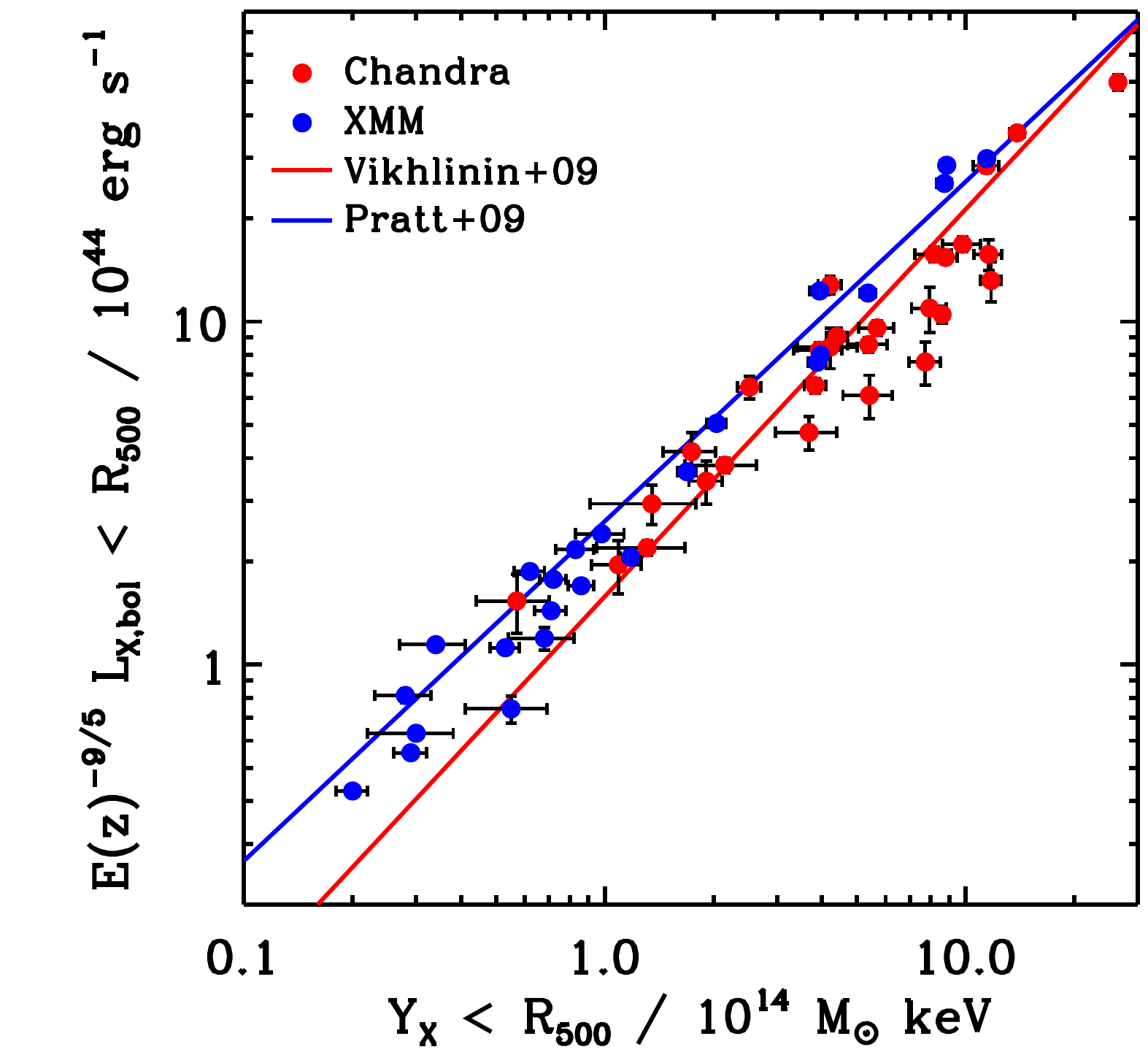}
\includegraphics[width=0.495\textwidth,keepaspectratio=true,clip=true]{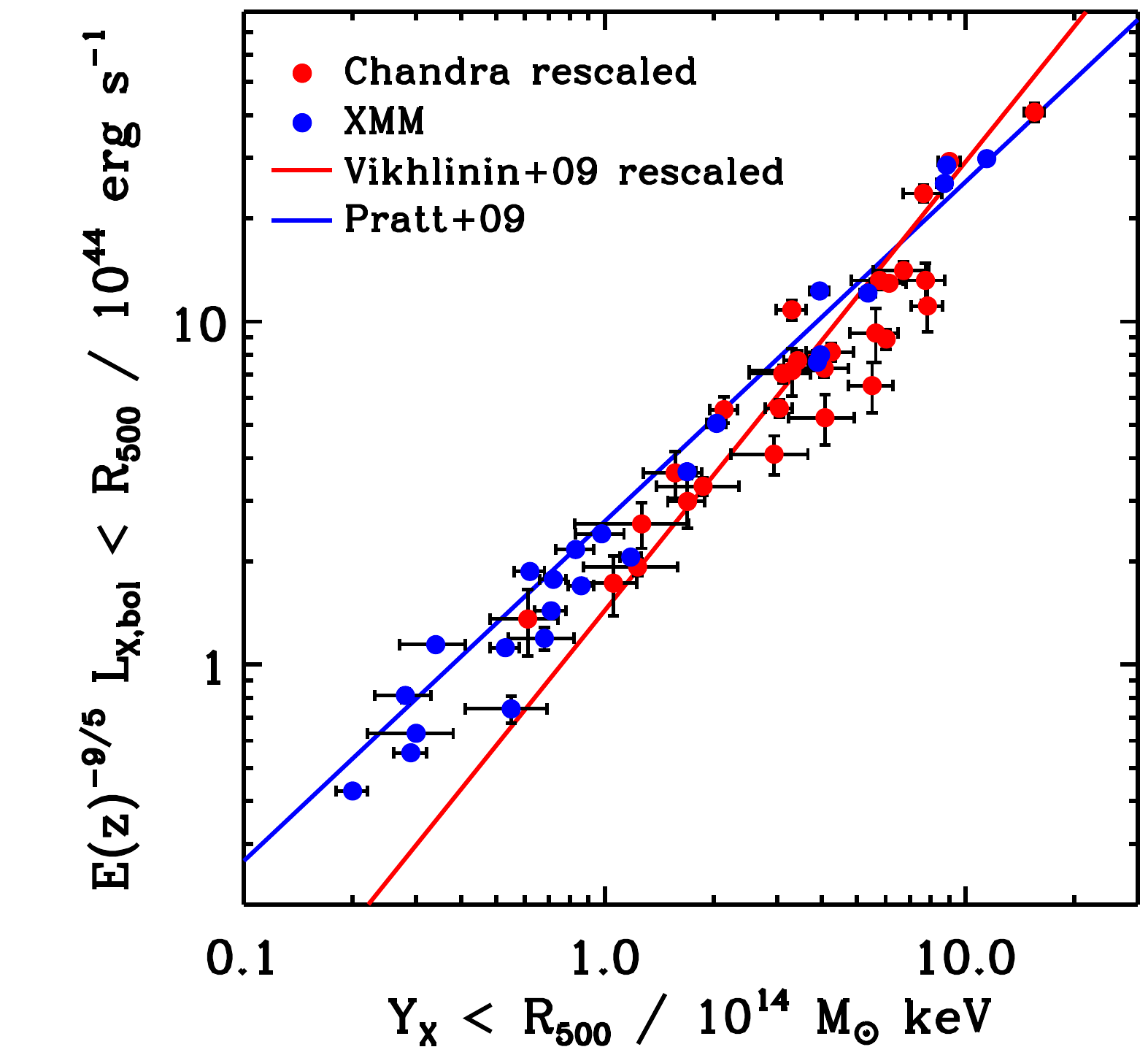}
 	\end{center}
	\caption{
{\sl Left}: the $L_X-Y_X$ relation for individual clusters from Table~\ref{t:xray}. The red line is the $L_X-Y_X$ relation \protect\citep{2009ApJ...692.1033V} based on \cha\ calibration, while the blue line \protect\citep{2009AA...498..361P} is based on \xmm\ calibration.
{\sl Right}: \cha\ data ($L_X$ and $Y_X$) and $L_X-Y_X$ relation are rescaled to \xmm\ with the in-house cross-calibration in Appendix~\ref{app:cc}. These clusters follow the normal $L_X-Y_X$ relation with less scatters than those in Fig.~\ref{fig:L-N}.
	}
	 \label{fig:L-Y}
\end{figure*}

\subsection{X-ray morphology parameters}
\label{ss:morph}
Besides the dynamical criterion based on the BCG/X-ray offset and cluster environment as presented in Table~\ref{t:xray}, we also apply several quantitative X-ray morphology parameters in Table~\ref{t:mor}: the centroid shift ($w$; \citealt{1993ApJ...413..492M}), the surface brightness concentration ($c_{\rm SB}$; \citealt{2008A&A...483...35S}), and the symmetry-peakiness-alignment (SPA; \citealt{2015MNRAS.449..199M}). Generally, the $w$, $s$, and $a$ measure the bulk asymmetry, while $c_{\rm SB}$ and $p$ judge the CC. The distribution of these parameters are shown in Fig.~\ref{fig:morph}.

\section{Discussion}
\label{s:dis}

\subsection{Scaling relation from optical, X-ray and SZ}
There is mismatch or discrepancy between optical and SZ: the expected SZ signals from the model
are higher than the observed ones for stacked optically selected clusters (e.g. \citealt{2011A&A...536A..12P}; \citealt{2013ApJ...767...38S}; \citealt{2017MNRAS.468.3347S}).
However, the scaling relations between SZ and X-ray are consistent with each other. For example, \cite{2011A&A...536A..10P} extracts the SZ signal individually for the Meta-catalogueue of X-ray detected Clusters of galaxies and find that the SZ measurements and
the X-ray based predictions are in excellent agreement. Similar results are found based on both X-ray selected clusters (\citealt{2011A&A...525A.139M}; \citealt{2011A&A...536A..12P}; \citealt{2013ApJ...767...38S}) and SZ selected clusters (\citealt{2011A&A...536A...9P}; \citealt{2011A&A...536A..11P}; \citealt{2011ApJ...738...48A}). Moreover even for optically selected clusters, \cite{2011A&A...536A..12P} compares the stacked SZ signal with the stacked ROSAT X-ray luminosity \citep{2008ApJ...675.1106R} around \max\ clusters in the same richness bins and finds the mean SZ signal and mean X-ray luminosity are consistent with the model predictions.

The consistency between SZ and X-ray observations is expected, because they are both from the same ICM. 
The problem is that optically selected clusters post a challenge for robust stacking of the X-ray or SZ signals.

\subsection{Possible solutions for the discrepancy}
As mentioned in Section~\ref{s:int}, there are some possible solutions for the discrepancy between the stacked \pla\ SZ fluxes and the model expectations for the optically selected clusters.
First, we focus on the ICM part to rule out some solutions.
The consistency between SZ and X-ray observations demonstrates that the calibration and contamination of the SZ signal are not significantly biased. The assumed pressure profile used to estimate the SZ signal \citep{2010AA...517A..92A} and the X-ray scaling relation are not significantly biased.
These richest \max\ clusters follow normal $L_X-T_X$ and $L_X-Y_X$ relations, which suggests that the fraction of X-ray-dark-but-optical-normal clusters is not significant (at least for massive ones), consistent with some previous studies (e.g. \citealt{2011A&A...536A..37A}).

Secondly, we focus on galaxies to discuss some remaining solutions. The miscentering (e.g. A1986, A1961) decreases the X-ray luminosity/SZ signal, as some flux moves outside of the aperture. \cite{2013ApJ...767...38S} demonstrated the effect of miscentering on decreasing SZ signal. They also proposed that the miscentering effect causes their lower measured SZ signal compared to \pla\ due to the finer resolution of {\em ACT}. However, as they point out, the miscentering distribution from their sample alone can only explain part of the discrepancy between optical and SZ, unless an unrealistic larger offset exists. Moreover, most miscentered clusters are merging or disturbed clusters with lower X-ray or SZ surface brightness, which makes them more difficult to be detected. The discrepancy of SZ signal is at a level of 10\% in our richness range \citep{2012ApJ...757....1B}. The projection (e.g. A750, A1319) will increase the optical richness. As the cluster mass-richness relation is close to a linear relation, projection causes the projected clusters to simply slide up and down the mass-richness relation, without deviating from it \citep{2017MNRAS.466.3103S}. However, the $L_X-M$ or $Y_{SZ}-M$ relation is a power-law relation with an index greater than one ($\sim 1.6$; \citealt{2014MNRAS.438...62R}). For example, if we have two clusters with $N_{200}=80$ ($M_{500|N=80}=5\times10^{14} M_{\odot}$, bolometric X-ray luminosity $L_{X|N=80}=11\times10^{44}{\rm erg\ s^{-1}}$) projected together, they will be detected as a N=160 cluster,  the corresponding expected mass and X-ray luminosity are $M_{500|N=160}=10\times10^{14} M_{\odot}$ and $L_{X|N=160}=43\times10^{44}{\rm erg\ s^{-1}}$, respectively. Though the mass is equal to the sum of two subclusters, the total X-ray luminosity is overestimated by a factor of $\sim2$ than the linear combination of $2L_{X|N=80}=22\times10^{44}{\rm erg\ s^{-1}}$. Thus, projection causes the expected X-ray luminosity or the SZ signal from the summed optical richness to be higher than the actual summed values. The projection fraction of samples extending to much lower richness is around 10\% \citep{2017MNRAS.466.3103S}, even higher for these richest \max\ clusters (Fig.~\ref{fig:L-N}). The above two effects can act together, especially in super clusters and large-scale filaments (Fig.~\ref{fig:ls}).
The contamination of low mass halos, whose true halo mass is far below the value suggested by the optical richness, would also dilute and reduce the mean mass of the sample.
We note that there is contamination of such low mass systems based on Fig.~\ref{fig:Ls-N-app}. These low mass halos are mostly blended systems with boosted richness affected by nearby large scale structure. Thus there is a mixture of halo masses at very high $N_{200}$, the clean and the blended, and the PDF of $M$ given $N_{200}$ (or $\lambda$) will be asymmetric with a low-mass tail.  A skew-normal or Hermite polynomial expansion \citep{2010ApJ...716..281S} are good alternatives to mixture modeling.  
Next, we roughly estimate the contamination fraction. There are 4 low $L_X$ systems beyond the $2\sigma$ line of $L_X-N$ relation from \cite{2014MNRAS.438...78R} in Fig.~\ref{fig:Ls-N-app} (and 7 beyond the $1.5\sigma$ line towards lower $L_X$ vs. 1 beyond the $1.5\sigma$ line towards higher $L_X$). Taking these numbers at face value, the contamination is 10\% - 15\%.

Thirdly, the discrepancy may be induced by the mass bias and covariance bias between the ICM and galaxy scaling relations. The mass bias is either from the X-ray HE mass or the lensing mass. \cite{2014MNRAS.438...78R} found the discrepancy problem could be solved by lowering the galaxy weak-lensing mass by 10\% while raising the X-ray mass by 21\%. The weak-lensing mass could be overestimated due to line-of-sight (LOS) contamination and triaxial halo (e.g. \citealt{2007MNRAS.380..149C}). The red-sequence cluster finding algorithm tends to find more prolate clusters (major axis aligned with LOS) than oblate clusters due to higher galaxy density and background contrast. This orientation bias can lead to 3-6\% overestimate of lensing mass \citep{2014MNRAS.443.1713D}. The covariance bias between $M_{\rm wl}$ and $N$ at fixed mass can also induce a 10\% overestimate of lensing mass (Eq.~12 of \citealt{2014MNRAS.438...78R} and Appendix~\ref{app:msr}). We compare the weaking-lensing mass with the mass estimated from the galaxy velocity dispersion \citep{2016MNRAS.460.3900F}, and CMB weak-lensing (\citealt{2017NatAs...1..795G}; \citealt{2018MNRAS.tmp..309B}). Indeed, the mean galaxy weak-lensing mass is higher than the mass from the other two methods, although uncertainty with the latter two relations is substantial.
 Meanwhile, the X-ray HE mass could be underestimated due to gas bulk motion and turbulence (e.g. \citealt{2007ApJ...668....1N}; \citealt{2009ApJ...705.1129L}). We also include the 25\% HE bias as solid lines in Figs.~\ref{fig:T-N}, \ref{fig:L-N}, and \ref{fig:Y-N}.  Compared with the dashed line of no mass bias gives a idea of how much the mass bias could reduce the discrepancy. Moreover, the covariance bias from the multivariate scaling relations (Appendix~\ref{app:msr}) leads an additional $\sim10\%$ correction.

\subsection{Can we further increase the robustness of the optical richness} 
\label{ss:improve}
From the ICM-galaxy scaling relations, we find that optically selected clusters have large scatter with more clusters biased to lower mass.
Both \cite{2011A&A...536A..12P} and \cite{2013ApJ...767...38S} selected a BCG-dominated subsample of their optically selected clusters, defined as the BCG luminosity ratio $L_{BCG}/(L_{tot}-L_{BCG})$, being larger than the average ratio for a given richness bin. They find that the BCG-dominated sample has a higher normalization, closer to the predicted relation.

\cite{2013A&A...557A..52P} compared the scaling relation between the SZ signal and the stellar mass for a large sample of locally brightest galaxies, analogous to a BCG-dominated sample. The relation is close to the self-similar prediction extending from rich clusters down to groups ($M_{500} \sim 2 \times 10^{13} M_{\odot}$), but with a normalization $\sim$ 20\% lower than the X-ray selected clusters. This discrepancy is mainly from the Malmquist bias from the X-ray sample and the miscentering from satellite contamination \citep{2013A&A...557A..52P}.
Meanwhile, \cite{2011ApJ...736...39H} and \cite{2015ApJ...808..151G} found consistent results, at least down to the group mass scale (but necessarily to the lower mass systems). Moreover, \cite{2014MNRAS.445..460G} found that the stacked SZ signal from radio selected sources is also consistent with the self-similar prediction. The SZ signal is mainly from the AGN hosted halos of giant galaxies instead of galaxy clusters or groups. The miscentering and projection problems are insignificant for such giant galaxies compared with the massive optically selected clusters.

We also select a BCG-dominated subsample with the \max\ and \red\ catalogues locating the same BCG. Fig.~\ref{fig:L-N-app} shows that the BCG-dominated clusters tend to be relaxed clusters and agree better with the model prediction than the full sample. Three of the four low luminosity outliers have the projection problems (see Fig.~\ref{fig:L-N}).
Therefore, the BCG-dominated subsample more closely correspond to the X-ray selected and SZ selected samples.

Another potentially useful information could improve the optical mass proxy is the optical luminosity of either member galaxies, or the BCG, or the intracluster light (e.g. \citealt{2007ApJ...666..147G}; \citealt{2008MNRAS.390.1157R}). Though the stars are a minor fraction of baryon in clusters, they are still scaled with the total halo mass. Moreover, the velocity dispersion from the upcoming Dark Energy Spectroscopic Instrument (DESI) survey will also increase the robustness of the optical richness.

\subsection{Dynamical state and cool core fraction}
As an optically selected sample, this sample also provides an opportunity to study the ICM dynamical state and CC fraction using the X-ray morphology parameters presented in \S~\ref{ss:morph}, without any ICM selection bias. 
Due to the diversity in the recent merger histories of individual clusters, which is further complicated by projection, 
the morphological parameters should be treated with caution. However, they provide useful statistical tools to characterise trends of properties in large cluster samples (e.g. \citealt{2010A&A...514A..32B}). We use $w$ and $c_{\rm SB}$ to compare the fractions of relaxed ($w\leq0.01$) and CC ($c_{\rm SB}\geq0.075$) clusters among optical, SZ, and X-ray selected samples (Table~\ref{t:fra}). Both the fractions of relaxed and CC clusters change as: optical $<$ SZ $<$ X-ray. Moreover, the combination of asymmetry and CC indicators provide a more rigorous definition for relaxed clusters, e.g. the SPA criterion ($s > 0.87,\ p > -0.82,\ {\rm and}\ a > 1.00$; \citealt{2015MNRAS.449..199M}). Only 1 out of 55 clusters in our sample (2\%) is close to the SPA criterion, compared with 57/361 (16\%) for an X-ray selected sample \citep{2015MNRAS.449..199M}. Simulations also tend to find less relaxed clusters compared with an X-ray selected sample (e.g. \citealt{2010A&A...514A..32B}).

In Table~\ref{t:relation}, we further compare the amplitudes (fix the slope to the model prediction) of the whole (all) sample, as well as the relaxed (R) and disturbed (M+P) subsamples, with the model prediction including 25\% mass bias. We find that the amplitude of whole sample is close to the model prediction, while the relaxed subsample is higher and the disturbed subsample is lower than the model. This fact indicates that cluster with different dynamical state may have different level of mass bias.

\begin{table*}
\centering 
\caption{X-ray morphological parameters}
\tabcolsep=0.4cm
\begin{tabular}{lccccc} \hline \hline
Cluster	&	$w$			&	$c_{\rm SB}$			&	$s$			&	$p$			&	$a$			\\ \hline
A2142	&	0.007	$\pm$	0.001	&	0.075	$\pm$	0.001	&	0.788	$\pm$	0.014	&	-0.921	$\pm$	0.005	&	1.159	$\pm$	0.017	\\
J150M	&	0.011	$\pm$	0.005	&	0.054	$\pm$	0.010	&	0.843	$\pm$	0.179	&	-1.259	$\pm$	0.323	&	0.870	$\pm$	0.308	\\
J150E	&	0.025	$\pm$	0.008	&	0.050	$\pm$	0.015	&	0.772	$\pm$	0.196	&	-1.596	$\pm$	0.412	&	0.663	$\pm$	0.284	\\
J150W	&	0.026	$\pm$	0.011	&	0.031	$\pm$	0.014	&	0.475	$\pm$	0.218	&	-1.694	$\pm$	0.697	&	0.280	$\pm$	0.240	\\
A1689	&	0.002	$\pm$	0.001	&	0.116	$\pm$	0.002	&	1.166	$\pm$	0.053	&	-0.829	$\pm$	0.018	&	1.423	$\pm$	0.035	\\
A1443	&	0.037	$\pm$	0.010	&	0.025	$\pm$	0.003	&	0.650	$\pm$	0.164	&	-1.501	$\pm$	0.142	&	0.789	$\pm$	0.090	\\
A781	&	0.046	$\pm$	0.002	&	0.023	$\pm$	0.002	&	0.446	$\pm$	0.023	&	-1.881	$\pm$	0.825	&	0.941	$\pm$	0.032	\\
A781M	&	0.007	$\pm$	0.002	&	0.029	$\pm$	0.004	&	1.172	$\pm$	0.150	&	-1.772	$\pm$	0.349	&	1.203	$\pm$	0.220	\\
A781E	&	0.009	$\pm$	0.003	&	0.020	$\pm$	0.005	&	1.181	$\pm$	0.216	&	-1.770	$\pm$	0.492	&	1.381	$\pm$	0.326	\\
A781W	&	0.010	$\pm$	0.003	&	0.073	$\pm$	0.007	&	0.966	$\pm$	0.151	&	-1.158	$\pm$	0.071	&	1.030	$\pm$	0.180	\\
A1986	&	0.019	$\pm$	0.005	&	0.047	$\pm$	0.010	&				&				&				\\
A1882E	&	0.024	$\pm$	0.005	&	0.051	$\pm$	0.004	&	0.575	$\pm$	0.091	&	-1.843	$\pm$	0.540	&	0.995	$\pm$	0.221	\\
A1882W	&	0.020	$\pm$	0.005	&	0.051	$\pm$	0.010	&	0.552	$\pm$	0.238	&	-1.674	$\pm$	0.274	&	0.137	$\pm$	0.234	\\
A1882M	&	0.012	$\pm$	0.005	&	0.271	$\pm$	0.041	&	0.883	$\pm$	0.150	&	-0.895	$\pm$	0.036	&	0.819	$\pm$	0.209	\\
A1882N	&	0.034	$\pm$	0.011	&	0.058	$\pm$	0.016	&				&				&				\\
A1758N	&	0.026	$\pm$	0.019	&	0.016	$\pm$	0.005	&	0.517	$\pm$	0.240	&	-1.604	$\pm$	0.281	&	1.438	$\pm$	0.090	\\
A1758S	&	0.015	$\pm$	0.001	&	0.041	$\pm$	0.002	&	0.497	$\pm$	0.034	&	-1.376	$\pm$	0.402	&	0.971	$\pm$	0.053	\\
A1760	&	0.029	$\pm$	0.005	&	0.037	$\pm$	0.007	&				&				&				\\
A1622	&	0.036	$\pm$	0.001	&	0.046	$\pm$	0.001	&				&				&				\\
A750E	&	0.007	$\pm$	0.001	&	0.075	$\pm$	0.003	&	1.021	$\pm$	0.098	&	-1.086	$\pm$	0.022	&	1.298	$\pm$	0.072	\\
A750W	&	0.042	$\pm$	0.004	&	0.056	$\pm$	0.006	&	0.178	$\pm$	0.056	&	-1.172	$\pm$	0.208	&	0.827	$\pm$	0.274	\\
A1682	&	0.034	$\pm$	0.006	&	0.039	$\pm$	0.005	&	0.429	$\pm$	0.072	&	-1.414	$\pm$	0.332	&	0.444	$\pm$	0.140	\\
A1246	&	0.011	$\pm$	0.004	&	0.047	$\pm$	0.007	&	0.671	$\pm$	0.214	&	-1.265	$\pm$	0.111	&	0.745	$\pm$	0.291	\\
A1961	&	0.026	$\pm$	0.008	&	0.051	$\pm$	0.013	&	0.564	$\pm$	0.124	&	-1.496	$\pm$	0.798	&	0.624	$\pm$	0.231	\\
A2034	&	0.015	$\pm$	0.002	&	0.027	$\pm$	0.001	&	0.904	$\pm$	0.039	&	-1.316	$\pm$	0.081	&	0.911	$\pm$	0.035	\\
A655	&	0.018	$\pm$	0.006	&	0.047	$\pm$	0.005	&	0.466	$\pm$	0.144	&	-1.427	$\pm$	0.147	&	0.774	$\pm$	0.236	\\
A1914	&	0.010	$\pm$	0.001	&	0.061	$\pm$	0.002	&	0.660	$\pm$	0.034	&	-1.000	$\pm$	0.014	&	1.352	$\pm$	0.069	\\
Z5247	&	0.056	$\pm$	0.013	&	0.029	$\pm$	0.006	&	0.530	$\pm$	0.149	&	-1.767	$\pm$	0.673	&	1.247	$\pm$	0.319	\\
A657	&	0.011	$\pm$	0.003	&	0.114	$\pm$	0.007	&	0.571	$\pm$	0.104	&	-1.073	$\pm$	0.083	&	0.638	$\pm$	0.187	\\
J229	&	0.012	$\pm$	0.003	&	0.061	$\pm$	0.006	&	1.148	$\pm$	0.195	&	-1.607	$\pm$	0.517	&	1.100	$\pm$	0.289	\\
A2051	&	0.012	$\pm$	0.001	&	0.050	$\pm$	0.003	&	0.718	$\pm$	0.074	&	-1.349	$\pm$	0.212	&	1.001	$\pm$	0.114	\\
A2051S	&	0.009	$\pm$	0.005	&	0.105	$\pm$	0.019	&	0.888	$\pm$	0.291	&	-1.353	$\pm$	0.264	&	1.084	$\pm$	0.455	\\
A2051N	&	0.015	$\pm$	0.005	&	0.069	$\pm$	0.009	&	0.480	$\pm$	0.138	&	-2.324	$\pm$	1.295	&	0.757	$\pm$	0.374	\\
A2050	&	0.011	$\pm$	0.001	&	0.052	$\pm$	0.002	&	0.777	$\pm$	0.050	&	-1.230	$\pm$	0.072	&	1.335	$\pm$	0.126	\\
A1423	&	0.007	$\pm$	0.004	&	0.101	$\pm$	0.008	&	1.005	$\pm$	0.111	&	-1.003	$\pm$	0.036	&	1.034	$\pm$	0.122	\\
A801	&	0.014	$\pm$	0.004	&	0.082	$\pm$	0.009	&	0.772	$\pm$	0.153	&	-1.141	$\pm$	0.066	&	0.863	$\pm$	0.186	\\
A773	&	0.007	$\pm$	0.002	&	0.045	$\pm$	0.003	&	1.147	$\pm$	0.115	&	-1.185	$\pm$	0.052	&	1.264	$\pm$	0.082	\\
A1576	&	0.017	$\pm$	0.004	&	0.048	$\pm$	0.005	&	0.805	$\pm$	0.097	&	-1.138	$\pm$	0.051	&	1.021	$\pm$	0.081	\\
A2631	&	0.030	$\pm$	0.004	&	0.033	$\pm$	0.003	&	0.636	$\pm$	0.083	&	-1.282	$\pm$	0.123	&	0.988	$\pm$	0.077	\\
A1703	&	0.014	$\pm$	0.003	&	0.047	$\pm$	0.003	&	0.830	$\pm$	0.166	&	-1.130	$\pm$	0.046	&	1.396	$\pm$	0.080	\\
A2219	&	0.007	$\pm$	0.001	&	0.027	$\pm$	0.001	&	1.008	$\pm$	0.073	&	-1.159	$\pm$	0.033	&	1.442	$\pm$	0.051	\\
A1319M	&	0.013	$\pm$	0.006	&	0.070	$\pm$	0.022	&				&				&				\\
A1319NW	&	0.022	$\pm$	0.009	&	0.089	$\pm$	0.018	&				&				&				\\
A1560	&	0.065	$\pm$	0.007	&	0.060	$\pm$	0.007	&				&				&				\\
J175N	&	0.023	$\pm$	0.014	&	0.043	$\pm$	0.014	&				&				&				\\
J175S	&	0.017	$\pm$	0.006	&	0.051	$\pm$	0.015	&				&				&				\\
J249SW	&	0.014	$\pm$	0.004	&	0.039	$\pm$	0.006	&	0.751	$\pm$	0.125	&	-1.586	$\pm$	0.399	&	1.288	$\pm$	0.218	\\
J249NE	&	0.009	$\pm$	0.004	&	0.072	$\pm$	0.010	&	0.907	$\pm$	0.249	&	-1.476	$\pm$	0.566	&	0.792	$\pm$	0.360	\\
A1201	&	0.023	$\pm$	0.001	&	0.100	$\pm$	0.002	&	0.539	$\pm$	0.048	&	-1.034	$\pm$	0.022	&	0.696	$\pm$	0.119	\\
A2009	&	0.003	$\pm$	0.001	&	0.163	$\pm$	0.007	&	1.141	$\pm$	0.141	&	-0.894	$\pm$	0.026	&	1.381	$\pm$	0.088	\\
A2111	&	0.043	$\pm$	0.014	&	0.040	$\pm$	0.004	&	0.495	$\pm$	0.108	&	-1.479	$\pm$	0.567	&	0.792	$\pm$	0.051	\\
A815	&	0.027	$\pm$	0.012	&	0.020	$\pm$	0.005	&				&				&				\\
Z1450	&	0.009	$\pm$	0.003	&	0.036	$\pm$	0.009	&	0.878	$\pm$	0.151	&	-1.525	$\pm$	0.158	&	0.799	$\pm$	0.240	\\
A1765	&	0.005	$\pm$	0.002	&	0.065	$\pm$	0.009	&	0.908	$\pm$	0.228	&	-1.604	$\pm$	0.629	&	0.897	$\pm$	0.400	\\
A1902	&	0.008	$\pm$	0.003	&	0.148	$\pm$	0.009	&	0.796	$\pm$	0.135	&	-0.939	$\pm$	0.020	&	0.897	$\pm$	0.172	\\
\hline
\end{tabular}
\begin{tablenotes}
\item {\sl Note.} The centroid shift ($w$ within $R_{500}$, \citealt{1993ApJ...413..492M}), the surface brightness concentration ($c_{\rm SB}$ in 40-400 kpc range; \citealt{2008A&A...483...35S}), and the symmetry-peakiness-alignment (SPA; \citealt{2015MNRAS.449..199M}).
\end{tablenotes} 
\label{t:mor}
\end{table*}

\begin{figure*}
 	\begin{center}
\includegraphics[width=0.495\textwidth,keepaspectratio=true,clip=true]{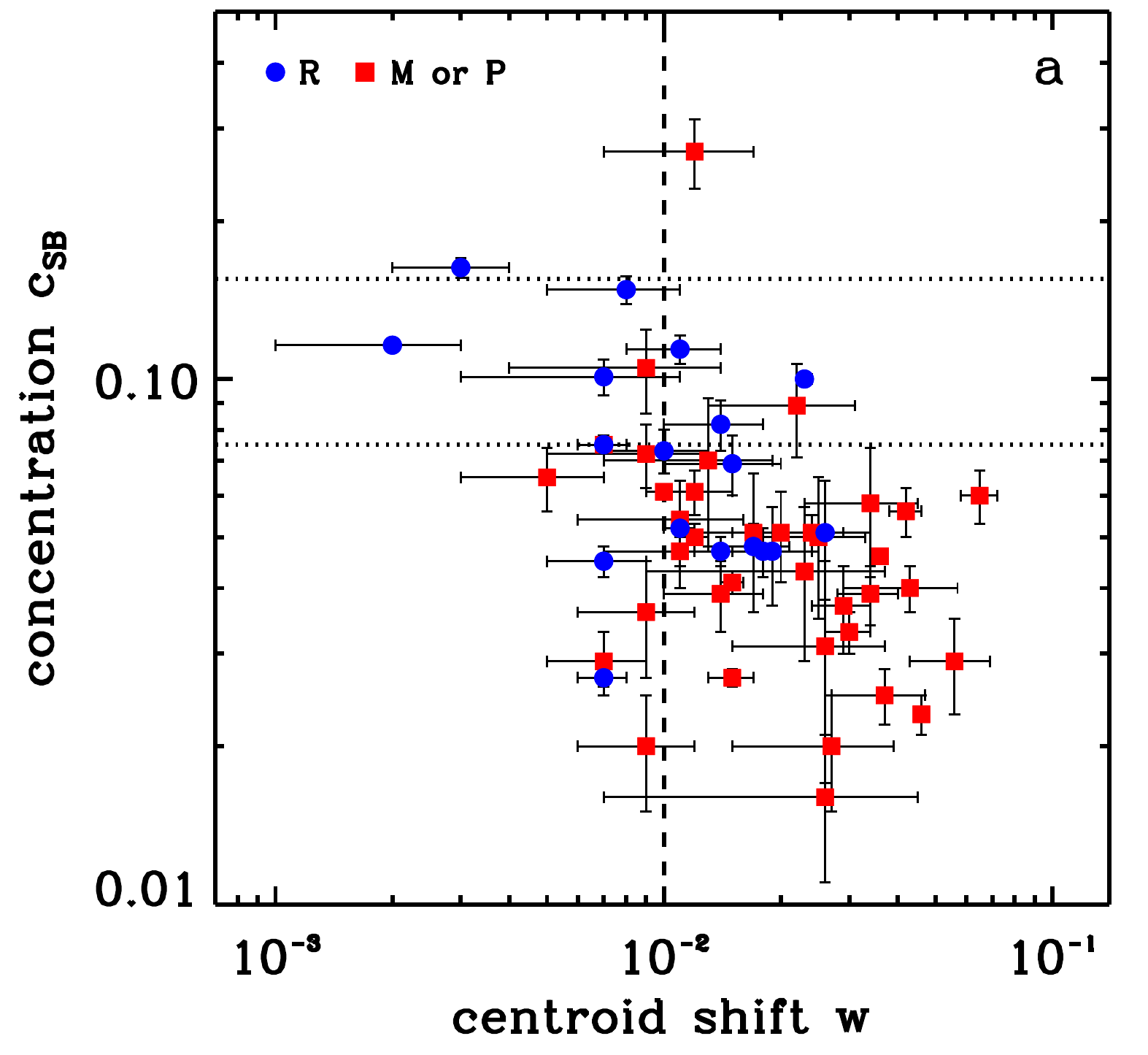} 	
\includegraphics[width=0.495\textwidth,keepaspectratio=true,clip=true]{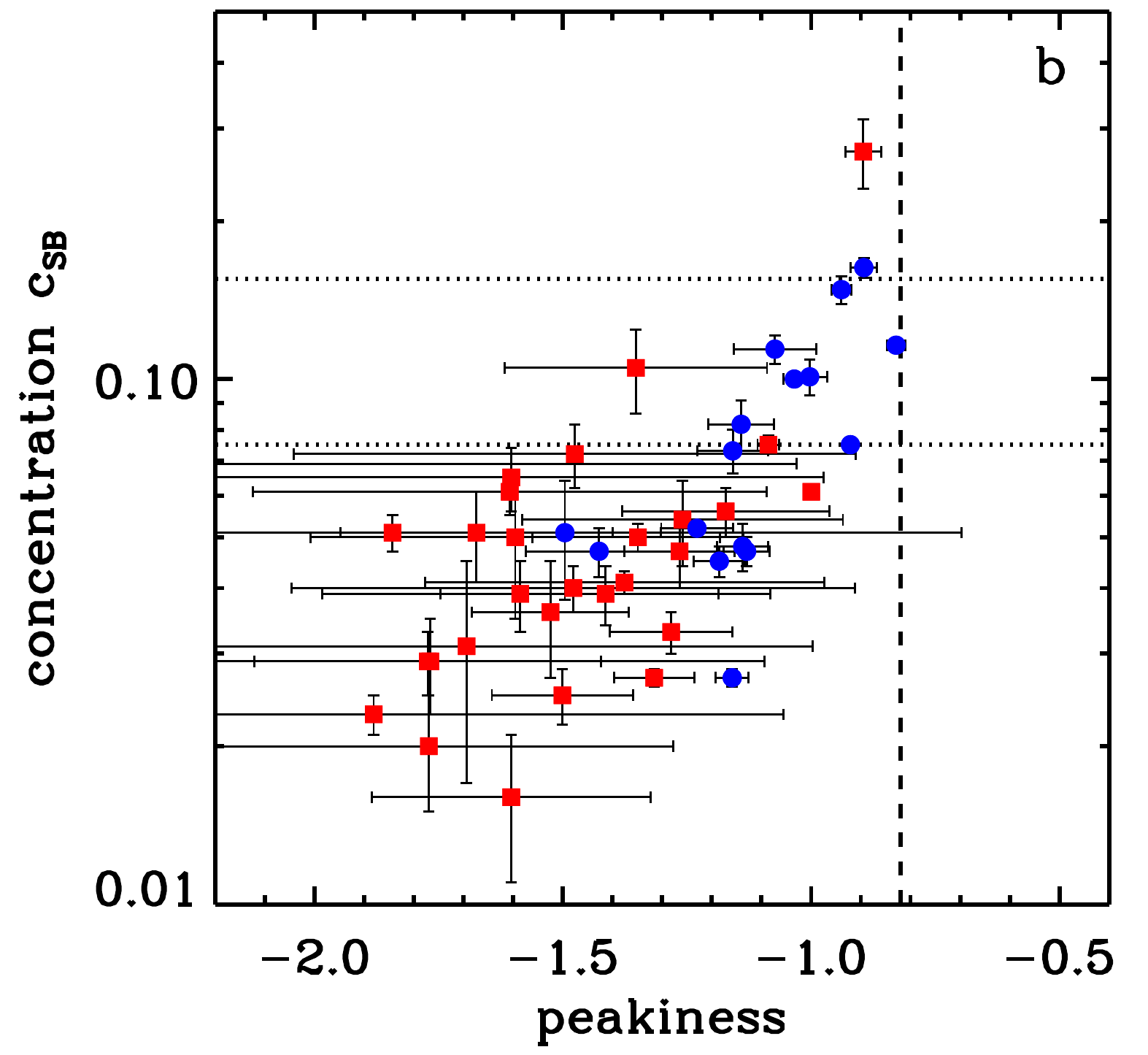}
\includegraphics[width=0.495\textwidth,keepaspectratio=true,clip=true]{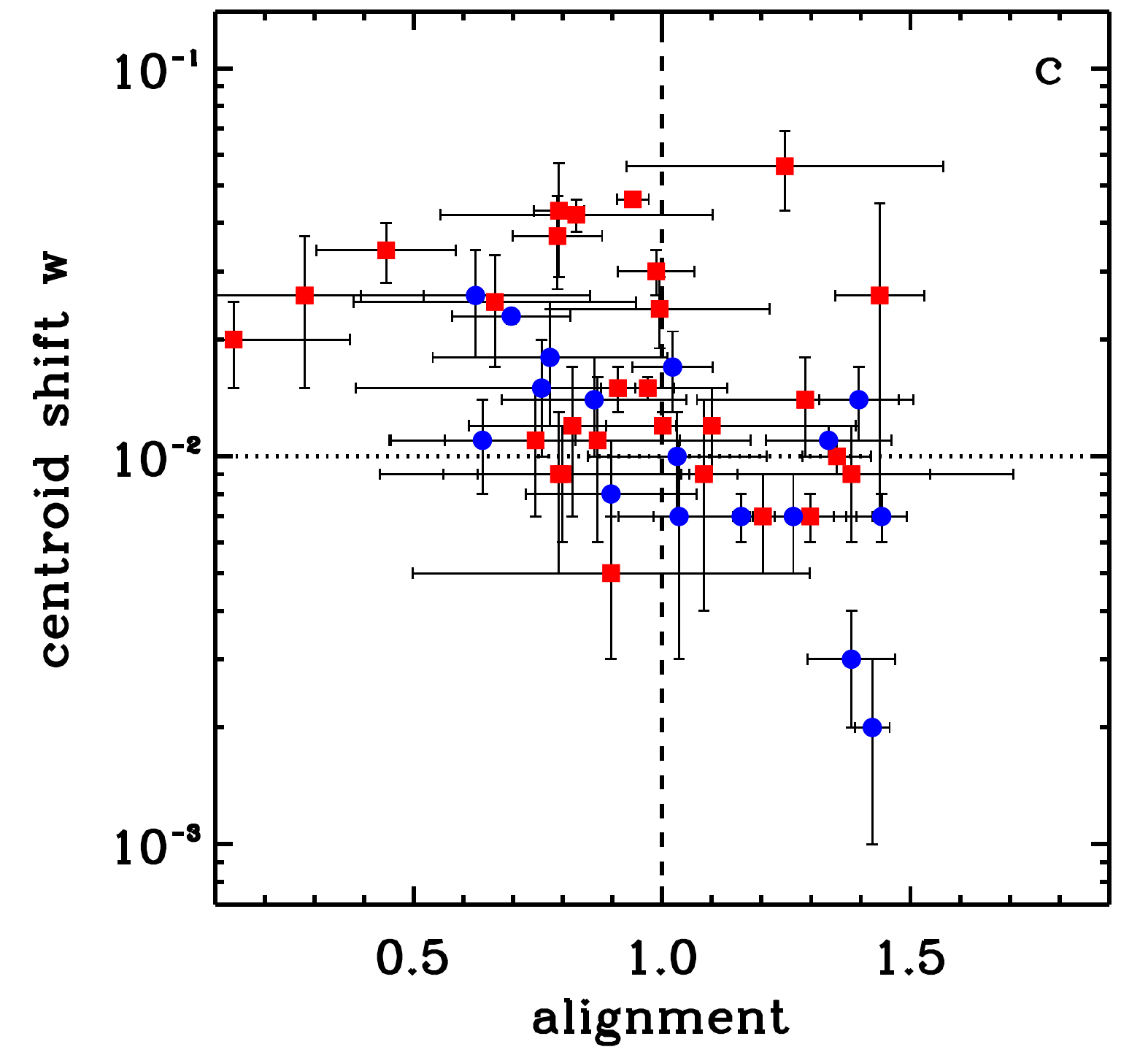}
\includegraphics[width=0.495\textwidth,keepaspectratio=true,clip=true]{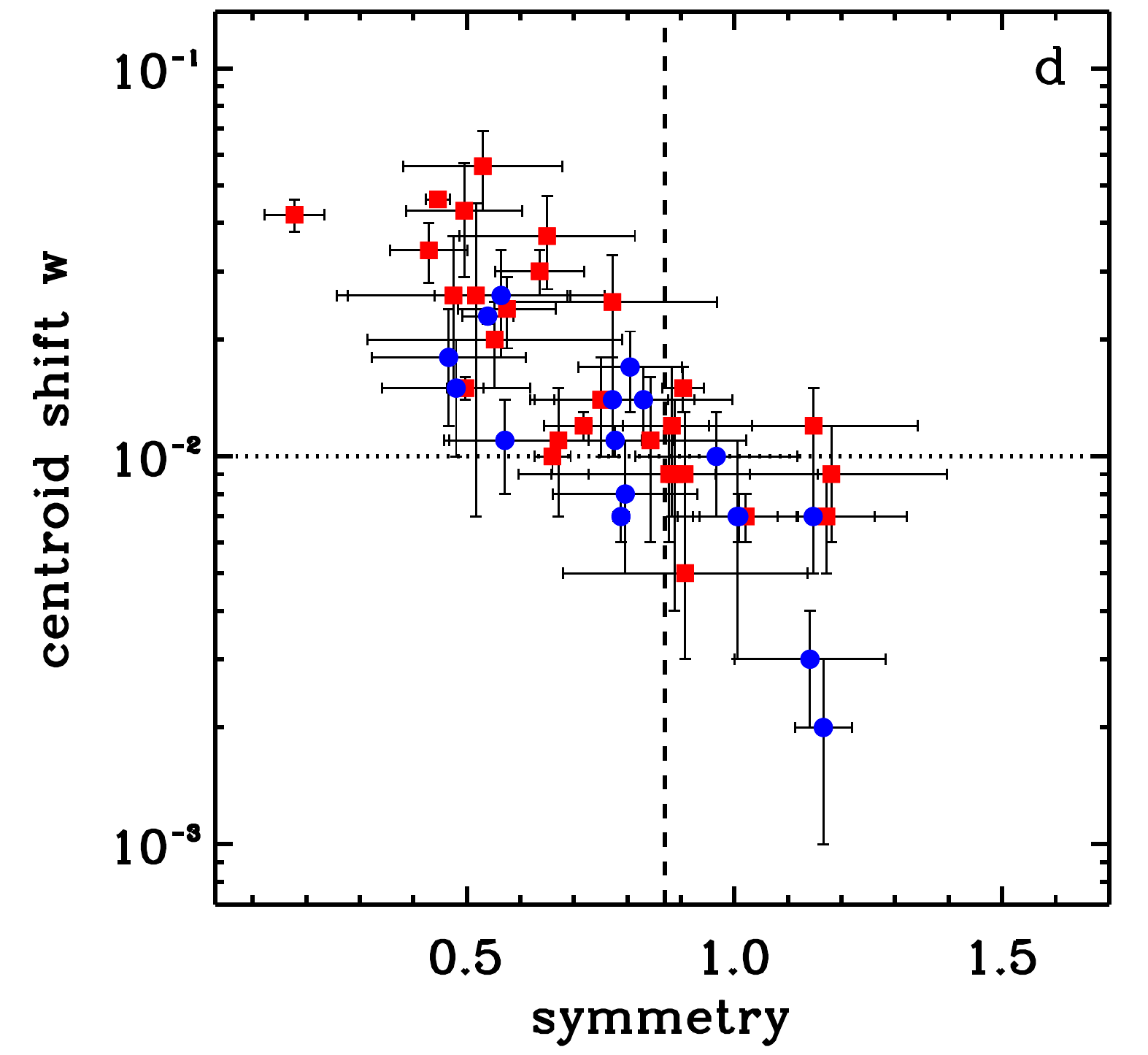}
 	\end{center}
	\caption{
The distribution of X-ray morphology parameters. Red squares are disturbed clusters with the same miscentering and projection problems as in Figs.~\ref{fig:L-N} and \ref{fig:Y-N}, while blue dots are relaxed clusters without miscentering and projection problems. The lines mark the classification threshold: $w \leq 0.01$ for relaxed clusters; $c_{\rm SB} \geq 0.075$ for moderate CC clusters and $c_{\rm SB} \geq 0.155$ for pronounced CC clusters; $s > 0.87,\ p > -0.82,\ {\rm and}\ a > 1.00$ for relaxed clusters. The CC indicators $c_{\rm SB}$ and $p$ are correlated with each other. The asymmetry indicators $w$, $s$, and $a$ are also correlated.
	}
	 \label{fig:morph}
\end{figure*}

\begin{table}
\caption{The fractions of relaxed and CC clusters}
\tabcolsep=0.22cm
\begin{tabular}{lccc} \hline \hline
	& optical	& SZ	&	X-ray \\	\hline
Relaxed ($w\leq0.01$) & 29\% (55$^a$) & 35\% (120$^b$) & 61\% (31$^c$) \\	
CC ($c_{\rm SB}\geq0.075$) & 22\% (55$^a$) & 36\% (164$^d$) & 61\% (100$^d$) \\	
\hline
\end{tabular}
\begin{tablenotes}
\item {\sl Note.}
$^a$This work from Table~\ref{t:mor}, the number in brackets is the total cluster number in the sample.  
$^b$\cite{2017ApJ...846...51L}.
$^c$\cite{2010A&A...514A..32B}.
$^d$\cite{2017ApJ...843...76A}.
This optically selected sample shows lower fractions of relaxed and CC clusters compared with SZ and X-ray selected samples.
\end{tablenotes} 
\label{t:fra}
\end{table}

\section{Conclusions}
\label{s:sum}

Great progress on galaxy clusters has been made in the last decade with X-ray, SZ and optical surveys.
The scaling relations between X-ray and SZ are consistent with each other for X-ray, SZ and even optically selected clusters, because both the X-ray emission and SZ signal are from the same ICM.
However, discrepancies emerge when we compare the ICM scaling relation based on X-ray and SZ data with the galaxy scaling relation based on optical data. 

In order to study the discrepancies, we directly compare the optical and X-ray scaling relations for a complete sample of 38 richest \max\ clusters. We list these factors contributing to the discrepancies: (1) miscentering, $\sim10\%$; (2) projection, $\sim10\%$; (3) contamination of low mass systems of optical selection, $\sim10\% - 15\%$; (4) hydrostatic mass bias, $\sim25\%$; (5) weak-lensing mass bias, $\sim10\%$; and (6) covariance bias, $\sim10\%$. These biases mix in some cases and can compensate with each other, but the dominant one is the mass bias. More studies are required to constrain these biases better. 

This sample offers insights into maxBCG mass selection $P(M|N)$ from the distribution of $P(X|N)$, where X is the low scatter X-ray mass proxy like $T_X$, $L_X$, and $Y_X$. In such top richest ($\sim$0.3\% of whole \max\ sample) range, a significant amount of blended halos with boosted richness mixes with the clean halos. However, all the blended halos are resolved by the X-ray imaging, as they follow the X-ray scaling relations calibrated from X-ray selected halos. The fraction of blended system could be reduce when including more information from optical data, e.g. whether hosts a dominant BCG or not.

This optically selected sample also provides an unbiased perspective to the ICM properties.
We find a rising fraction of relaxed or CC clusters from optical ($\sim26\%$), to SZ ($\sim36\%$), and to X-ray ($\sim61\%$) selected samples.
Moreover, the disturbed subsample shows higher mass bias than the relaxed subsample. 

Optical surveys and algorithms are very successful and efficient for finding clusters, 
more works need to be done to better understand the halo selection properties of optical catalog and the mass bias,
with the aid of simulations and mock catalog, before we implement the resulting scaling relations to study cosmology.

\section*{ACKNOWLEDGEMENTS}
We thank the referee, Gus Evrard, for important comments and suggestions. We thank Andrea Morandi for his early work on the \cha\ data.
Support for this work was provided by the National Aeronautics and Space Administration grants NNX16AH32G and NNX16AH26G. Support for this work was also provided by the National Aeronautics and Space Administration through \cha\ Award Number GO4-15119B and GO4-15115X issued by the \cha\ X-ray Center, which is operated by the Smithsonian Astrophysical Observatory for and on behalf of the National Aeronautics Space Administration under contract NAS8-03060. ER is supported by DOE grant DE-SC0015975 and by the Sloan Foundation, grant FG-2016-6443. NS acknowledges support from NSF grant 1513618. DN acknowledges support from NSF grant AST-1412768. This research has made use of data and/or software provided by the High Energy Astrophysics Science Archive Research Center (HEASARC), which is a service of the Astrophysics Science Division at NASA/GSFC and the High Energy Astrophysics Division of the Smithsonian Astrophysical Observatory.

\twocolumn
\appendix
\section{Multivariate scaling relations}
\label{app:msr}
Due to the slope or asymmetry of the mass function and the mass variance, the mean of the correlated multivariate scaling relation is not equal to the naive `plug in' expectation (e.g. \citealt{2011ARA&A..49..409A}; \citealt{2014MNRAS.441.3562E}).
\cite{2014MNRAS.441.3562E} model is aimed at describing the underlying massive halo population. The $P(N_{\rm gal}|M)$ kernel can be separated into intrinsic halo scatter, $P(N_{\rm int}|M)$, and a LOS/noise component, $P(N_{\rm gal}|N_{\rm int}, M)$.  Asymmetry in the latter PDF is expected from non-linear clustering.  This generates a corresponding skewness in the mass selection, $P(M|N_{\rm gal})$. The selection of the 38 (0.3\% of whole sample) richest \max\ clusters will exacerbate the skewness relative to a lower-richness selection. However, as a first order approximation and largely for illustrative purposes here, we assume the conditional probability is a log-normal distribution.

Following \cite{2014MNRAS.438...62R}, we define $m={\rm ln}(M/M_0)$ and the mean distribution of log-observable, e.g. temperature $t={\rm ln}(T_X/T_{X, 0})$, is
\begin{equation}
\avg{t|m} = a_{t|m} + s_{t|m} \, m,
\end{equation}
where $a_{t|m}$ is the amplitude and $s_{t|m}$ is the slope of scaling relation.
Using Bayes Theorem, we can convert the $T_X-M$ relation to $M-T_X$ relation with the Eq.~A5 of \cite{2014MNRAS.438...62R},
\begin{equation}
\avg{m|t} =[\frac{t-a_{t|m}}{s_{t|m}}]-\beta\sigma_{m|t}^2,
\end{equation}
The first term in square brackets is the naive expected mean from Eq. A1, while the second term is the Eddington bias correction. $\beta$ is the slope of the halo mass function (${\rm d}n/{\rm d\ ln}\ M \propto M^{-\beta}$) and $\sigma_{m|t}$ is the scatter in $m$ at fixed $t$. Assuming the  \cite{2008ApJ...688..709T} mass function and the {\em WMAP9}  cosmology \citep{2013ApJS..208...19H}, in the typical mass range of this sample ($0.6-13.3\times10^{14}M_{\odot}$), $\beta = 1.4-4.5$.

{\bf $\pmb{L_X-N}$ relation} is derived from the $L_X-M$ relation (\citealt{2009AA...498..361P}; $L_1-M_Y$ MB in their Table 2) and the $M-N$ relation (\citealt{2009ApJ...699..768R}; \citealt{2017MNRAS.466.3103S}) using Eq.~A13 of \cite{2014MNRAS.438...62R},
\begin{equation}
\avg{l|n} = a_{l|m} + s_{l|m}(\avg{m|n}+r\beta\sigma_{m|l}\sigma_{m|n}).
\end{equation}
Thus the amplitude and slope of $L_X-N$ relation are
\begin{equation}
a_{l|n} = [a_{l|m} + s_{l|m}a_{m|n}]+r_{l,n|m}\beta s_{l|m}\sigma_{m|l}\sigma_{m|n},
\end{equation}
\begin{equation}
s_{l|n}=s_{l|m}s_{m|n}.
\end{equation}
The scatter of $L_X-N$ relation is estimated using Eq.~A14 of \citep{2014MNRAS.438...62R}:
\begin{equation}
\sigma_{l|n}^2=s_{l|m}^2(\sigma_{m|n}^2+\sigma_{m|l}^2-2r_{l,n|m}\sigma_{m|n}\sigma_{m|l}).
\end{equation}
On the amplitude, the term in square brackets is the naive `plug in' value. The other term is the covariance bias, $r_{l,n|m}$ is the correlation coefficient between $l$ and $n$ at fixed $m$. We note the binned masses of \cite{2009ApJ...699..768R} and \cite{2017MNRAS.466.3103S} measure the ${\rm ln}\avg{M|N}$ rather than $\avg{{\rm ln\ }M|N}$, and ${\rm ln}\avg{M|N}= \avg{{\rm ln\ }M|N}+\frac{1}{2}\sigma_{m|n}^2$ assuming a log-normal distribution, thus the amplitude is related as $\tilde{a}_{m|n}=a_{m|n}+\frac{1}{2}\sigma_{m|n}^2$. We quote their $\tilde{a}_{m|n}$ and get $a_{m|n}$ by subtracting $\frac{1}{2}\sigma_{m|n}^2$.
To get an order of magnitude estimate for the bias term, we set $\beta=2.8$.
To the first order, the hot gas - galaxy correlation coefficient is zero, which results in no bias.
However, recent studies suggest negative hot gas - galaxy correlation and we also take $r_{l,n|m}=-0.5$ (\citealt{2017arXiv171104922F} \& private communication with Gus Evrard) to examine its impact. We also apply $\sigma_{m|l}=0.28$ \citep{2012MNRAS.426.2046A} and $\sigma_{m|n}=0.45$ \citep{2009ApJ...699..768R} or $\sigma_{m|\lambda}=0.25$ \citep{2014ApJ...783...80R}, which yields a correction term $\sim -0.34$ ($\sim -0.17$ for $\lambda$), corresponds to a $28\%$ ($15\%$) down offset for the amplitude of $L_X$. If we set $\beta$ in the range of $1.4-4.5$, the down offset is $15\%-42\%$ ($8\%-23\%$ for $\lambda$) instead. Fig.~\ref{fig:L-N-app} compares the difference between the `plug in' method and the bias corrected method with different $\beta$ values for the \max\ sample.
In order to compare with previous publications, especially the ones from \rosat, we also present the $L_X $ in 0.1-2.4 keV in Fig.~\ref{fig:Ls-N-app}. The $L_{0.1-2.4\ \rm keV}$ is converted from $L_{X,bol}$ using an {\tt apec} model with metallicity fixed to $0.3Z_{\odot}$. The black line is the predicted $L_X-N$ relation from the $L_X-M$ relation (\citealt{2009AA...498..361P}; $L[0.1-2.4]-M_Y$ MB in their Table B2) and the $M-N$ relation (\citealt{2009ApJ...699..768R}) using Eqs.~A4 and A5. The purple line is from the preferred $L_X-N$ relation (\citealt{2014MNRAS.438...78R}; in their Table 4).

\begin{figure}
 	\begin{center}
\includegraphics[width=0.495\textwidth,keepaspectratio=true,clip=true]{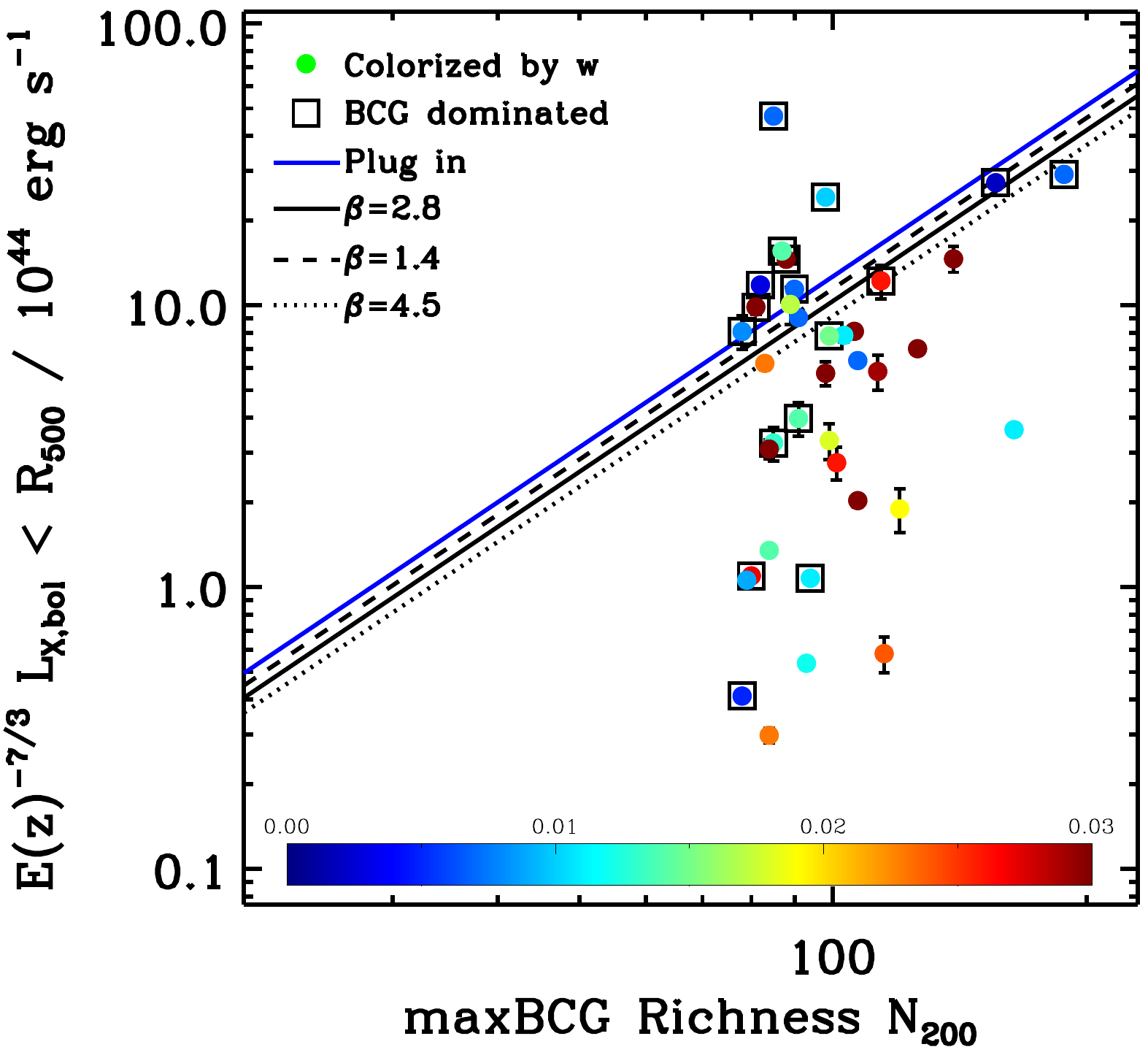}
 	\end{center}
	\caption{
		Comparison of the `plug in' $L_X-N_{200}$ relation (blue solid line) with the bias corrected ones. The `plug in' relation simply combines the X-ray $L_X-M$ relation (\citealt{2009AA...498..361P}) and the weak-lensing $M-N_{200}$ relation \citep{2009ApJ...699..768R}. The black solid, dashed, and dotted lines show bias corrected relation for $\beta$ of 2.8, 1.4, and 4.5, respectively. The dots are colorized by the centroid shift $w$. The black boxes represent a BCG dominated subsample with the \max\ and \red\ locating the same galaxy as the BCG (see Section~\ref{ss:improve}). The BCG dominated clusters tend to be relaxed clusters and follow more closely to the model prediction of solid line.
	}
	 \label{fig:L-N-app}
\end{figure}

{\bf $\pmb{T_X-N}$ relation} is derived from the $M-T_X$ relation (\citealt{2009ApJ...693.1142S}; Tier $1+2+$clusters in their Table 6) and the $M-N$ relation (\citealt{2009ApJ...699..768R}; \citealt{2017MNRAS.466.3103S}). From Eq.~A13 of \cite{2014MNRAS.438...62R}, we find
\begin{equation}
\avg{t|n} = a_{t|m} + s_{t|m}(\avg{m|n}+r_{t,n|m}\beta\sigma_{m|t}\sigma_{m|n})
\end{equation}
The $M-T_X$ relation  is related to the $T_X-M$ relation from Eq.~A2,
\begin{equation}
a_{t|m}=-\frac{a_{m|t}+\beta\sigma_{m|t}^2}{s_{m|t}},\ s_{t|m}=\frac{1}{s_{m|t}},
\end{equation}
substitute $a_{t|m}$ and $s_{t|m}$ into Eq.~A7, we get the amplitude and slope of $T_X-N$ relation as
\begin{equation}
a_{t|n} = [\frac{a_{m|n}-a_{m|t}}{s_{m|t}}]+\frac{\beta\sigma_{m|t}}{s_{m|t}}(r_{t,n|m}\sigma_{m|n}-\sigma_{m|t}),
\end{equation}
\begin{equation}
s_{t|n}=\frac{s_{m|n}}{s_{m|t}}.
\end{equation}
Similar to the scatter of $L_X-N$ relation, but substitute $s_{t|m}=1/{s_{m|t}}$,
\begin{equation}
\sigma_{t|n}^2=\frac{1}{s_{m|t}^2}(\sigma_{m|n}^2+\sigma_{m|t}^2-2r_{t,n|m}\sigma_{m|n}\sigma_{m|t}).
\end{equation}
We assume $r_{t,n|m}=-0.5$ (\citealt{2017arXiv171104922F} \& private communication with Gus Evrard) and $\sigma_{m|t}$=0.20 \citep{2006ApJ...650..128K} to estimate the bias, which yields a correction term of $\sim -0.13$, corresponding to a $12\%$ down offset for $n$ or $\lambda$.

{\bf $\pmb{Y_X-N}$ relation} is derived from the $M-Y_X$ relation (\citealt{2010AA...517A..92A}; their Eq.~2) and the $M-N$ relation (\citealt{2009ApJ...699..768R}; \citealt{2017MNRAS.466.3103S}) as:
\begin{equation}
a_{x|n} = [\frac{a_{m|n}-a_{m|x}}{s_{m|x}}]+\frac{\beta\sigma_{m|x}}{s_{m|x}}(r_{x,n|m}\sigma_{m|n}-\sigma_{m|x}),
\end{equation}
\begin{equation}
s_{x|n}=\frac{s_{m|n}}{s_{m|x}}.
\end{equation}
\begin{equation}
\sigma_{x|n}^2=\frac{1}{s_{m|x}^2}(\sigma_{m|n}^2+\sigma_{m|x}^2-2r_{x,n|m}\sigma_{m|n}\sigma_{m|x}).
\end{equation}
Assuming $r_{x,n|m}=-0.5$ (\citealt{2017arXiv171104922F} \& private communication with Gus Evrard) and $\sigma_{m|x}$=0.087 \citep{2010AA...517A..92A}, we derive a correction term of $\sim -0.12$, corresponding to a $11\%$ down offset for $n$ or $\lambda$.

{\bf $\pmb{Y_X-N}$ relation} is also from the $Y_{SZ}-Y_X$ relation (\citealt{2014MNRAS.438...62R}; M10 data set in their Table 1) and the $Y_{SZ}-N$ relation (\citealt{2014MNRAS.438...78R}; in their Table 4) as:
\begin{equation}
a_{x|n} = [\frac{a_{sz|n}-a_{sz|x}}{s_{sz|x}}]+\frac{\beta\sigma_{sz|x}}{s_{sz|x}}(r_{x,n|sz}\sigma_{sz|n}-\sigma_{sz|x}),
\end{equation}
\begin{equation}
s_{x|n}=\frac{s_{sz|n}}{s_{sz|x}},
\end{equation}
We assume $r_{x,n|sz}=0$. The scatter of $Y_X-N$ relation is estimated using Eq.~A14 of \citep{2014MNRAS.438...62R}:
\begin{equation}
\sigma_{x|n}^2=s_{x|m}^2[\sigma_{m|n}^2+\sigma_{m|x}^2-2r_{x,n|m}\sigma_{m|n}\sigma_{m|x}].
\end{equation}
The resultant $\sigma_{x|n}=0.69$ is very close to the $\sigma_{sz|n}=0.70$ (\citealt{2014MNRAS.438...78R}; in their Table 4).

The binned $Y_X$ presented as green bowties in Fig.~\ref{fig:Y-N} is $\avg{{\rm ln\ }Y_X|N}$, which is evaluated from ${\rm ln}\avg{Y_X|N}-\frac{1}{2}\sigma_{x|n}^2$ assuming a log-normal distribution. When we present the expected $Y_X$ inferred from \pla\ stacking $Y_{SZ}$, we include additional corrections listed as below.
(1) From stacking ${\rm ln}\avg{Y_{SZ}|N}$ (\citealt{2011A&A...536A..12P}; in their Table 1) to $\avg{{\rm ln\ }Y_{SZ}|N}$, $-\frac{1}{2}\sigma_{sz|n}^2=-22\%$. (2) Aperture-induced correction due to covariance, as the \pla\ $Y_{SZ}$ is measured within $R_{500}$, which is based on \max\ $N_{200}$. From Eq.~A4 and replacing $L_X$ with $Y_{SZ}$, the amplitude is over-biased with a factor of $r_{sz,n|m}\beta s_{sz|m}\sigma_{m|sz}\sigma_{m|n}$ with $r_{sz,n|m}=0.47$ \citep{2012MNRAS.426.2046A}, which is different from the case of $Y_{SZ}$ measured from $R_{500}$ independent of $N_{200}$. In that case, $r_{sz,n|m}$ should be 0 or even negative, suggested by a negative hot gas - galaxy correlation when we derive the predicted multivariate scaling relation. The amplitude bias of $r_{sz,n|m}\beta s_{sz|m}\sigma_{m|sz}\sigma_{m|n}$ is further divided by a factor of 2, as \pla\ measurements are template-amplitude fits rather than cylindrically integrated $Y_{SZ}$ measurements and the inner radii weight more than the cylindrical integration. The final correction is $-5\%$ \citep{2014MNRAS.438...78R}. (3) Miscentering correction at a level of $10\%$ in the richness range we present \citep{2012ApJ...757....1B}. (4) Eddington bias correction based on Eq.~A2, as we convert $Y_{SZ}$ to $Y_{X}$ using $Y_{SZ}-Y_X$ relation (\citealt{2014MNRAS.438...62R}; M10 data set in their Table 1; $\sigma_{x|sz}=\sigma_{sz|x}/s_{sz|x}=0.15$) with a correction of $-\beta\sigma_{x|sz}^2=-6\%$.
(5) Aperture-induced correction due to hydrostatic or weak-lensing mass bias is ignored, because it is typically small relative with the mass bias itself.

In summary, we list all the cited and derived relations in Table~\ref{t:relation}.

\begin{figure}
 	\begin{center}
\includegraphics[width=0.495\textwidth,keepaspectratio=true,clip=true]{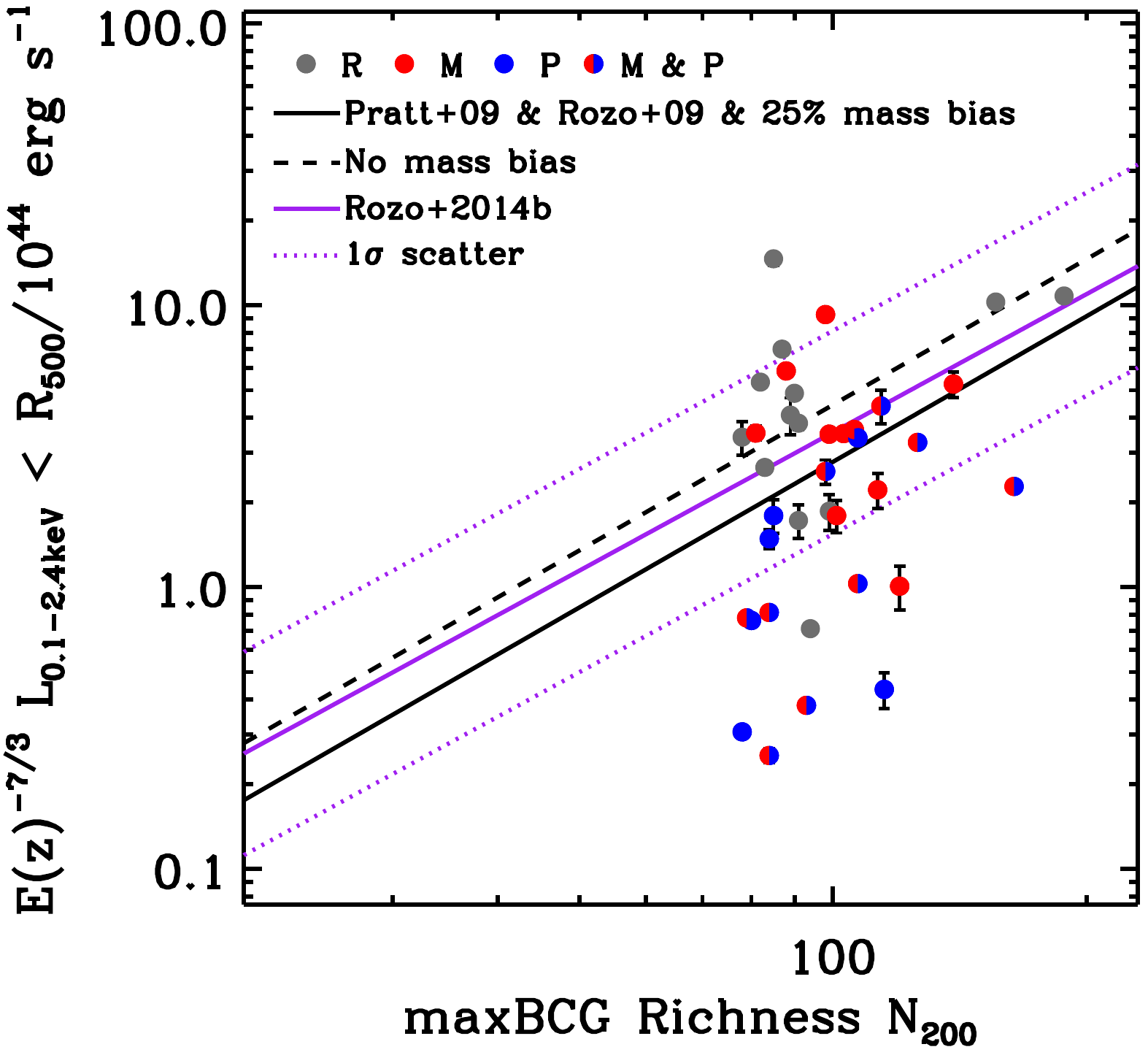}
 	\end{center}
	\caption{
The same plot as Fig.~\ref{fig:L-N}, but for soft band 0.1-2.4 keV luminosity. The purple solid line with $1\sigma$ dashed lines are from \citet{2014MNRAS.438...78R}.
	}
	 \label{fig:Ls-N-app}
\end{figure}

\begin{table*}
\caption{Scaling relation}
\tabcolsep=0.4cm
\begin{tabular}{lcccc} \hline \hline
Relation & Amplitude & Slope & Scatter & Sample\\
\hline
\multicolumn{5}{|c|}{$M-X$ relation from literature}\\
$L_X-M_{500}$ & $a_{l|m}=-0.68\pm0.08$ & $s_{l|m}=1.90\pm0.11$ & $\sigma_{m|l}^a=0.28$ & \cite{2009AA...498..361P}\\
$M_{500}-T_X$ & $a_{m|t}=-1.34\pm0.06$ & $s_{m|t}=1.65\pm0.04$ & $\sigma_{m|t}^b=0.20$ & \cite{2009ApJ...693.1142S}\\
$M_{500}-Y_X$ & $a_{m|x}=0.84\pm0.02$ & $s_{m|x}=0.56\pm0.02$ & $\sigma_{m|x}=0.09$ & \cite{2010AA...517A..92A}\\
\hline
\multicolumn{5}{|c|}{$M-N$ relation from literature}\\
$M_{500}^c-N_{200}$ & $a_{m|n}=0.86\pm0.12$ & $s_{m|n}=1.06\pm0.11$ & $\sigma_{m|n}=0.45\pm0.19$ & \cite{2009ApJ...699..768R}\\
$M_{500}^d-\lambda$ & $a_{m|n}=0.66\pm0.07$ & $s_{m|n}=1.30\pm0.10$ & $\sigma_{m|n}=0.25\pm0.05$ & \cite{2017MNRAS.466.3103S}\\
\hline
\multicolumn{5}{|c|}{$X-N$ relation from model}\\
$T_X-N_{200}$ & $a_{t|n}=1.02\pm0.09$ & $s_{t|n}=0.64\pm0.07$ & $\sigma_{t|n}=0.35\pm0.11$ & \max\ (Fig.~\ref{fig:T-N} left)\\
$T_X-\lambda$ & $a_{t|n}=0.93\pm0.06$ & $s_{t|n}=0.79\pm0.06$ & $\sigma_{t|n}=0.24\pm0.03$ & \red\ (Fig.~\ref{fig:T-N} right)\\
$L_X-N_{200}$ & $a_{l|n}=0.08\pm0.28$ & $s_{l|n}=2.01\pm0.24$ & $\sigma_{l|n}=1.22\pm0.33$ & \max\ (Fig.~\ref{fig:L-N} left)\\
$L_X-\lambda$ & $a_{l|n}=-0.16\pm0.16$ & $s_{l|n}=2.47\pm0.24$ & $\sigma_{l|n}=0.87\pm0.10$ & \red\ (Fig.~\ref{fig:L-N} right)\\
$Y_X-N_{200}$ & $a_{x|n}=-0.61\pm0.22$ & $s_{x|n}=1.89\pm0.21$ & $\sigma_{x|n}=0.90\pm0.33$ & \max\ (Fig.~\ref{fig:Y-N} left)\\
$Y_X-\lambda$ & $a_{x|n}=-0.93\pm0.13$ & $s_{x|n}=2.32\pm0.20$ & $\sigma_{x|n}=0.54\pm0.09$ & \red\ (Fig.~\ref{fig:Y-N} right)\\
\hline
\multicolumn{5}{|c|}{$X-N$ relation from data}\\
$T_X-N_{200}$ & $a_{t|n}=1.05\pm0.16$ & $s_{t|n}=0.64\ ({\rm fixed})$ & $\sigma_{t|n}=0.51\pm0.03$ & \max\ (all)\\
$T_X-N_{200}$ & $a_{t|n}=1.31\pm0.28$ & $s_{t|n}=0.64\ ({\rm fixed})$ & $\sigma_{t|n}=0.46\pm0.10$ & \max\ (R)\\
$T_X-N_{200}$ & $a_{t|n}=0.92\pm0.28$ & $s_{t|n}=0.64\ ({\rm fixed})$ & $\sigma_{t|n}=0.55\pm0.08$ & \max\ (M+P)\\
$L_X-N_{200}$ & $a_{l|n}=0.01\pm0.16$ & $s_{l|n}=2.01\ ({\rm fixed})$ & $\sigma_{l|n}=1.18\pm0.02$ & \max\ (all)\\
$L_X-N_{200}$ & $a_{l|n}=0.73\pm0.28$ & $s_{l|n}=2.01\ ({\rm fixed})$ & $\sigma_{l|n}=0.95\pm0.05$ & \max\ (R)\\
$L_X-N_{200}$ & $a_{l|n}=-0.37\pm0.20$ & $s_{l|n}=2.01\ ({\rm fixed})$ & $\sigma_{l|n}=1.13\pm0.02$ & \max\ (M+P)\\
$Y_X-N_{200}$ & $a_{x|n}=-0.56\pm0.17$ & $s_{x|n}=1.89\ ({\rm fixed})$ & $\sigma_{x|n}=1.13\pm0.02$ & \max\ (all)\\
$Y_X-N_{200}$ & $a_{x|n}=-0.10\pm0.28$ & $s_{x|n}=1.89\ ({\rm fixed})$ & $\sigma_{x|n}=0.98\pm0.05$ & \max\ (R)\\
$Y_X-N_{200}$ & $a_{x|n}=-0.85\pm0.22$ & $s_{x|n}=1.89\ ({\rm fixed})$ & $\sigma_{x|n}=1.14\pm0.03$ & \max\ (M+P)\\
\hline
\end{tabular}
\begin{tablenotes}
\item {\sl Note.} 
The scaling relation takes the form $\avg{{\rm ln}\ \psi}=a+s {\rm ln}\ (\chi/\chi_{0})$, they are evaluated at $z=0.23$, the median redshift of the \max\ cluster sample. The units are $10^{14} M_{\odot}$ for $M_{500}$, $10^{44}$ ergs s$^{-1}$ for $L_X$, keV for $T_{X}$, and $10^{14} M_{\odot}$ keV for $Y_X$, all the amplitude of $M-X$ relations are transfer to the values at unit pivot, while the pivot value of $M-N$ and $X-N$ relations is at richness = 40. A mass bias of $b = 0.25$ is included when derive th $X-N$ relation, through adding the amplitude of $M-N$ relation with a value of ln $(1-b)$.
$^a$ $\sigma_{m|l}=0.28$ is from \citet{2012MNRAS.426.2046A}.
$^b$ $\sigma_{m|t}=0.20$ is from \citet{2006ApJ...650..128K}.
$^c$ The binned masses of \citet{2009ApJ...699..768R} and \citet{2017MNRAS.466.3103S} measure the ${\rm ln}\avg{M|N}$ rather than $\avg{{\rm ln\ }M|N}$, and $\tilde{a}_{m|n}=a_{m|n}+\frac{1}{2}\sigma_{m|n}^2$. We quote their $\tilde{a}_{m|n}$ and get $a_{m|n}$ by subtracting $\frac{1}{2}\sigma_{m|n}^2$. 
$^d$ \citet{2017MNRAS.466.3103S} measure $M_{200m}-\lambda$ relation, we convert $M_{200m}$ to $M_{500c}$ with a typical ratio of 1.67.
We note that the $X-N$ amplitude of `all' sample is close to the model prediction with 25\% mass bias, while the `R' sample is higher and the `M+P' sample is lower than the model. This may indicate that the dynamical state of cluster affects the level of mass bias.  
\end{tablenotes} 
\label{t:relation}
\end{table*}

\section{\cha\ and \xmm\ cross-calibration}
\label{app:cc}
The X-ray data of our sample are from \cha\ and \xmm. There are some cross-calibration issues between these two instruments reported by the International Astronomical Consortium for High Energy Calibration (IACHEC, e.g. \citealt{2015A&A...575A..30S}). 
We use six clusters in our sample with both the \cha\ and \xmm\ data and spanning a wide temperature range to do the in-house cross-calibration. 
We first measure the X-ray properties such as $T_X$, $L_X$ (bolometric), and $Y_X$ individually from six clusters and independently from \cha\ and \xmm.
All the X-ray properties are derived with the same procedures as detailed in Section~\ref{s:obs}. Note that $R_{500}$ can be different for the \cha\ data and the \xmm\ data because of the different temperatures.

Fig.~\ref{fig:cross-T} compares the temperatures from \cha\ and \xmm, with \cha\ temperatures systematically higher than \xmm's. We then fit the relation with a power-law function at $T_X$ = 2 - 10.5 keV as

\begin{equation}
	\frac{kT_{XMM}}{\rm 1\ keV} = a_{T} (\frac{kT_{Ch}}{\rm 1\ keV})^{s_{T}},\ a_{T}=1.15\pm0.09,\ s_{T}=0.83\pm0.04.
\end{equation}
Similarly, we have the power-law relations on $L_{X}$ and $Y_{X}$:
\small
\begin{equation}
	\frac{L_{XMM}}{\rm 10^{44}\ erg\ s^{-1}} = a_{L} (\frac{L_{Ch}}{\rm 10^{44}\ erg\ s^{-1}})^{s_{L}},\ a_{L}=0.90\pm0.01,\ s_{L}=0.98\pm0.01.
\end{equation}
\begin{equation}
	\frac{Y_{XMM}}{10^{14}\ M_{\odot}\ \rm keV} = a_{Y} (\frac{Y_{Ch}}{10^{14}\ M_{\odot}\ \rm keV})^{s_{Y}},\ a_{Y}=0.98\pm0.08,\ s_{Y}=0.84\pm0.04.
\end{equation}
\normalsize

\begin{figure}
 	\begin{center}
\includegraphics[width=0.495\textwidth,keepaspectratio=true,clip=true]{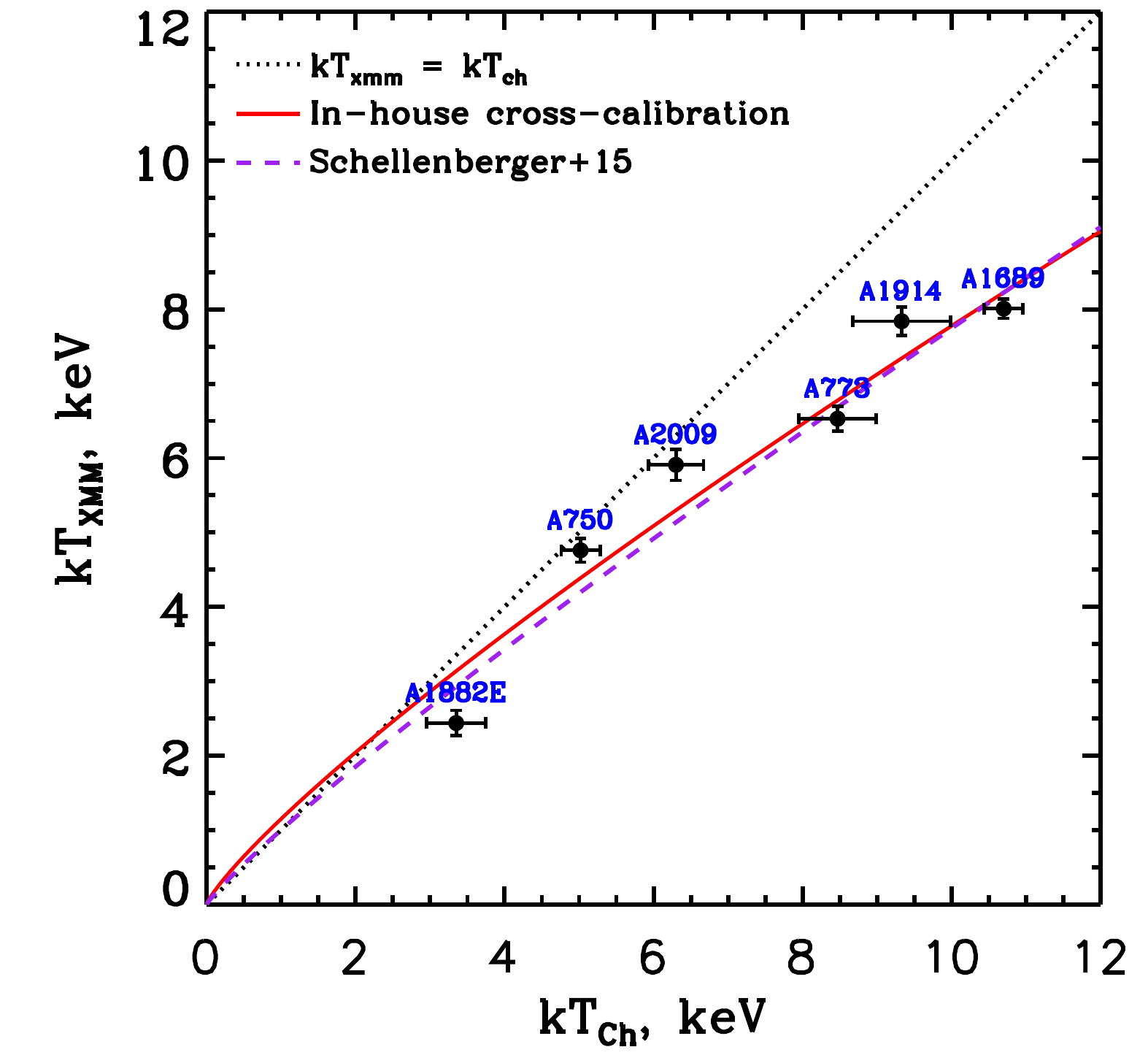}
 	\end{center}
	\caption{
		Comparison of temperatures from \cha\ and \xmm\ for six clusters in our sample. Temperatures are derived from 0.15-0.75 $R_{500}$ for \cha\ and \xmm\ respectively, with $R_{500}$ determined iteratively from the $M-T_X$ relation. The dotted line is the line of equality. The red line is our powerlaw best-fit relation (see Appendix~\ref{app:cc}) used in this work, while the purple dashed line is from \protect\cite{2015A&A...575A..30S} ($a_{T}=1.00$ and $s_{T}=0.89$), with older \cha\ and \xmm\ calibrations than what we used.
	}
	 \label{fig:cross-T}
\end{figure}

\section{Features of individual clusters}
Here we briefly comment clusters in the sample (in order of the \max\ richness $N_{200}$), with emphasis on the substructure and dynamical state.

\begin{figure}
 	\begin{center}
\includegraphics[width=0.495\textwidth,keepaspectratio=true,clip=true]{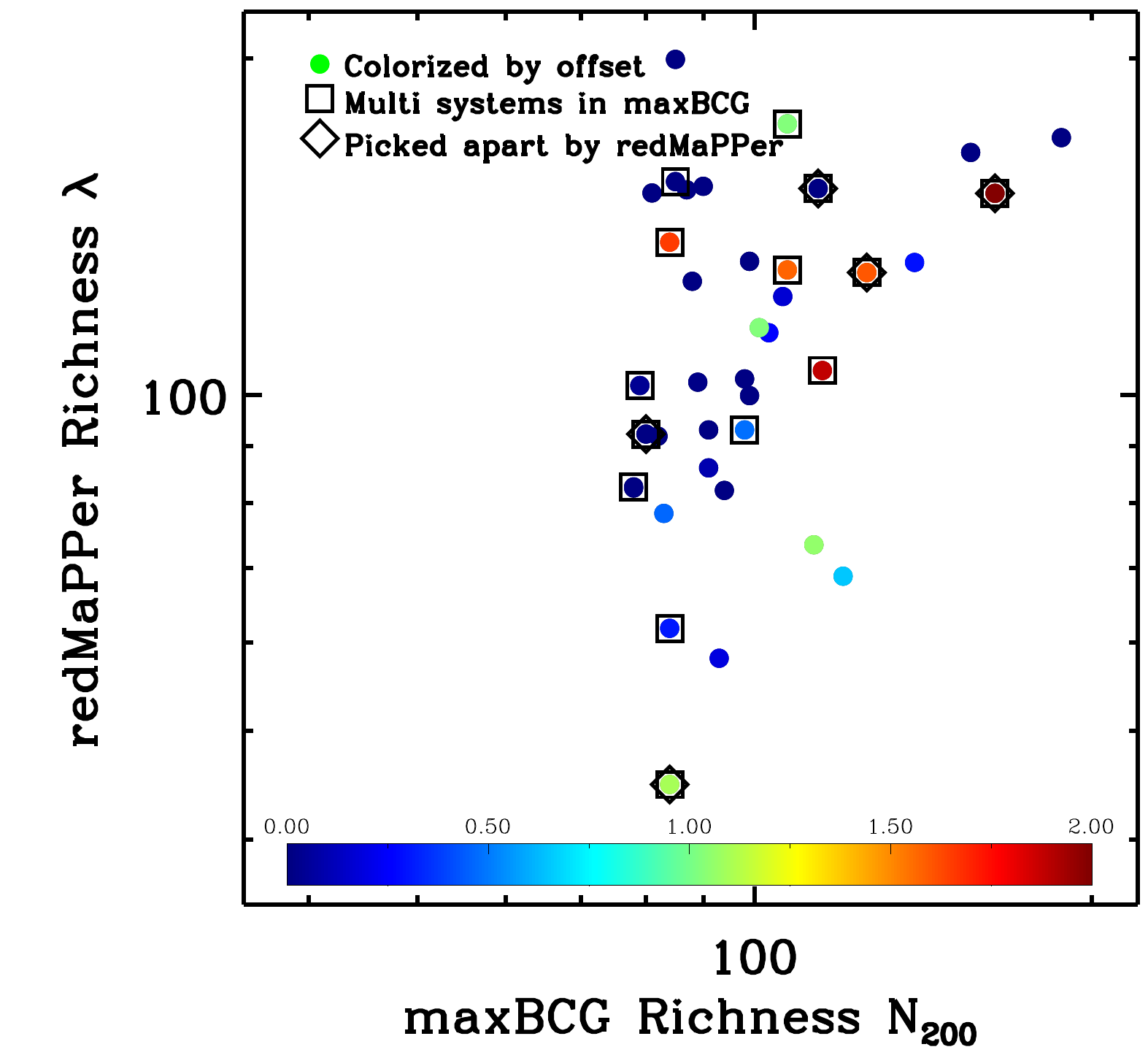}
 	\end{center}
	\caption{
		Comparison of richness from \max\ and \red\ of the sample. The dots are colorized by the centroid offset between \max\ and \red, in units of Mpc. The box shows the \max\ cluster with multiple X-ray clumps. The diamond shows the same \max\ cluster, but picked apart as multi clusters by \red. The clusters with larger centroid offset are more likely to be multi cluster systems.
	}
	 \label{fig:richness}
\end{figure}

\label{app:id}
\noindent {\bf A2142} is the richest \max\ cluster.
There is an ongoing merger as indicated by multiple cold fronts (\citealt{2000ApJ...541..542M}; \citealt{2013A&A...556A..44R}), radio halos (\citealt{2017A&A...603A.125V}), and group-scale substructures \citep{2011ApJ...741..122O}. However, its global X-ray emission appears relaxed and both \max\ and \red\ positions agree with each other.

\noindent {\bf J150} is the 2nd richest \max\ cluster. One may expect J150 as a massive ($\sim1.1\times10^{15}M_{\odot}$) and high temperature ($kT \sim 10$ keV) cluster inferred from its optical richness. However from the \XMM\ data, we find a $\sim$5 Mpc filament interlaced with three $kT \sim 3$ keV clusters (J150E, J150M and J150W) and one group (J150EE; $kT \sim 1.3$ keV). The \max\ center is on J150E. \red\ identifies a cluster centered on J150W and a group on J150EE. However, the \pla\ SZ center is located at the X-ray brightest cluster (J150M). Thus this large-scale filament suffers problems of miscentering and projection when selected as a single optical cluster.
We then compare the optical richness inferred from $T_X$ with the one in optical catalog.
Based on the mass derived from the X-ray temperature of different components, we estimate that the total \max\ richness for three clusters and the group is 82 (vs. 164 from \max), and the total \red\ richness for three clusters is 105.5 (vs. 151.4 from \red). Thus, this system is a particularly rich one in optical. Less X-ray luminous groups in this large-scale structure and projection contamination may also contribute.

\noindent {\bf A1689} has a large concentration parameter and its gravitational lensing mass is higher than the X-ray mass by a factor of 2, which could be explained by its triaxial shape with the major axis nearly orientated along the line of sight (e.g. \citealt{2011ApJ...729...37M}; \citealt{2013MNRAS.428.2241S}; \citealt{2015ApJ...806..207U}). The triaxiality may be induced by a merger along the line of sight as indicated by the asymmetric temperature distribution \citep{2004ApJ...607..190A} and the diffuse radio emission \citep{2011A&A...535A..82V}. However in the plane of sky, A1689 appears relaxed based on the X-ray morphology. The X-ray peak is also consistent with both \max\ and \red\ centers.

\noindent {\bf A1443} is a merging cluster from the X-ray E-W elongation and the diffuse radio emission \citep{2015MNRAS.454.3391B}. \max\ and \red\ selected different galaxies as the BCG of the cluster.

\noindent {\bf A781} in fact contains two large filaments projected on the sky, one $\sim$2 Mpc filament (A781, A781M, A781N; $z \sim$ 0.29) and another $\sim$7 Mpc filament (A781E, A781W; $z \sim$ 0.43). All these five clusters, including the X-ray faint A781N ($kT$=1.3 keV, $L_X=1.6\times10^{43}\ \rm erg\ s^{-1}$ in a $2^{\prime}$ radius), are confirmed by weak-lensing detections \citep{2008ApJ...673..163S}. 
A radio halo and a candidate radio relic was also discovered to be associated with the hottest cluster A781 (\citealt{2011A&A...529A..69G}). The \max\ only detects the brightest A781, while the \red\ detects A781E but mixes A781 with A781M.

\noindent {\bf A1986} is not detected by \pla. The \red\ center is on the X-ray peak. The mass estimated from the \red\ richness ($\sim 3.9\times10^{14}M_{\odot}$) is more consistent with the mass suggsted by the X-ray temperature ($\sim 2.4\times10^{14}M_{\odot}$). The \max\ center is $\sim$ 640 kpc from the X-ray center and the \max\ richness is biased high by $\sim 2$.

\noindent {\bf A1882} is in a $\sim$ 3 Mpc filament at $z \sim 0.14$ with at least four groups (A1882E, A1882M, A1882W, A1882N, $kT \sim 1.4 - 2.5$ keV), which are corresponding to four concentrations of galaxy distribution \citep{2013ApJ...772..104O}. The merger is considered to be at an early stage from the large projected separation and the dearth of evidence for a recent major interaction in X-ray data \citep{2013ApJ...772..104O}. The \max\ and \red\ centers are on A1882E and A1882W respectively. 
There is also no \pla\ SZ detection in this region.

\noindent {\bf A1758} is composed of A1758N and A1758S separated by $\sim 2$ Mpc in projection with similar redshift ($z \sim 0.28$) and X-ray luminosity. A1758N is in the late stage of a merger of two 7 keV subclusters near the plane of sky, while A1758S is in the early stage of a merger of two 5 keV subclusters close to the line of sight from detailed multi-wavelength observations (e.g. \citealt{2004ApJ...613..831D}; \citealt{2009A&A...507.1257G}; \citealt{2011A&A...529A..38D}; \citealt{2017MNRAS.466.2614M}). These two systems are most likely gravitationally bound and will eventually merge into a $\sim 12$ keV cluster \citep{2004ApJ...613..831D}. The X-ray mass of A1758N and A1758S is $11.7\times10^{14}M_{\odot}$ and $4.7\times10^{14}M_{\odot}$ respectively, which can be compared with the masses suggested by their \red\ richness values ($11.1\times10^{14}M_{\odot}$ for A1758N and $2.7\times10^{14}M_{\odot}$ for A1758S). The \max\ only identifies A1758N with an optical mass of $7.2\times10^{14}M_{\odot}$.

\noindent {\bf A1760} is divided into two clusters by \red\ and the total richness of these two clusters is compatible with \max's.

\noindent {\bf A1622} is composed of two clusters as shown in the X-ray image. Only the X-ray properties of the northern one can be constrained from the shallow \cha\ data. The \max\ and \red\ identified different galaxies as the BCG. Optical richness values from both \max\ and \red\ suggest mass values of 2 - 3 times higher than the mass estimated from the X-ray temperature. 

\noindent {\bf A750} is composed of A750E/A750W at $z \sim 0.16$. A750W is possibly falling into A750E, as indicated by the highly disturbed X-ray morphology and the large offset of two X-ray peaks. Both \max\ and \red\ mix these two clusters as one with different galaxies as the BCG. 

\noindent {\bf A1682} is a merging cluster as shown by the disturbed X-ray morphology. Both galaxy distribution and weak-lensing mass map show two peaks \citep{2002ApJS..139..313D} coincident with the X-ray peaks.
The diffuse radio emission is complex with possibly one halo and two relics (e.g. \citealt{2013A&A...551A.141M}). The \max\ and \red\ identified different galaxies as the BCG.

\noindent {\bf A1246} has been observed to the virial radius with {\em Suzaku} \citep{2014PASJ...66...85S}. The X-ray mass within $R_{500}$ ($5.5\times10^{14}M_{\odot}$) from the $M-T_X$ relation of the \cha\ data is consistent with the mass derived from the {\em Suzaku} data ($4.3\times10^{14}M_{\odot}$; \citealt{2014PASJ...66...85S}). The \max\ and \red\ identified different galaxies as the BCG.

\noindent {\bf A1961} is a poor cluster ($2.7\times10^{14}M_{\odot}$) from its X-ray temperature. However, the optical richness values from \max\ and \red\ suggest mass values of $\sim$3 times higher. The \max\ and \red\ identified different galaxies as the BCG. The miscentering of \max\ is large, $\sim$ 1 Mpc.

\noindent {\bf A2034} is a merging cluster with the merger axis along S-N, as indicated by the northern shock \citep{2014ApJ...780..163O} and the diffuse radio emission \citep{2009A&A...507.1257G}. The complex dymanics is also shown by the galaxy distribution \citep{2014ApJ...780..163O} and weak-lensing mass distribution \citep{2008PASJ...60..345O}. The \max\ and \red\ results are very similar.

\noindent {\bf A655} hosts a dominated cD galaxy at the center, identified by both \max\ and \red. However, the optical richness values from \max\ and \red\ suggest mass values of 2 - 3 times higher than that from the X-ray temperature.
The tentacle-like outskirts based on the X-ray morphology may suggest connection with other large-scale filaments and infalling galaxy groups as also suggested by \cite{2016ApJ...824...69P}.

\noindent {\bf A1914} is a merging cluster as indicated by two substructures along the NE-SW direction from the galaxy distribution \citep{2013MNRAS.430.3453B} and the weak-lensing mass distribution \citep{2008PASJ...60..345O}, as well as the diffuse radio emission \citep{2003A&A...400..465B}.
Both \max\ and \red\ identified the same BCG, 175 kpc south to the X-ray peak.

\noindent {\bf Z5247} is a merging cluster as indicated by the disturbed X-ray morphology. There are two X-ray peaks, corresponding to two substructures from the galaxy distribution and the weak-lensing mass distribution \citep{2002ApJS..139..313D}. The cluster also hosts a radio relic and a candidate radio halo \citep{2015A&A...579A..92K}. The \max\ and \red\ identified different galaxies as the BCG.

\noindent {\bf A657} is not detected by \pla. While \max\ and \red\ identified the same BCG close to the X-ray peak, the optical richness values from \max\ and \red\ suggest mass values of 3 - 4 times higher than that from the X-ray temperature.

\noindent {\bf J229} is a $\sim$2.5 keV system without \pla\ SZ detection. The \max\ and \red\ identified different galaxies as the BCG. The optical richness values from \max\ and \red\ suggest mass values of 2 - 4 times higher than that from the X-ray temperature.

\noindent {\bf A1423} is a relaxed cluster from the smooth X-ray morphology, galaxy distribution and weak-lensing mass distribution \citep{2002ApJS..139..313D}.

\noindent {\bf A801} appears relaxed in X-rays. The richness values from \max\ and \red\ suggest a cluster mass similar to that derived from its X-ray temperature.

\noindent {\bf A773} is a merging cluster as shown by evidence such as two X-ray peaks in the center along NE-SW, asymmetric X-ray temperature distribution \citep{2004ApJ...605..695G}, two peaks of galaxy distribution and their velocity distribution \citep{2007A&A...467...37B}, two peaks in the weak-lensing mass distributions \citep{2002ApJS..139..313D}, and diffuse radio emission \citep{2001A&A...376..803G}.

\noindent {\bf A1576} is a disturbed cluster as indicated by its lopsided X-ray morphology and the presence of multiple peaks in galaxy distribution and weak-lensing mass distribution \citep{2002ApJS..139..313D}.

\noindent {\bf A2631} is classified as a disturbed cluster based on the multiple morphology parameters \citep{2010ApJ...721L..82C}. However, the weak-lensing map shows only one single peak \citep{2010PASJ...62..811O} and there is not any significant extended radio emission \citep{2008A&A...484..327V}.

\noindent {\bf A1703} is a relaxed, unimodal cluster from the strong-lensing model \citep{2009A&A...498...37R}.

\noindent {\bf A2219} is a merging cluster with the main merger axis along NW-SE direction, as shown by a series shocks and a possible cold front \citep{2017MNRAS.464.2896C}, two luminous BCGs in the cluster center and substructure in galaxy distribution \citep{2004A&A...416..839B} and a diffuse radio halo \citep{2003A&A...400..465B}. 

\noindent {\bf A1319} is composed of three clusters at similar redshifts, A1319M, A1319NW and A1319SW.
Both \max\ and \red\ only identified one cluster centered on the BCG of A1319M.

\noindent {\bf A1560} is a merging cluster with two subclusters. The \max\ center is on A1560SW and the \red\ center is on A1560NE.

\noindent {\bf J175} is in a complex field with both foreground and background sources. X-ray emission mainly shows two clusters separated by $\sim 2.5$ Mpc in projection along the N-S direction, J175N at $z_{\rm spec}=0.117$ and J175S at $z_{\rm spec}=0.119$, which is also confirmed by \red. 
However, the \max\ cluster is centered on a luminous galaxy at $z_{\rm spec}=0.117$, nearly midway between J175N and J175S. The \max\ center is very close to the foreground galaxy group HCG 58 (\citealt{1982ApJ...255..382H}; \citealt{2009AJ....138..295F}) that hosts the brightest X-ray source in the \XMM\ field, the X-ray AGN of NGC~3822.
The X-ray diffuse luminosity from this \max\ region, excluding the NGC~3822 AGN, other point sources and diffuse emission from J175N/J175S is only $\sim 3.3 \times10^{42} {\rm erg\ s^{-1}}$ at the \max\ cluster redshift. There is no \pla\ SZ source in this region. There is also a faint background cluster at $z_{\rm spec}=0.280$, detected by both \max\ and \red.

\noindent {\bf J249} is a cluster pair with comparable X-ray temperatures. There is no \pla\ SZ detection. The \max\ and \red\ identified different galaxies in J249SW as the BCG.

\noindent {\bf A1201} is a merging cluster with the merger axis along NW-SE from the lopsided X-ray morphology. It hosts two cold fronts and an offset remnant core with a stripped tail \citep{2012ApJ...752..139M}. Substructures are identified from the spatial and velocity distribution of member galaxies \citep{2009ApJ...692..702O}. The strong-lensing arc indicates large mass elongation \citep{2003ApJ...599L..69E}. \max\ and \red\ identified different galaxies as the BCG.

\noindent {\bf A2009} appears relaxed from the smooth X-ray morphology.

\noindent {\bf A2111} is a merging cluster from the lopsided X-ray morphology and early X-ray observations (\citealt{1997MNRAS.288..702W}; \citealt{1999MNRAS.307...67H}). The member galaxy and weak-lensing mass distribution shows the same elongation as the X-rays in the NW-SE direction (\citealt{2002ApJS..139..313D}). No significant diffuse radio emission has been detected \citep{2008A&A...484..327V}.

\noindent {\bf A815} is a merging cluster from the disturbed X-ray morphology. A815N and A815S each has a luminous galaxy on its X-ray peak. There is no \pla\ SZ detection.
There is a group to the east detected by the \red\ at a similar redshift. Based on the \XMM\ data, we derive $kT$=1.1 keV and $L_X=1.5\times10^{43}\ {\rm erg\ s^{-1}}$ within $R_{500}$ that is determined from \red's richness.

\noindent {\bf Z1450} is a merging cluster from the asymmetric X-ray morphology. The \max\ and \red\ identified different galaxies as the BCG. The optical richness values from \max\ and \red\ suggest mass values of 5 - 7 times higher than that from the X-ray temperature. There is no \pla\ SZ detection.
There is a group to the SW at a similar redshift. The X-ray emission of this region is dominated by two bright point sources, which leaves insufficient amount of data to constrain the gas properties of this group.

\noindent {\bf A1765} is a low mass ($1.0\times10^{14}M_{\odot}$) cluster from the X-ray temperature ($kT=2.2$ keV), while the optical richness values from \max\ and \red\ suggest mass values of $\sim$ 5 times higher. There is no \pla\ SZ detection.
There is a group to the east at a similar redshift. Based on the \XMM\ data, we derive $kT=1.5$ keV, $L_X=4.3\times10^{42}\ {\rm erg\ s^{-1}}$ within a 1.5$^{\prime}$ radius.

\noindent {\bf A1902} appears relaxed. Masses from optical and X-ray are consistent.
  
\begin{figure*}
 	\begin{center}
\includegraphics[width=0.246\textwidth,keepaspectratio=true,clip=true]{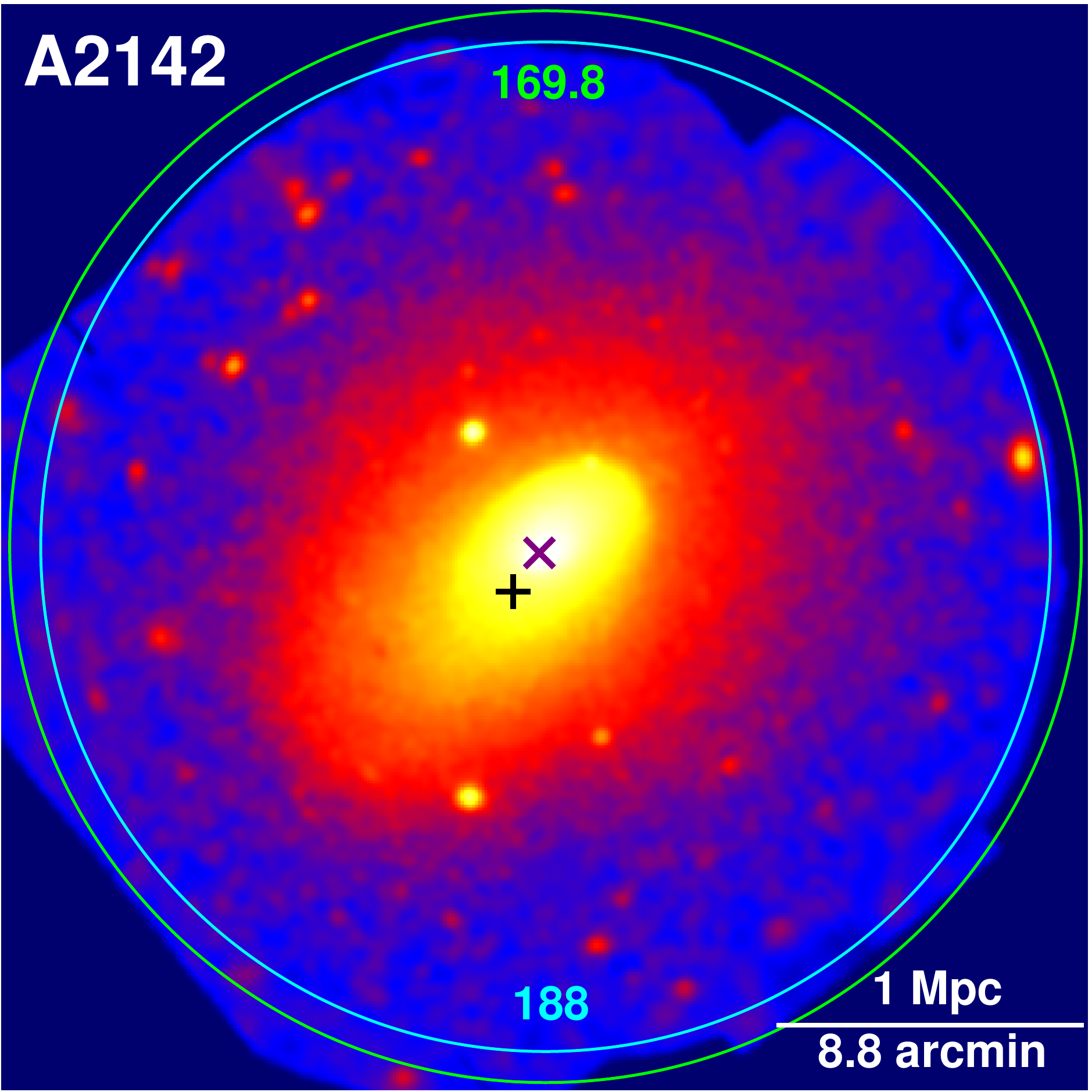}
\includegraphics[width=0.246\textwidth,keepaspectratio=true,clip=true]{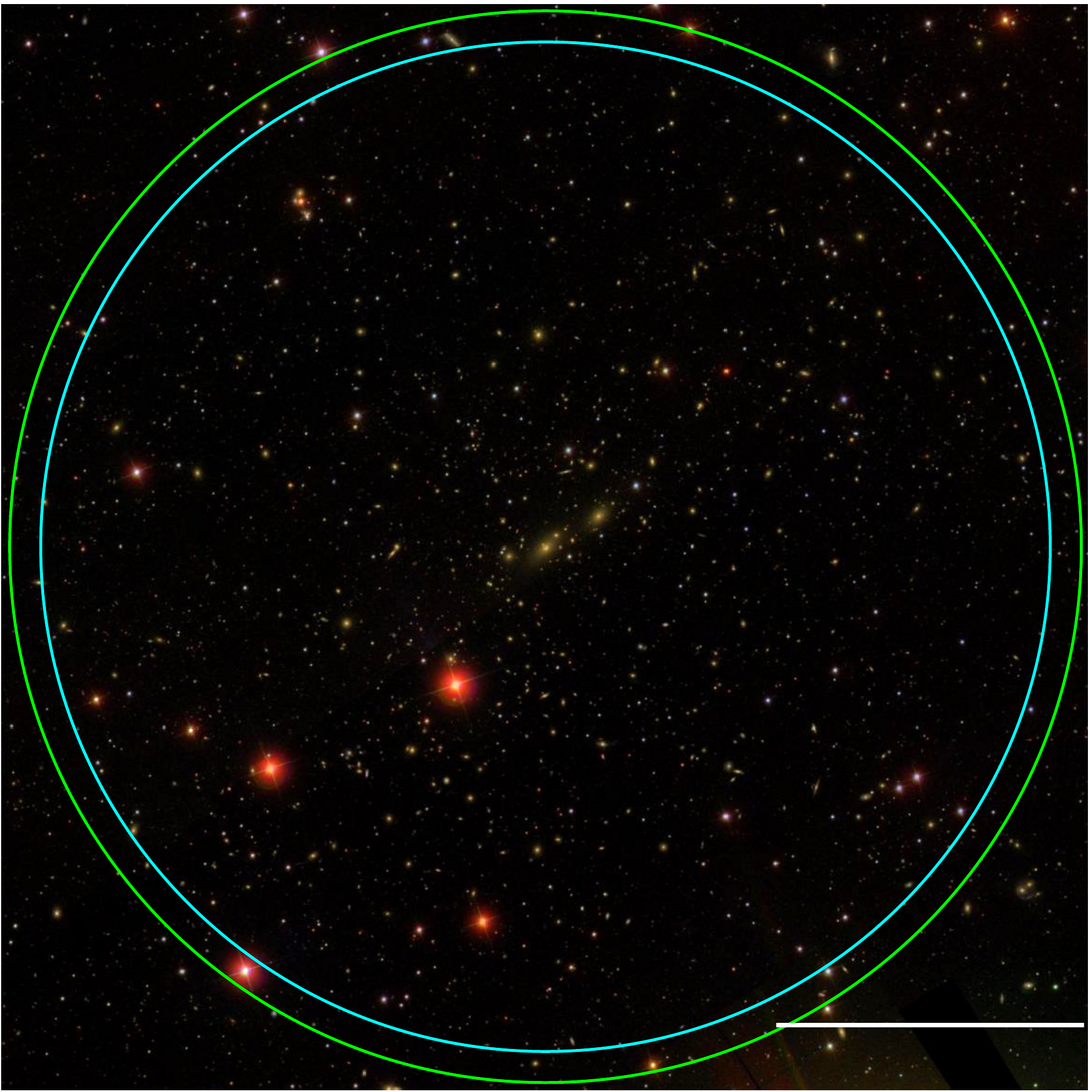}
\includegraphics[width=0.246\textwidth,keepaspectratio=true,clip=true]{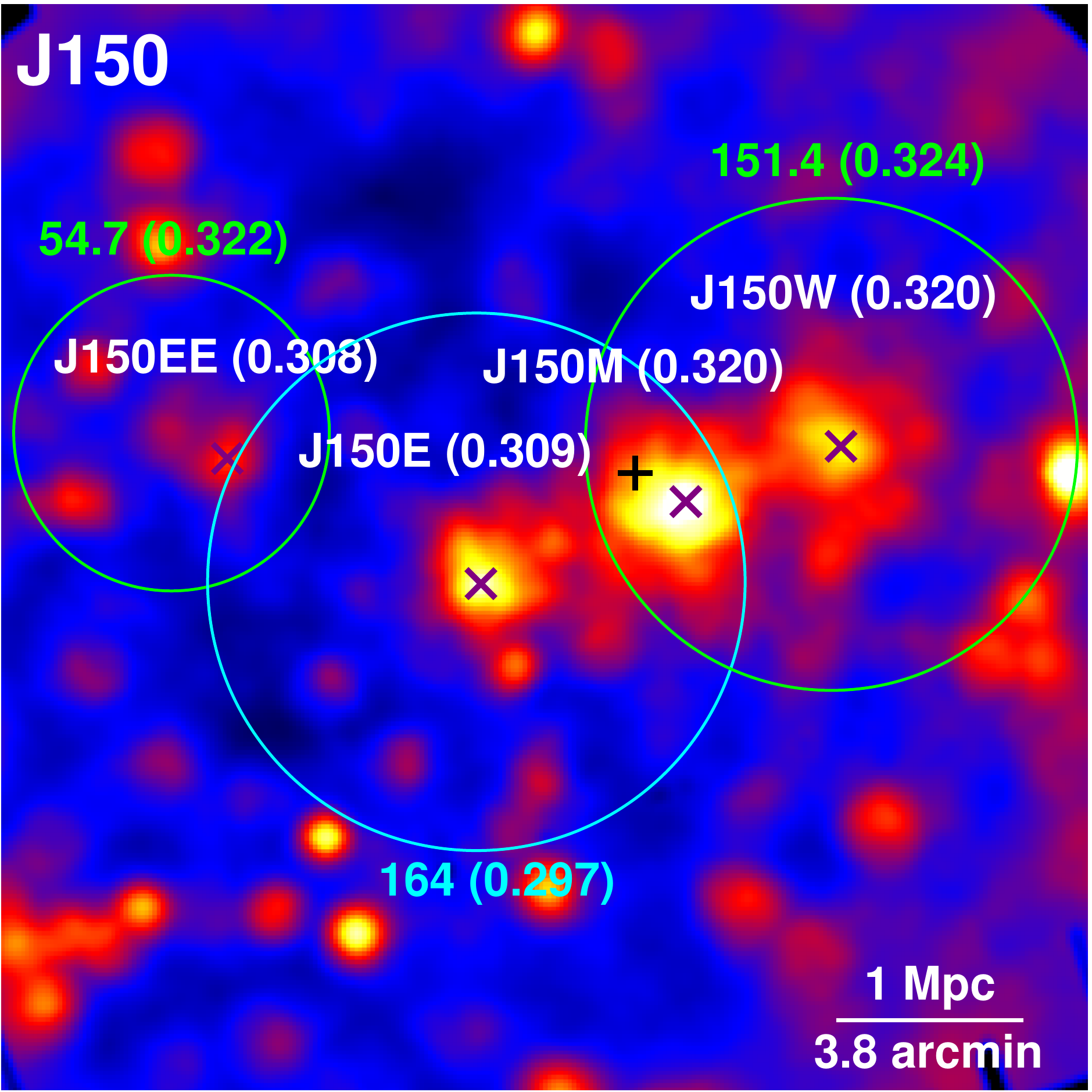}
\includegraphics[width=0.246\textwidth,keepaspectratio=true,clip=true]{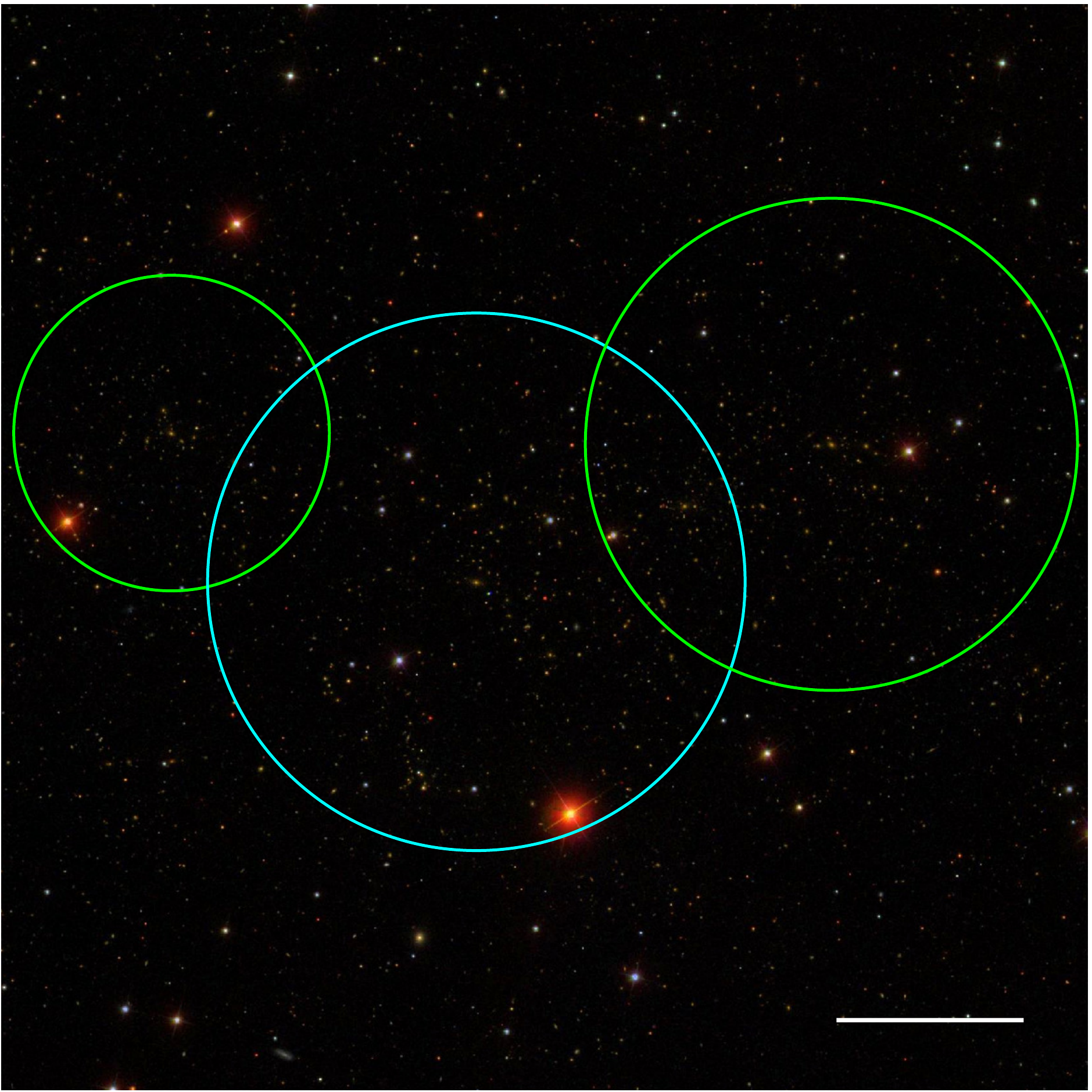}
\includegraphics[width=0.246\textwidth,keepaspectratio=true,clip=true]{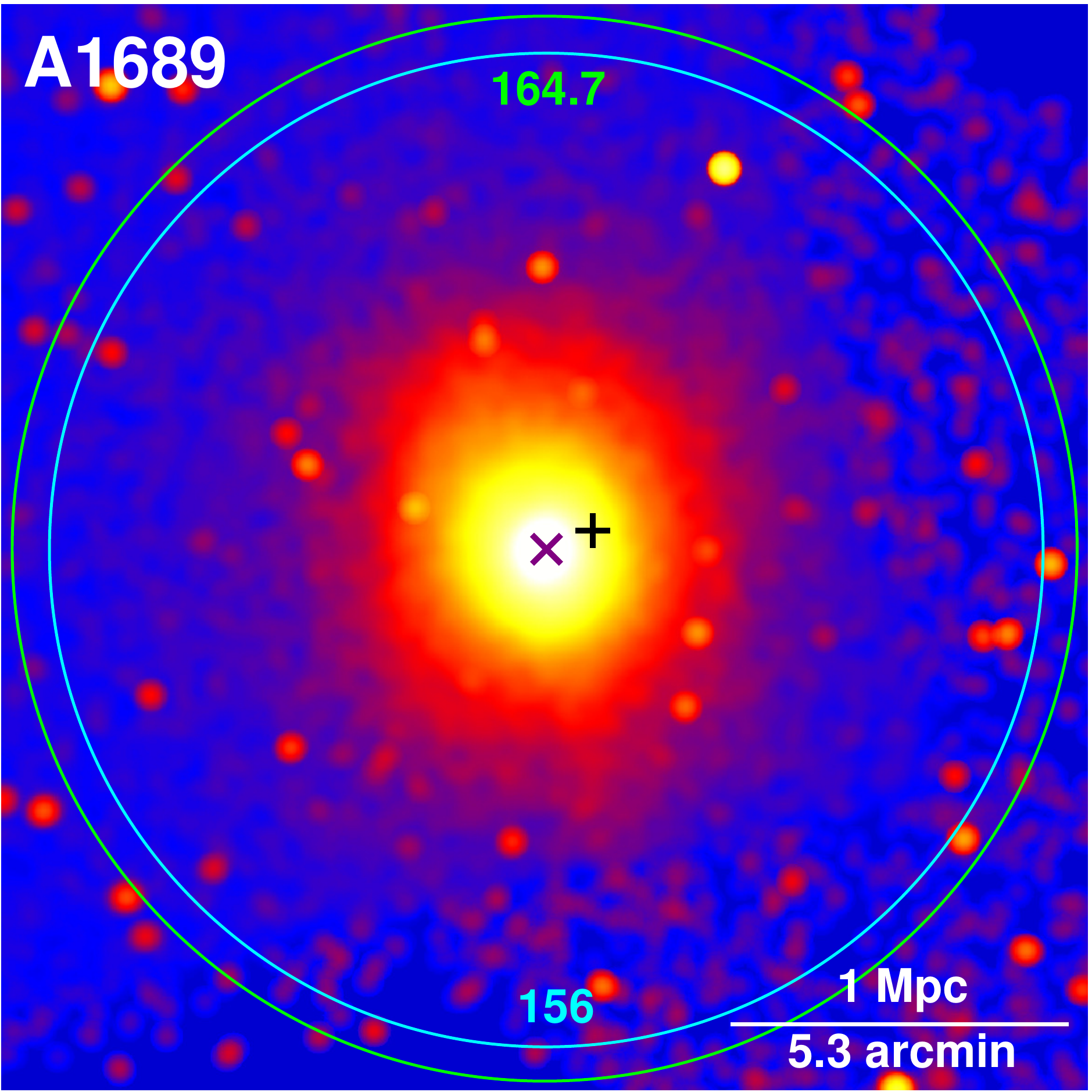}
\includegraphics[width=0.246\textwidth,keepaspectratio=true,clip=true]{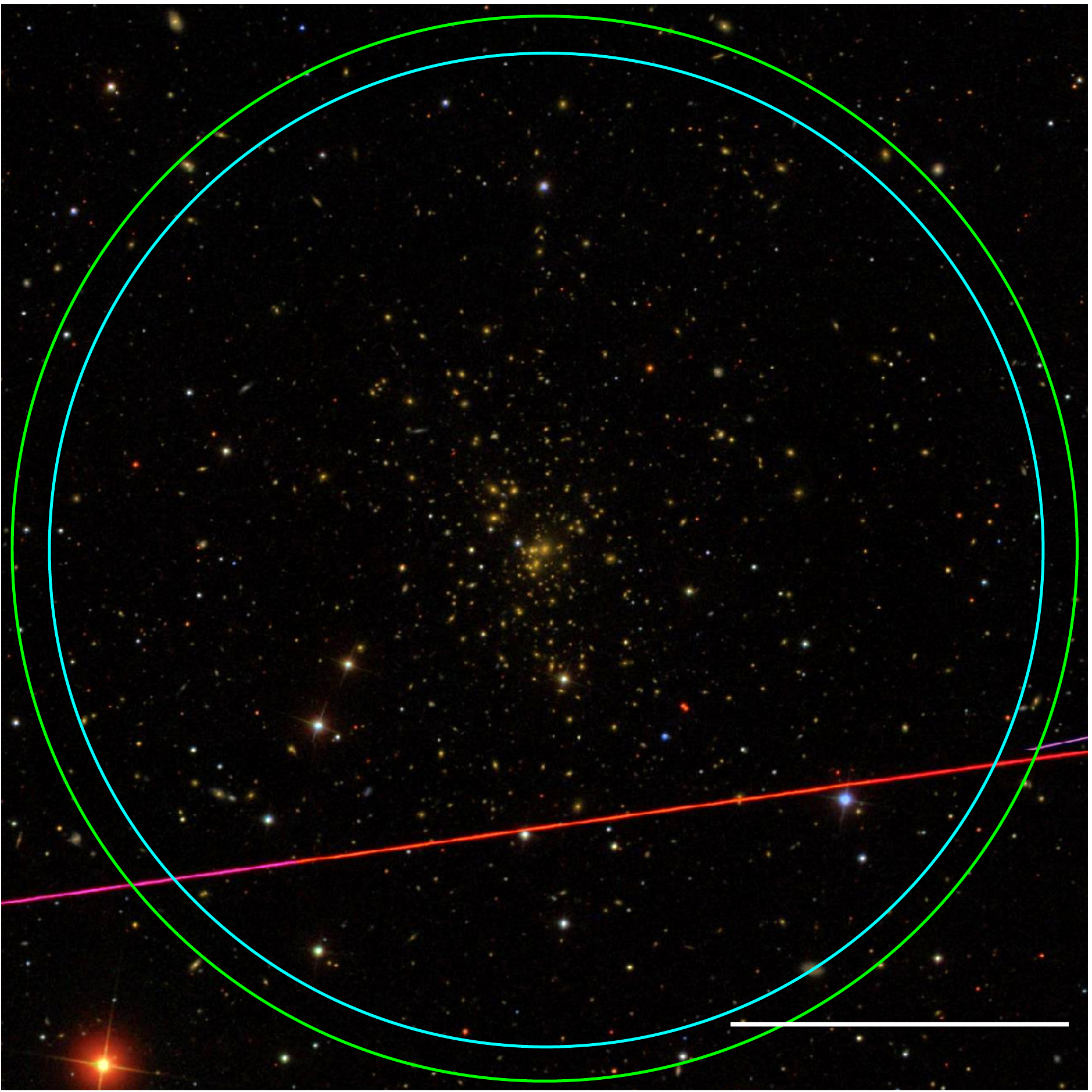}
\includegraphics[width=0.246\textwidth,keepaspectratio=true,clip=true]{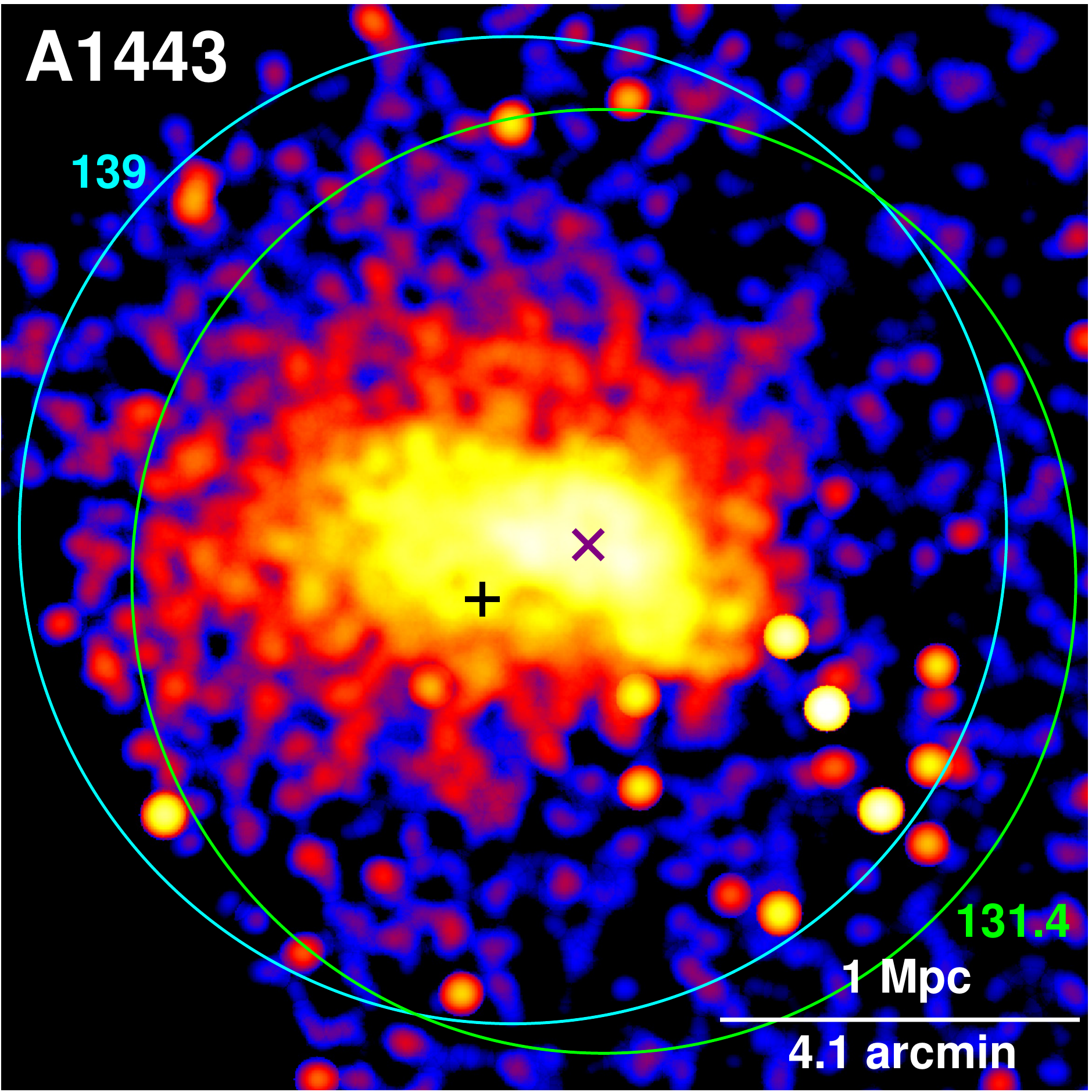}
\includegraphics[width=0.246\textwidth,keepaspectratio=true,clip=true]{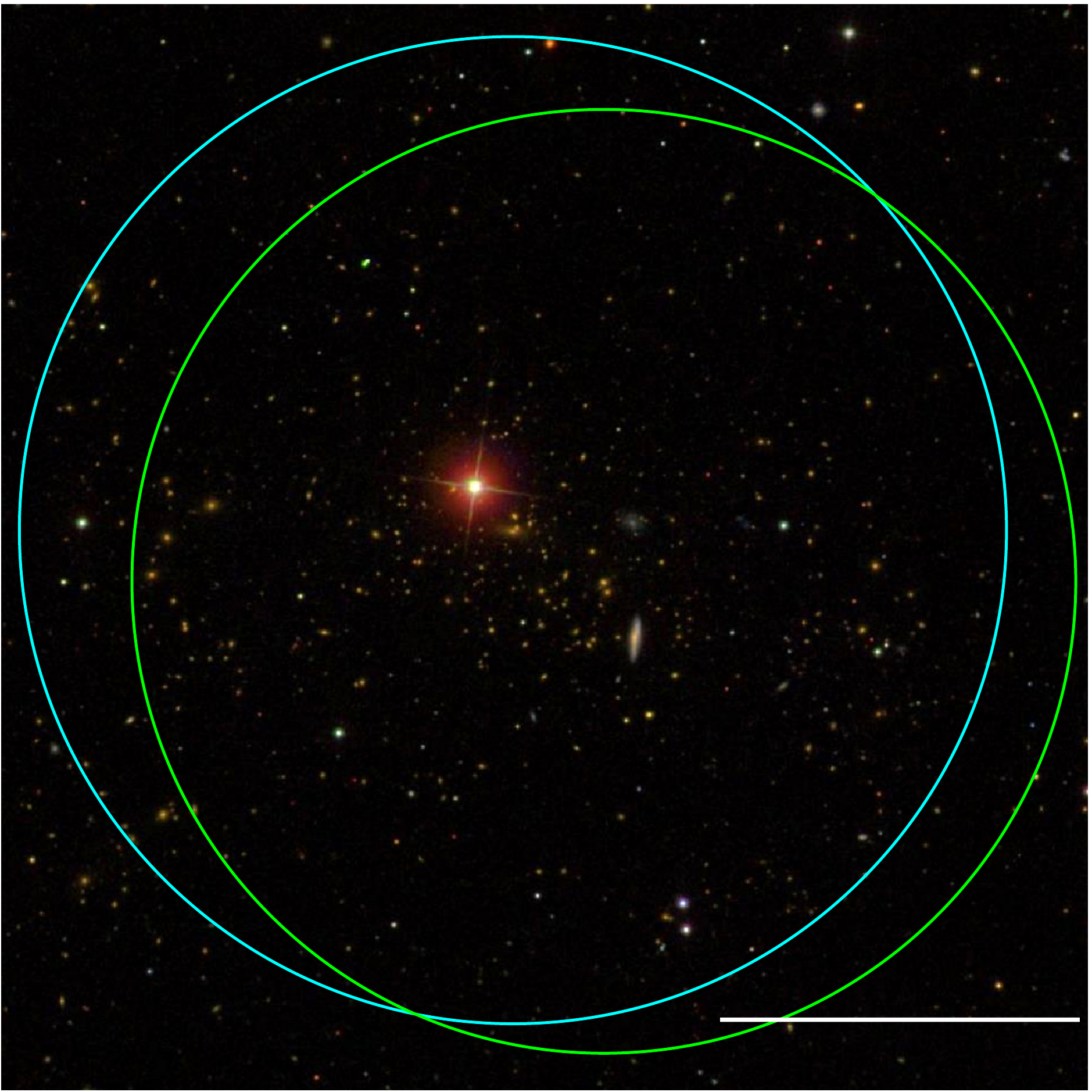}
\includegraphics[width=0.246\textwidth,keepaspectratio=true,clip=true]{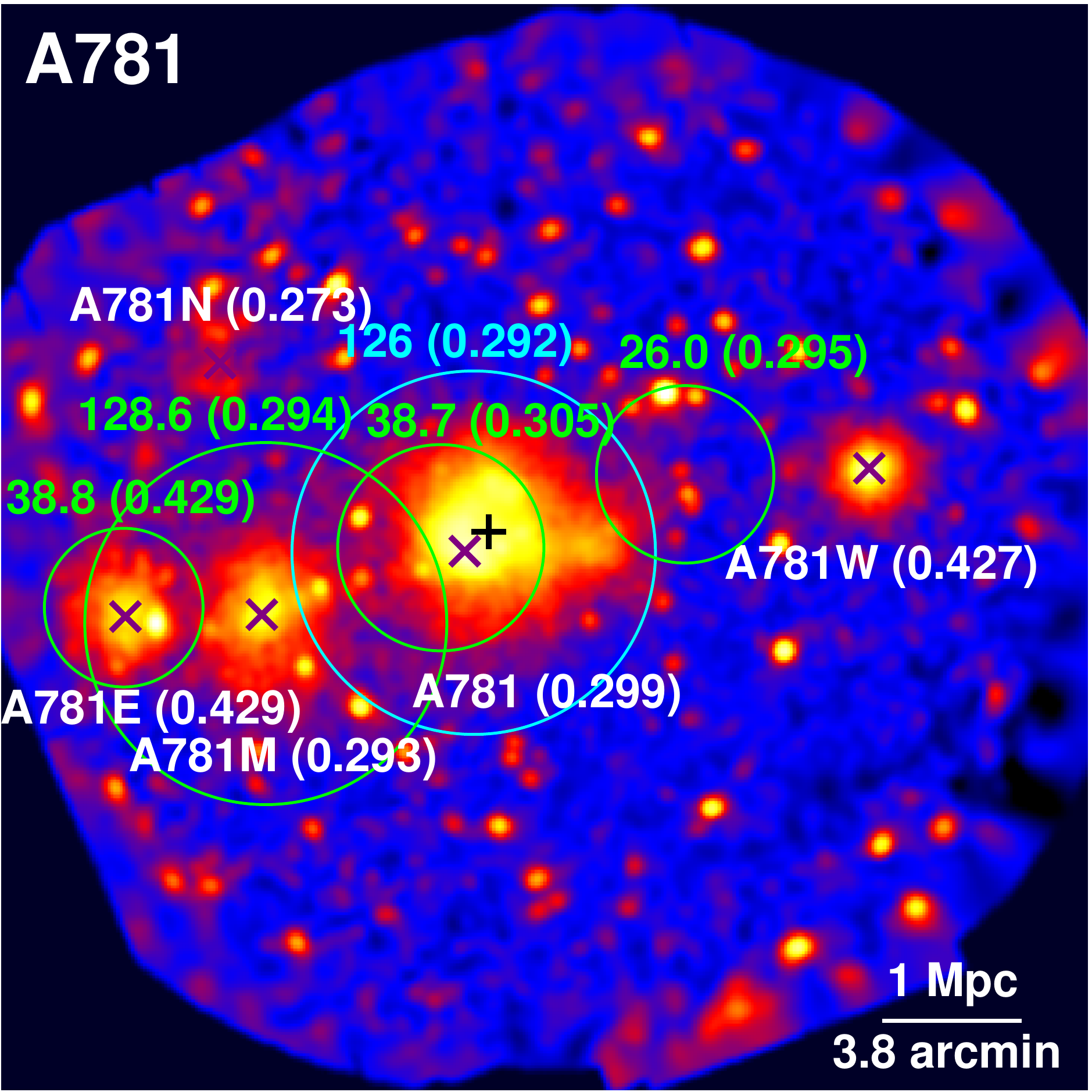}
\includegraphics[width=0.246\textwidth,keepaspectratio=true,clip=true]{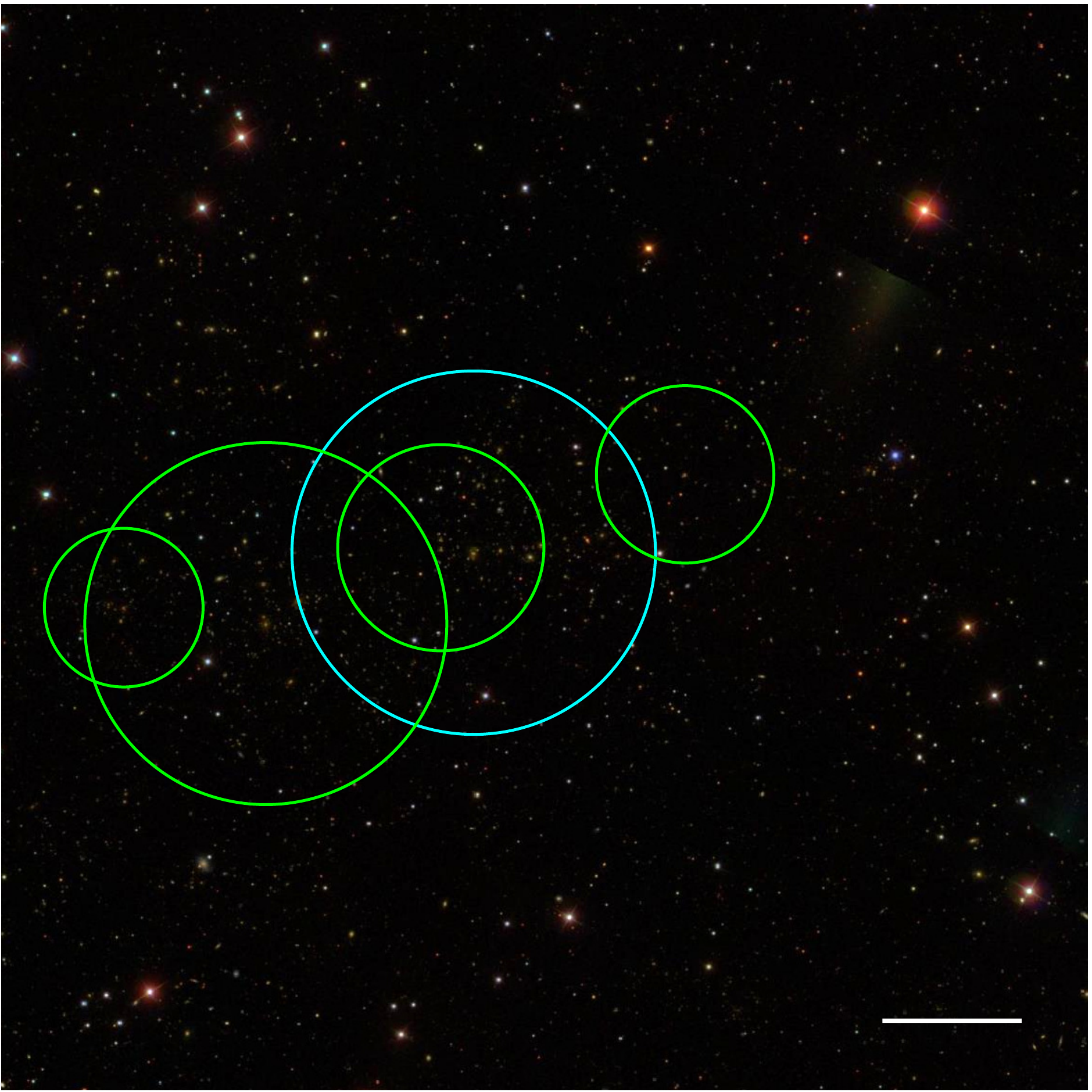}
\includegraphics[width=0.246\textwidth,keepaspectratio=true,clip=true]{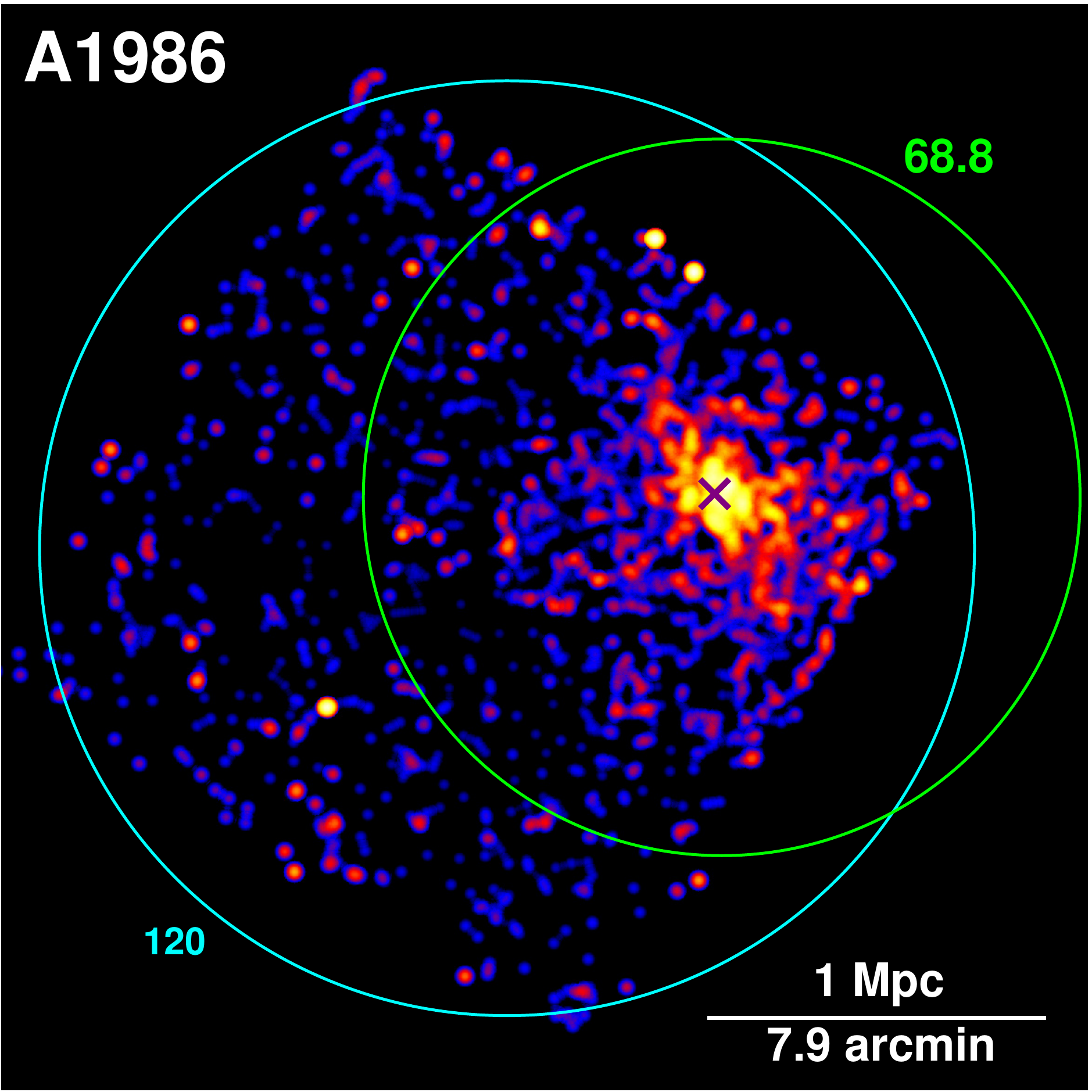}
\includegraphics[width=0.246\textwidth,keepaspectratio=true,clip=true]{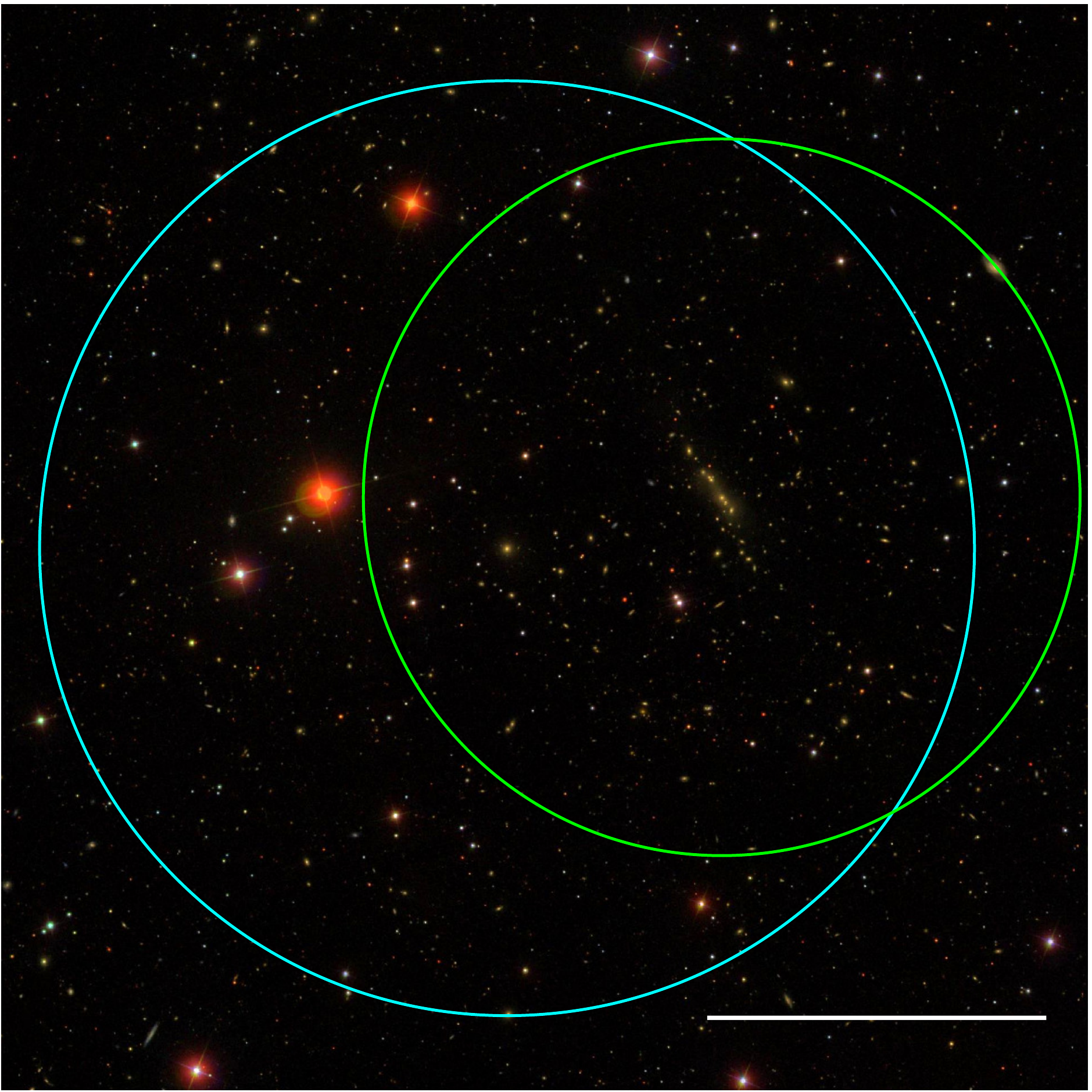}
\includegraphics[width=0.246\textwidth,keepaspectratio=true,clip=true]{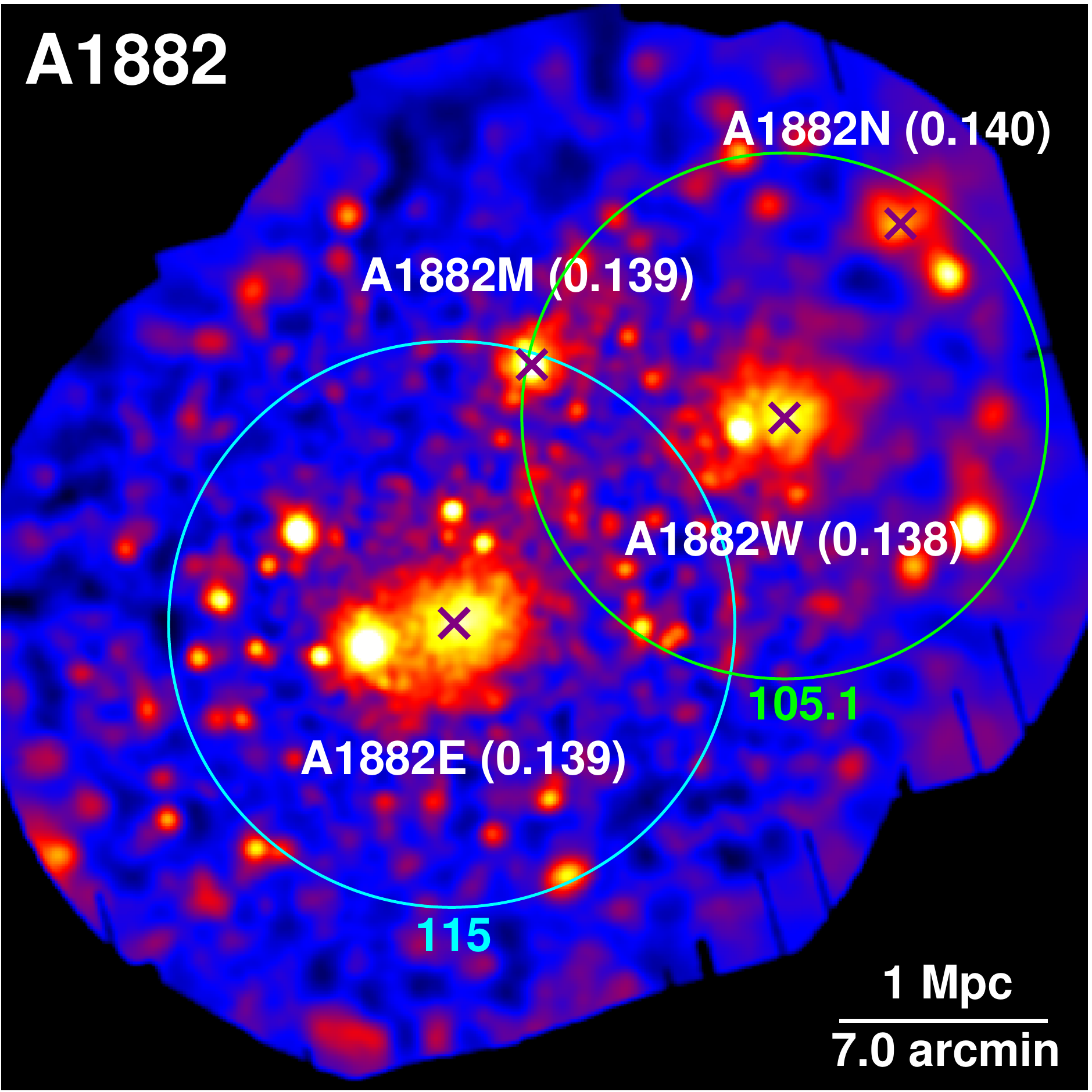}
\includegraphics[width=0.246\textwidth,keepaspectratio=true,clip=true]{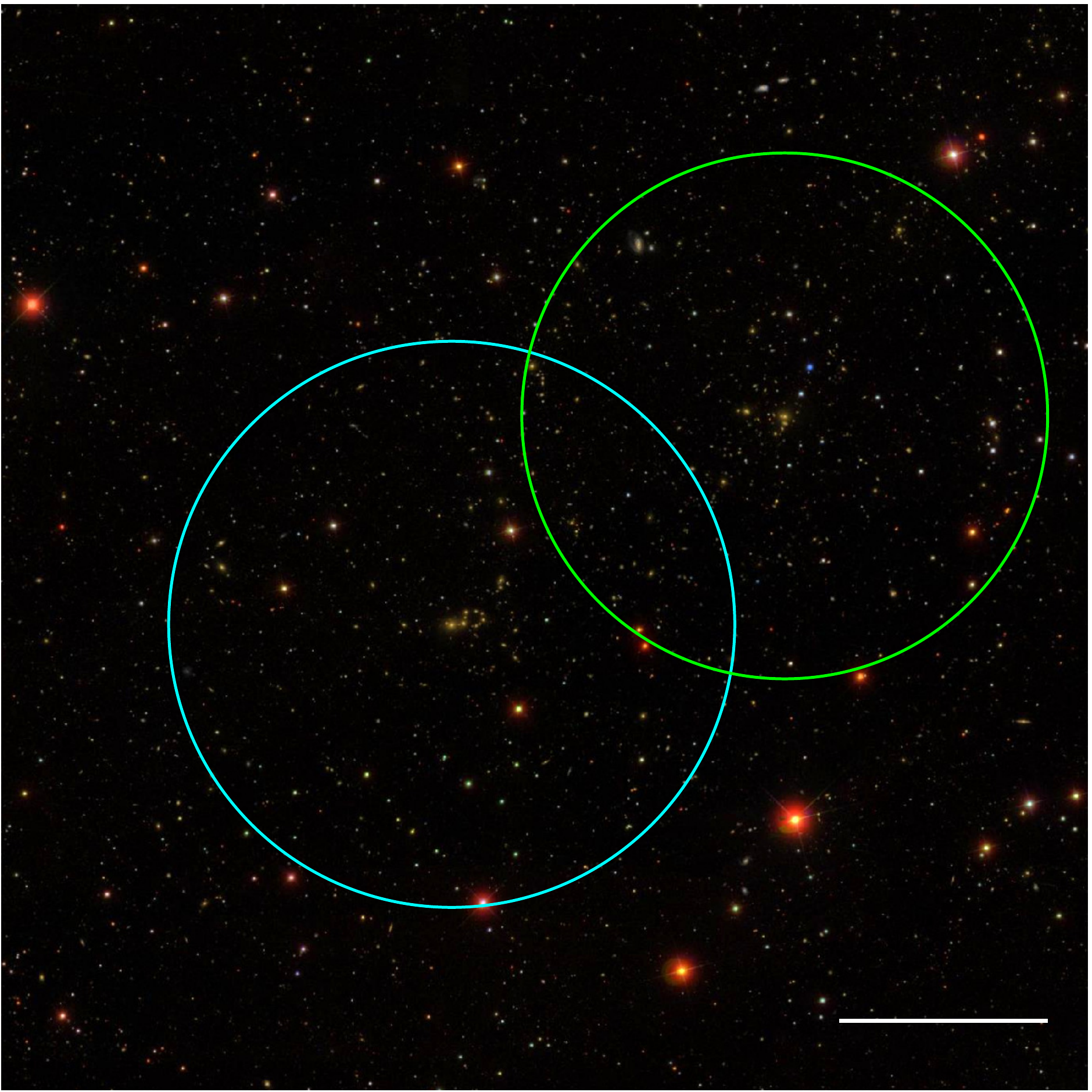}
\includegraphics[width=0.246\textwidth,keepaspectratio=true,clip=true]{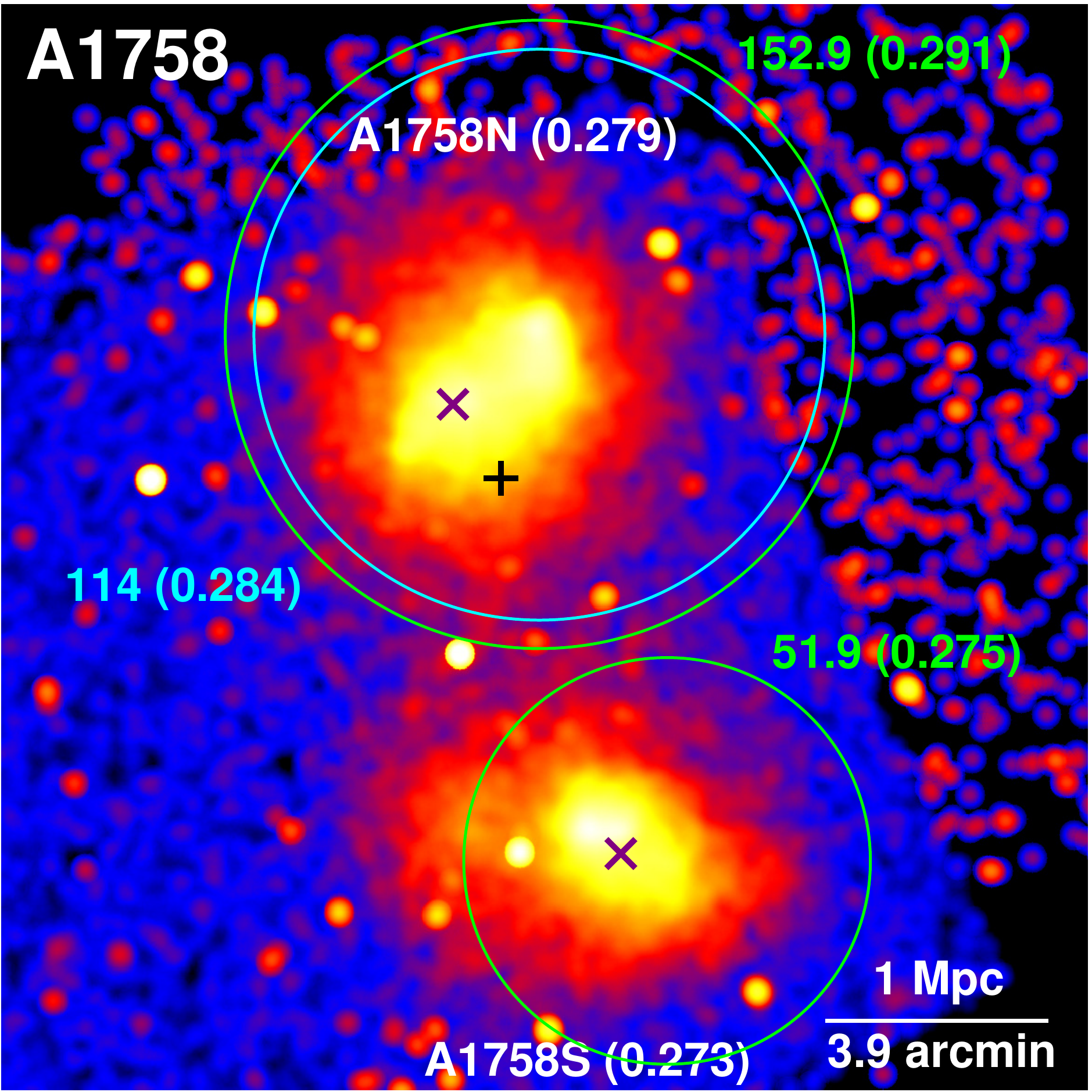}
\includegraphics[width=0.246\textwidth,keepaspectratio=true,clip=true]{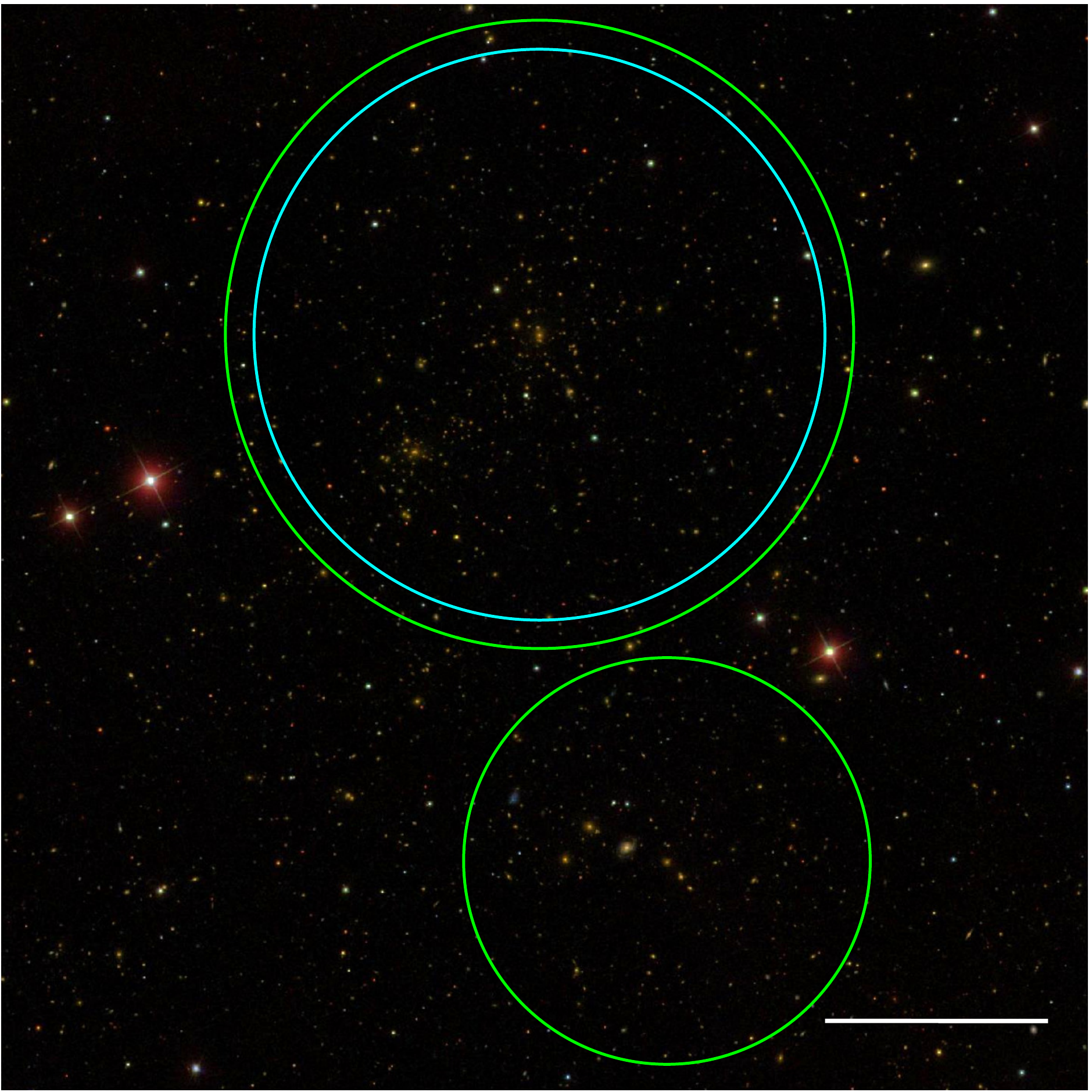}
\includegraphics[width=0.246\textwidth,keepaspectratio=true,clip=true]{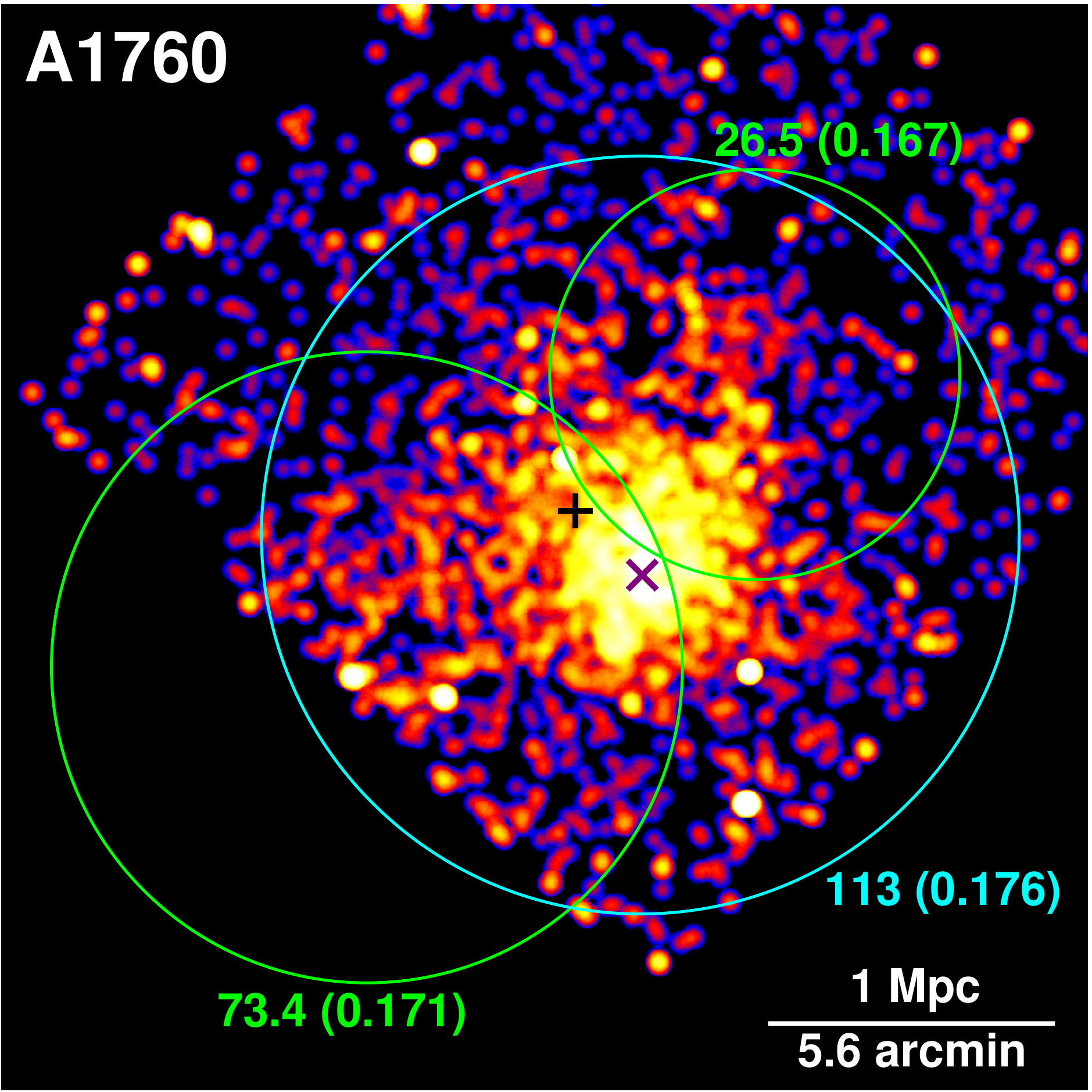}
\includegraphics[width=0.246\textwidth,keepaspectratio=true,clip=true]{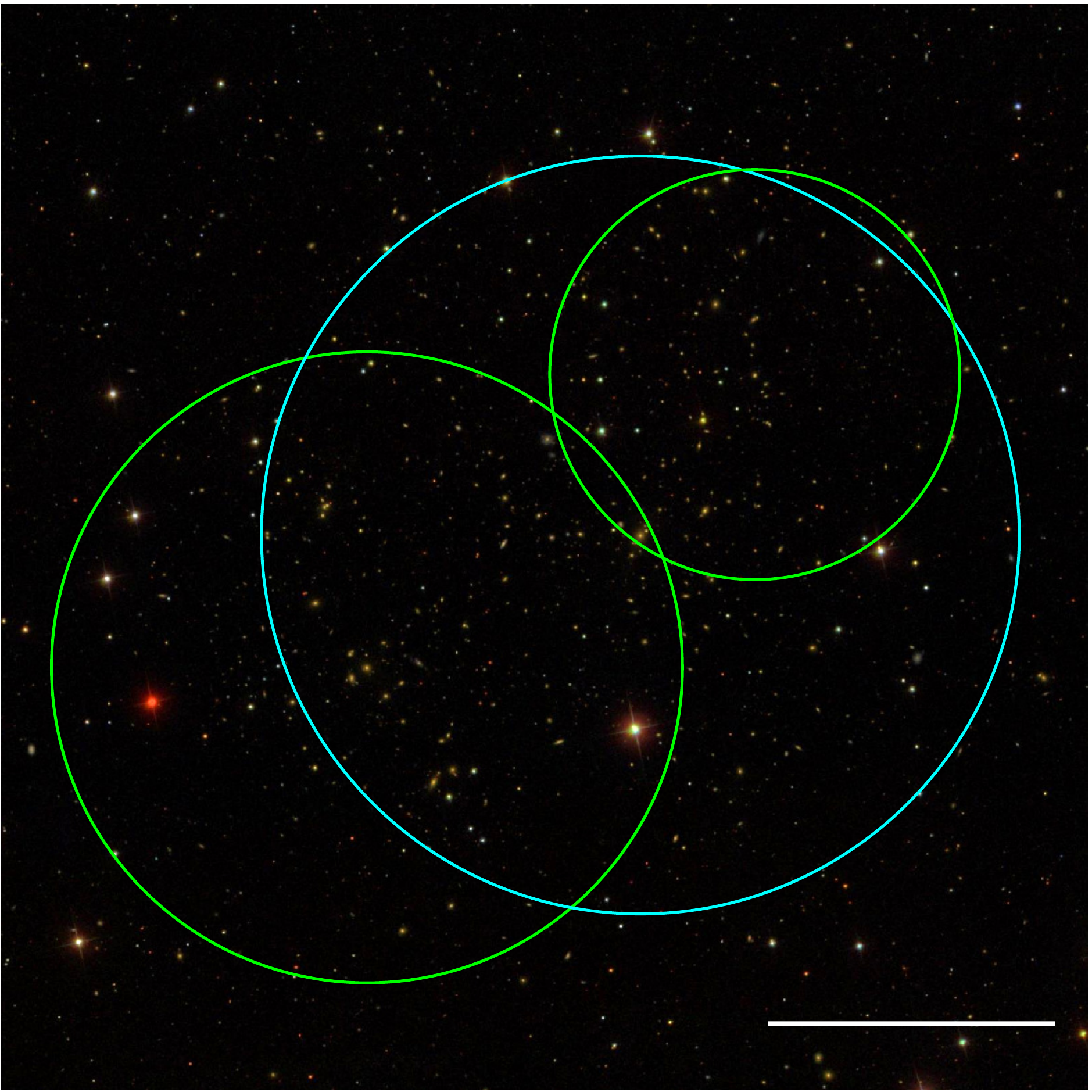}
\includegraphics[width=0.246\textwidth,keepaspectratio=true,clip=true]{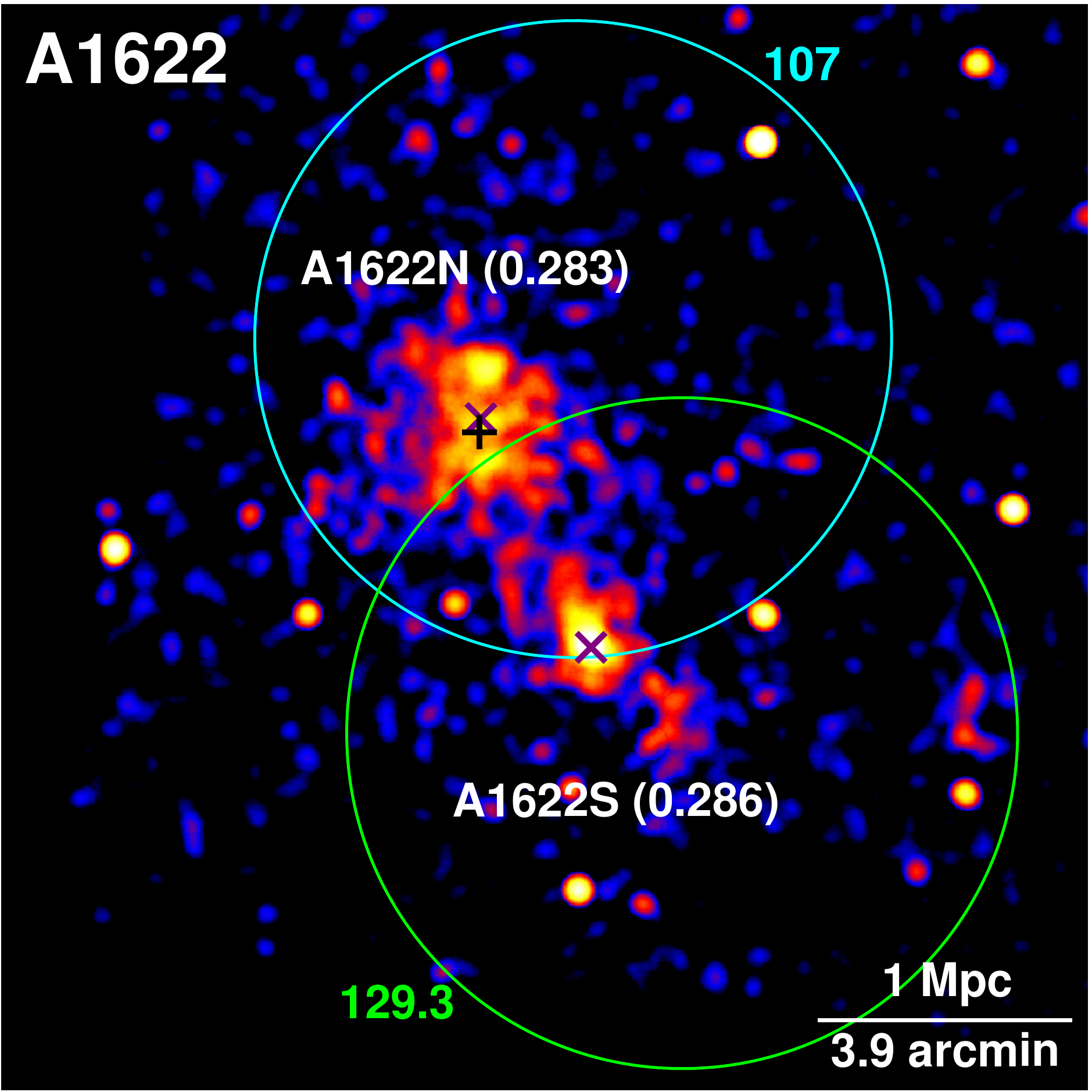}
\includegraphics[width=0.246\textwidth,keepaspectratio=true,clip=true]{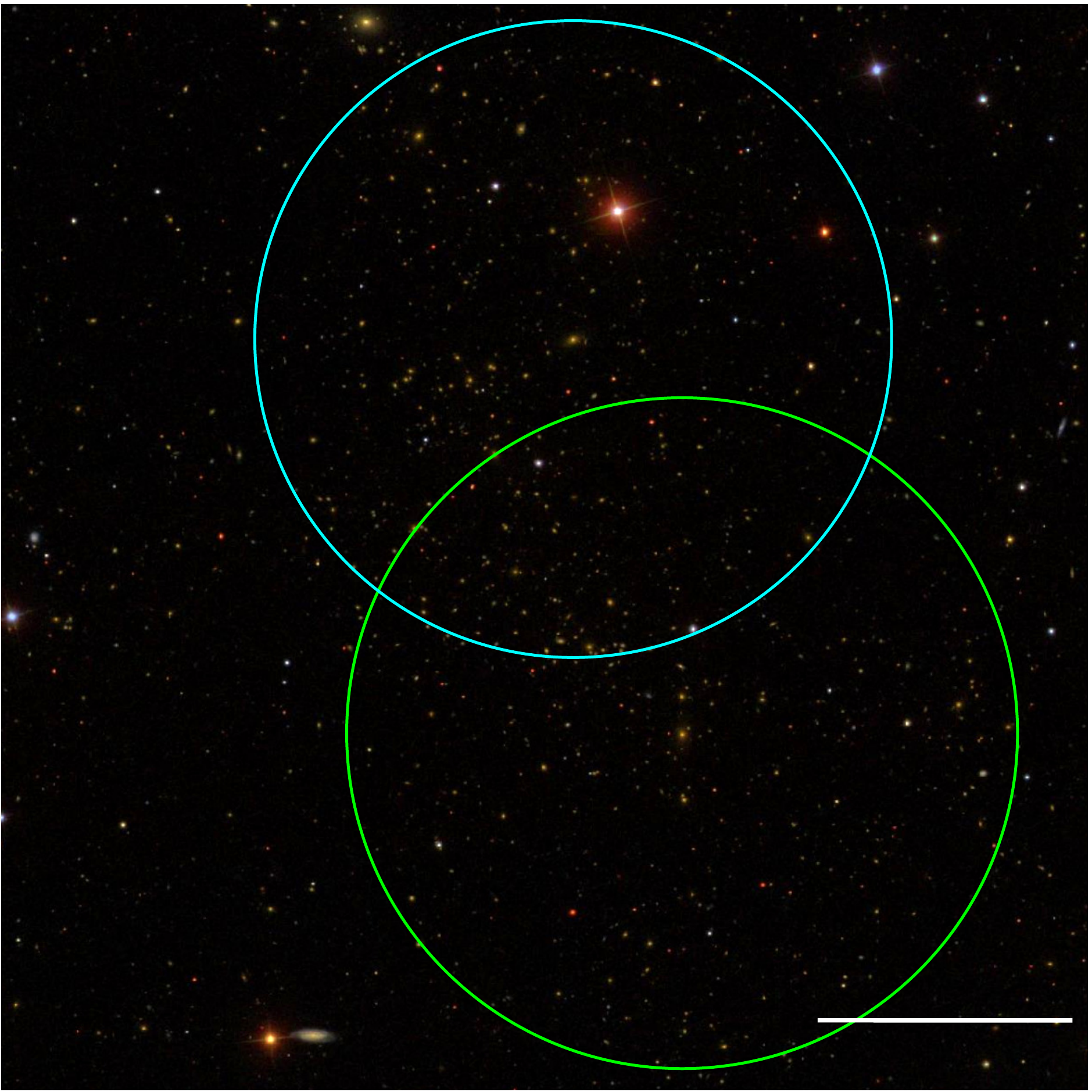}
 	\end{center}
\end{figure*}

\begin{figure*}\ContinuedFloat
 	\begin{center}
\includegraphics[width=0.246\textwidth,keepaspectratio=true,clip=true]{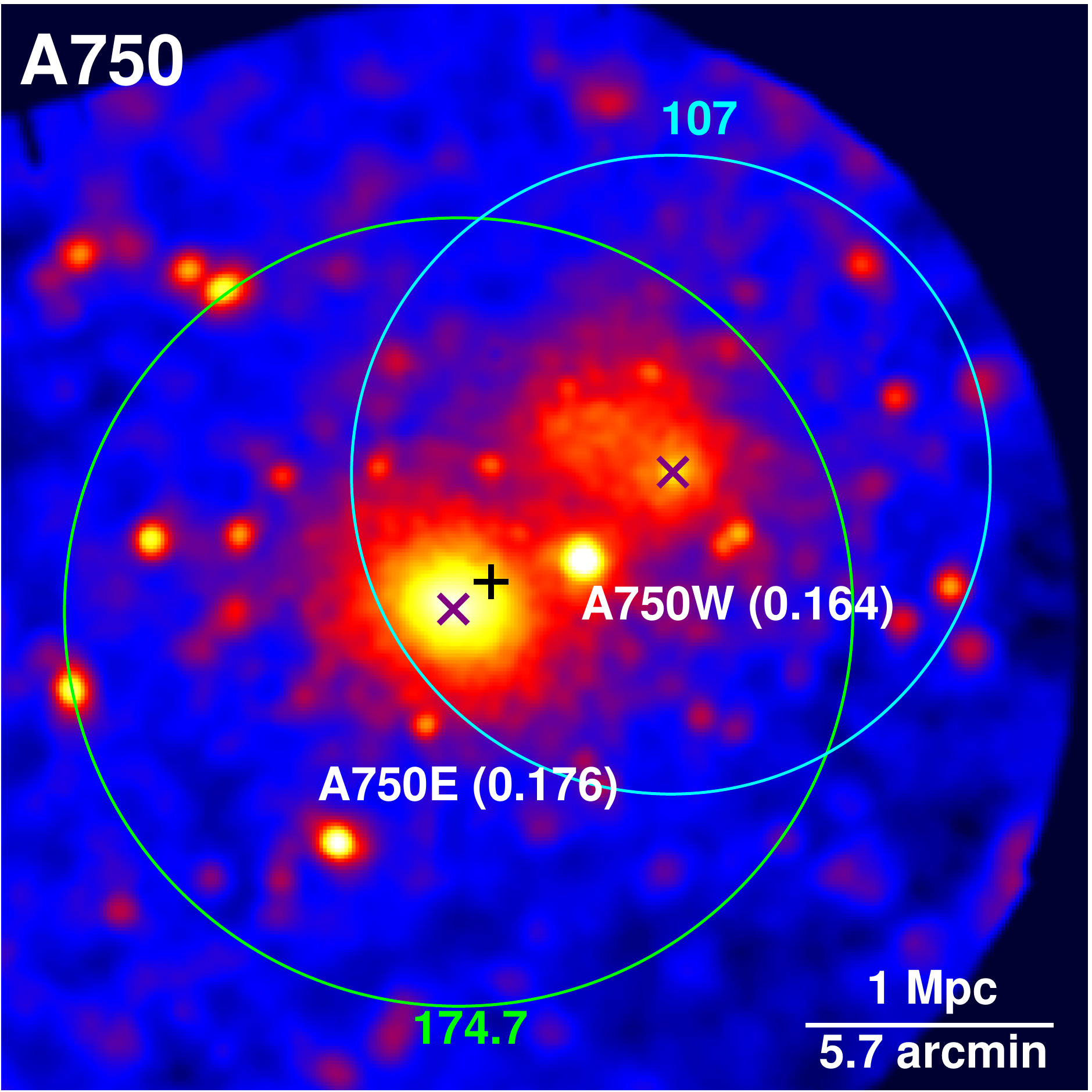}
\includegraphics[width=0.246\textwidth,keepaspectratio=true,clip=true]{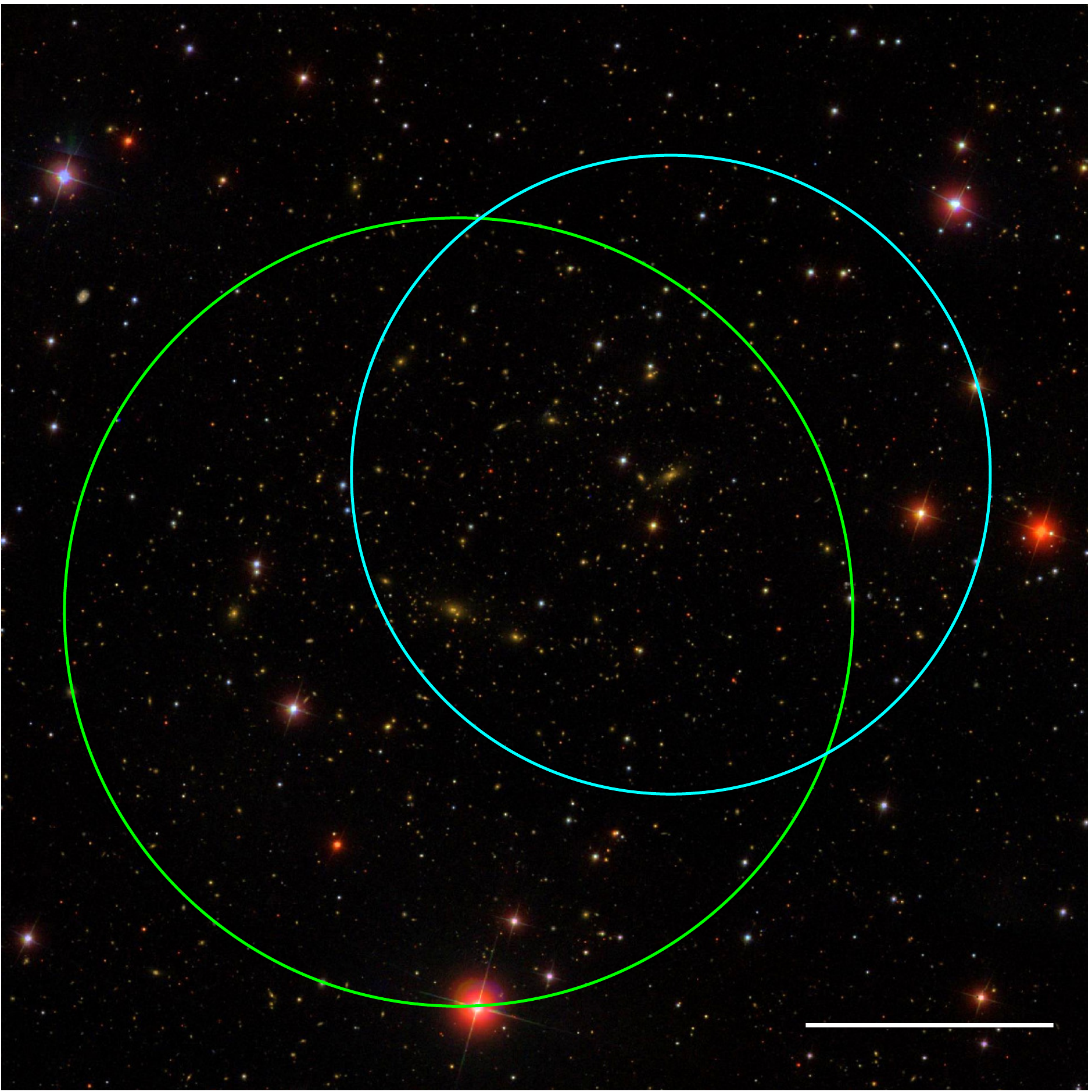}
\includegraphics[width=0.246\textwidth,keepaspectratio=true,clip=true]{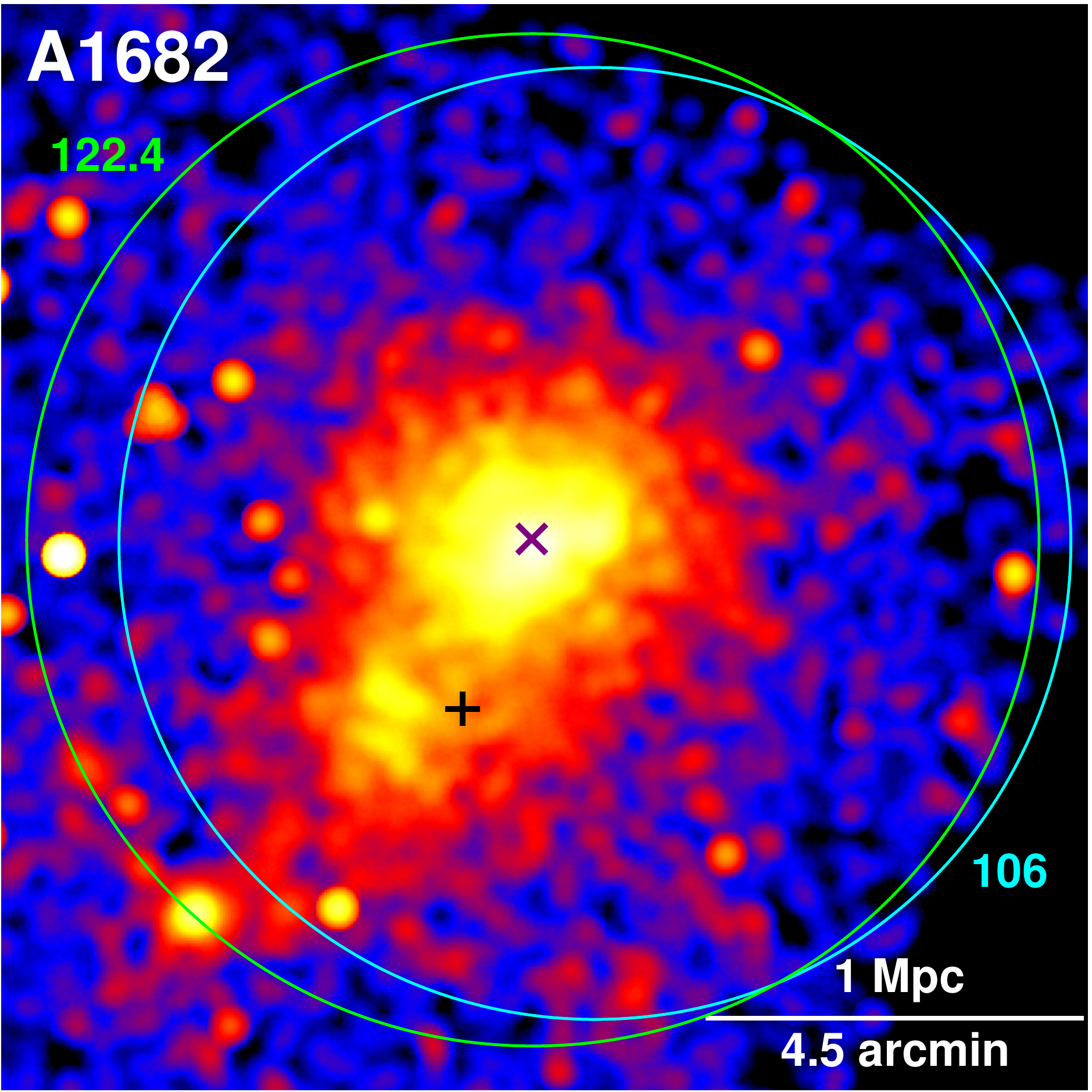}
\includegraphics[width=0.246\textwidth,keepaspectratio=true,clip=true]{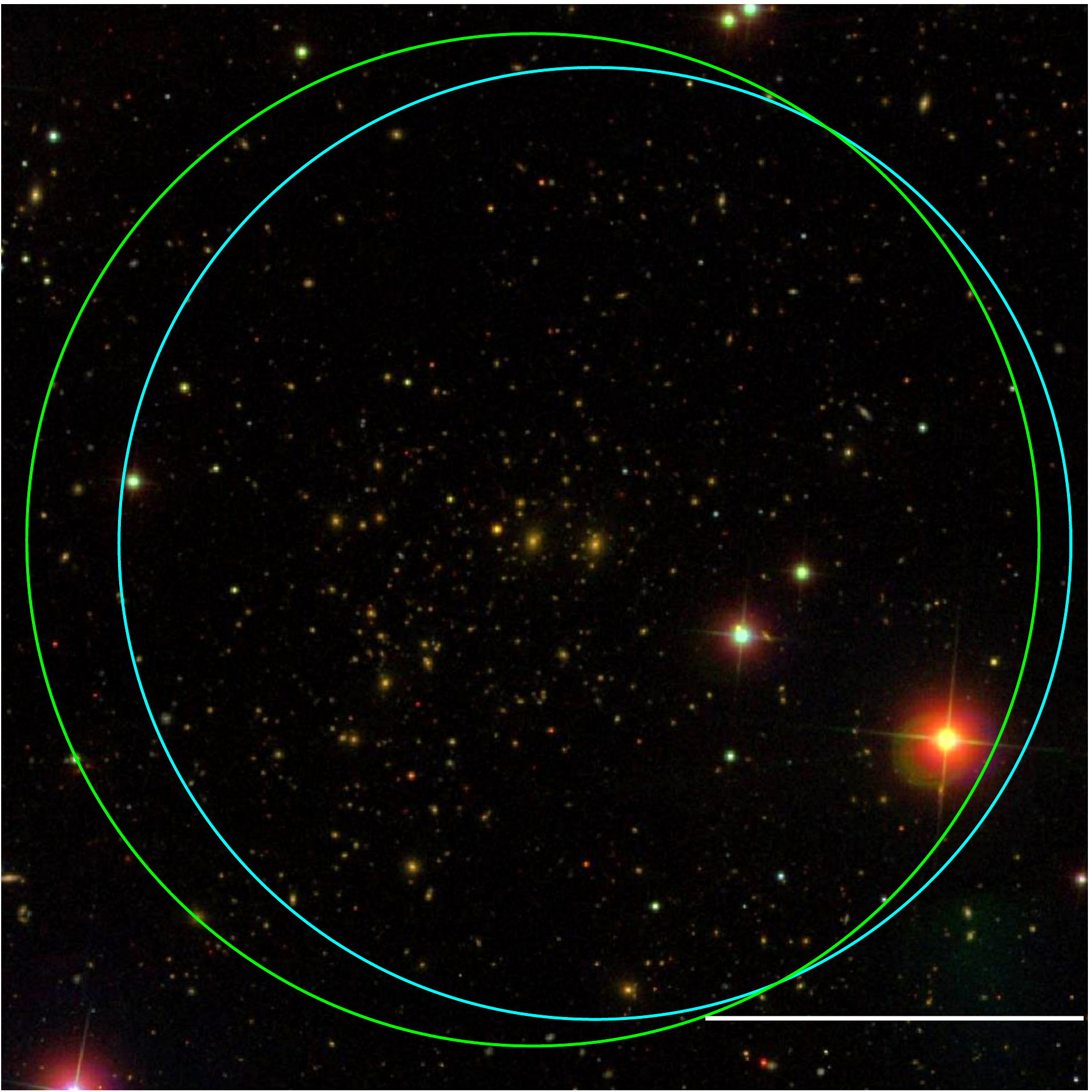}
\includegraphics[width=0.246\textwidth,keepaspectratio=true,clip=true]{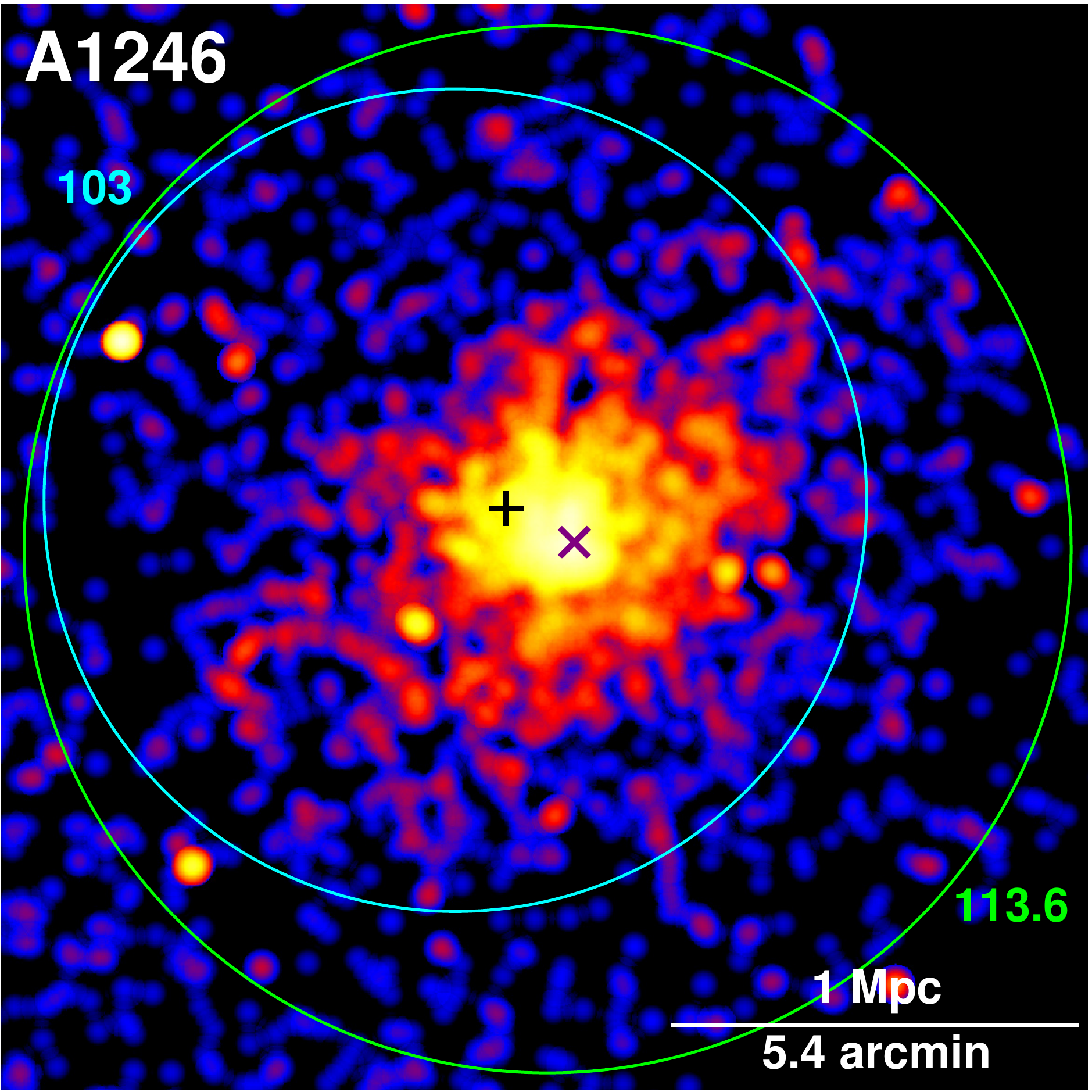}
\includegraphics[width=0.246\textwidth,keepaspectratio=true,clip=true]{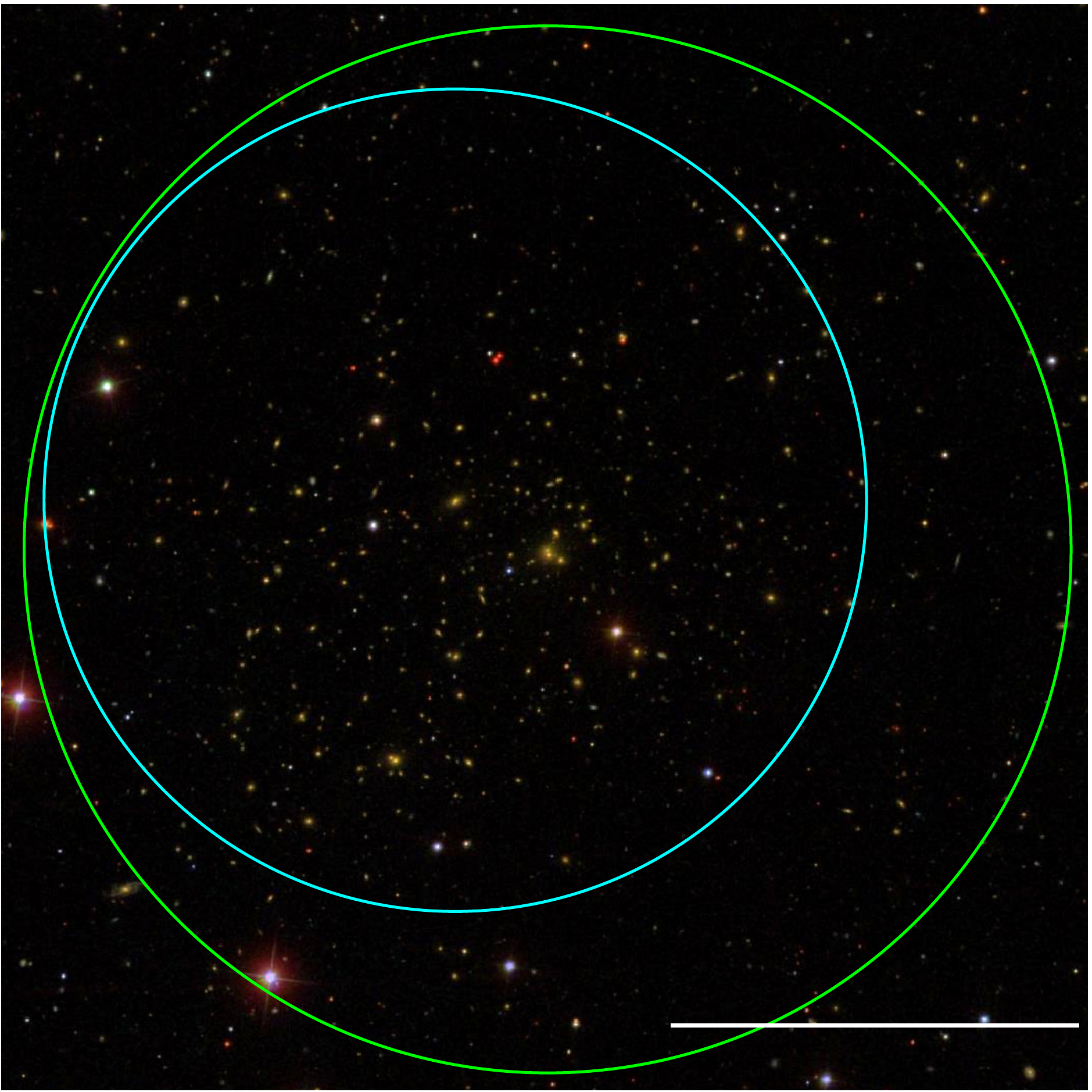}
\includegraphics[width=0.246\textwidth,keepaspectratio=true,clip=true]{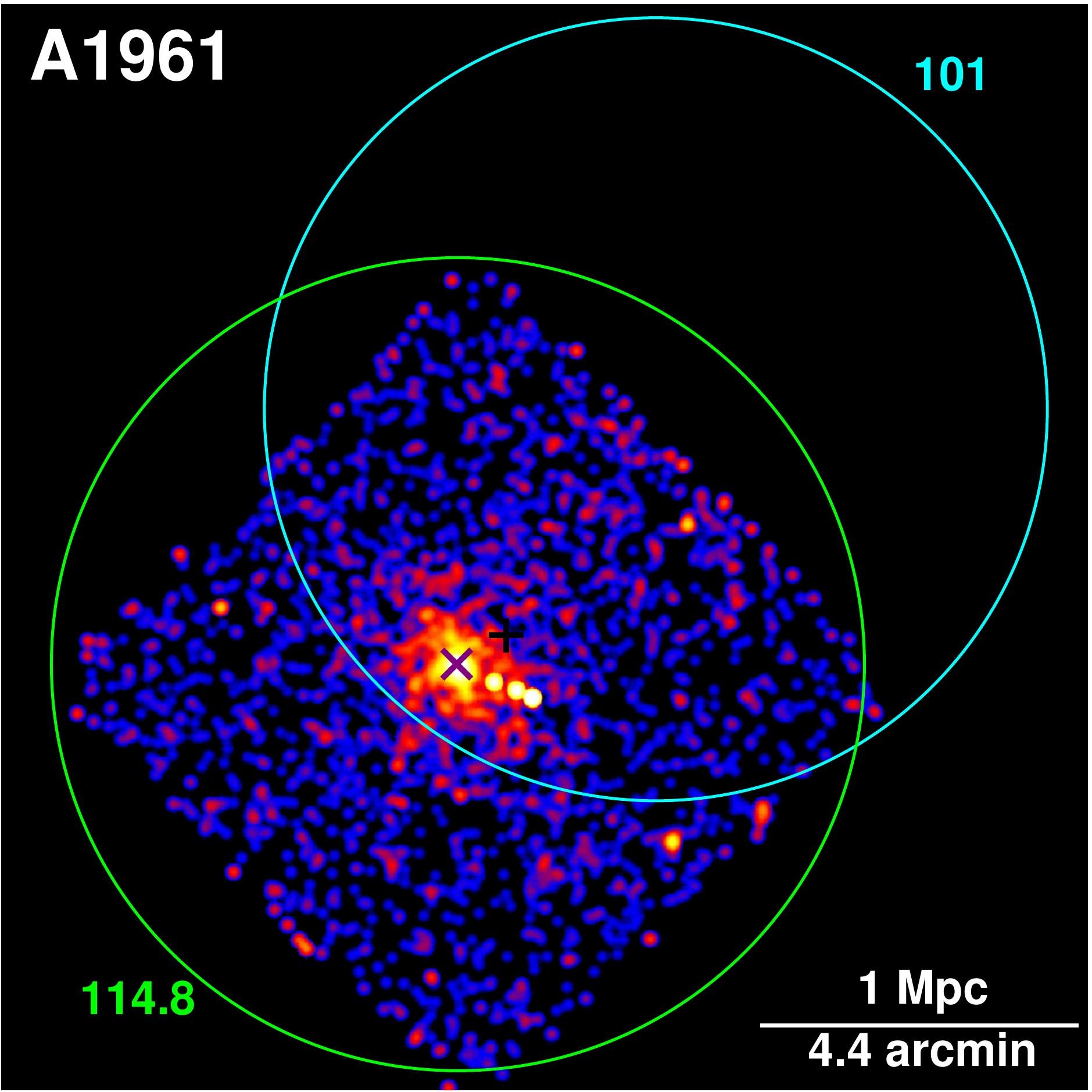}
\includegraphics[width=0.246\textwidth,keepaspectratio=true,clip=true]{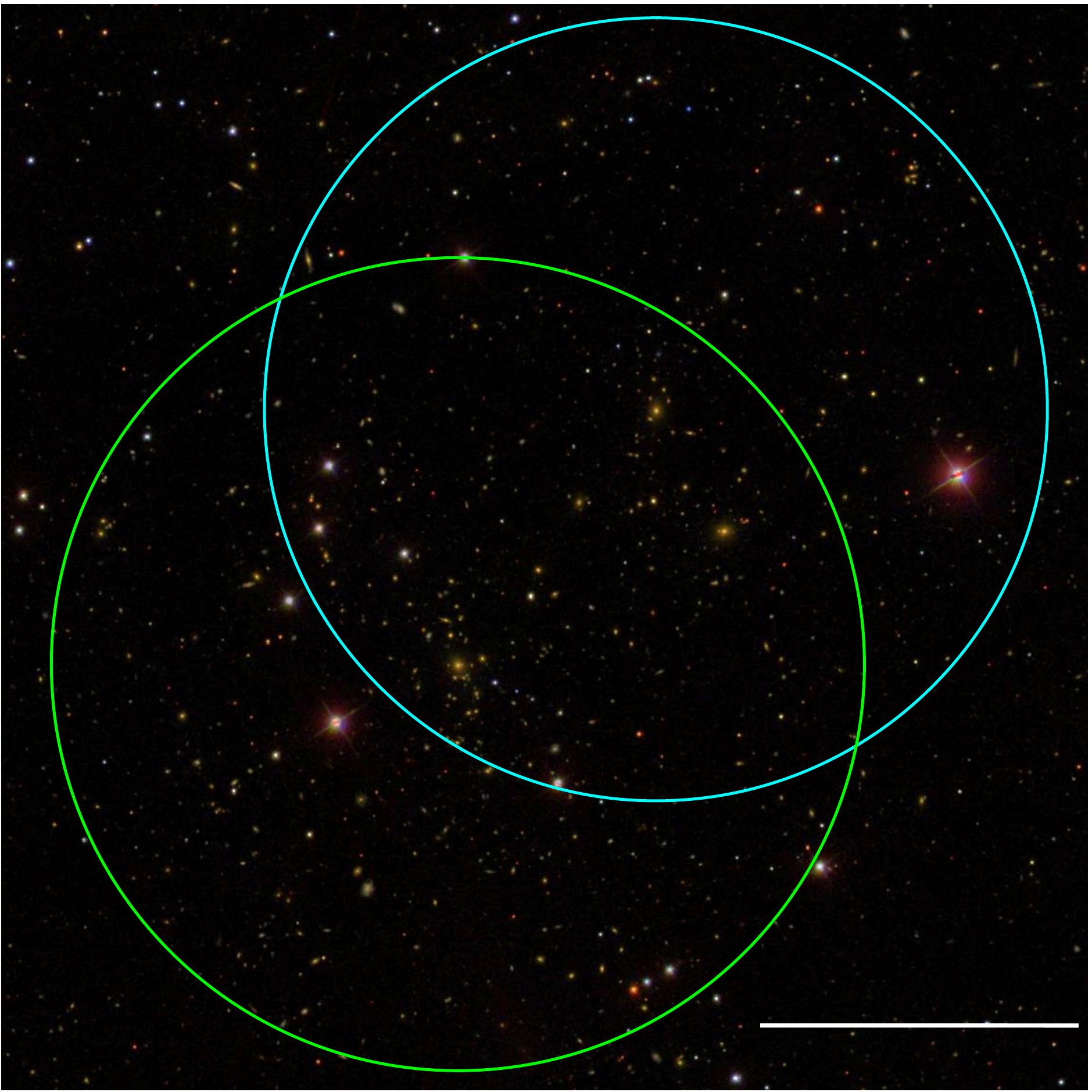}
\includegraphics[width=0.246\textwidth,keepaspectratio=true,clip=true]{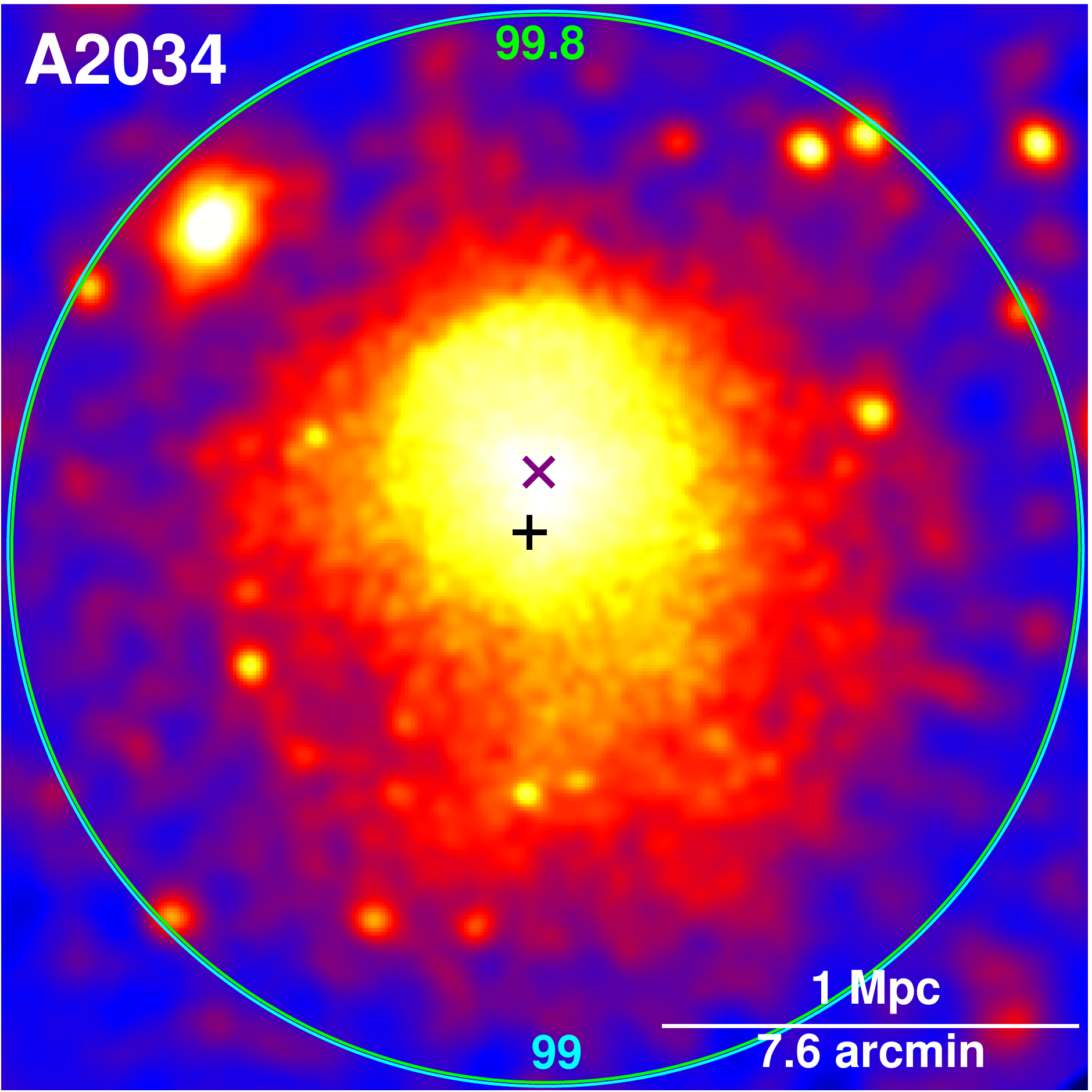}
\includegraphics[width=0.246\textwidth,keepaspectratio=true,clip=true]{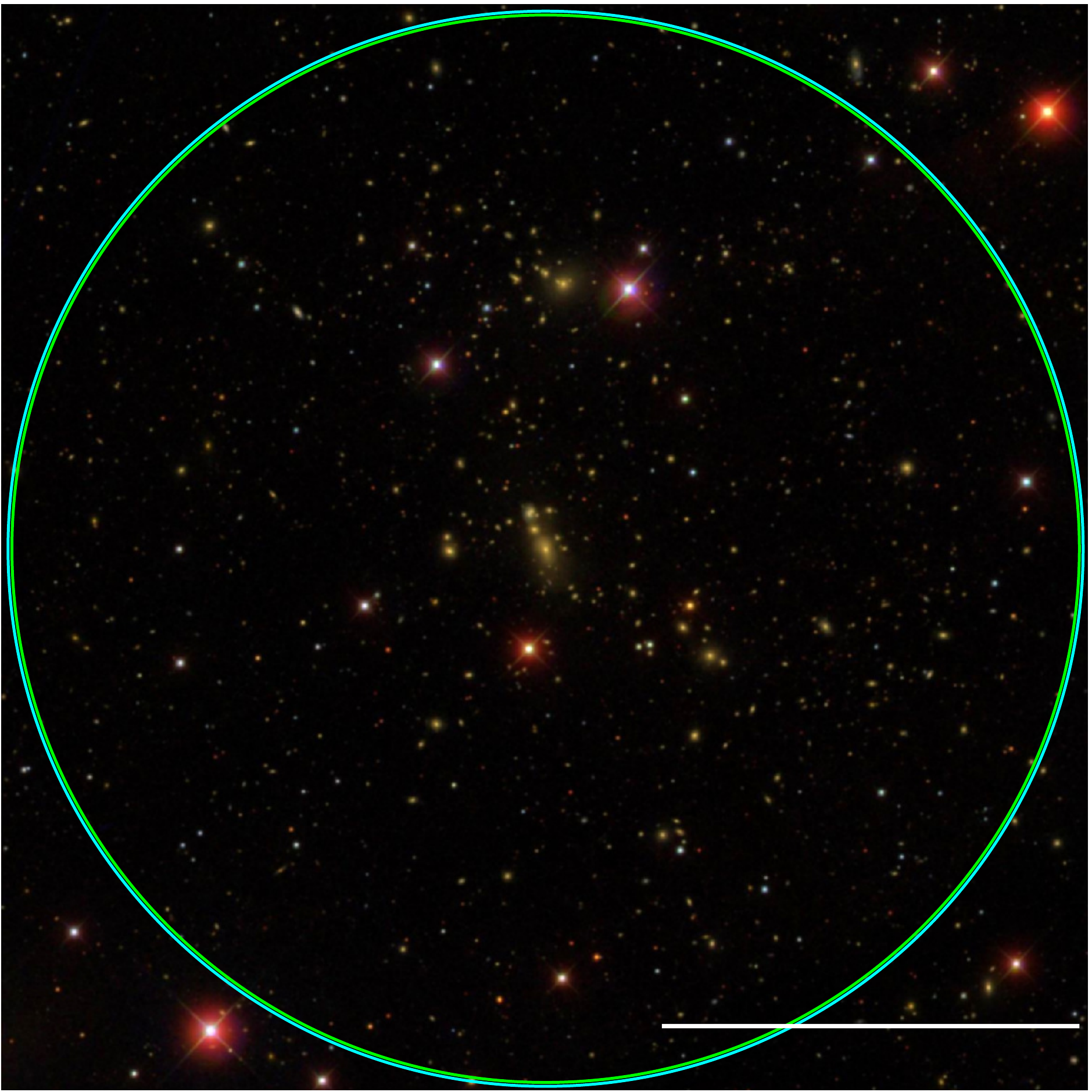}
\includegraphics[width=0.246\textwidth,keepaspectratio=true,clip=true]{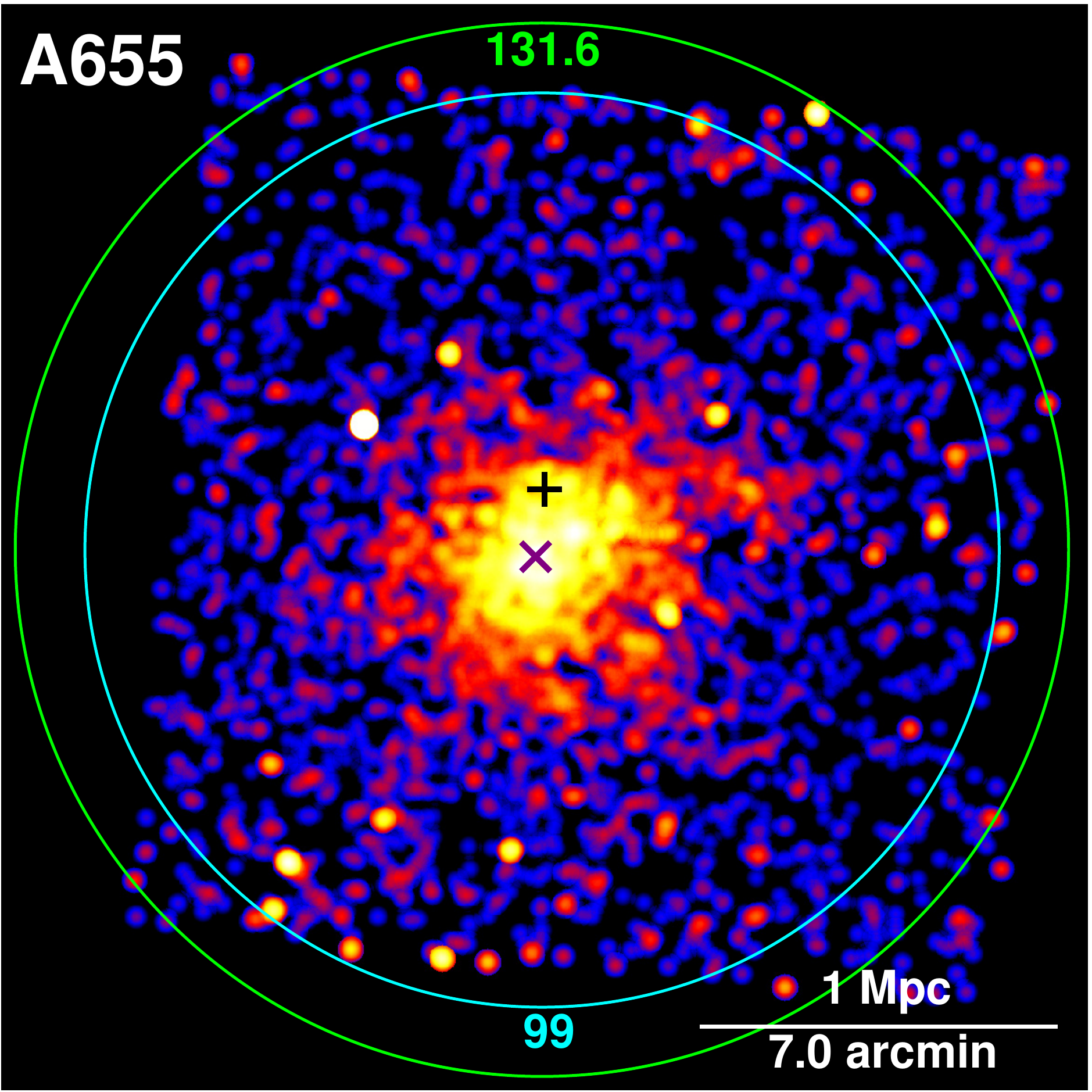}
\includegraphics[width=0.246\textwidth,keepaspectratio=true,clip=true]{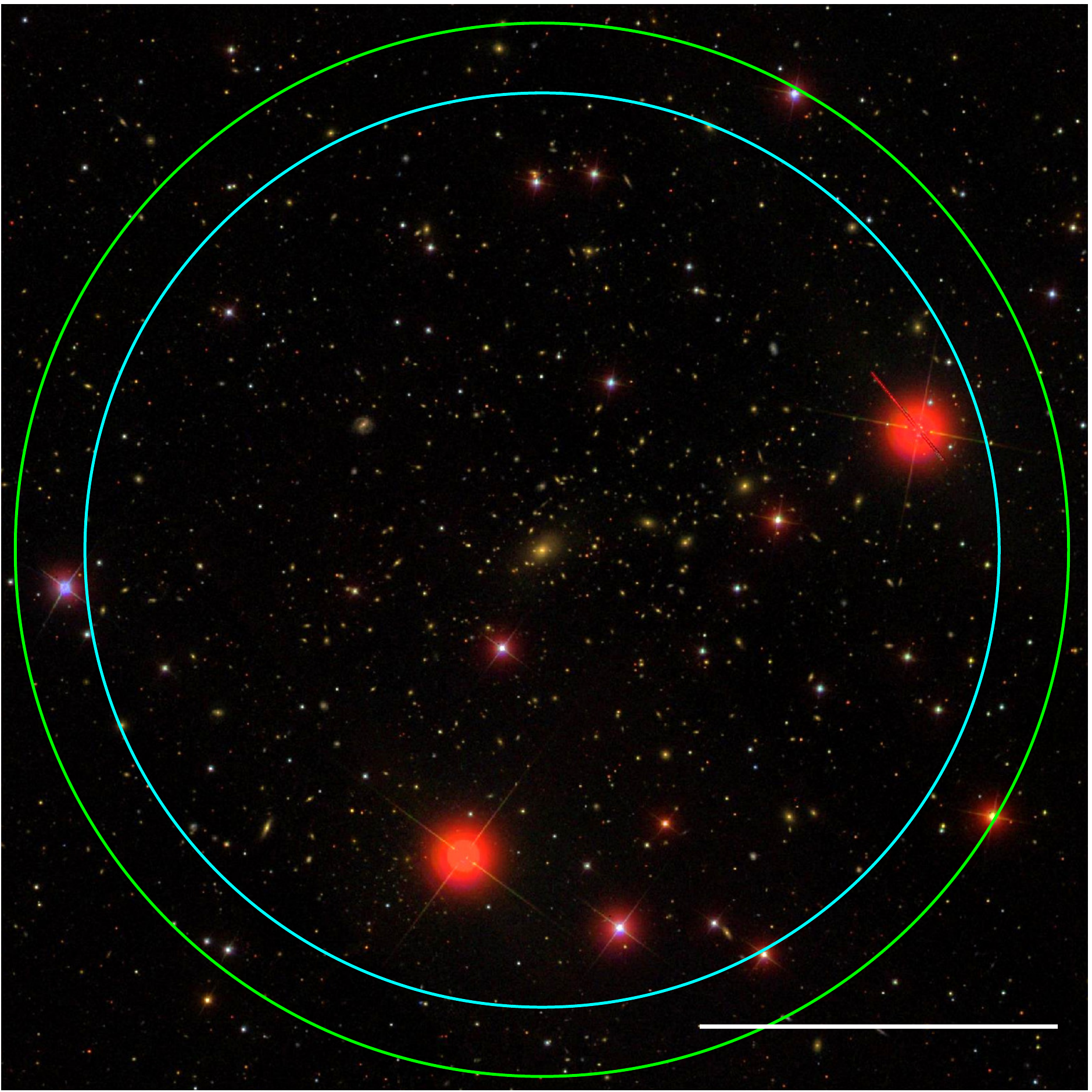}
\includegraphics[width=0.246\textwidth,keepaspectratio=true,clip=true]{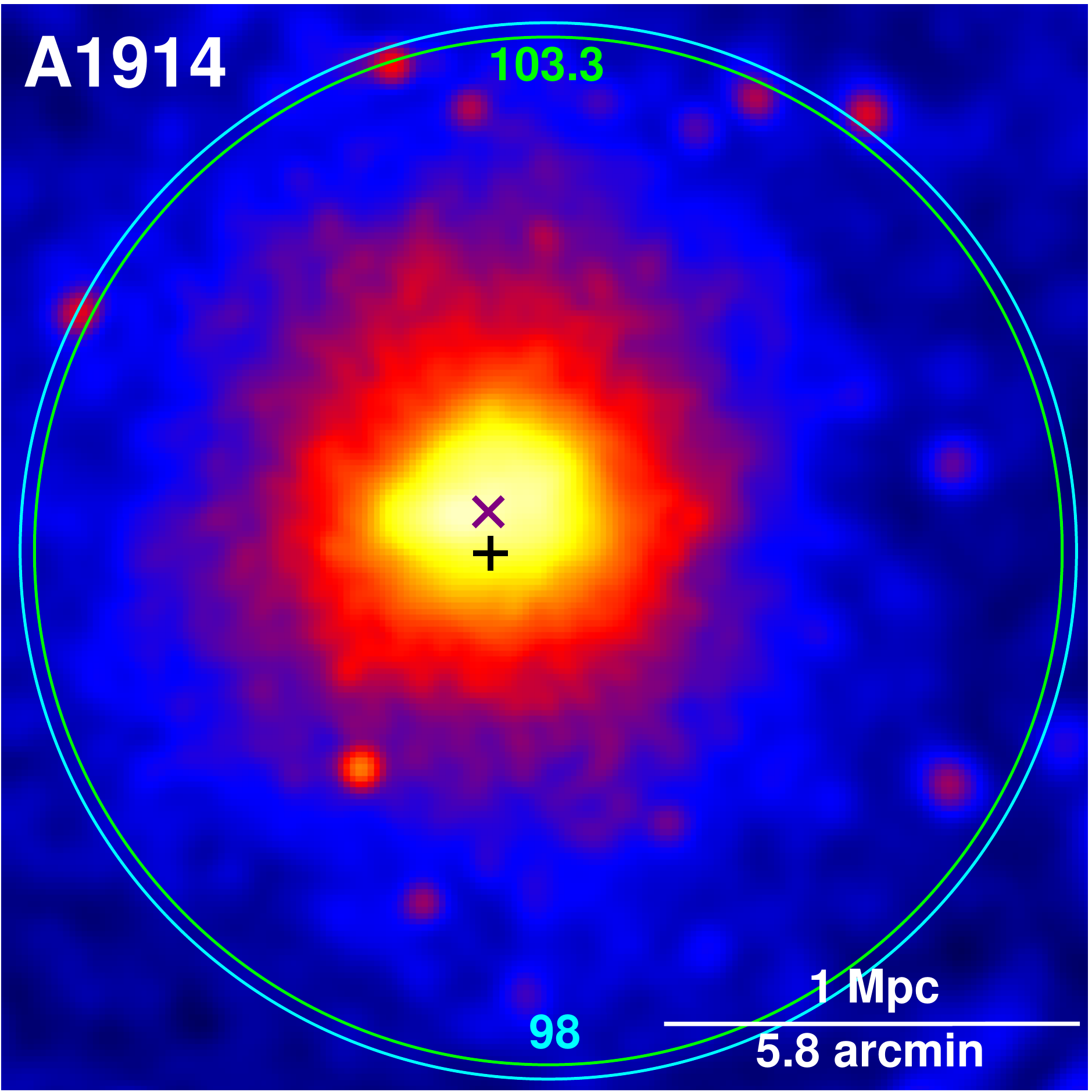}
\includegraphics[width=0.246\textwidth,keepaspectratio=true,clip=true]{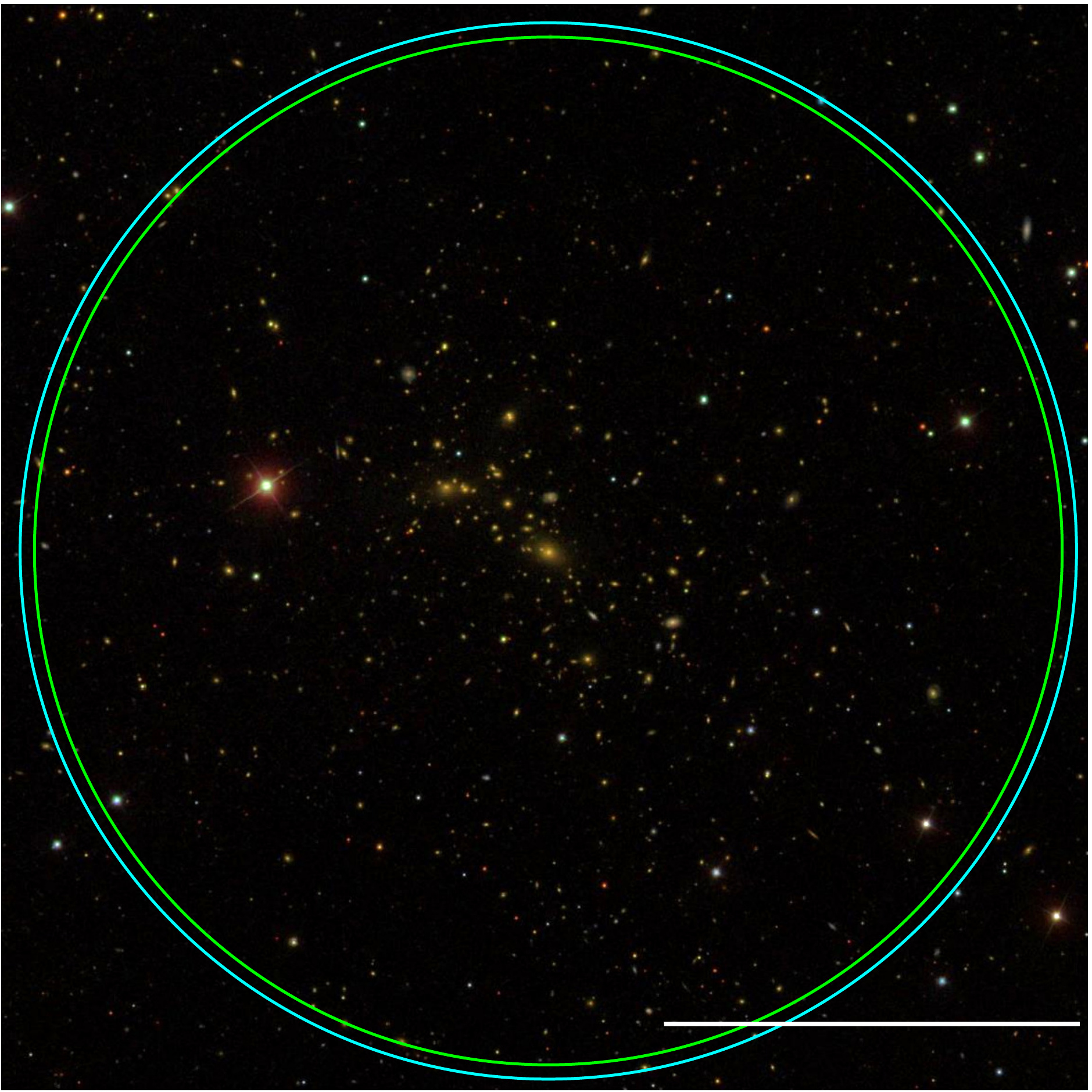}
\includegraphics[width=0.246\textwidth,keepaspectratio=true,clip=true]{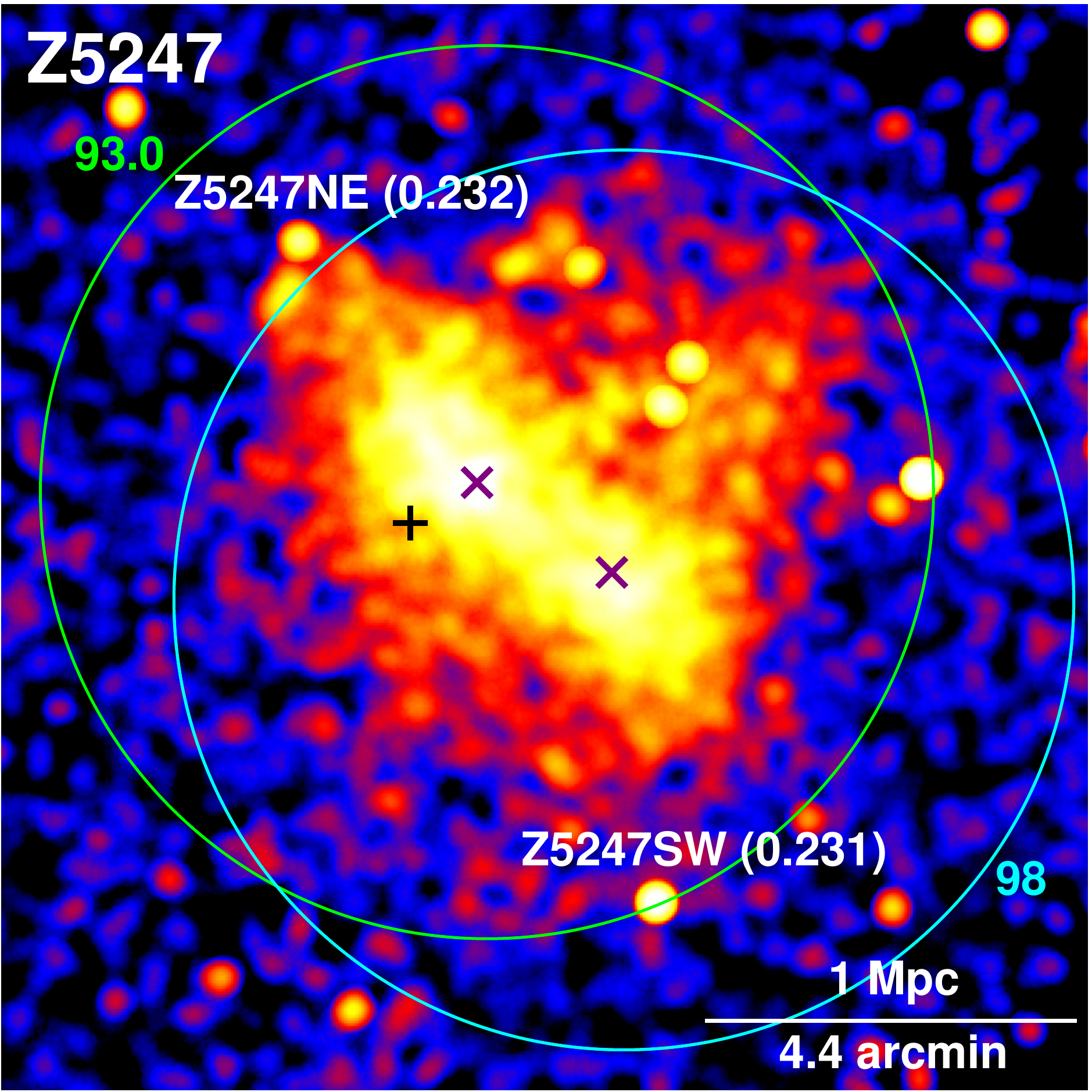}
\includegraphics[width=0.246\textwidth,keepaspectratio=true,clip=true]{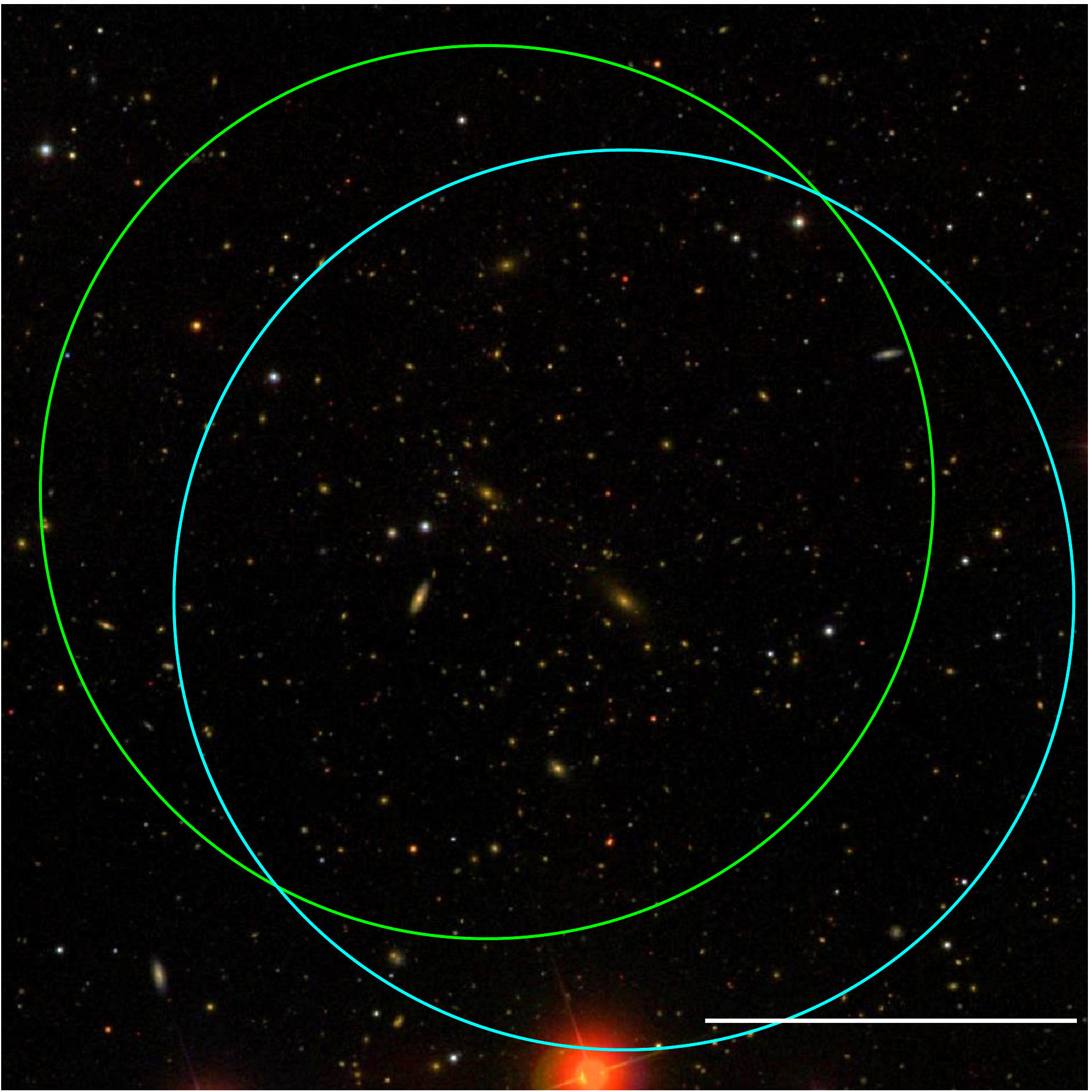}
\includegraphics[width=0.246\textwidth,keepaspectratio=true,clip=true]{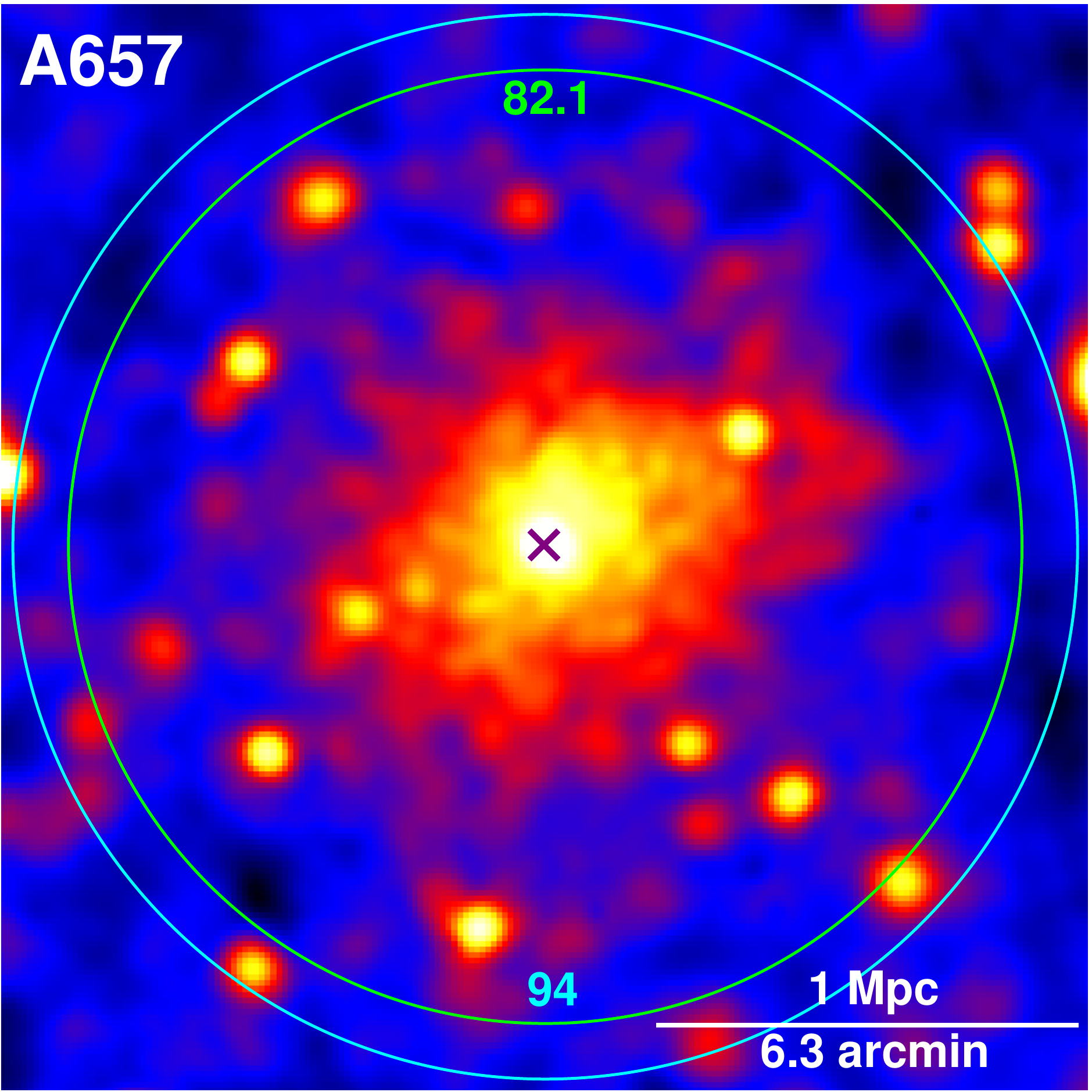}
\includegraphics[width=0.246\textwidth,keepaspectratio=true,clip=true]{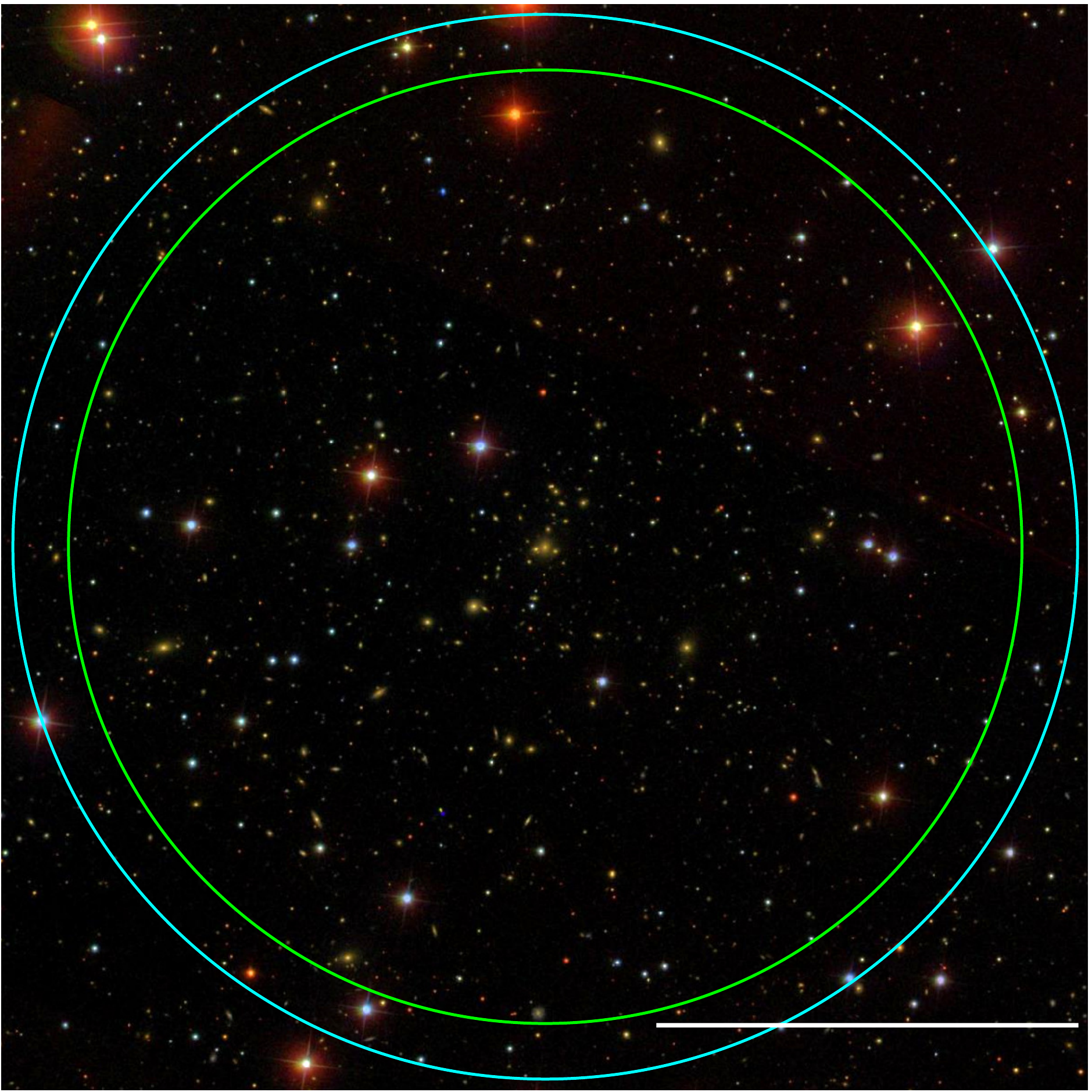}
\includegraphics[width=0.246\textwidth,keepaspectratio=true,clip=true]{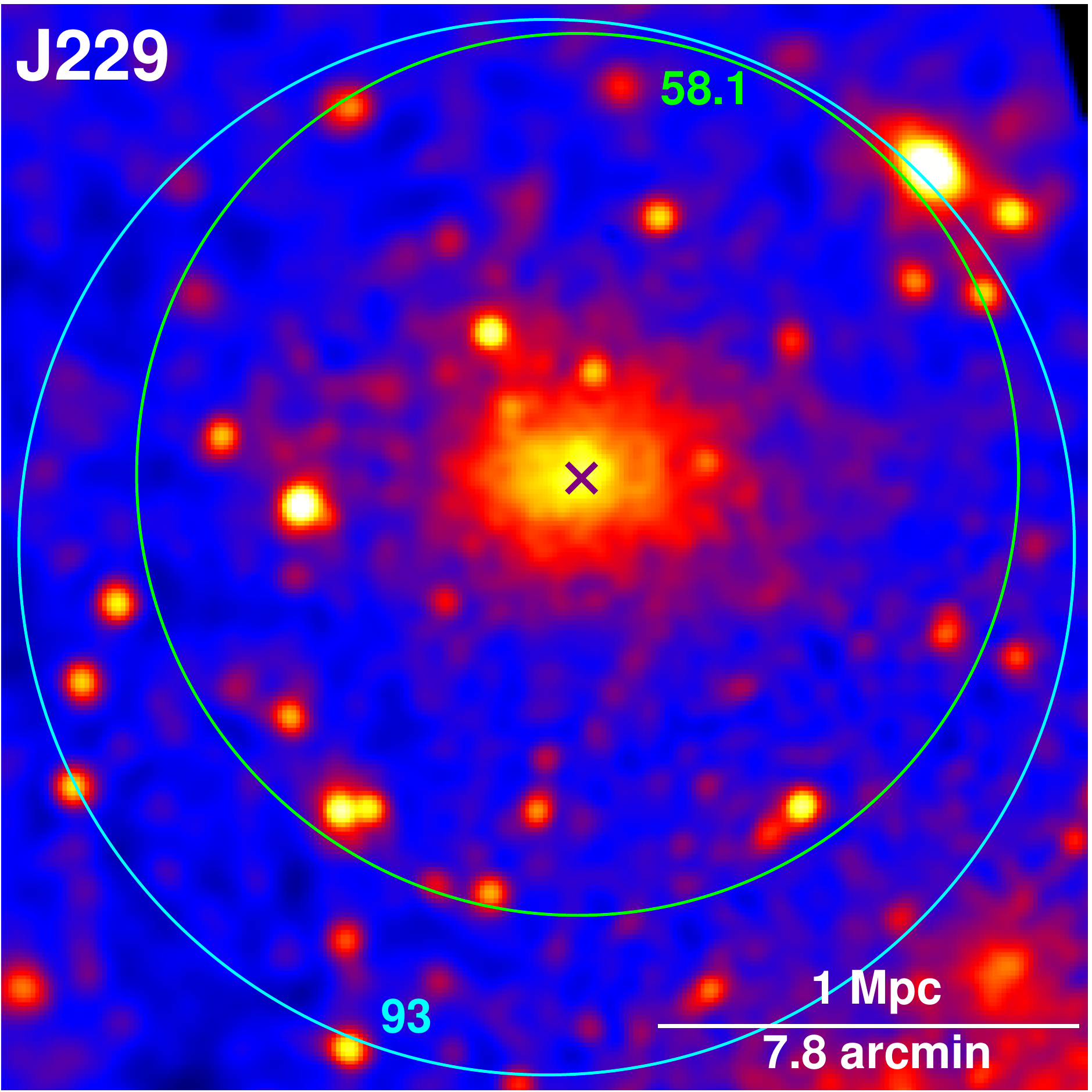}
\includegraphics[width=0.246\textwidth,keepaspectratio=true,clip=true]{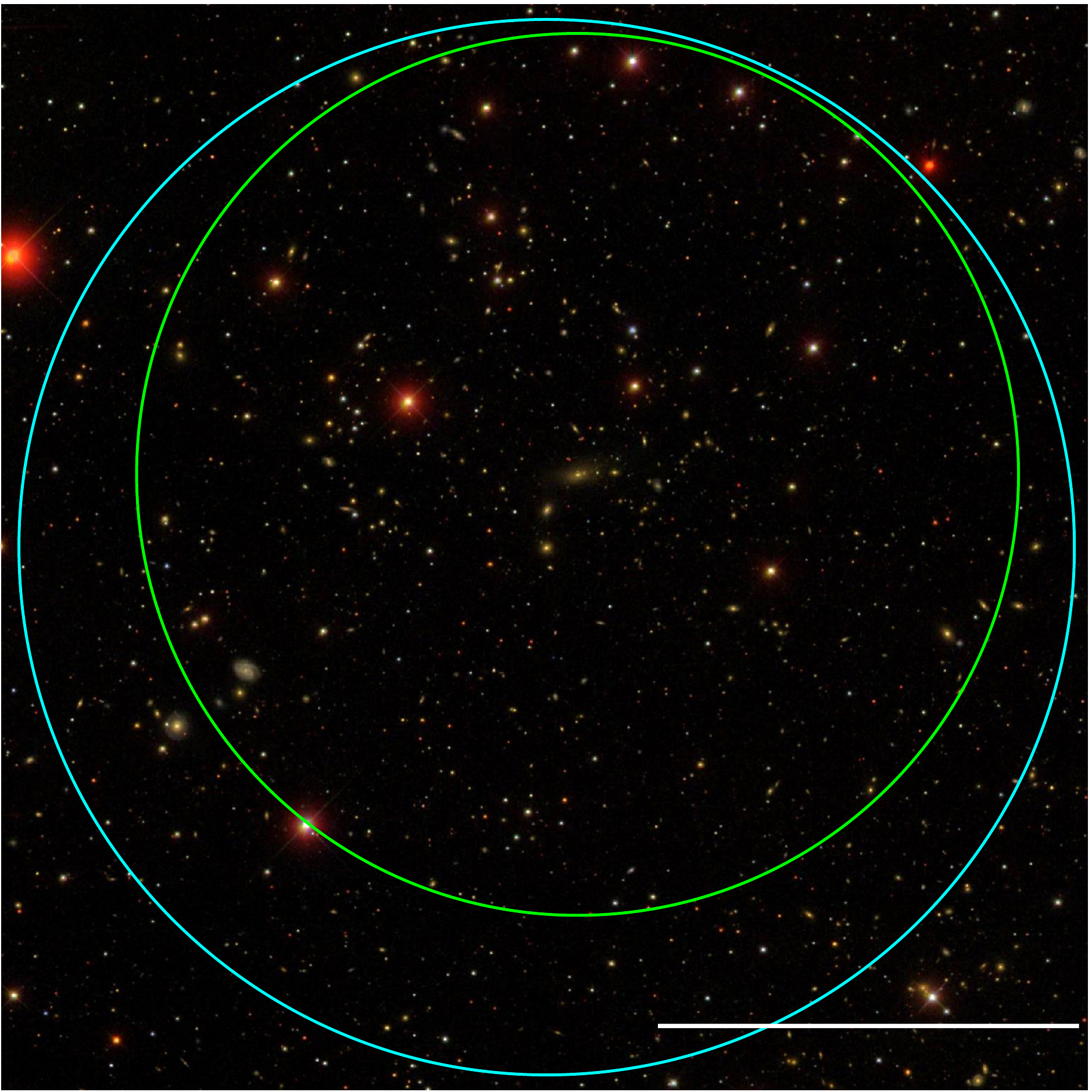}
 	\end{center}
\end{figure*}

\begin{figure*}\ContinuedFloat
 	\begin{center}
\includegraphics[width=0.246\textwidth,keepaspectratio=true,clip=true]{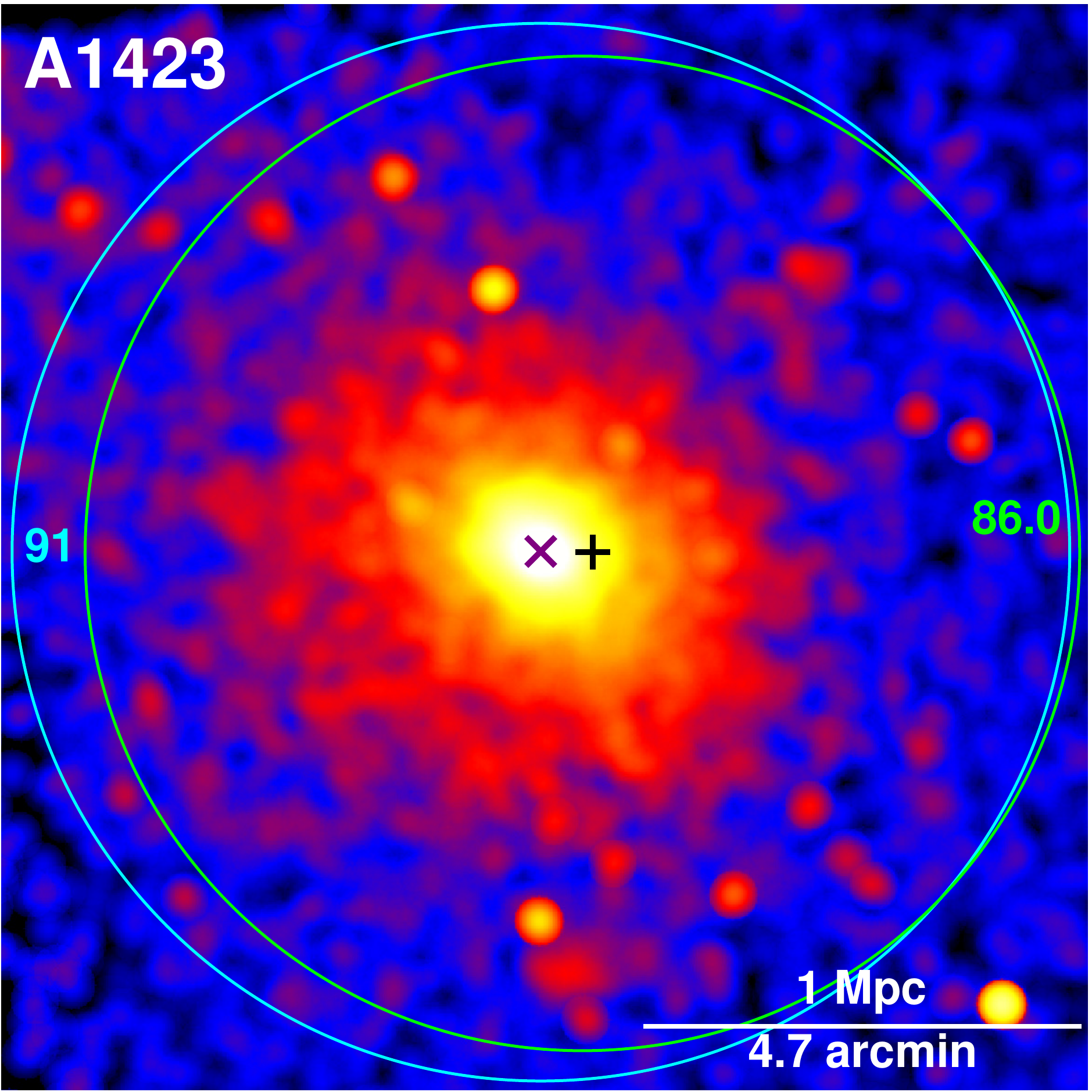}
\includegraphics[width=0.246\textwidth,keepaspectratio=true,clip=true]{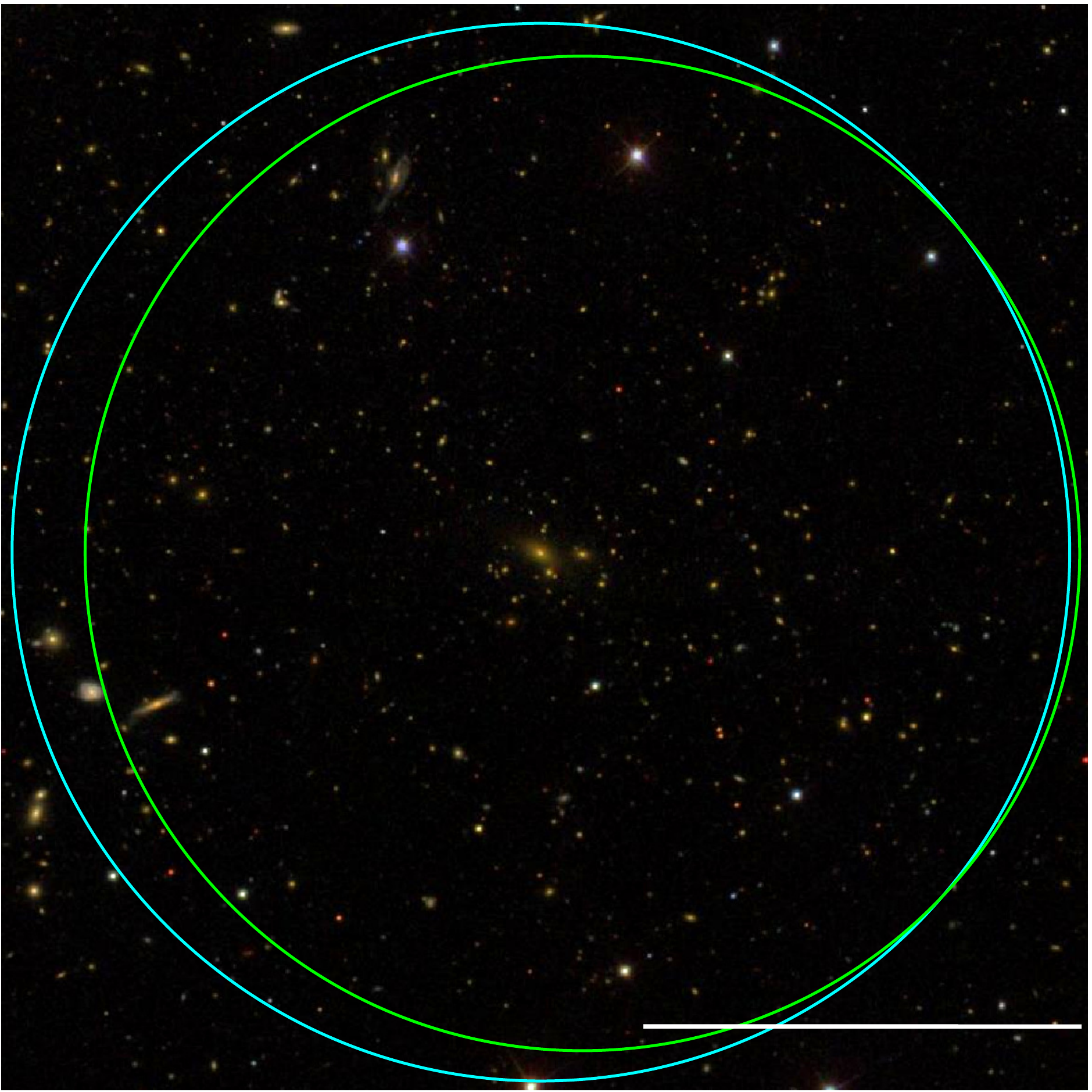}
\includegraphics[width=0.246\textwidth,keepaspectratio=true,clip=true]{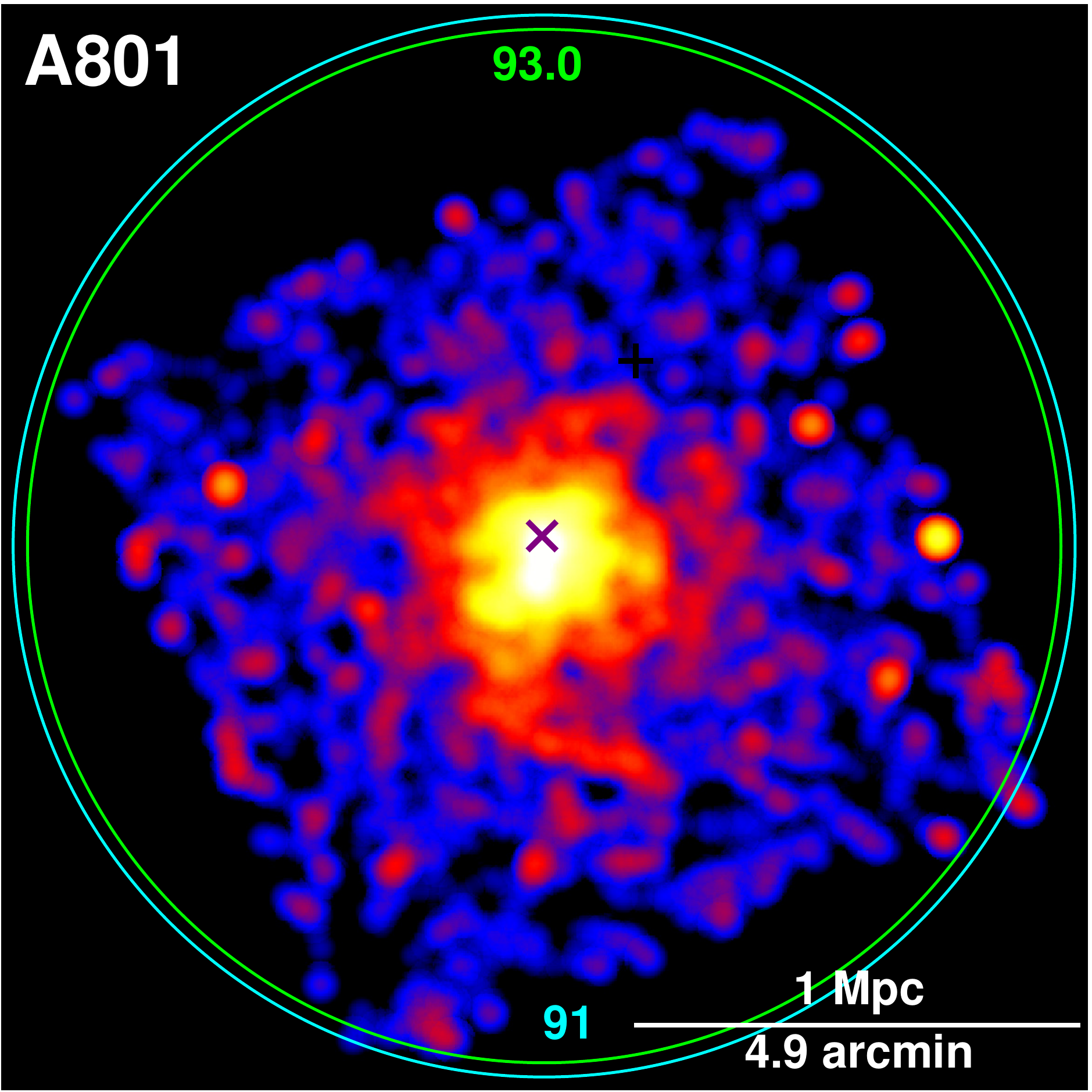}
\includegraphics[width=0.246\textwidth,keepaspectratio=true,clip=true]{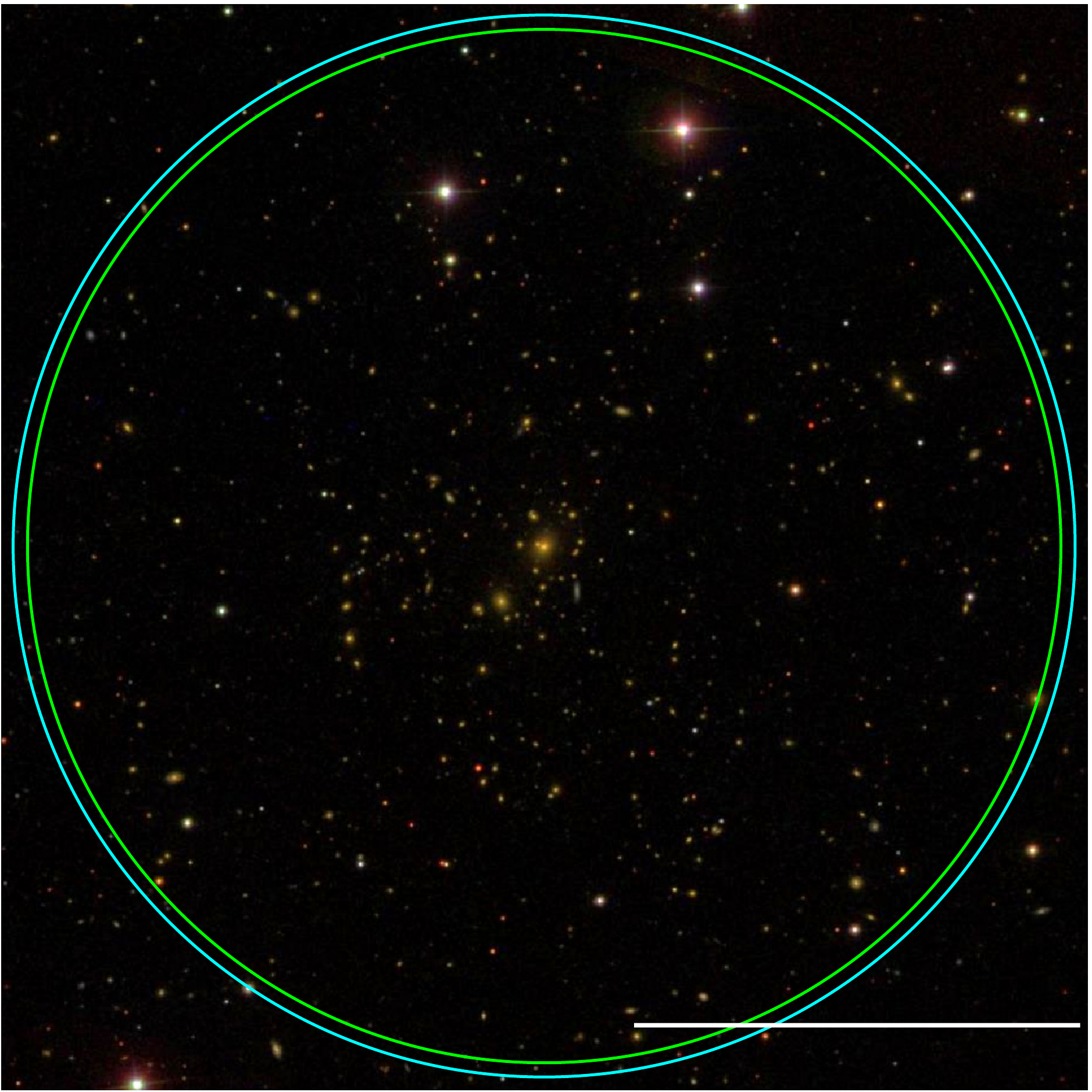}
\includegraphics[width=0.246\textwidth,keepaspectratio=true,clip=true]{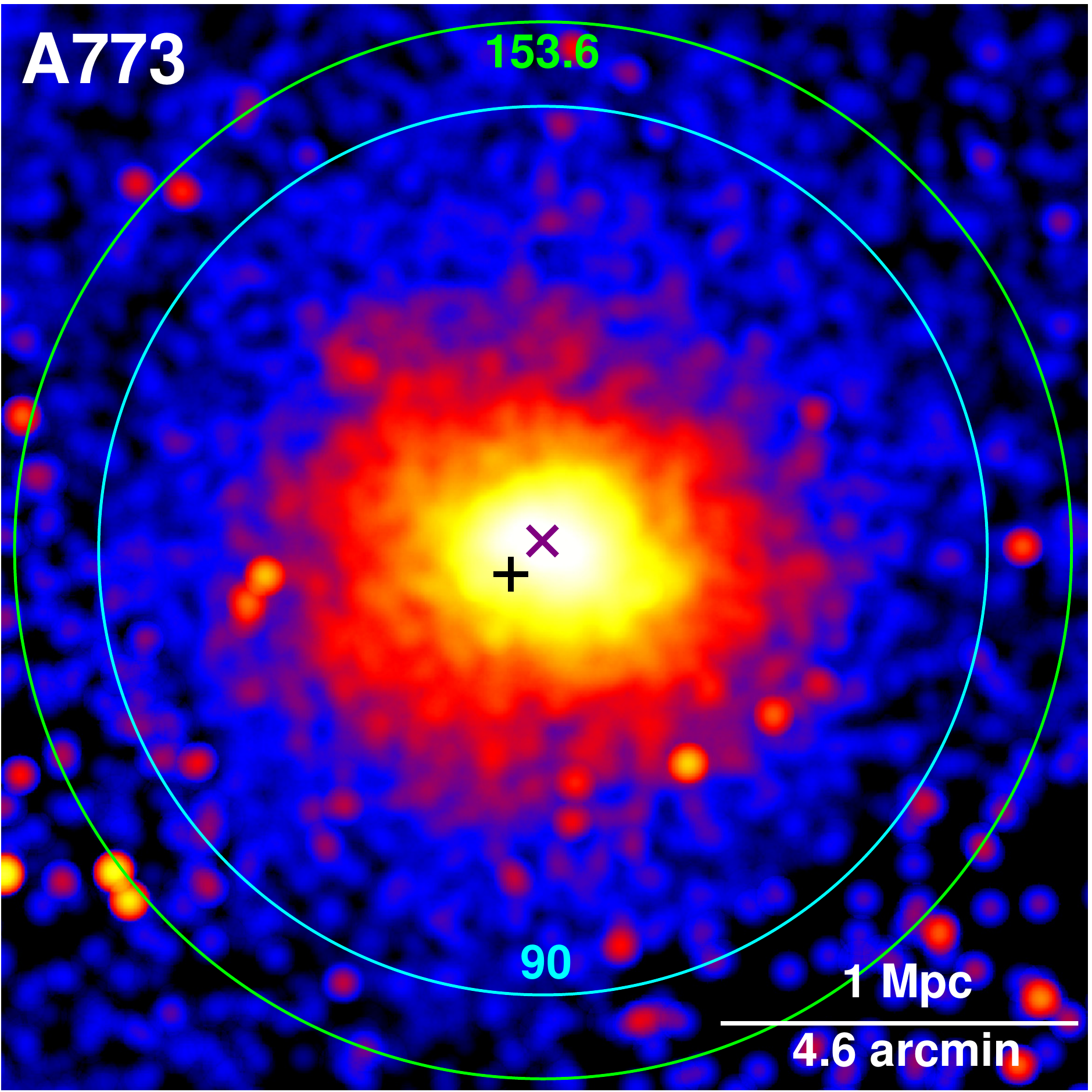}
\includegraphics[width=0.246\textwidth,keepaspectratio=true,clip=true]{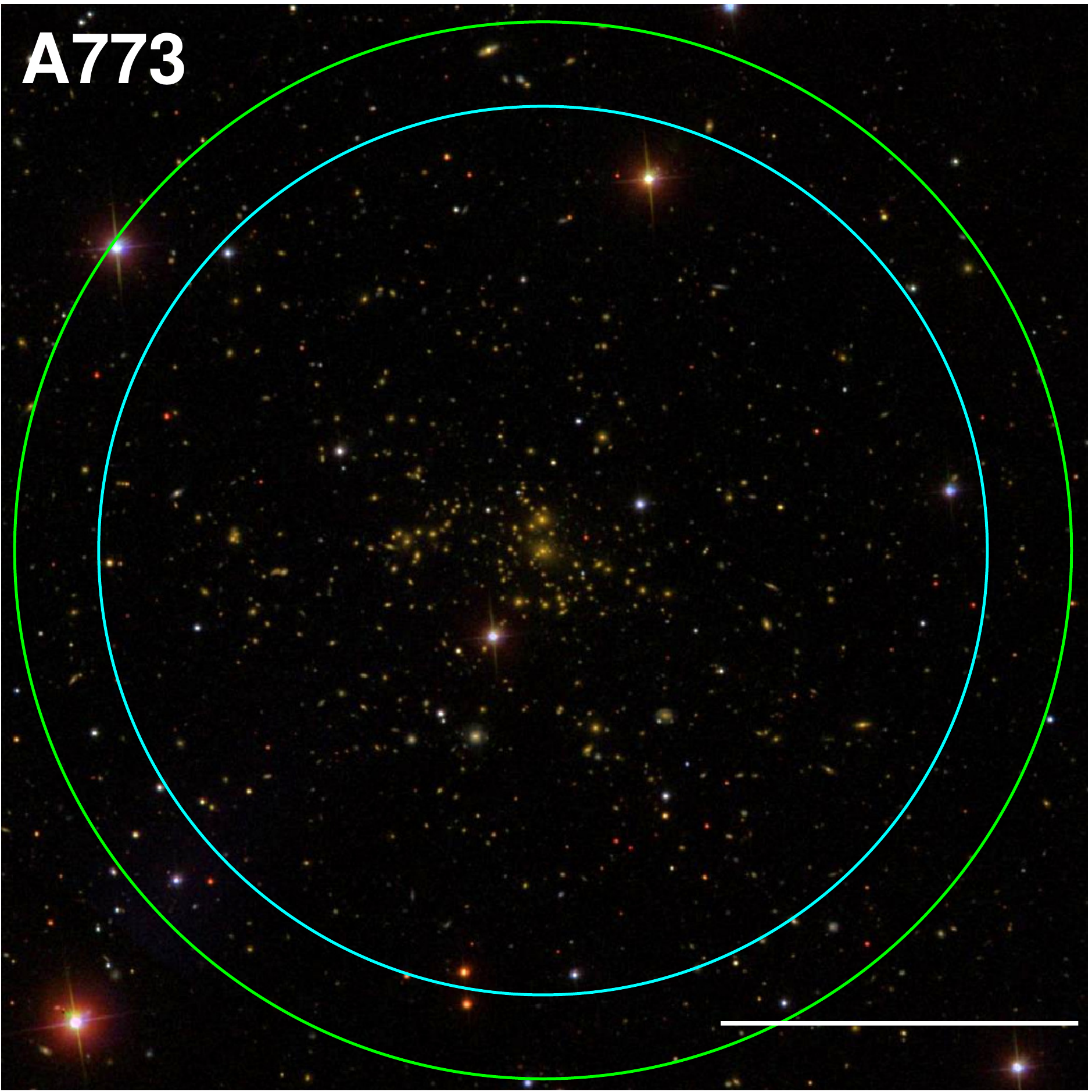}
\includegraphics[width=0.246\textwidth,keepaspectratio=true,clip=true]{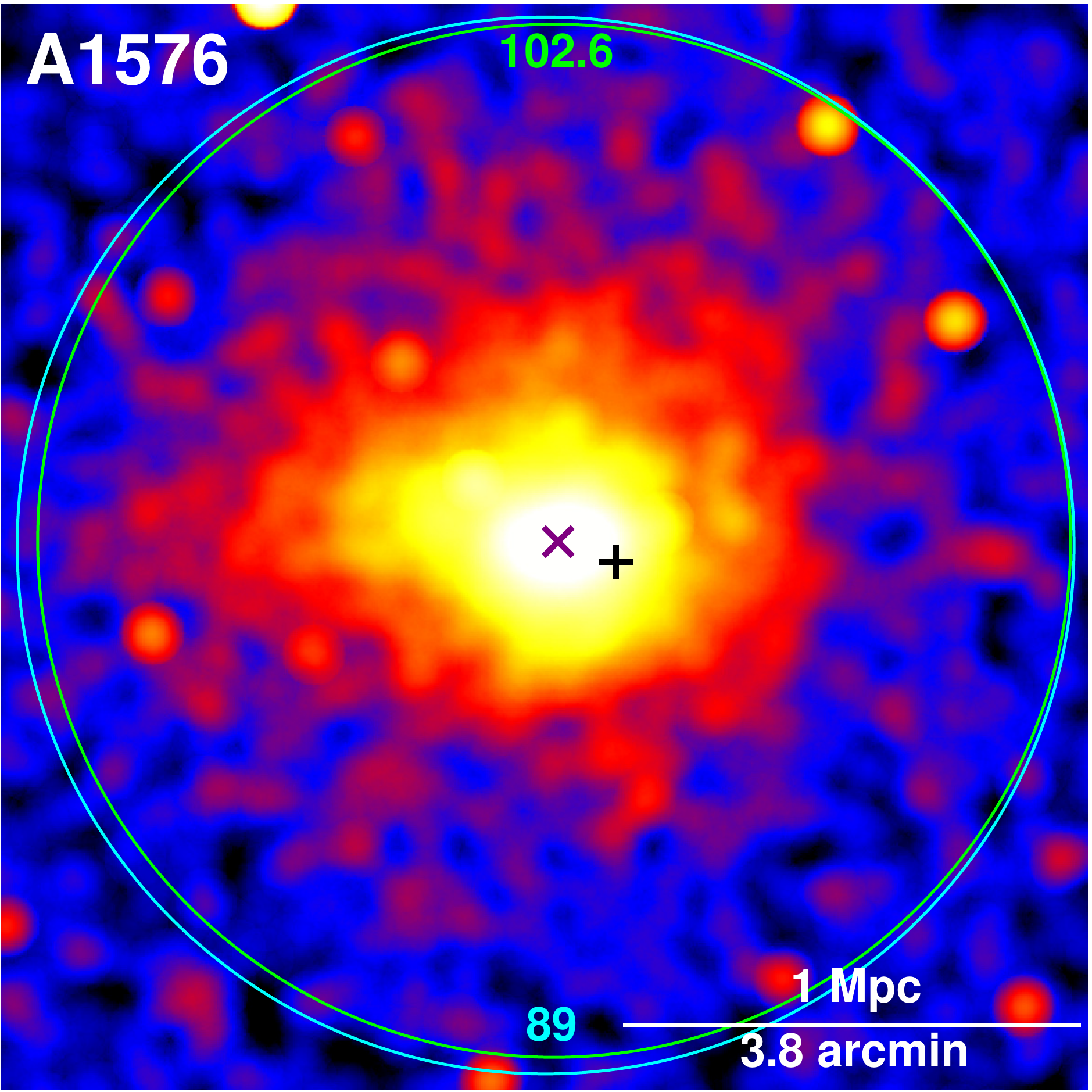}
\includegraphics[width=0.246\textwidth,keepaspectratio=true,clip=true]{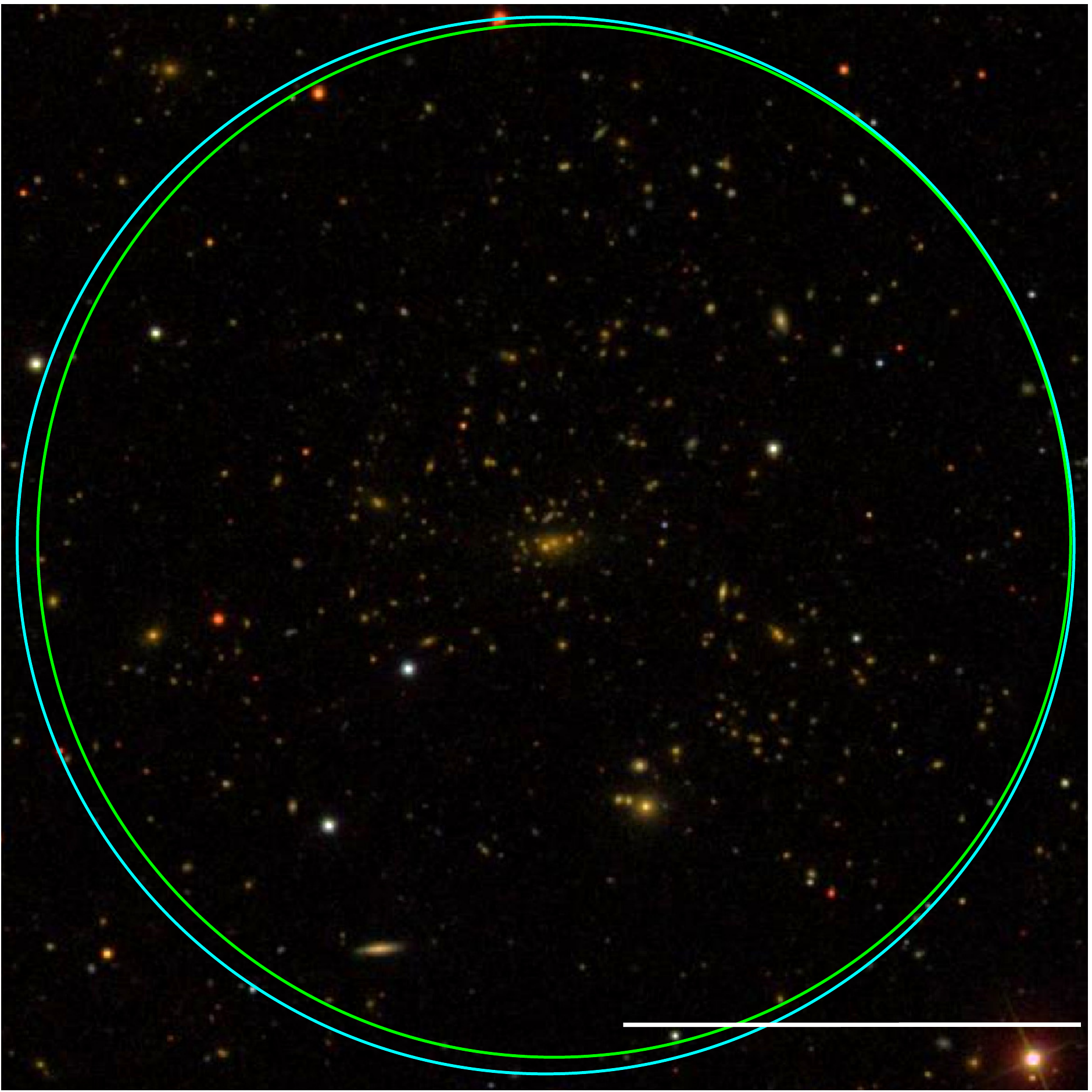}
\includegraphics[width=0.246\textwidth,keepaspectratio=true,clip=true]{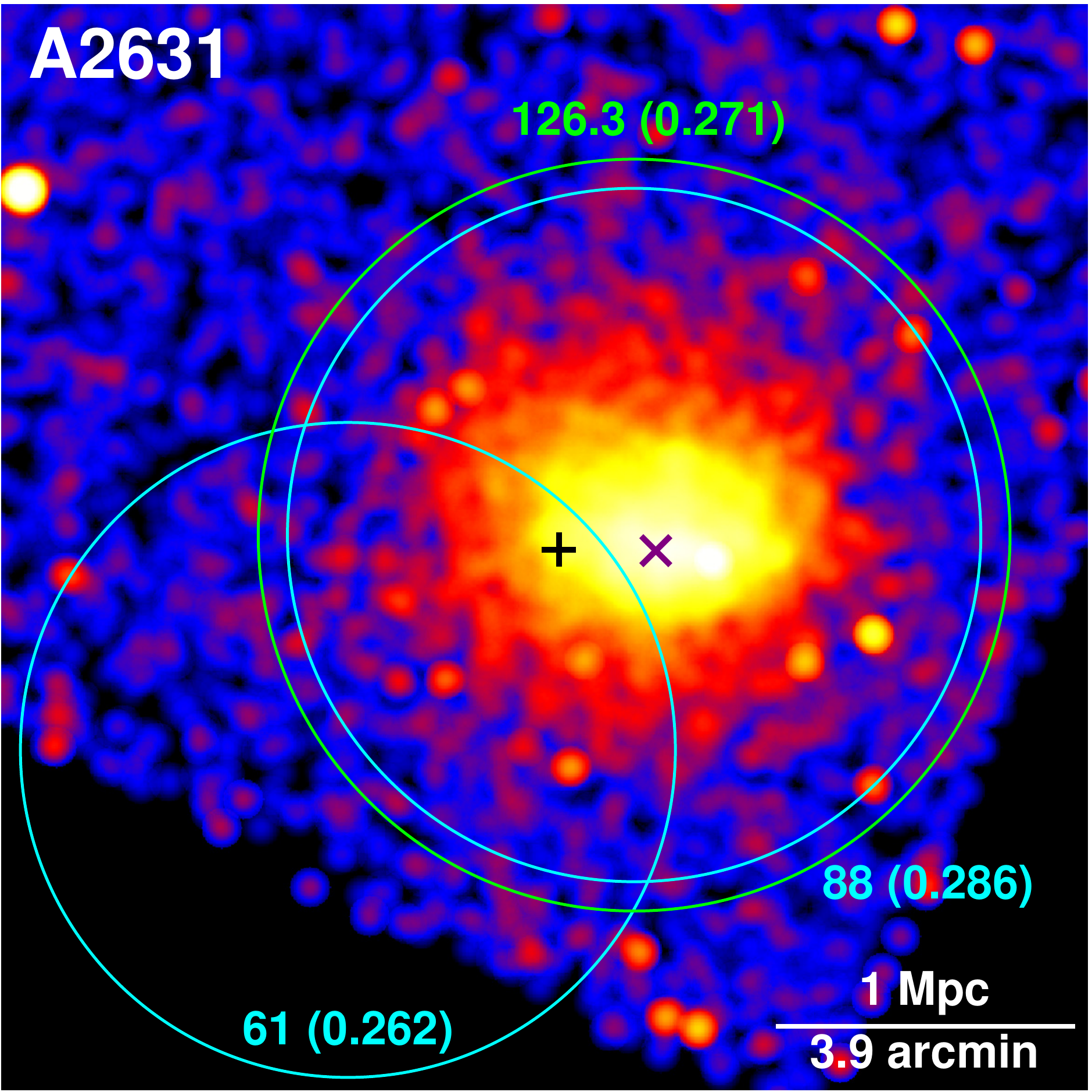}
\includegraphics[width=0.246\textwidth,keepaspectratio=true,clip=true]{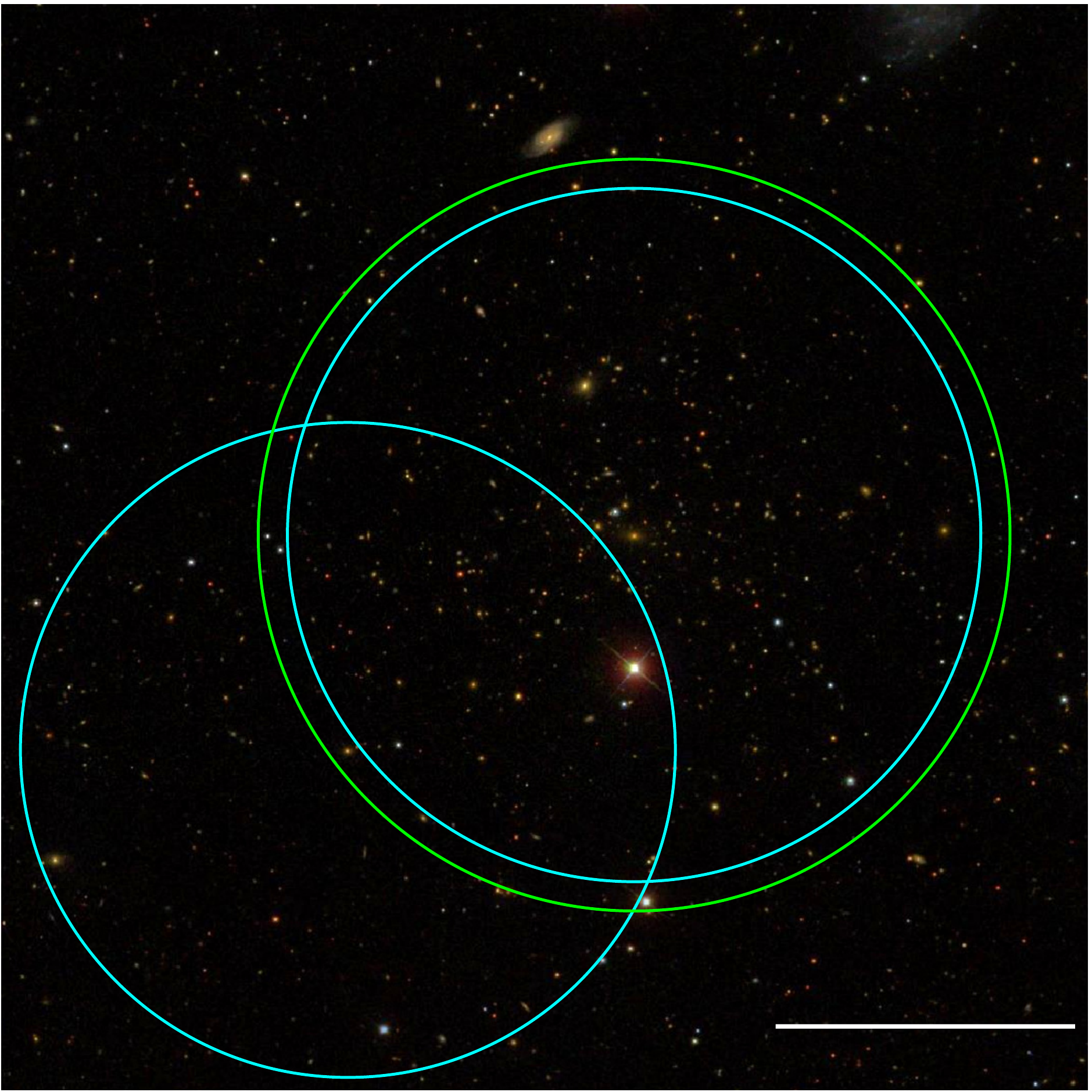}
\includegraphics[width=0.246\textwidth,keepaspectratio=true,clip=true]{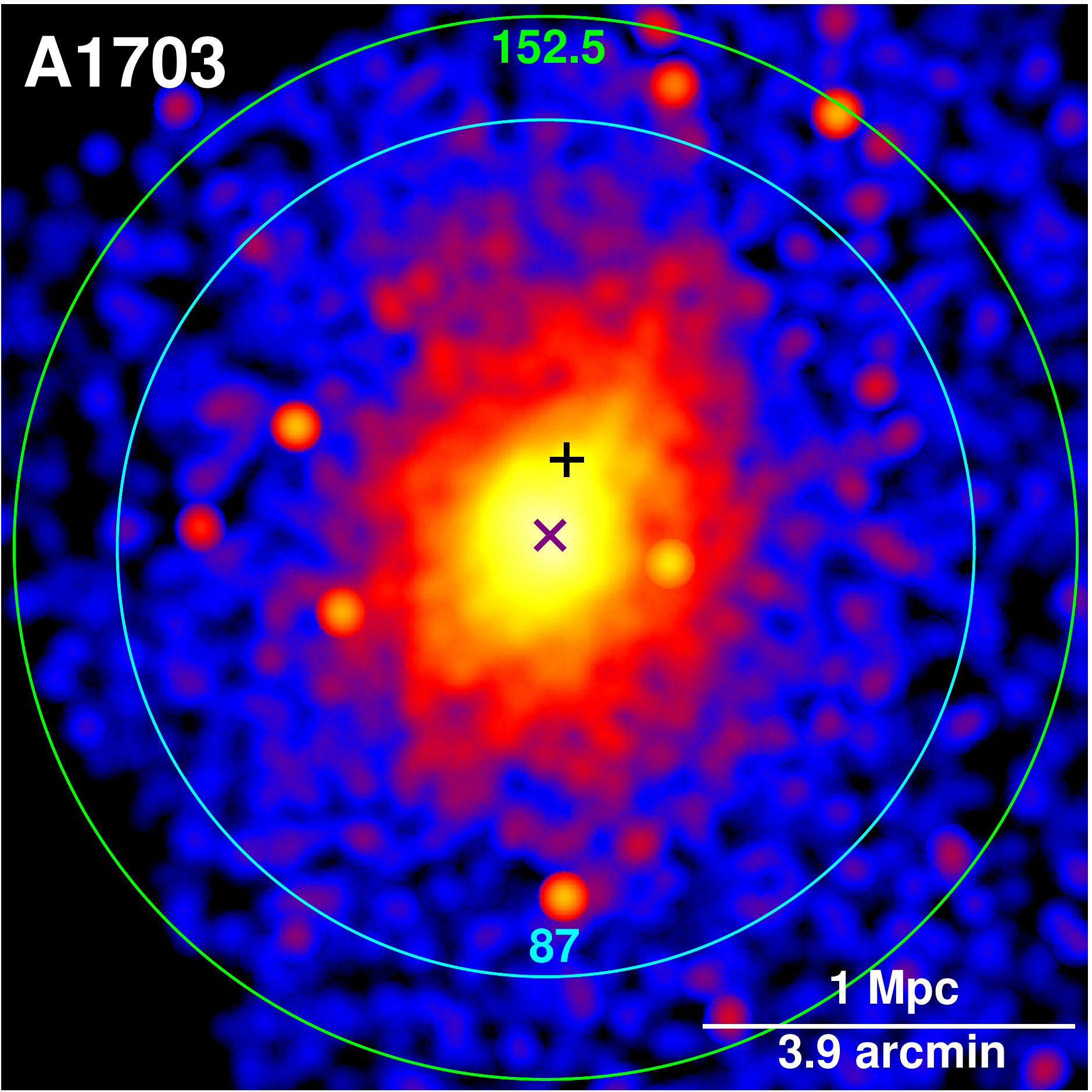}
\includegraphics[width=0.246\textwidth,keepaspectratio=true,clip=true]{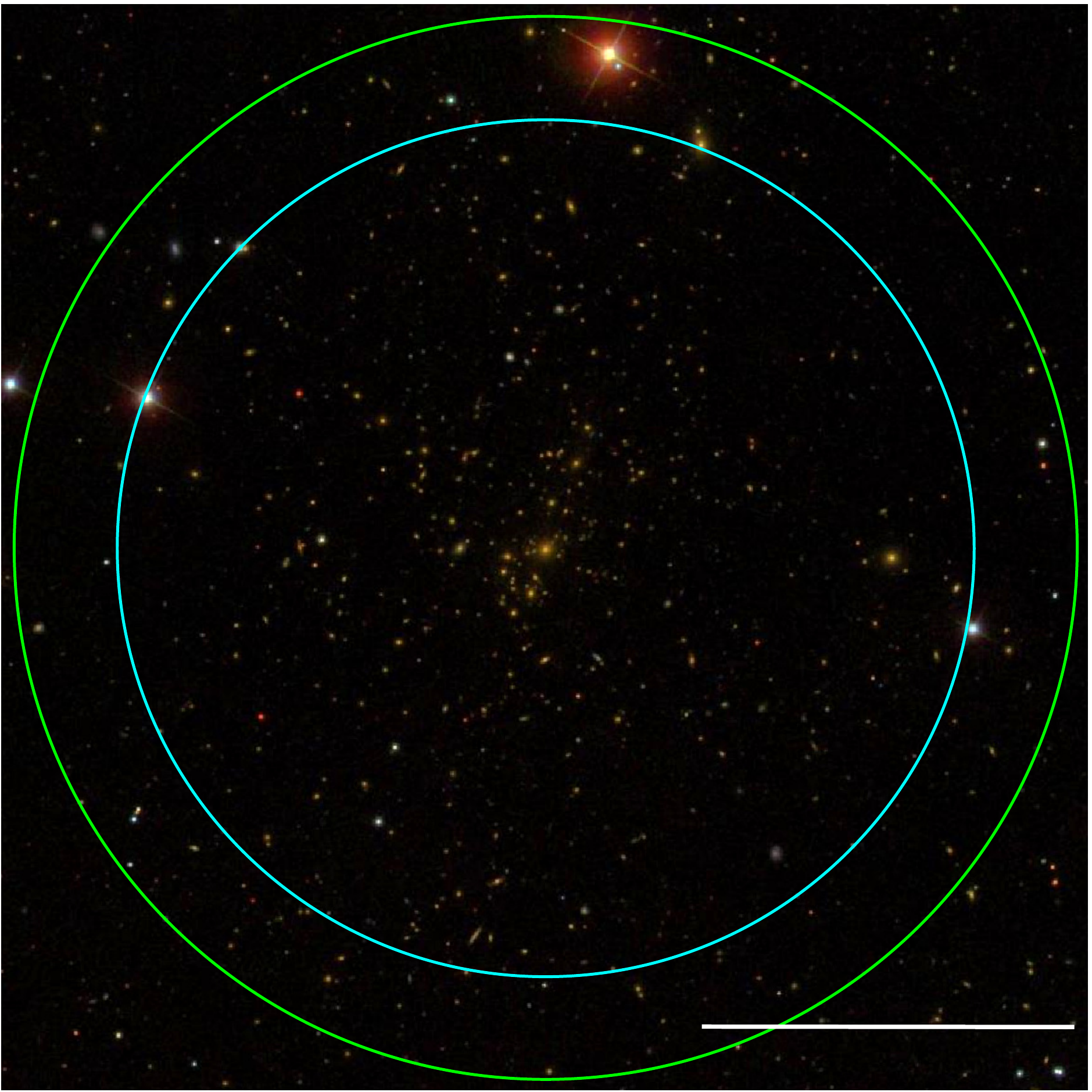}
\includegraphics[width=0.246\textwidth,keepaspectratio=true,clip=true]{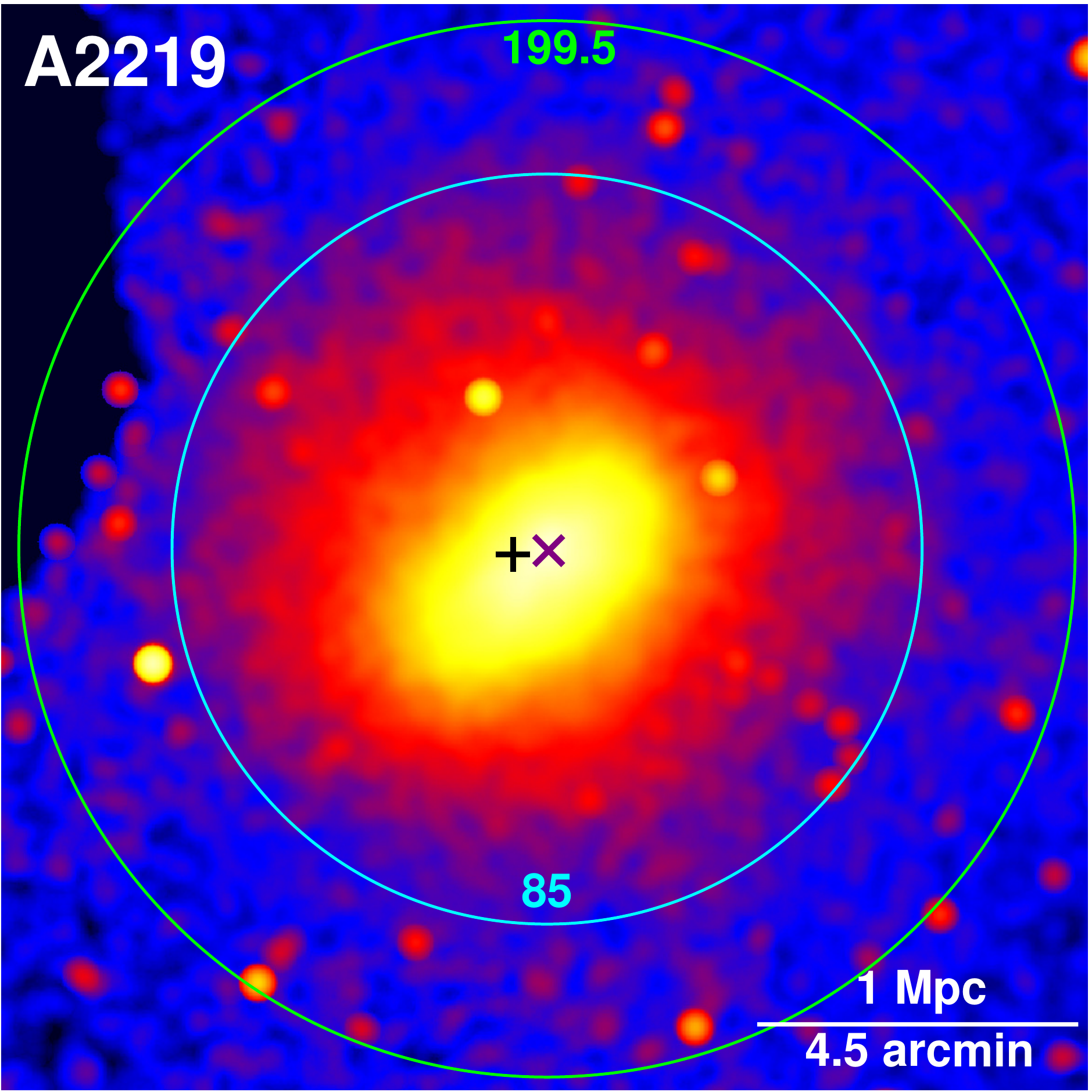}
\includegraphics[width=0.246\textwidth,keepaspectratio=true,clip=true]{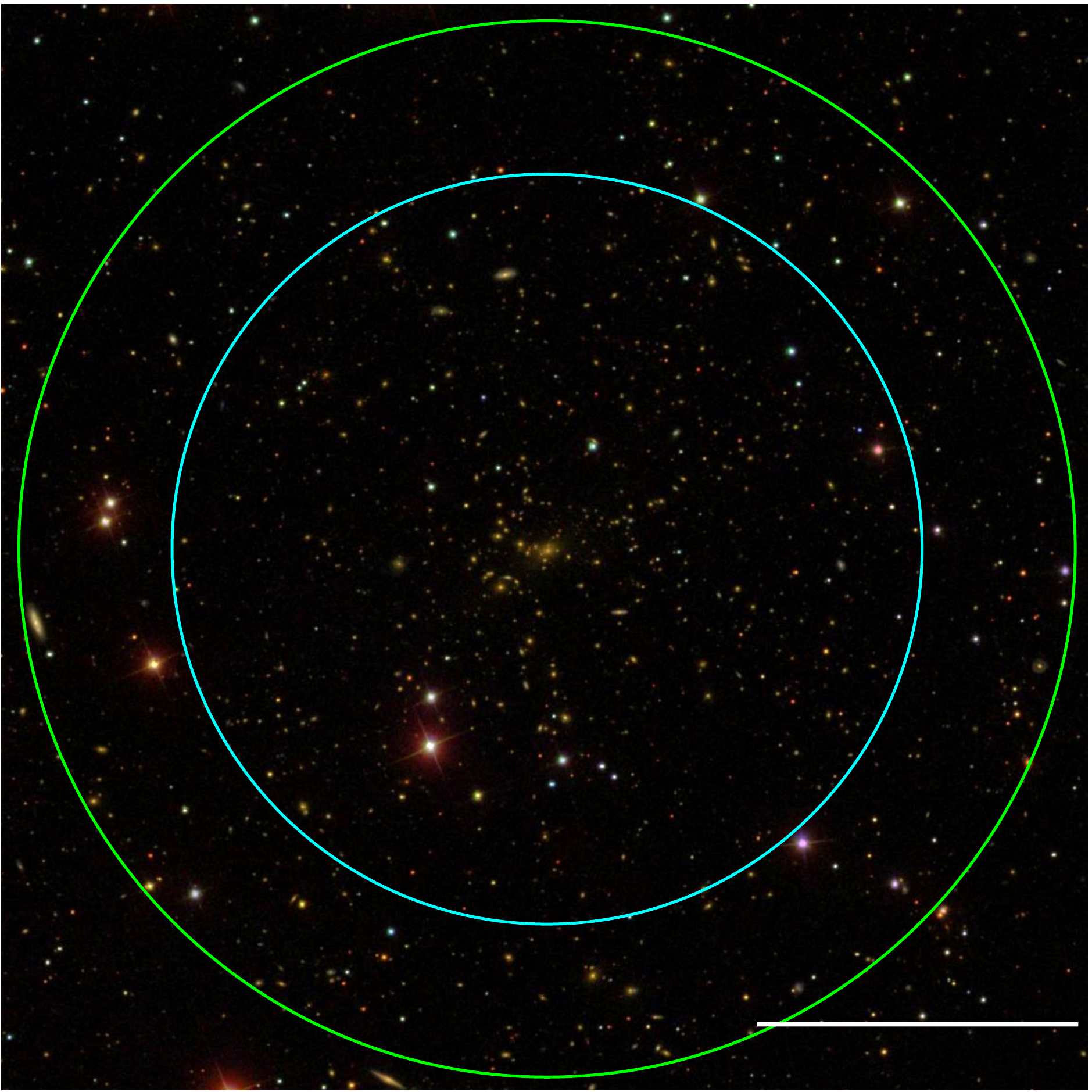}
\includegraphics[width=0.246\textwidth,keepaspectratio=true,clip=true]{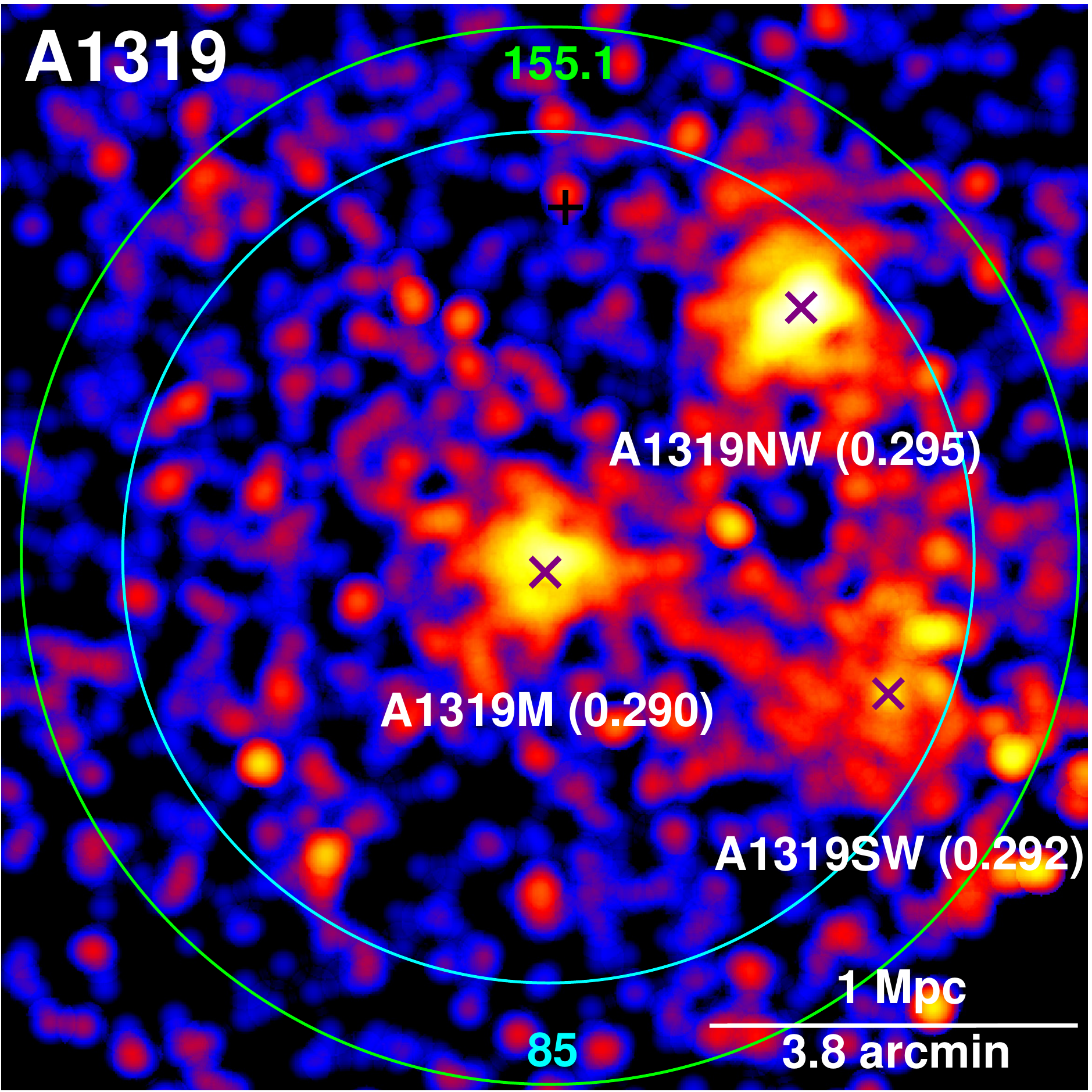}
\includegraphics[width=0.246\textwidth,keepaspectratio=true,clip=true]{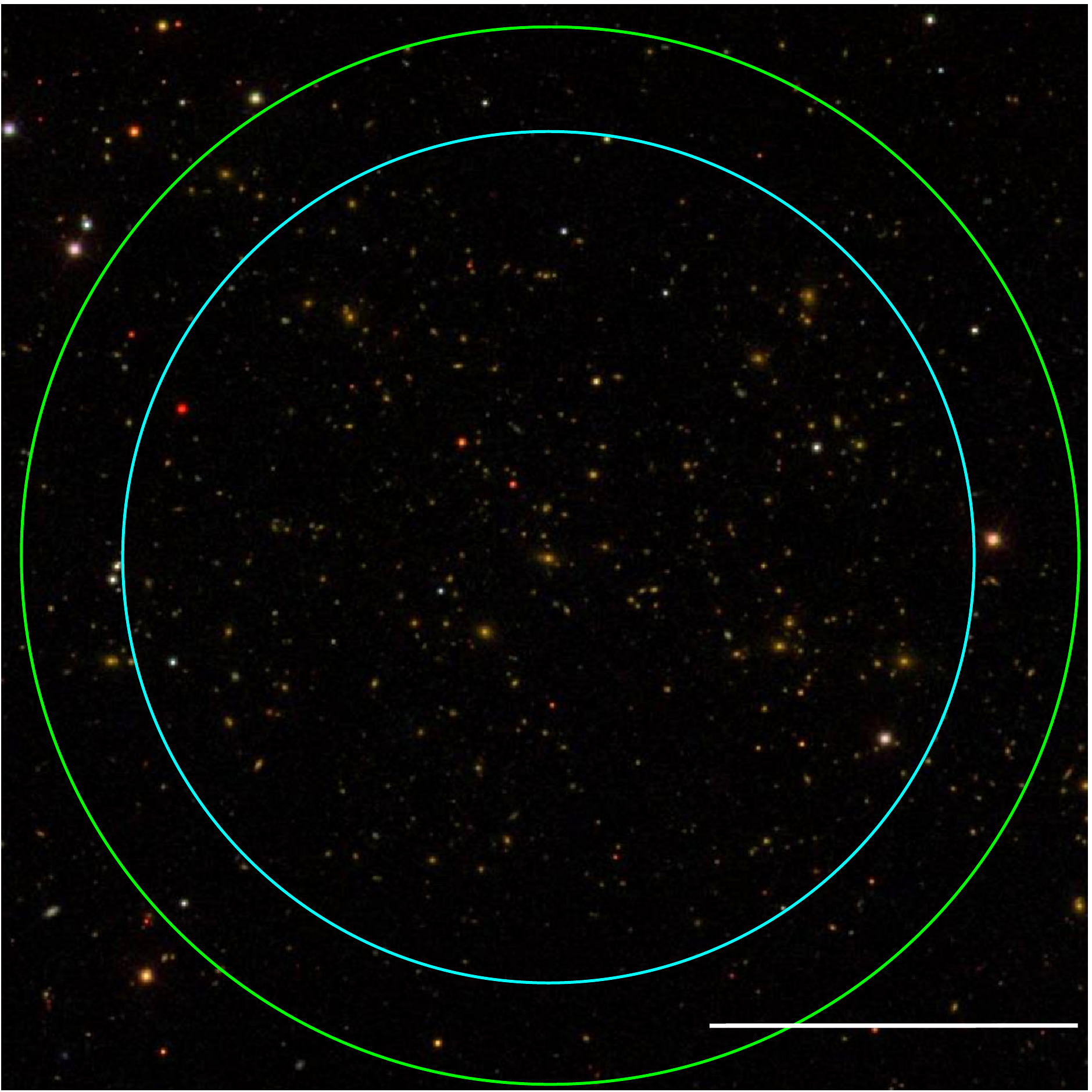}
\includegraphics[width=0.246\textwidth,keepaspectratio=true,clip=true]{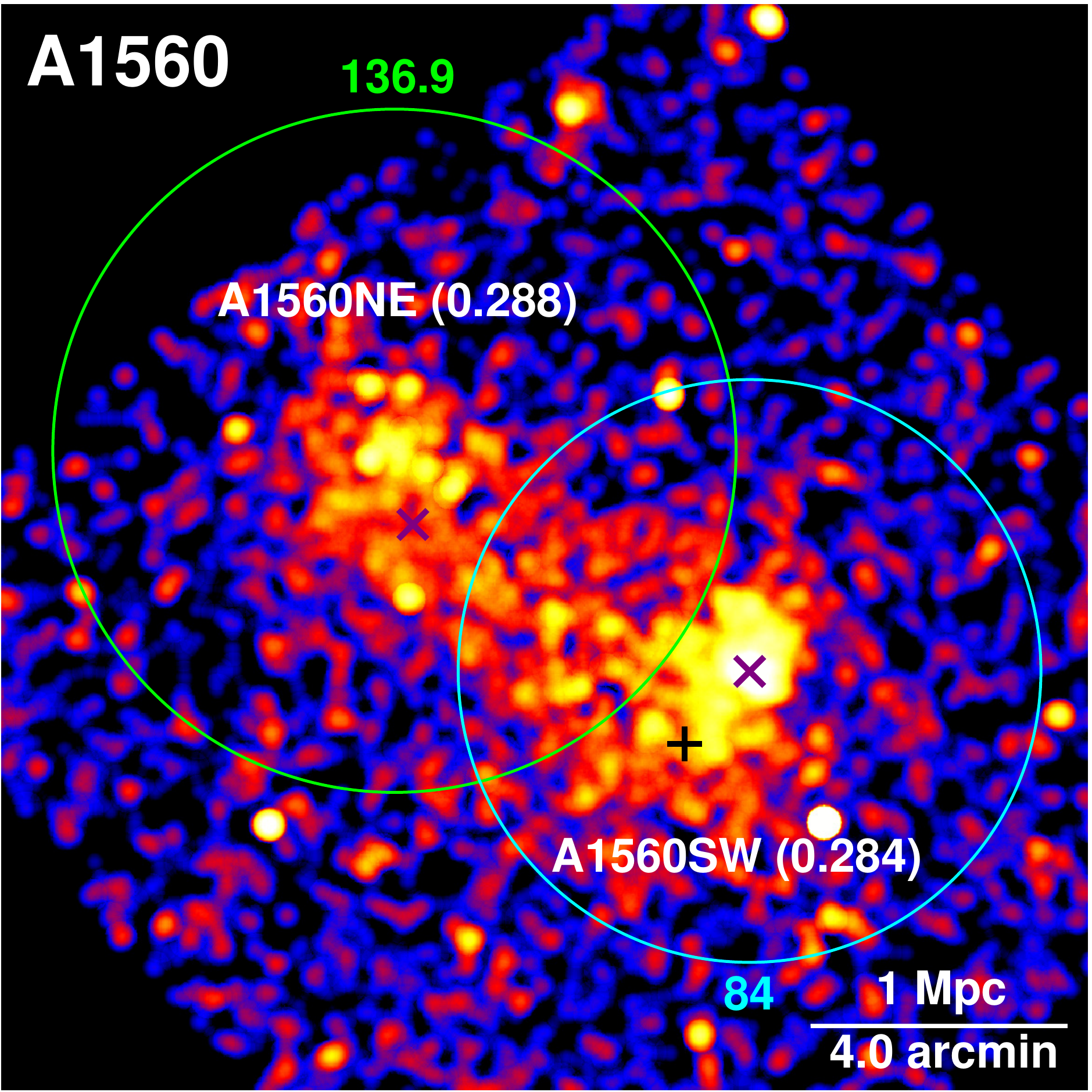}
\includegraphics[width=0.246\textwidth,keepaspectratio=true,clip=true]{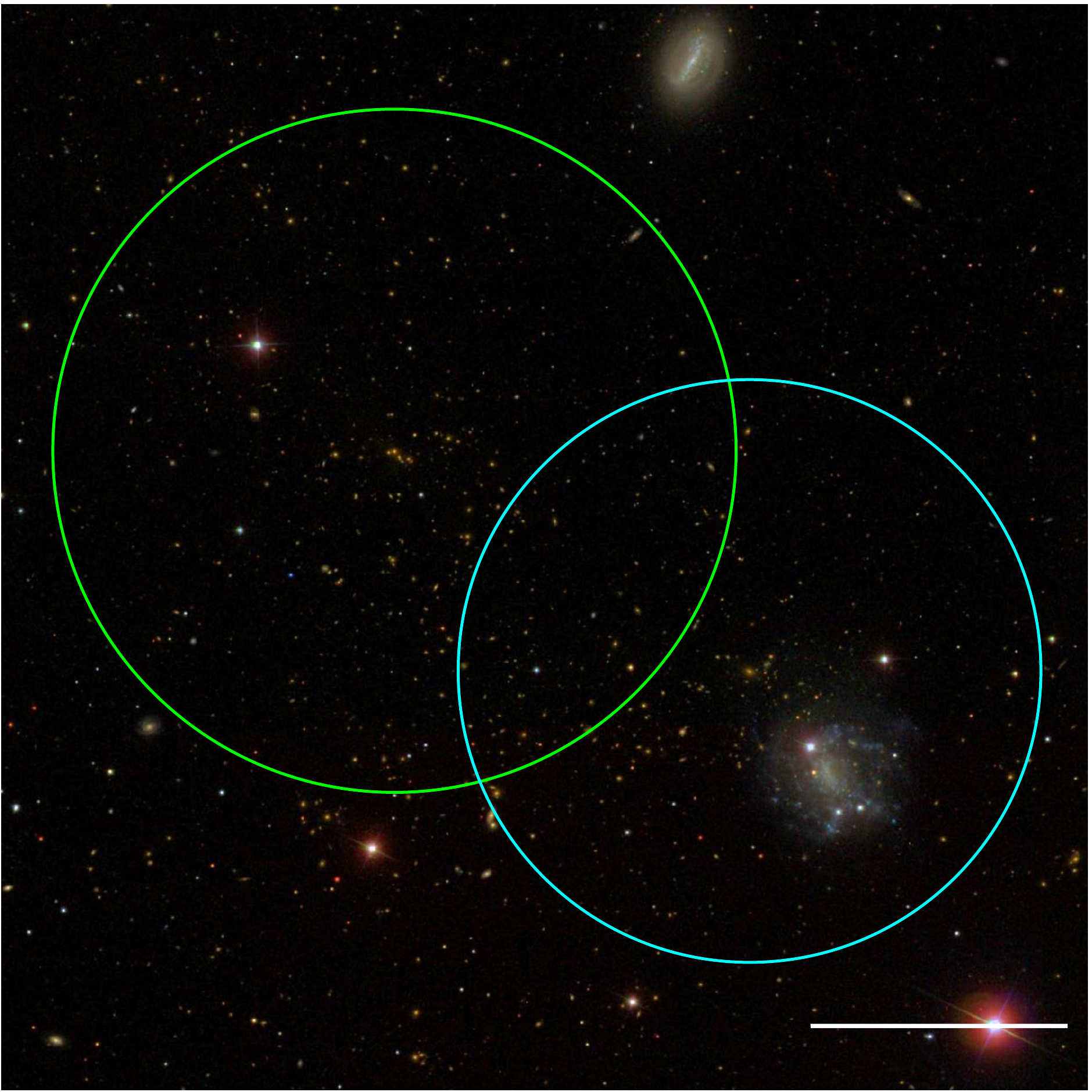}
\includegraphics[width=0.246\textwidth,keepaspectratio=true,clip=true]{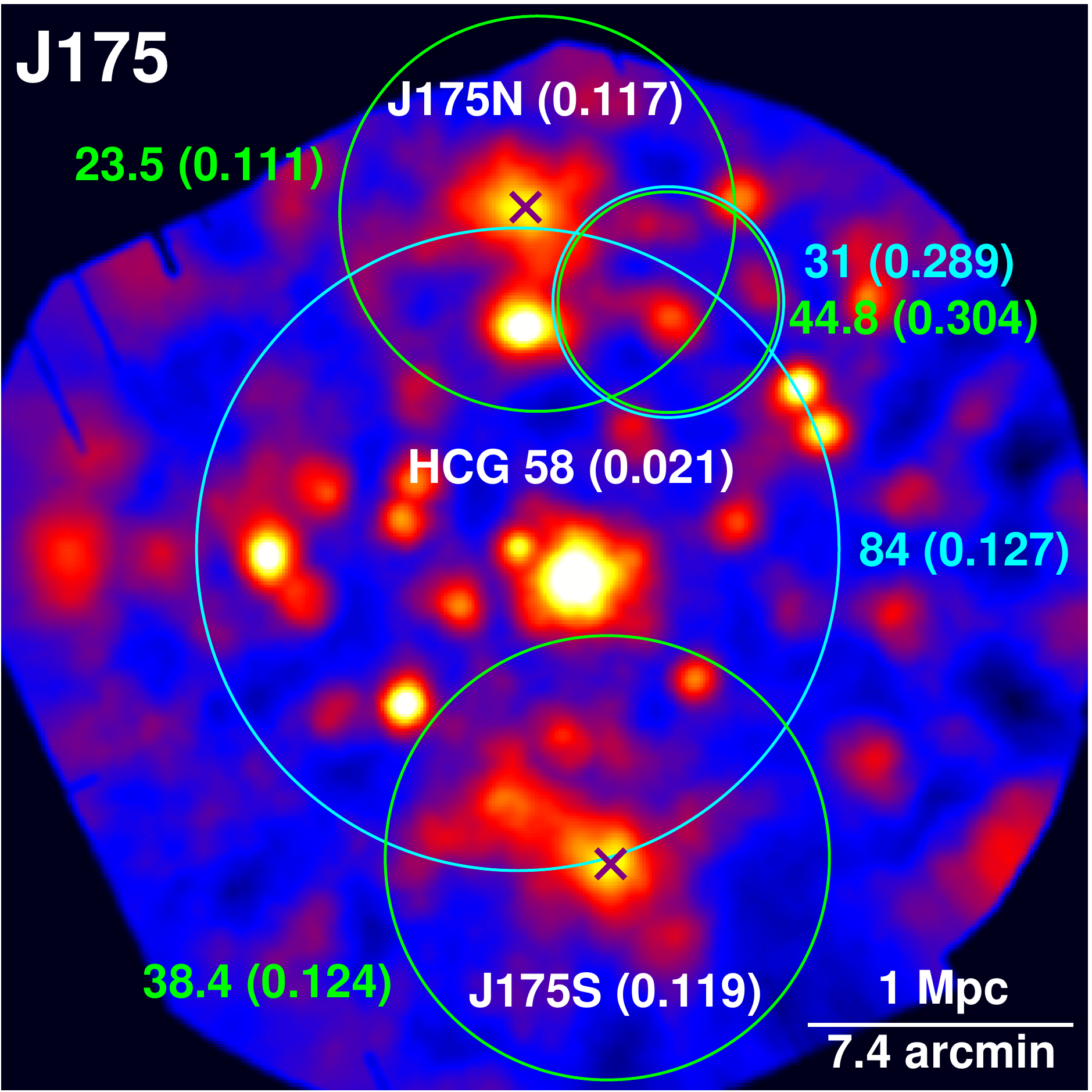}
\includegraphics[width=0.246\textwidth,keepaspectratio=true,clip=true]{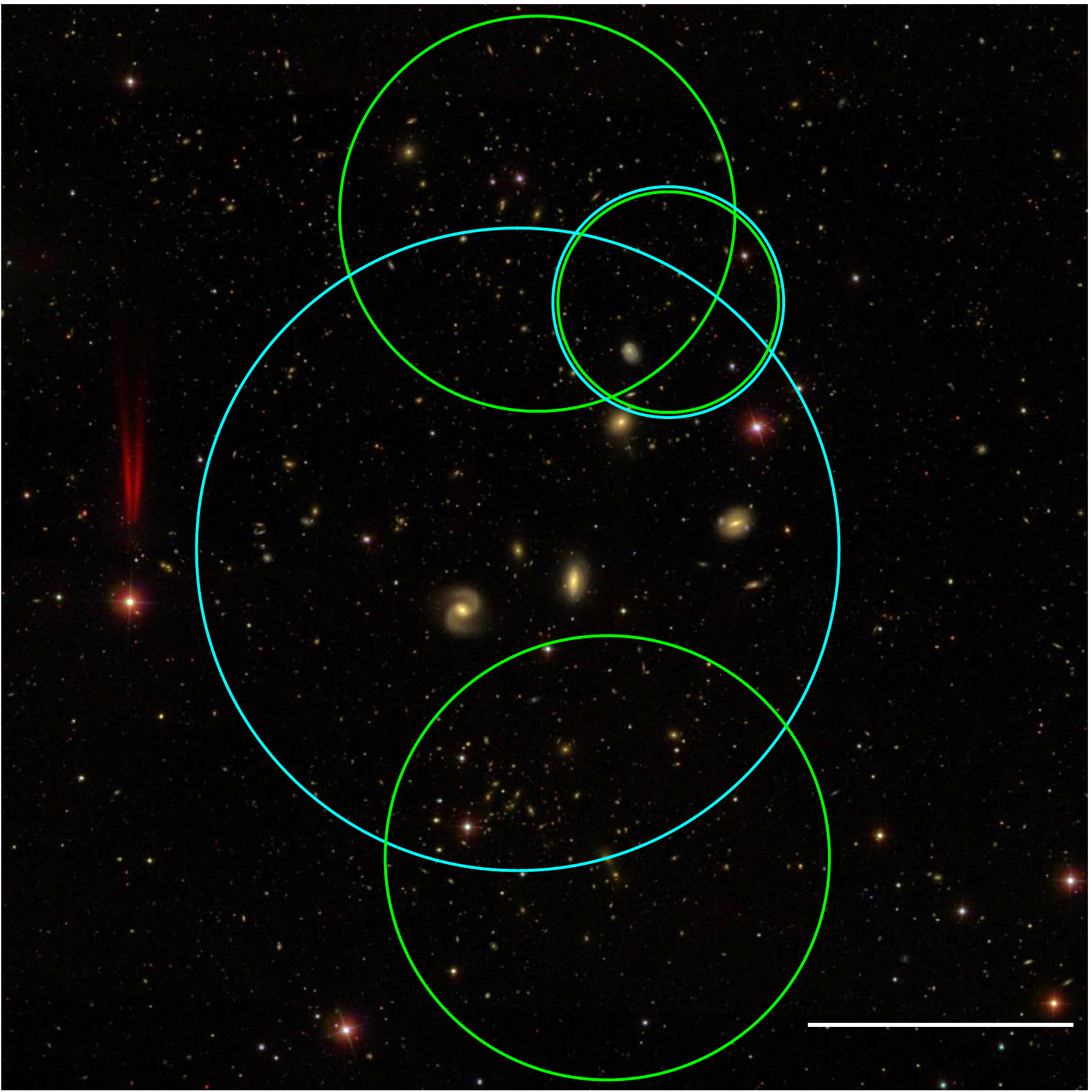}
 	\end{center}
\end{figure*}

\begin{figure*}
 	\begin{center}
\includegraphics[width=0.246\textwidth,keepaspectratio=true,clip=true]{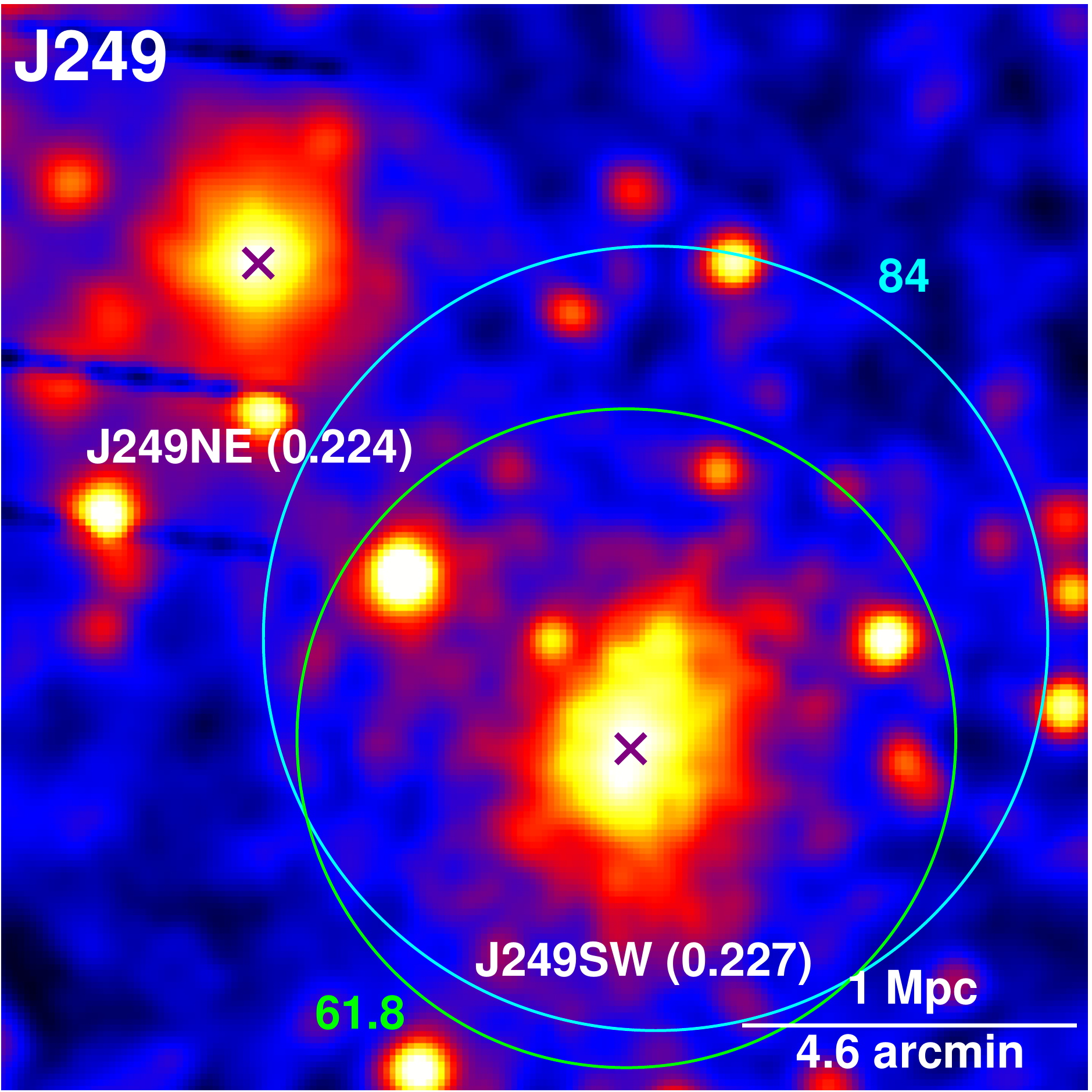}
\includegraphics[width=0.246\textwidth,keepaspectratio=true,clip=true]{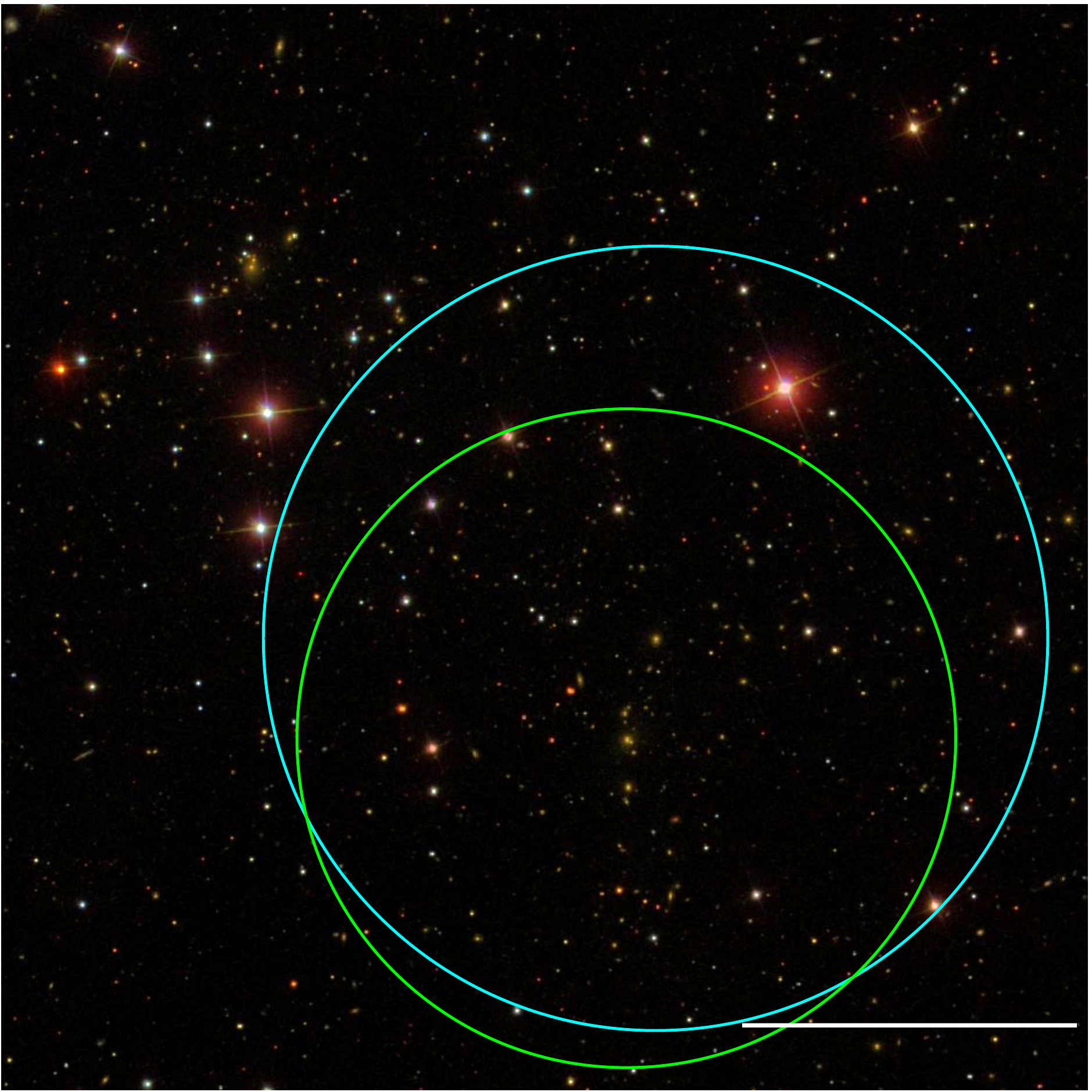}
\includegraphics[width=0.246\textwidth,keepaspectratio=true,clip=true]{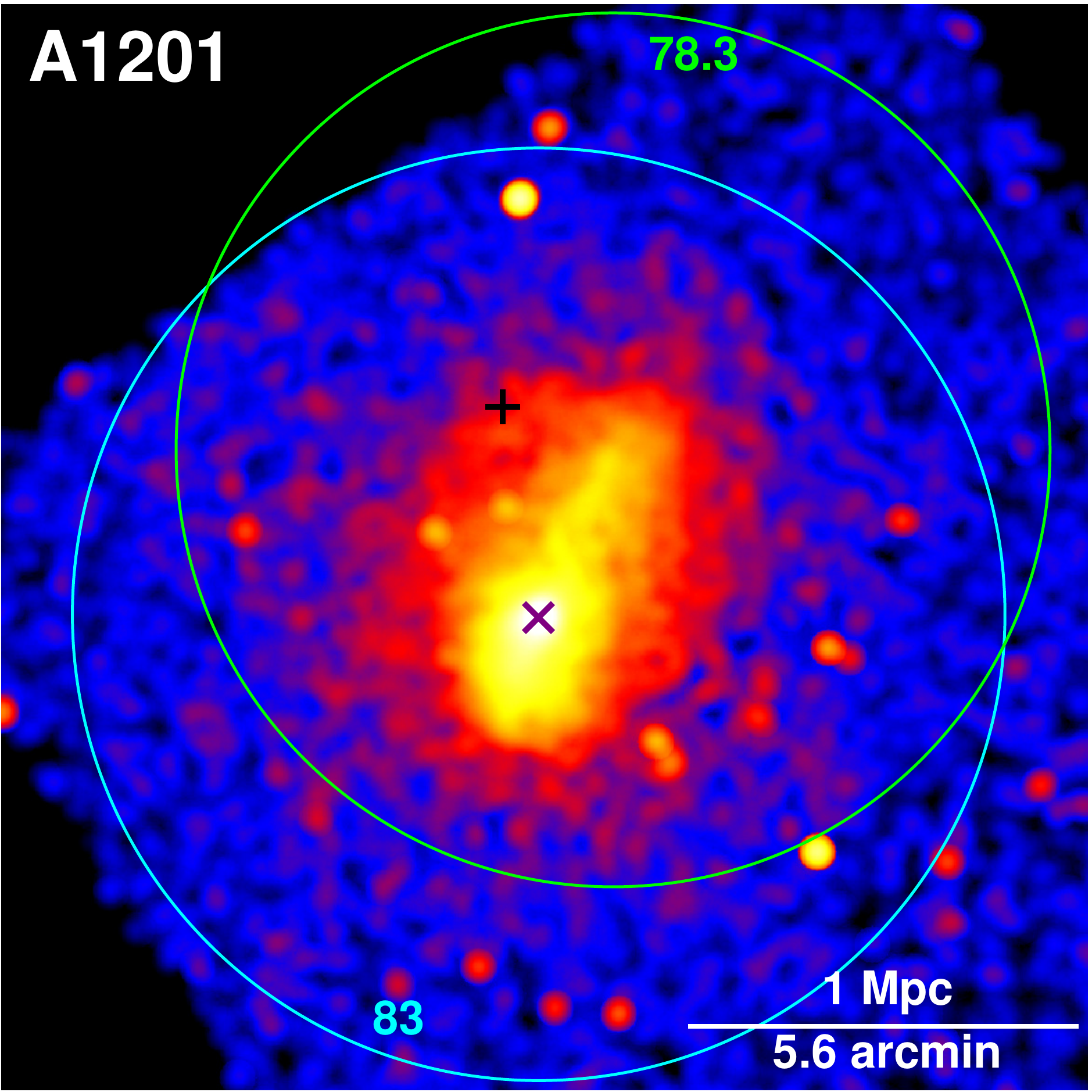}
\includegraphics[width=0.246\textwidth,keepaspectratio=true,clip=true]{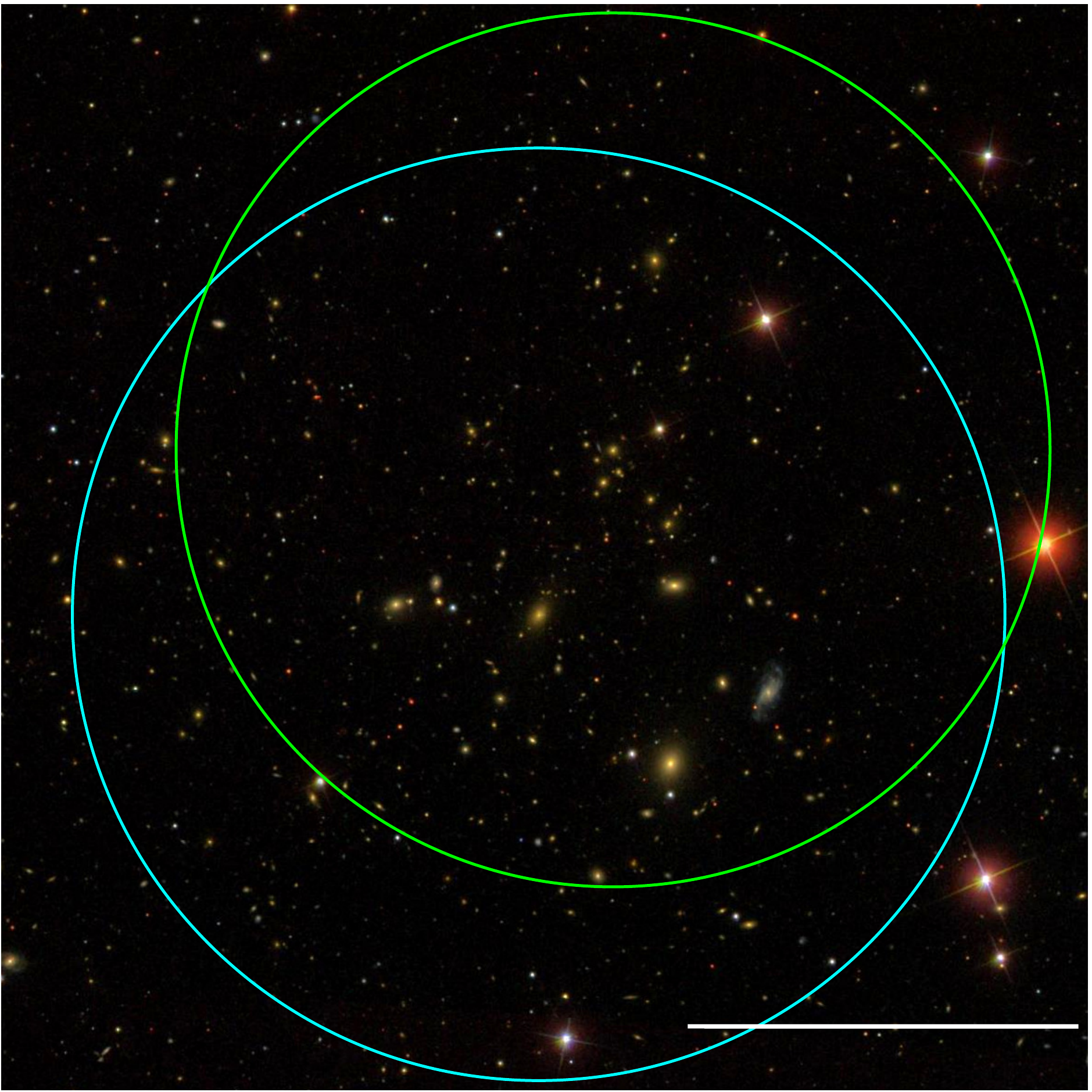}
\includegraphics[width=0.246\textwidth,keepaspectratio=true,clip=true]{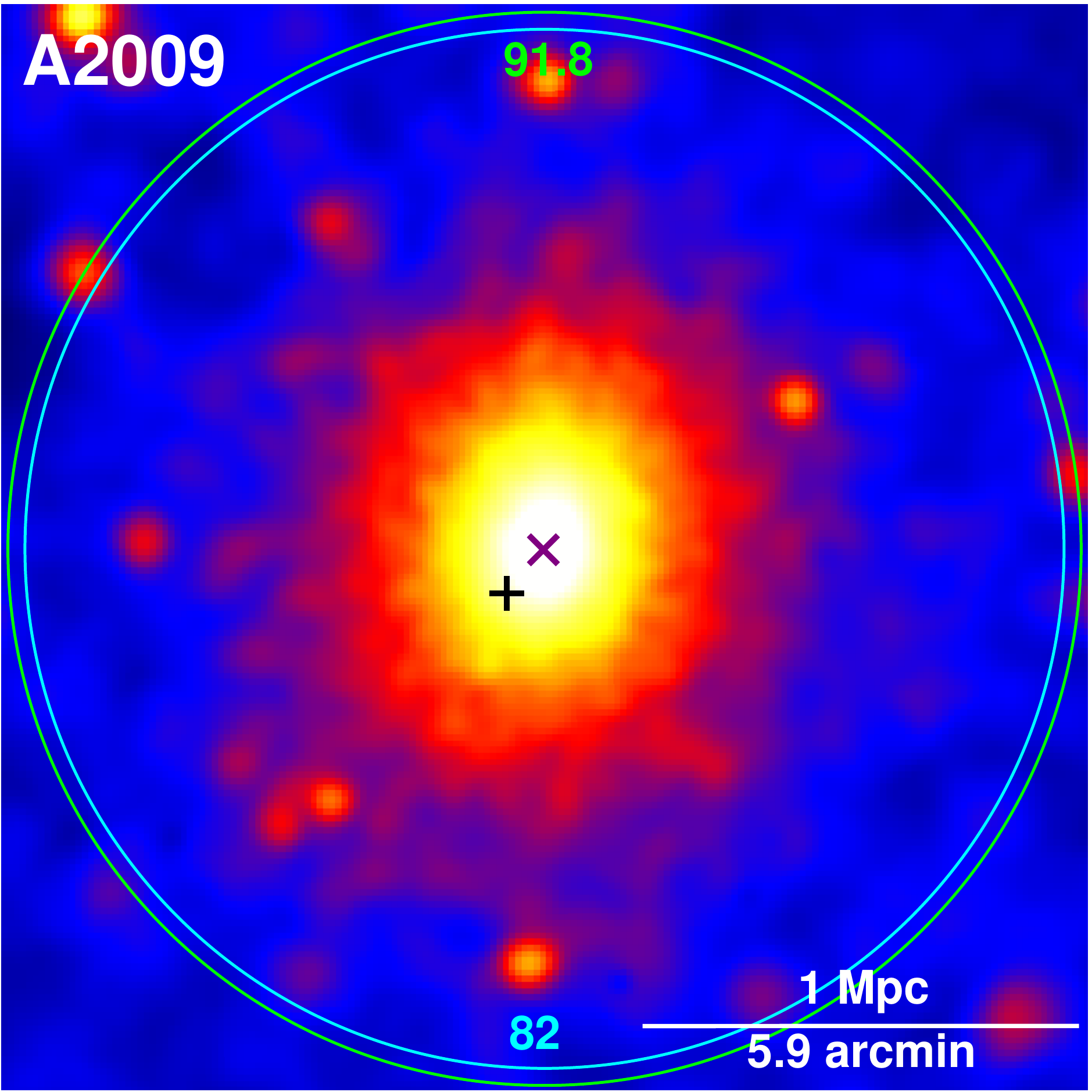}
\includegraphics[width=0.246\textwidth,keepaspectratio=true,clip=true]{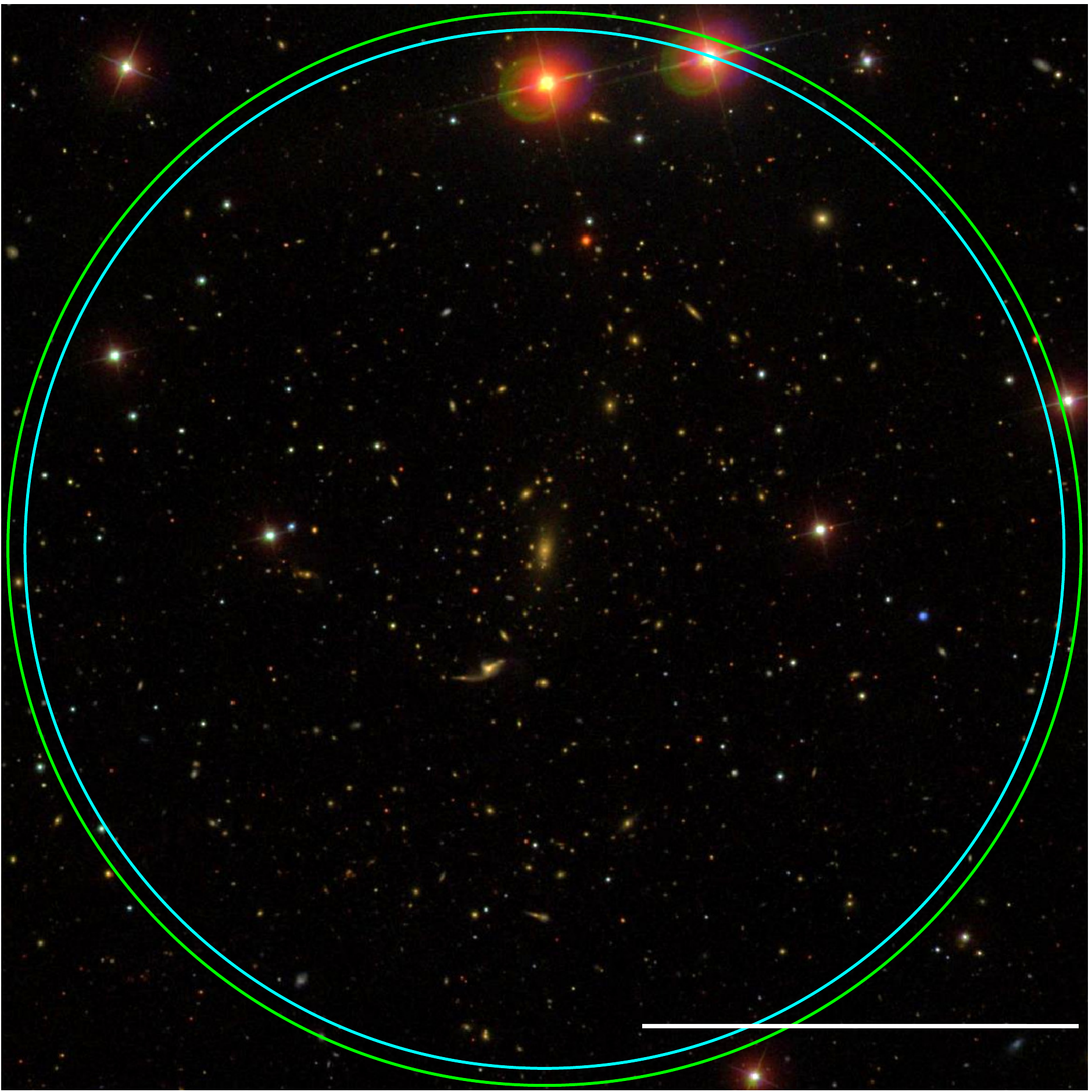}
\includegraphics[width=0.246\textwidth,keepaspectratio=true,clip=true]{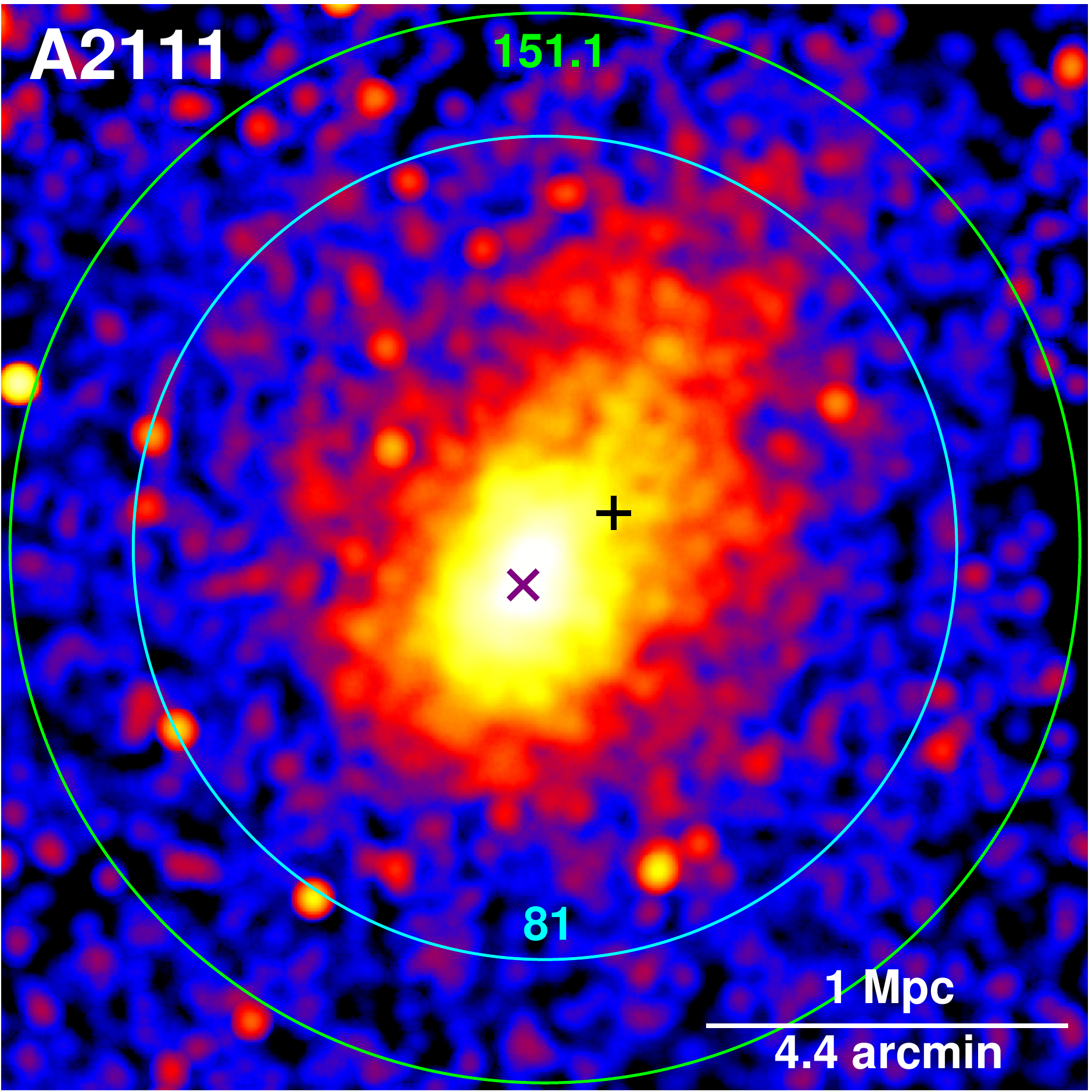}
\includegraphics[width=0.246\textwidth,keepaspectratio=true,clip=true]{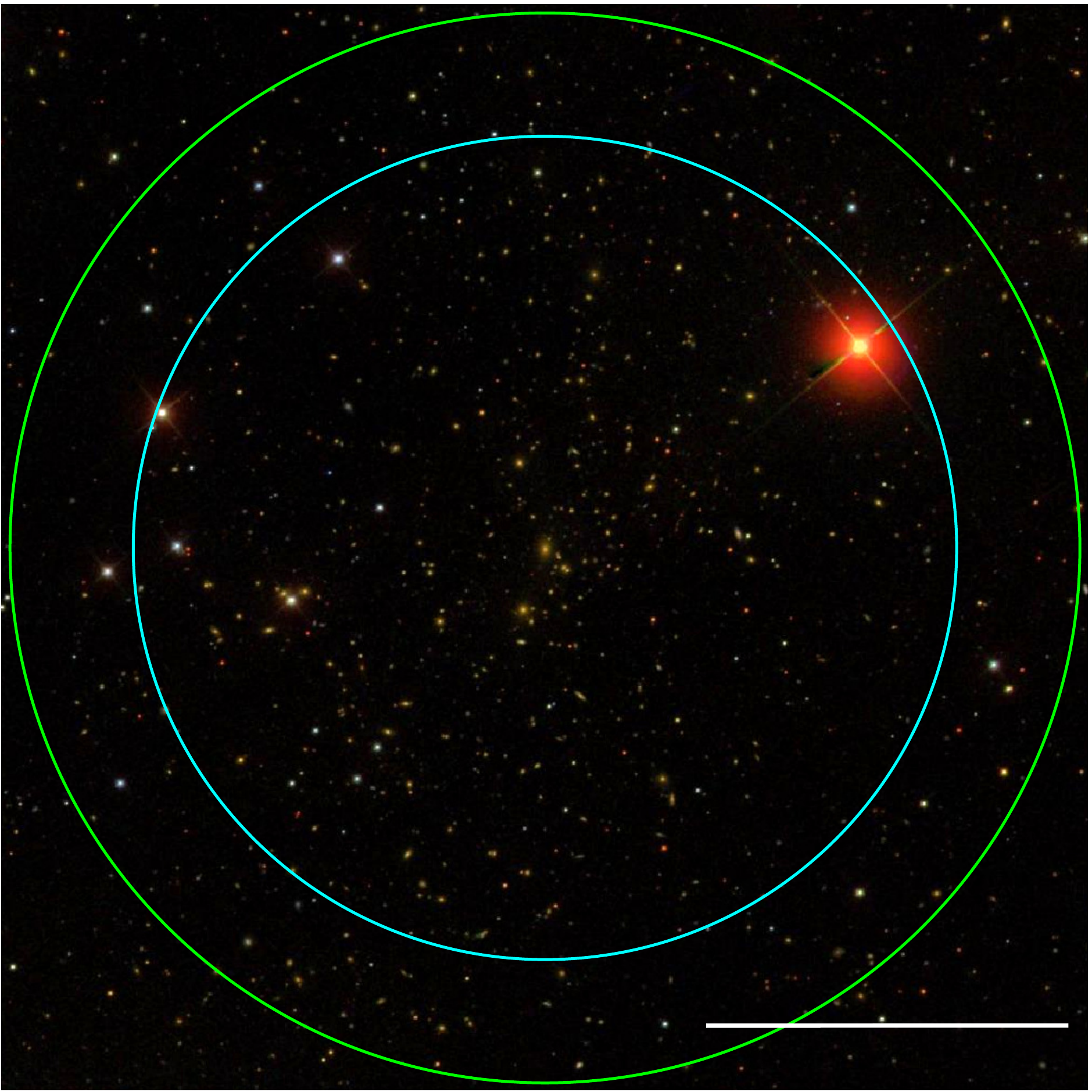}
\includegraphics[width=0.246\textwidth,keepaspectratio=true,clip=true]{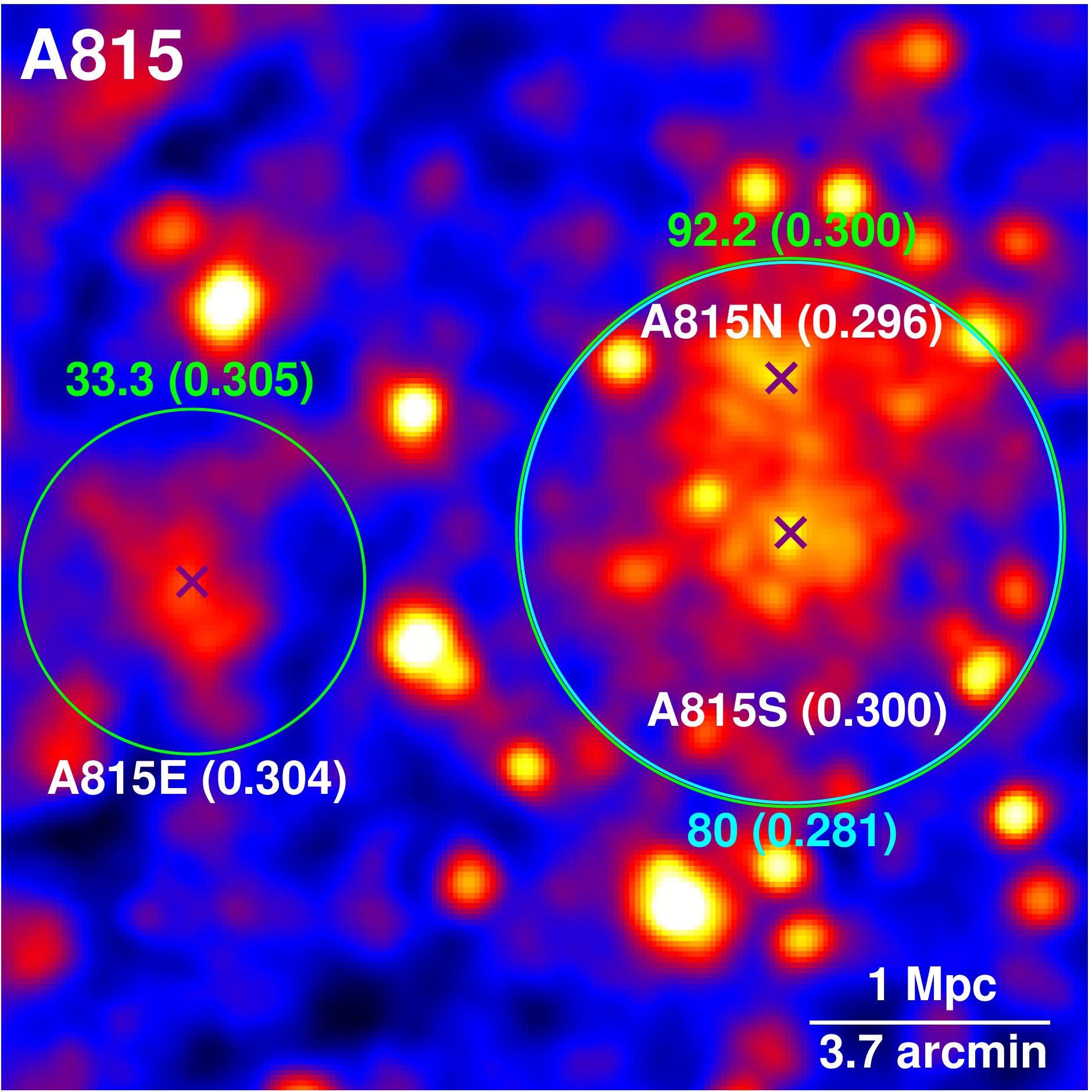}
\includegraphics[width=0.246\textwidth,keepaspectratio=true,clip=true]{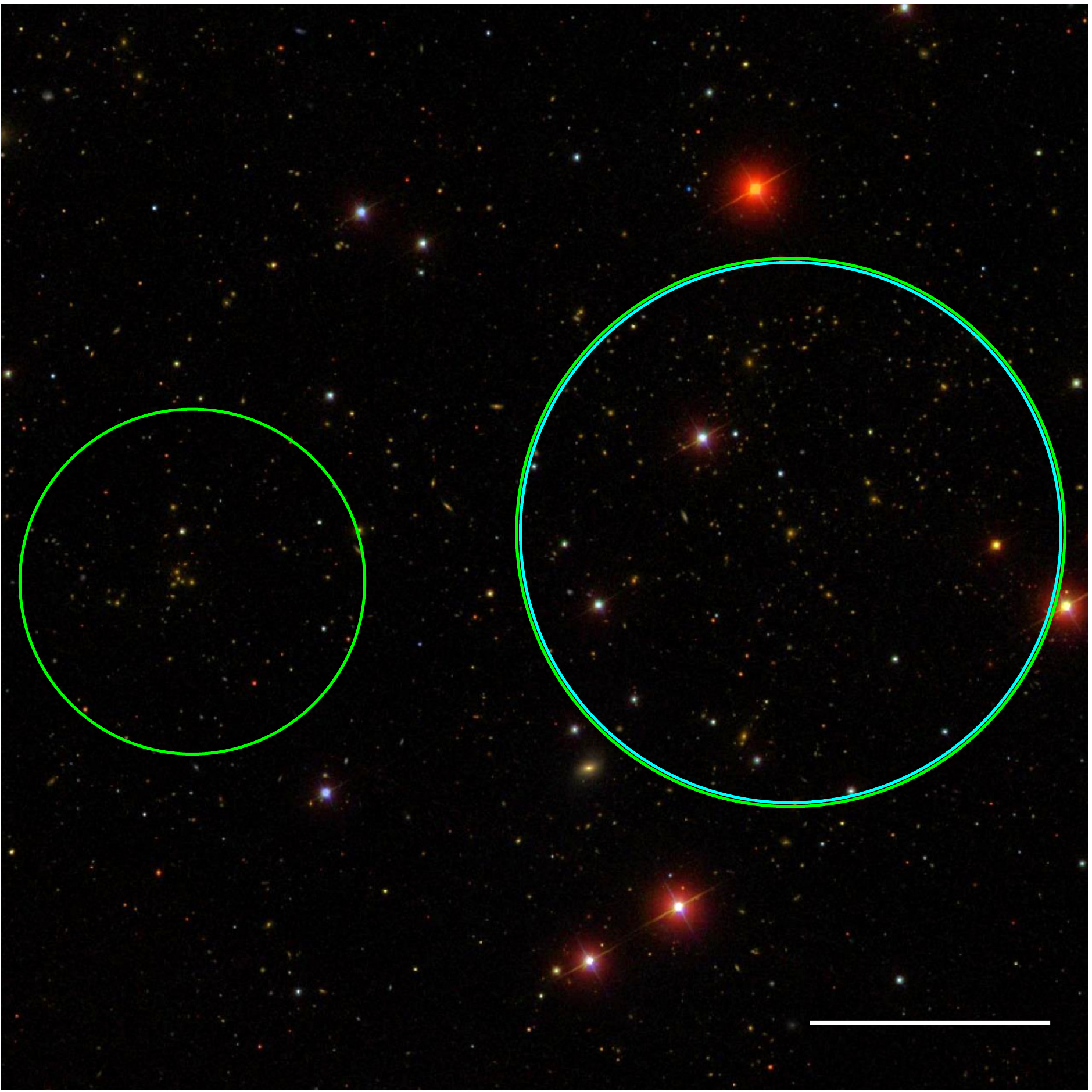}
\includegraphics[width=0.246\textwidth,keepaspectratio=true,clip=true]{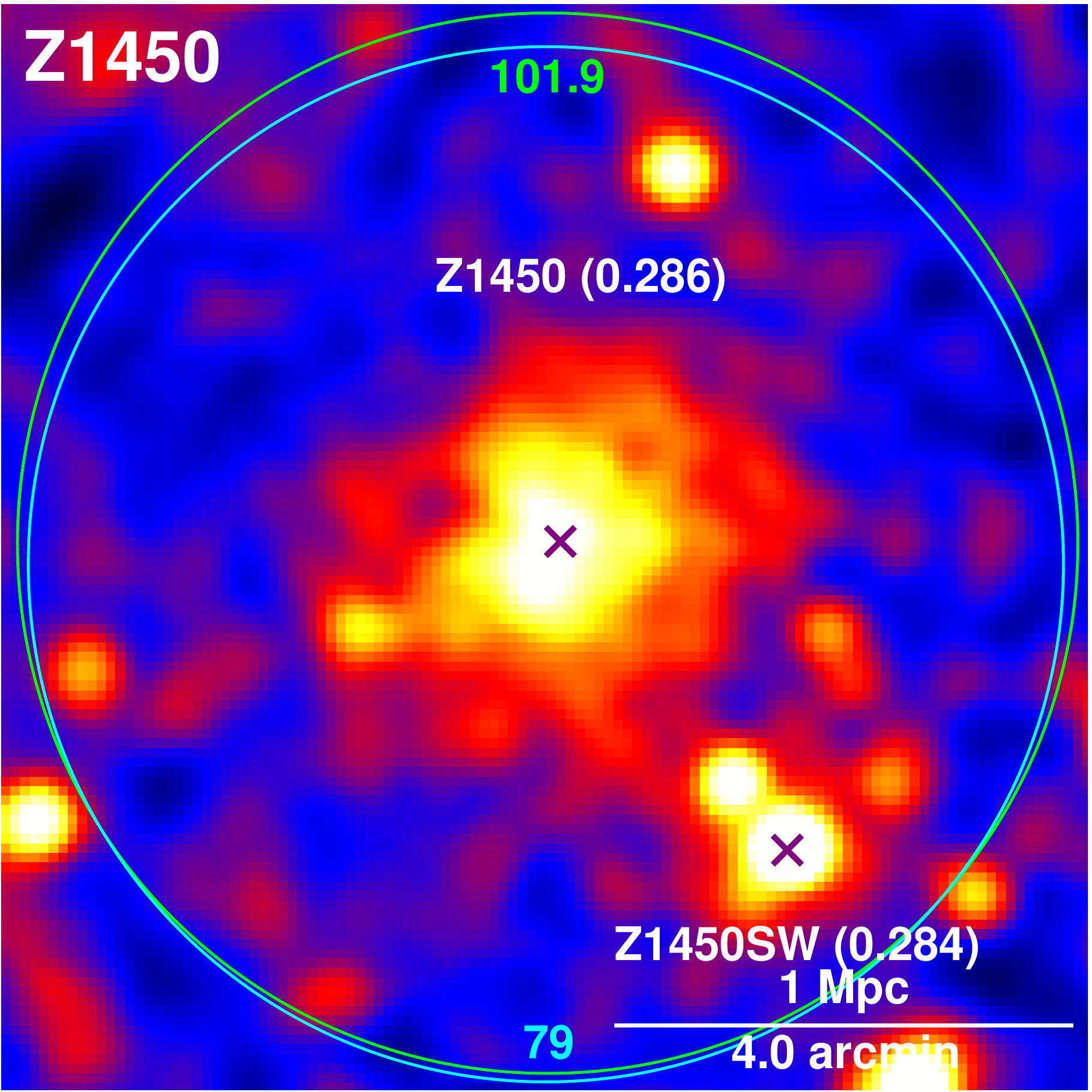}
\includegraphics[width=0.246\textwidth,keepaspectratio=true,clip=true]{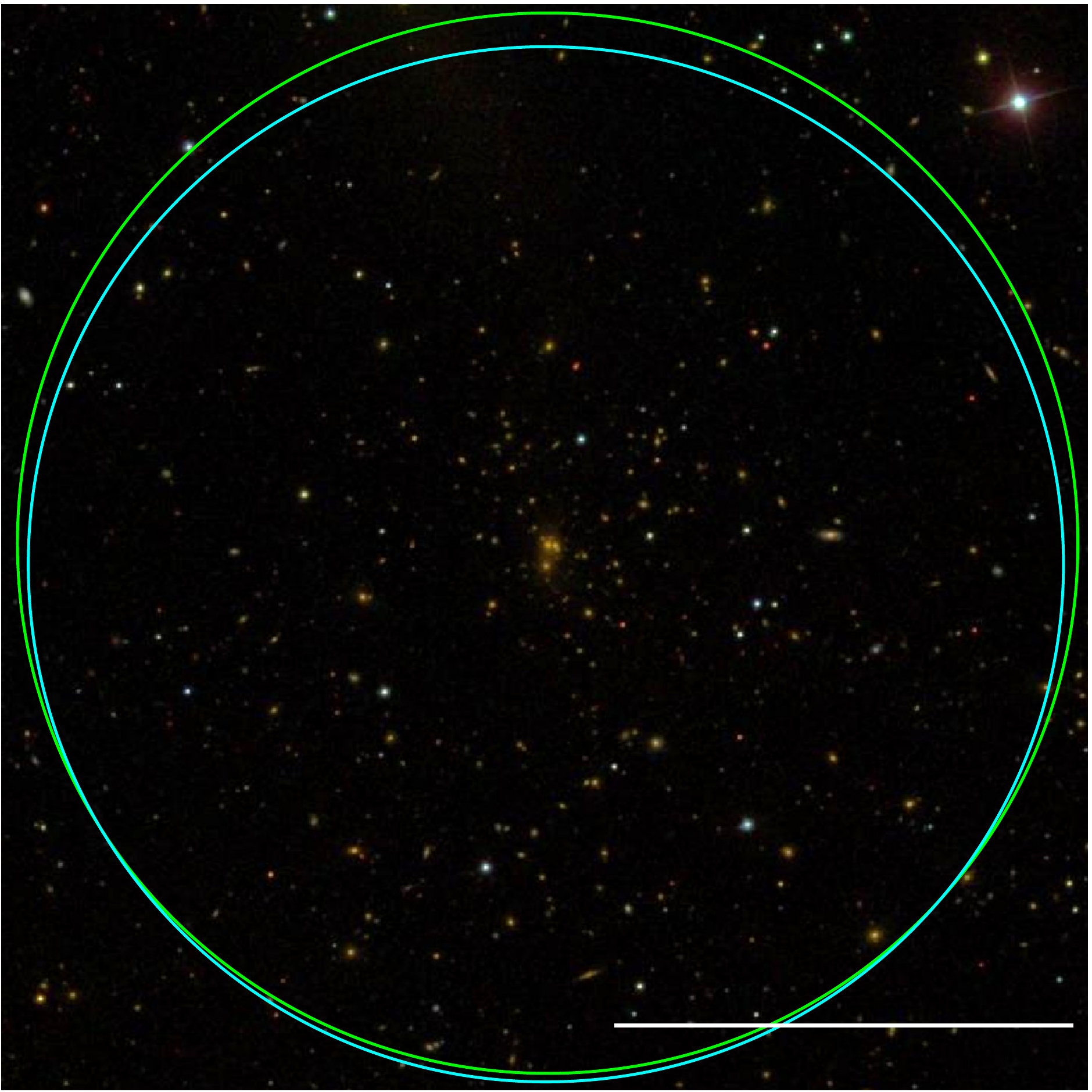} 	
\includegraphics[width=0.246\textwidth,keepaspectratio=true,clip=true]{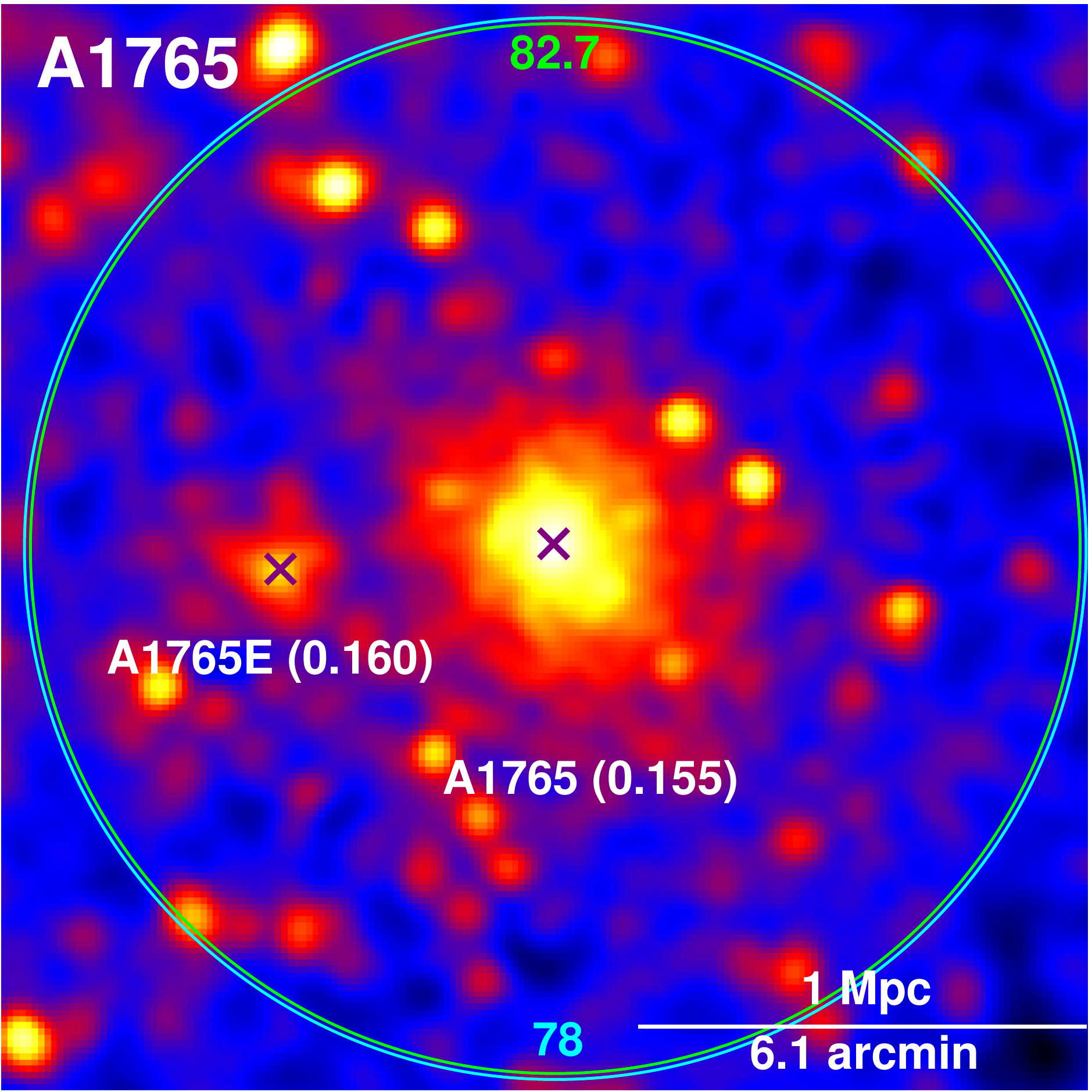}
\includegraphics[width=0.246\textwidth,keepaspectratio=true,clip=true]{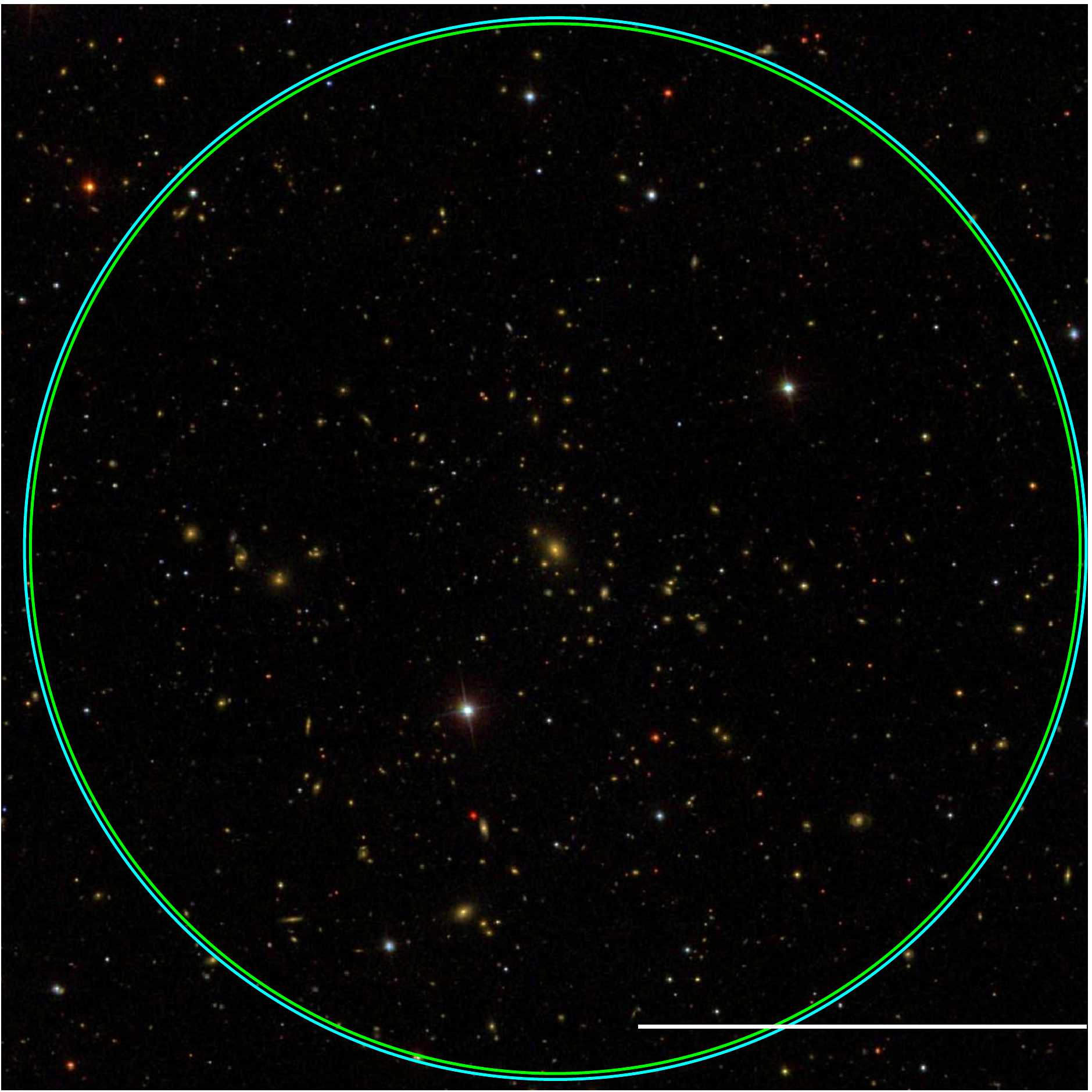}
\includegraphics[width=0.246\textwidth,keepaspectratio=true,clip=true]{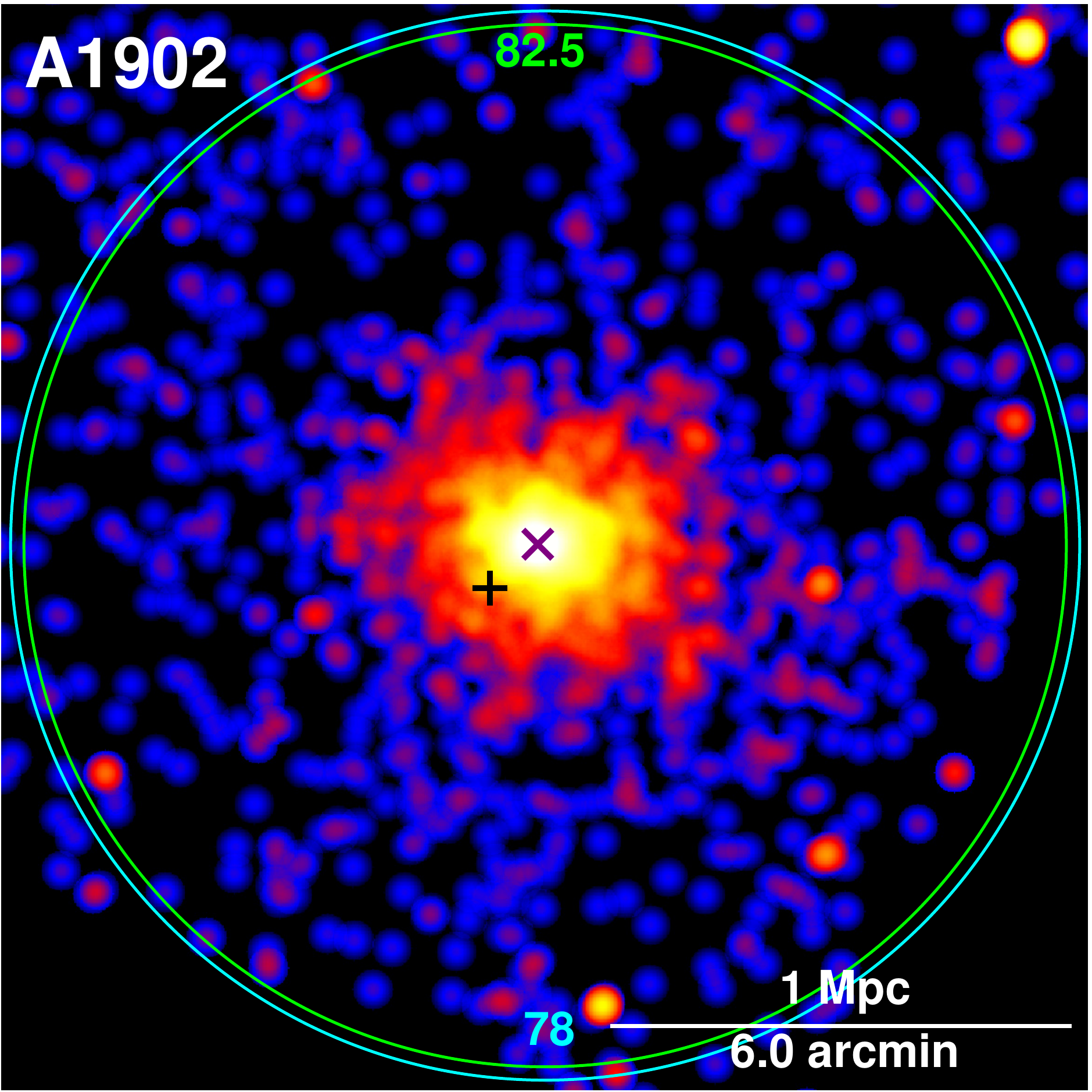}
\includegraphics[width=0.246\textwidth,keepaspectratio=true,clip=true]{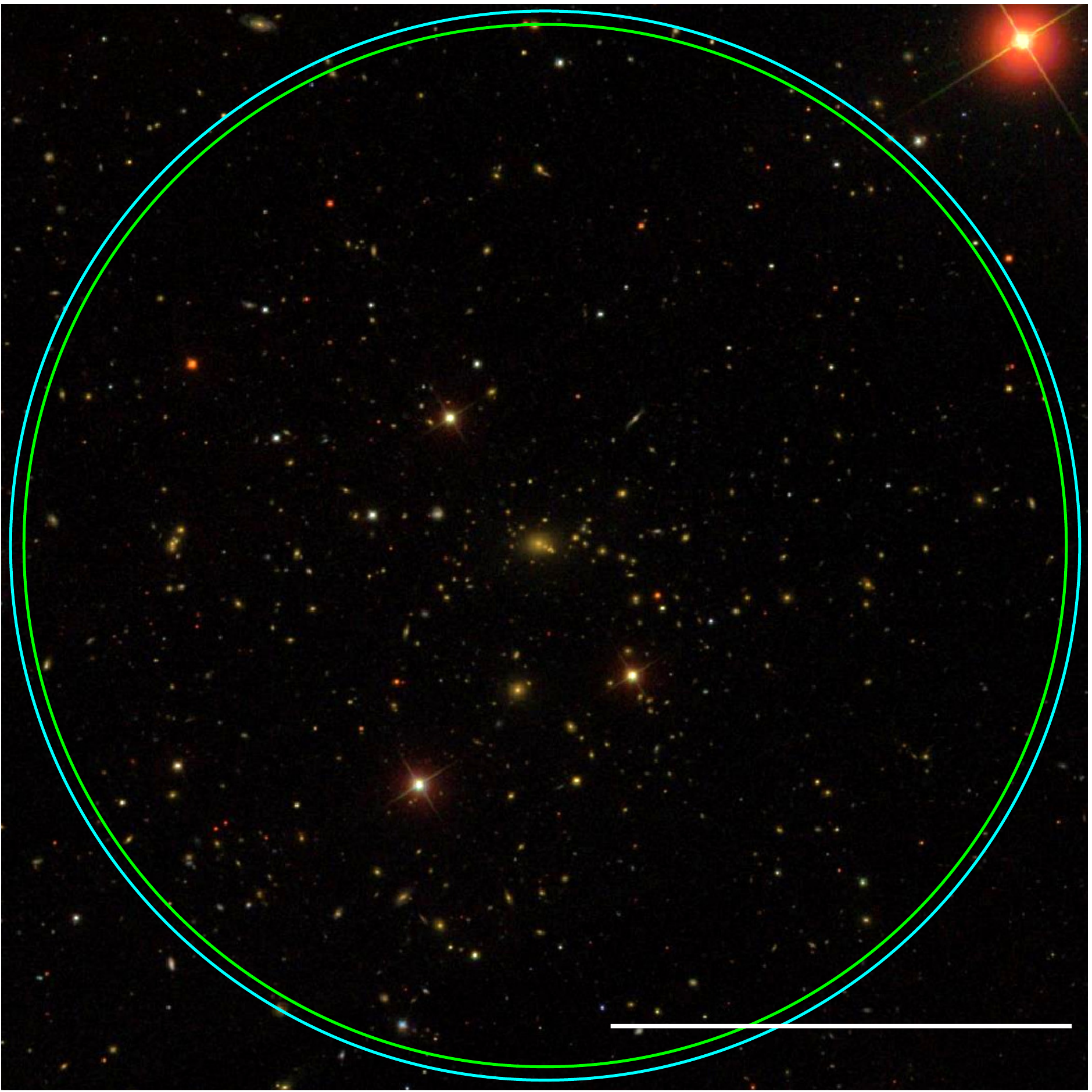}
 	\end{center}
	\caption{
\cha\ 0.7-2 keV/\xmm\ 0.5-2 keV images of the 38 richest \max\ clusters (in order of the \max\ richness). All images are instrumental background subtracted, exposure corrected and smoothed. The side-by-side images are {\em SDSS} RGB images with the same FOV. Purple ${\pmb \times}$ marks the X-ray peak, while black ${\pmb +}$ marks the SZ center from PSZ2 catalogue \citep{2016A&A...594A..27P}. The centers of cyan and green circles are from the \max \citep{2007ApJ...660..239K} and \red \citep{2014ApJ...785..104R} catalogue respectively, while the number marks the optical richness and the circles are $R_{500}$ from the mass-richness relation (\citealt{2009ApJ...699..768R} for \max\ and \citealt{2017MNRAS.466.3103S} for \red). For multiple cluster systems, we also show the name and the spectroscopic redshift from \sdss\ for each X-ray cluster, as well as the photometric redshifts for the optical cluster from the optical catalogues.
}
	 \label{fig:smp}
\end{figure*}

\section{Large-scale structures}
We also searched for large-scale structures around each \max\ cluster in our sample, by looking at the \max\ and \red\ catalogues, and examining \cha\ and \xmm\ data archives in a 1 degree radius (corresponding to $\sim 13$ Mpc at $z=0.23$).
Besides merging clusters like J150, A781, A1882 and A1319 as detailed in \ref{app:id}, there are two additional large-scale structures, as shown in Fig.~\ref{fig:ls}.

{\bf J229-LS} contains at least four clusters (A2050, J229, A2051, A2051S; $z \sim 0.12$) in a $\sim$10 Mpc filament. The X-ray properties of these clusters are also listed in Table~\ref{t:xray}. Note that A2051N is a background cluster at $z \sim 0.38$. Interestingly, the poor cluster J229 is identified as a rich optical cluster, while the hotter cluster A2051 is missed by both \max\ and \red.

{\bf A2219-LS} is composed of both A2219 and J249 in a $\sim$ 6 Mpc filament at $z \sim 0.23$. Both clusters are in this sample with similar \max\ richness, while J249 is actually composed of two subclusters from the X-ray data.

\begin{figure*}
 	\begin{center}
\includegraphics[width=0.495\textwidth,keepaspectratio=true,clip=true]{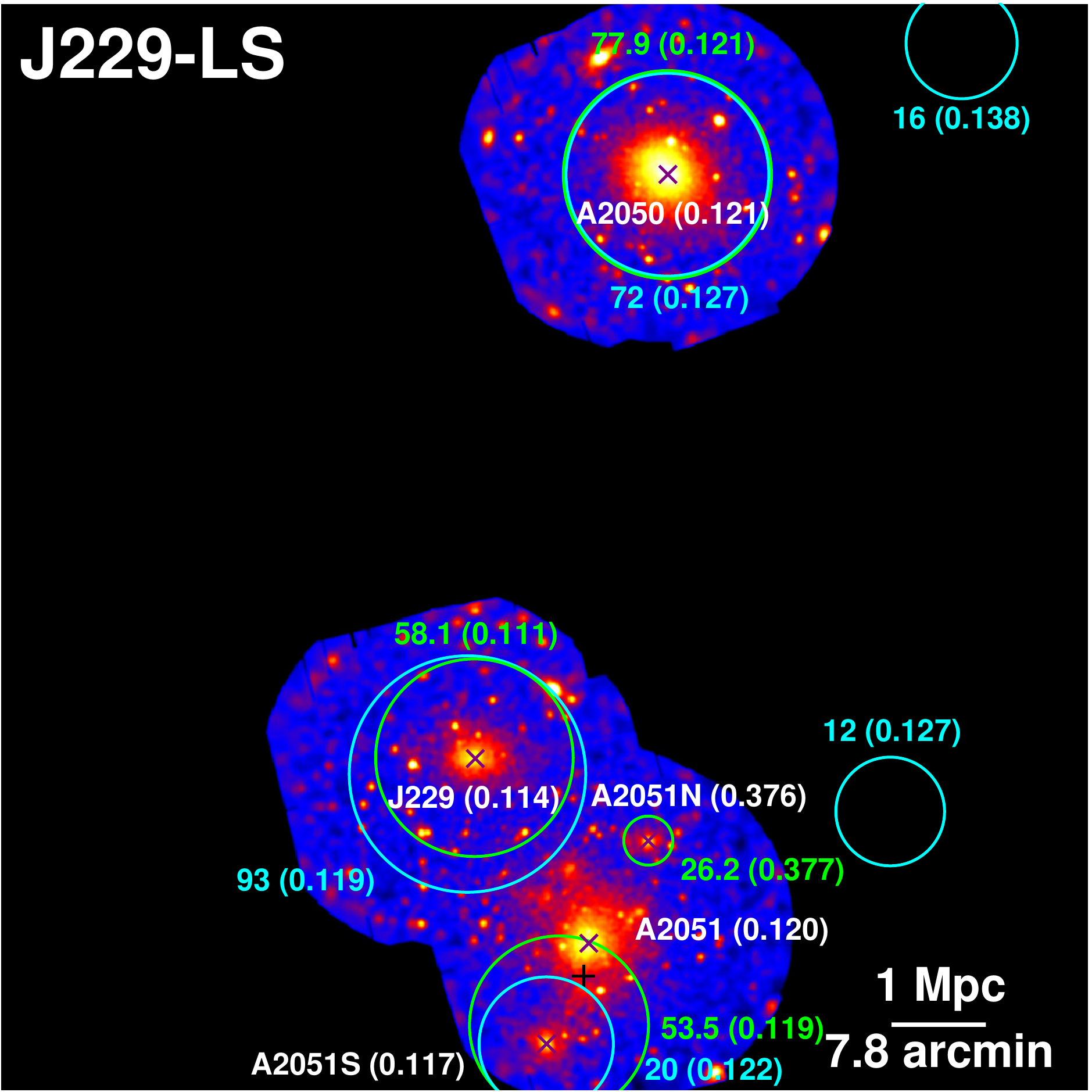}
\includegraphics[width=0.495\textwidth,keepaspectratio=true,clip=true]{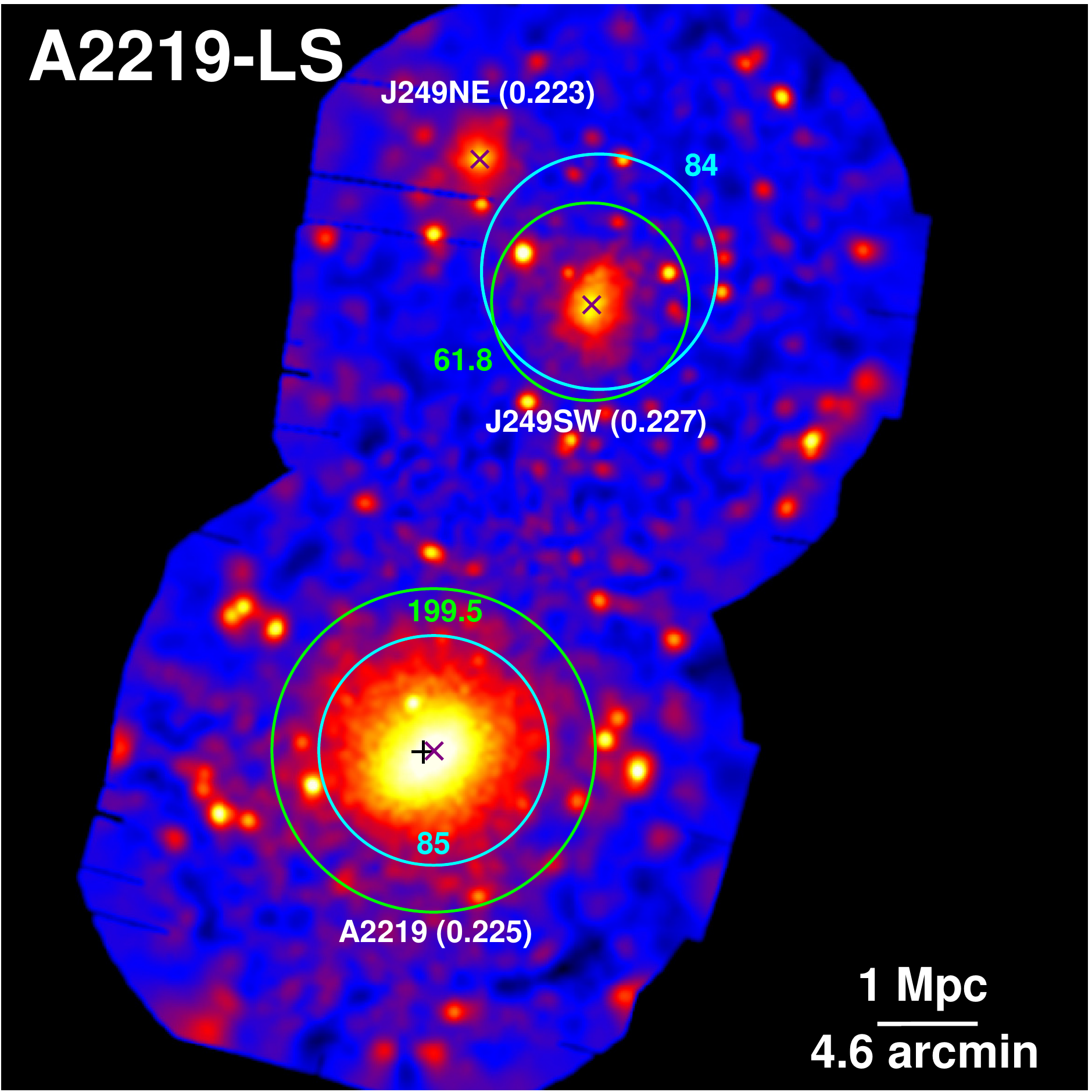}
 	\end{center}
	\caption{Large-scale structures identified around \max\ clusters in our sample (the marks are the same as in Fig.~\ref{fig:smp}).
}
	 \label{fig:ls}
\end{figure*}

\bibliography{ms}

\end{document}